\documentclass[11pt]{book}
\usepackage[top=3cm, bottom=3cm, width=13.5cm, centering]{geometry}

\usepackage{fancyhdr}
\usepackage{amsmath}
\usepackage{amsfonts}
\usepackage{amssymb}
\usepackage{amsthm}
\usepackage{latexsym}
\usepackage{listings}
\usepackage{makeidx}
\usepackage{graphicx}
\usepackage{pdfpages}

\makeindex

\numberwithin{equation}{section}

\newtheorem{crule}{cr}
\newtheorem{flse}{flse}
\newtheorem{dsj}{dsj}

\newtheorem{axi}{axi}
\newtheorem{ord}{ord}

\newtheorem{axint}{axdi}
\newtheorem{axdia}{axdia}
\newtheorem{axoc}{axoc}

\newtheorem{axarr}{axa}
\newtheorem{smlt}{smlt}
\newtheorem{dm}{dim}
\newtheorem{orda}{orda}

\newtheorem{axmtx}{axm}
\newtheorem{dmm}{dimm}

\newtheorem{definition}{Definition}[section]

\newcommand{\be}{\begin{equation*}}
\newcommand{\ee}{\end{equation*}}
\newcommand{\ben}{\begin{equation}}
\newcommand{\een}{\end{equation}}
\newcommand{\bal}{\begin{aligned}}
\newcommand{\eal}{\end{aligned}}
\newcommand{\ba}{\begin{eqnarray}}
\newcommand{\ea}{\end{eqnarray}}

\newcommand{\la}{\langle}
\newcommand{\ra}{\rangle}

\newcommand{\m}{\mathcal}

\newcommand{\mb}{\mathbb}

\newcommand{\lra}{\longrightarrow}
\newcommand{\lla}{\longleftarrow}

\begin{document}



\pagestyle{empty}
\newlength{\drop} 
\drop=0.1\textheight 
\rule{\textwidth}{1pt}\par 
\vspace{2pt}\vspace{-\baselineskip} 
\rule{\textwidth}{0.4pt}\par 
\vspace{\drop} 
{\centering 
{{\Huge A Formal System:}\\[0.5\baselineskip]
{\Huge Rigorous Constructions of}\\[0.5\baselineskip]
{\Huge Computer Models}}\\[0.5\baselineskip]
\vspace{0.25\drop} 
\rule{0.3\textwidth}{0.4pt}\par 
\vspace{100mm}
\hspace{90mm}{\text{G. Pantelis}}\par 
\rule{\textwidth}{0.4pt}\par 
\vspace{2pt}\vspace{-\baselineskip} 
\rule{\textwidth}{1pt}\par} 

\frontmatter

\parindent0pt
\parskip10pt
\pagestyle{plain}

\chapter*{Preface}

This book draws upon a number of converging ideas that have emerged over recent decades from various researchers involved in the construction of computer models.
These ideas are challenging the dominant paradigm where a computer model is constructed as an attempt to provide a discrete approximation of some continuum theory.

For reasons discussed in the first chapter, there is an argument that supports a departure from this paradigm towards the construction of discrete models based on simple deterministic rules.
Although still limited in their use in the sciences, these models are producing results that show promise and cannot be easily dismissed.
But one can take this one step further and argue that such discrete models not only provide alternative tools for simulation but in themselves can be used as a new language that describe real world systems.

The question arises as to how one can build a solid foundation for validating such discrete models, both as a simulation tool as well as a language that describe the laws that govern the application.
It appears that these two aspects of the model are highly linked and rely heavily upon a single overriding property, namely that of computability.

Encouraged by current trends in theoretical physics, we are particularly interested in dynamical systems that model the flow and interaction of information.
The state variables of such systems can only take on a finite number of assigned integer or rational values and are subject to the law of conservation of information. 
Thus there is high degree of compatibility with the discrete model and the machine upon which it is to be executed.
It seems plausible then that the laws that govern the computability of any model based on a finite dynamical system can be directly linked to the allowable computational operations on the machine itself.

Another important issue is that a computer model based on a finite dynamical system may involve algorithms that are not readily expressible in the notation of conventional mathematics.
This suggests a new paradigm in which a real world system is best described by a language of algorithms and programs rather than a language based on any conventional mathematical representation.
If this is the case then one should attempt to construct a language that is both simple enough to be adopted by those whose background is not rooted in the computer sciences and yet be powerful enough to be employed as a tool of analysis at a sufficiently high level.  

This book is primarily aimed at students and researchers in the mathematical sciences who have little or no knowledge of subjects in the computer sciences, although some experience with programming will be helpful. 
Specific discrete models will not be discussed in great detail since the focus is directed towards the basic operations of finite state arithmetic on a real world classical computer.
A simple language based on functional programs will be constructed for the purposes of analysis. 
From such a study it is hoped will emerge the basic theoretical tools that will lay down the foundations for both the construction and rigorous validation of this class of computer models.

Chapter \ref{ci} outlines the motivation behind the material in this book.
Chapters \ref{cps}-\ref{appl} are largely dedicated to the construction of the formal language based on functional programs.
When constructing a new language the reader will be bombarded with many definitions before the language can actually be used in analysis.
This cannot be avoided and the reader will need to make some effort to acquire some understanding of these definitions and the motivations behind them.
Therefore the material contained in Chapters \ref{cps}-\ref{appl} require some patience and perseverance on behalf of the reader.

In Chapters \ref{int}, \ref{array} and \ref{mtx} a number of basic properties of scalar, array and matrix arithmetic on a deterministic machine with finite memory will be derived.
The reader should be alerted to the subtleties that underly the way in which the axioms are structured and how they differ from the conventional axioms of fields and rings.  
Although the focus is on basic algebraic identities and inequalities, all proofs are provided for completion.
The reader may wish to go through some of these proofs to get a feel of how the formal language works.
When the reader is satisfied that they have a sufficient understanding of the process the reader may then wish to skim through the remaining proofs of these chapters.

In Chapter \ref{cfds} we examine the computability of finite dynamical systems based on a single scalar variable.
Here we extend the theory of Chapter \ref{int} to include discrete interval enclosures.
In Chapter \ref{cpcap} we reflect on the formal language that has been constructed in Chapters \ref{cps}-\ref{appl} and where it can be placed in the wider context of the conventional formal methods of proof theory.
In Chapter \ref{chap_ds} an attempt is made to bring together all of the ideas of the previous chapters to the final goal of formal constructions of computer models based on dynamical systems on lattices.
The book concludes with Chapter \ref{cfsis} where future directions are explored that address some of the unresolved issues that surround formal systems and their role in the scientific method.

\parskip0pt

\tableofcontents

\mainmatter

\parindent0pt
\parskip10pt

\pagestyle{fancy}
\renewcommand{\chaptermark}[1]{\markboth{#1}{}}
\renewcommand{\sectionmark}[1]{\markright{\thesection\ #1}}
\fancyhf{} 
\fancyhead[LE,RO]{\bfseries\thepage}
\fancyhead[LO]{\bfseries\rightmark}
\fancyhead[RE]{\bfseries\leftmark}
\renewcommand{\headrulewidth}{0.5pt}
\renewcommand{\footrulewidth}{0pt}
\addtolength{\headheight}{0.5pt} 
\fancypagestyle{plain}{%
\fancyhead{} 
\renewcommand{\headrulewidth}{0pt} 
}

\chapter{Introduction.}\label{ci}

\section{Continuous versus discrete.}

Debates surrounding continuous versus discrete mathematics arise in many sub-disciplines of the mathematical sciences.
Where this debate is of paramount importance can be found in the general area of computer modeling.
While we will not be discussing any particular computer model in great detail, it is appropriate that we at least start with some background to this topic since it largely represents the motivation behind much of the discourse presented throughout this book.

Computer modeling has become an important component of many scientific studies.
However, despite its widespread use in the sciences, computer modeling still has a reputation of taking on aspects of an art rather than an exact science.
The reasons for this reputation are somewhat historic.
While this field has provided many useful results and enhanced the insights into a wide area of scientific research there still remain weaknesses in the validation of computer models at a theoretical level.

The difficulties largely originate from two, not entirely unrelated, sources.
The controversy surrounding the existence of the real numbers, $\mb{R}$, is a philosophical debate that has been around since antiquity.
More recently, the existence of infinite sets, that are the basis of many mathematical abstractions, is also one that is contested in the philosophical arena.
We may include within this debate the existence of the rationals, $\mb{Q}$, and the integers, $\mb{Z}$.
From the perspective of raw computations on a real world computer there is no philosophical issue.
A machine neither recognizes an infinitesimal nor infinite sets in general.

Pure mathematicians are motivated by structures that have an elegant representation in the platonic world of ideal forms and have very little interest in real world applications.
In this realm the construction of the reals along with other mathematical abstractions involving infinite sets cannot be readily dismissed.
Nevertheless, the extensive products of the efforts of pure mathematicians over the years, that were largely motivated by theoretical interest, have been found to be useful in the sciences.

It is widely accepted that mathematics is the language of science but with this come some unexpected compromises.
Mathematical structures are abstract constructions.
In order to maintain some consistency, the physical system under consideration often needs to be idealized to suit the language that is employed to describe it.
As with the construction of the language upon which it is based, the constructions of models of the physical world under such idealizations can lead to theories that take on the appearance of elegance.
Unfortunately, the abstractions embedded in the descriptive language of contemporary mathematics can also lead to complications.

Following in the footsteps of many other scientists, computer modelers have often adopted the language of continuous mathematics without question.
The constraints of the finite resources of the computer and its inability to recognize the reals and infinite sets in general have led to the acceptance of the computer as a tool of approximation.
This in turn has led to the emergence of the discipline of numerical analysis that is largely dedicated to providing a rigorous foundation for approximation theory.

As an example, let us look at the wide application area of hydrodynamics.
In particular we want to focus on the traditional procedures for model validation.
We are concerned here with validation in the theoretical sense and this should not be confused with model validation that involves comparing simulation results with empirical data.

Hydrodynamics is largely based on conservation laws that are expressed in the form of second order partial differential equations (PDEs).
The most widely used hydrodynamic computer models are based on some kind of discrete system of equations that are meant to represent an approximation of the continuum model.
We can view this as a map that transforms a system of PDEs to a system of difference equations (DEs),
\be
\text{Continuum model (PDEs)} \longrightarrow \text{Discrete model (DEs)}
\ee
Under this map the continuum model is regarded as the template that represents the exact description of the physical system being modeled.
The discrete model attempts to approximate the continuum model by employing some type of discretization scheme.
These include finite difference methods, spectral methods and other variants of discretization.
Associated with any discretization scheme is a characteristic spatial resolution, $\Delta x$, and in the case of time dependent problems a temporal resolution, $\Delta t$. 
The map associated with the continuum to discrete model is regarded as valid if all of the following conditions are met. 

\begin{itemize}

\item The continuum model is well posed. The classical notion of a well posed system is based on the existence and uniqueness of a solution along with continuous dependence on the initial data.

\item Existence and stability of the solution of the discrete system of equations.

\item Consistency. The discrete system of equations converge to the continuum equations in the limit, $\Delta x \rightarrow 0$, $\Delta t \rightarrow 0$.

\item Convergence. The solution of the discrete system converges to the solution of the continuum equations in the limit, $\Delta x \rightarrow 0$, $\Delta t \rightarrow 0$.

\end{itemize}

There is an enormous wealth of mathematical theories that have been developed to address these items.
To name a few are finite difference methods that are often based on Taylor expansions, weak or generalized formulations of PDEs and solution methods and stability analysis of difference equations. 
Despite the extensive theoretical work on the subject, except for some special cases, a complete rigorous theoretical validation is rarely achievable.

The above conditions establish validity on theoretical grounds but do not entirely address the practical aspects of machine computation.
The exact solutions of the theoretical discrete system will not necessarily coincide with that of the discrete system implemented on the machine.
This is because we are forced to employ floating point arithmetic.
Thus round off errors introduce another source of complication that needs to be considered if a complete model validation is demanded.    

Matters become even worse when we find that often it is difficult to establish that the continuum model itself is well posed.
A case in point is the Navier-Stokes equation upon which fluid mechanics is based.
It is currently unknown whether there exist solutions in general, weak or strong, that satisfy typical boundary and initial data. 
Despite this, it is standard practice to formally establish convergence of the solution of the discrete equations to the exact solution of the Navier-Stokes equation, the existence of the latter being unknown. 

Firstly, it is worthwhile examining the origins of the continuum model.
The most fundamental continuum theories are based on the Euler equations of an ideal fluid.
Here the laws of fluid flow are derived largely from the principle properties of a continuum.
In applications the Euler's equations are replaced by equations that include terms associated with viscosity, heat transfer and other phenomena that may be deemed important for the particular problem being considered.
This class of semi-empirically based continuum equations of hydrodynamics are derived from a particle model through the application of Boltzmann equations that employ continuous distributions.
We can represent this as a map under the action of Boltzmann statistics 
\be
\text{Particle model} \longrightarrow \text{Continuum model}
\ee
Thus the continuum model itself is derived from a microscopic scale discrete system.
The combination of the maps given above is then a two step procedure of discrete to continuous back to discrete.
The properties of the two discrete systems differ in that the Boltzman equations employ continuous mathematics whereas the computer model is fully discrete. 

In light of these observations one may ask whether it is not better to compare the computer model directly with the kinetic particle model.
Both are discrete systems and the existence of solutions of the formulation associated with the particle model is much more tractable.
Fortunately, this notion has, in some sense, been around since the advent of the first computers.
Cellular automata have been used to model many complex systems from applications in the biological sciences to networks of information flow.
In more recent decades has emerged the use of cellular automata in the area of hydrodynamics.
The popularity of cellular automata in this application area has waxed and wained over the years but the results that they produce cannot be readily dismissed and are worthy of continued examination. 

Cellular automata in fluid dynamics are discrete rule based algorithms that attempt to mimic the particle model.
As such they can be directly translated into a computer program.
The fluid medium is discretized into a lattice and within each cell of the lattice there are only two possible states, $[0~1]$.
The dynamics of the system is governed by a collection of simple deterministic rules of cell pair interactions. 
The principle and defining feature of such systems is the conservation of information that reflects the physical law of conservation of mass.

For cellular automata to be effective the spatial domain needs to discretized into a very large number of small cells.
But even with the smallest refinement that is of practical use on even the most powerful computers, the characteristic size of the cells is still very much larger than the mean free path of the particles that it is meant to simulate.
This suggests that allowing only two states per cell may be inadequate. 

Issues of scale inconsistencies also arise in the formulation of continuum models.
In the early days of computational fluid dynamics it was found that the discrete models based on the Navier-Stokes equation did not perform well for large Reynolds numbers.
(Large Reynolds number flows are associated with the onset of hydrodynamic instabilities leading to turbulence.)
The earliest attempts to remedy this situation involved the introduction of a turbulence viscosity term that was identical in form to the molecular viscosity term.
While there were some improvements in simulation results, the introduction of the turbulence viscosity constant appeared to be inadequate to capture observed flows where a high degree of accuracy was required.
This led to the area of large eddy simulation models where the constant turbulence viscosity coefficient was replaced by a variable usually based on a function of the deformation tensor (see for instance \cite{liu01}).
The so called Smagorinsky model \cite{smag} introduced in the 1960s remains the most popular model whose variants are still in use to this day.

The problem with turbulent fluid models based on large eddy simulation is their dependence on artificial parameters that need to be readjusted for each specific application.
The presence of these artificial parameters is an indication that the continuum model exhibits some type of inconsistency.
One can identify this inconsistency as arising from the fact that the continuum model is ill defined in the sense that it fails be scale invariant.
The important properties that would be demanded from a reformulation is that it be scale invariant and be independent of artificial parameters that require tuning to specific applications.

One way to do this is to accept that the dependent variables of the continuum equations are filtered variables that must not only be dependent on space and time but on a new independent variable associated with scale.
One introduces the following conditions that define consistency based on scale invariance \cite{pan99}-\cite{pan09}.\index{scale invariance}

\begin{itemize}

\item The macroscopic formulation is described by a system of equations that represent conservation laws of the filtered variables and contain residual terms that capture all of the dissipative and dispersive effects that are associated with microscopic scale fluctuations.
The macroscopic scale formulation must be form invariant with respect to scale.

\item The dependent variables of the macroscopic formulation must also satisfy the filter equations that are expressed as second order partial differential equations rather than integrals.
The Gaussian filter is a convenient filter to use for this purpose.
The filter equations provide a continuous relationship between the filtered variables with respect to scale.

\item In the limit of increasing spatial resolution the residual terms vanish and the macroscopic equations collapse to the fully resolved continuum equations. 

\end{itemize}

Close examination of the conditions for consistency of scale invariance suggests that a continuum formulation can be removed altogether. 
An important observation is that the scale parameter is proportional to the square of the desired characteristic spatial resolution, $\Delta x$.
Since this can be directly associated with the spatial resolution of the discretization that is employed in the computer model it appears reasonable to consider the possibility that one could discard the continuum model altogether and adapt the above conditions for consistency based on scale invariance to an entirely discrete formulation.

To this end we must rely on the working hypothesis that objects of the physical world, at all scales, can be defined in terms of information so that all knowledge of the objects, including their state of motion, are contained in finite state vectors.
Under this regime the conservation laws of the continuum theories are replaced by laws that govern the conservation of information in some form.
Such a rule based algorithm involves finite state arithmetic as is reflective of a real world deterministic machine.
Here there is a major philosophical shift in that the discrete model is both the computer model and the defining template that represents the physical system being simulated.
In this paradigm the traditional notions of model validation by way of consistency, convergence and stability of a discrete computer model versus a continuum theory are bypassed by the single notion of computability of a program built upon sequentially ordered statements.
By pursuing this path it will eventually become apparent that, by necessity, the discourse will be predominantly transferred from the language of mathematical equations to the language of rule based algorithms and computer programs.

It should be stressed that the issue being discussed here is one of a choice of the most efficient language that can be employed to model the physical world.
Whether we possess a language that is rich enough to allow us to completely describe the physical world will remain a controversial issue.
We can, however, be encouraged by current trends in theoretical physics where there is an increasing tendency towards formulating physical laws in a language based on information flow, where physical objects and their dynamical state are represented by finite state vectors.

Information physics is related to the general area of digital physics of which early proponents include Zuse \cite{zuse} and Jaynes \cite{jay01}-\cite{jay02}.
More recent proponents of information and digital physics include Wolfram \cite{wolf} who examines the computational universe at the most fundamental level.
Allied to this subject are the deeper meta-mathematical investigations of Chaitin (see for example \cite{chait01}-\cite{chait10}) who is credited as the initiator of the subject of algorithmic information theory.
Of particular note is Chaitin's interest in the work dating back to Leibniz \cite{leib} who, amongst being the earliest known discoverer of binary arithmetic, appears to have explored early notions of complexity and how complexity can be employed to ultimately construct a formal definition of what actually constitutes a scientific theory.
Some aspects of these works, along with the more controversial views of Zeildberger on discrete versus continuous mathematics (see for example \cite{zeil01}-\cite{zeil02}), are highly influential in some of the ideas presented throughout this book. 

Nonetheless, continuum theories have so far served us very well, providing insights in many branches of scientific endeavor.
But their limitations in providing closed form solutions for many complex systems, and hence the need to introduce discrete approximations along with their inherent problems, are increasingly becoming recognized.
In a future where greater detail of solutions to complex systems is demanded alternatives should be investigated.
A language based on programs and its association with discrete mathematics appear to provide a good candidate for such an alternative.     

It will be premature here to embark on a detailed review of methodologies associated with the construction of fully discrete models.
Before we can do this we must first reassess the very foundations of basic arithmetic on a deterministic machine with finite memory.
Indeed, the axioms that dictate the basic rules of machine arithmetic will play a significant role in defining the laws that govern the construction of the discrete model.
If we are to seriously take the discrete model as the defining language that describes a real world application then it is not unreasonable to expect that the conventional physical laws will emerge as manifestations of the allowable finite state computations of that system. 
In this book we will explore some of these ideas, starting with the most elementary laws that govern machine arithmetic.
From such a study it is hoped will emerge a platform upon which a formal and rigorous approach to computer modeling can be constructed.     

There are two aspects to the work presented in this book.
The first is the introduction of an inference scheme based on the so called \emph{program extension rule}.
This formal system is not so much meant to replace the traditional formal schemes of proof theory but rather as an adaptation of them to efficiently deal with a preferred language.
The language that we will choose in this book is one based on representing formal statements as functional programs.
This language will be employed as an alternative to methods of analysis using traditional abstract mathematical structures. 
The second aspect of the book is concerned with the construction of axioms that directly address the constraints imposed by a deterministic machine with finite memory and the most elementary operations of arithmetic that can be performed on it.

\section{Machine arithmetic.}

Our main objective is to develop a formal language from which computer models can be constructed in such a way that computability is guaranteed at each step of the model's construction.
In the previous section we discussed discrete models based on simple deterministic rules.
As we shall see in a later chapter, for practical application it will be necessary to reformulate these models on lattices that represent larger scales.
It follows that such macroscopic scale formulations can be expressed in terms of algorithms that rely heavily on numerical computations.

It should be kept in mind that our goal for establishing computability will not be entirely restricted to the avoidance of underflows and overflows of the numerical computations.
We start with the hypothesis that the fully discrete model, and hence the operational parameters that characterize the machine upon which the model is to be executed, are in some sense reflective of the underlying structure of the real world system.
Under this hypothesis we are raising the status of the conditions of computability of a model by associating them with the laws that govern the dynamics of the real world system.

In any attempt to construct a tool for the validation of programs largely based on numerical computation one first looks to the basic foundations of arithmetic starting with the axioms of rings and fields (see for example \cite{burk}, \cite{mclane}).
Unfortunately, when encountering machine arithmetic one will eventually observe a departure from the elementary rules of arithmetic upon which one has been accustomed.
To explain some aspects of these departures one may delve deeper into analysis through topics such as modulo arithmetic and finite fields \cite{mclane}, but these too fall short of addressing many of the problems that are encountered when dealing with machine computations.

One promising approach that provides rigor through direct numerical computations can be found in interval arithmetic (see for example \cite{moor}, \cite{ale}).
This has found wide use in computations attempting to approximate continuum theories by way of floating point arithmetic.
For discrete based models where integer or fixed precision rational solutions are desired, we can define discrete interval arithmetic in a similar way.
There is, however, a significant difference in the way that interval methods are employed when dealing with fully discrete systems.

To tackle this problem in its entirety one soon finds the need to investigate topics in a much wider area, many of which are found in the realm of the computer sciences.
In particular, the initial motivation of program verification evolves into an area involving inference methods in a more general sense.  

Traditional studies of computers and computation often start by constructing a theoretical model that reflects some properties of real world computers.
Such examples can be found in Turing machines along with abstractions of programming languages themselves such as lambda calculus \cite{baren} leading to the study of logic and the important link between proofs and programs.
The latter in turn leads one into the subject of proof theory.
This is a wide area of study of which an excellent coverage can be found in \cite{buss}.

The formal systems in the general area of proof theory were primarily developed to address important theoretical problems in logic and were not optimally designed for practical implementation in a machine environment. 
The approach taken here is to construct a formal language such that the rules of inference are dictated not so much by an external abstract theory of logic and computation but rather by the constraints dictated by finite memory storage and allowable operations on a real world computer.
As a consequence there will be a need to abandon some of the expressiveness of formal systems found in current proof theory.  
Motivated by a more practical approach to program verification, the language is presented in a form that is less abstract than traditional studies of theoretical computers and functional programing based on the lambda calculus.

These methods will be described in the context of the software package VPC\index{VPC} (Verification of Program Computation) in its current phase of development.
While the source code of VPC will not be presented here, an effort will be made throughout this book to describe its functionality in sufficient detail so that the reader will be equipped to construct their own version if they so desire.

In the construction of our formal language the following properties are of primary importance.

\begin{itemize}
	\item \emph{Simplicity}. The language should be simple and accessible to those of various backgrounds outside of the computer sciences.
	
	\item \emph{Analysis.} The simplicity of the language should not compromise its power to be employed as a tool of analysis at a sufficiently high level.
	
	\item \emph{Proof assistance.} As a language based on functional programs it should be readily implemented on a machine platform.
	As such automated procedures can be constructed that assist in the generation of proofs.
	This assistance comes in the form of, (i) generating on screen real time constructions of formulations that remove the laborious and error prone task of writing down symbols on paper and (ii) a step by step guidance of valid options in a proof construction.
	
	\item \emph{Compatibility.} A language that is specially designed to address the issues of computability on a real world computer and its connection with the practical implementation of models based on finite state arithmetic.
	It is also advantageous to converse in a language that closely resembles the actual code that will ultimately represent the computer model.
	
	\item \emph{Expressiveness.} The language will largely deal with objects as subtypes of strings that are immediately recognized by the machine.
	Consequently, it will be a low level language that will lack the expressiveness found in standard formal systems of proof theory.
	However, it should possess the properties that it can be used as a primitive upon which theories demanding higher levels of abstractions can be built.   
	
	\item \emph{Improvements.} The language should be flexible enough to be open for future developments that enhance both its scope for analysis and automated procedures.
\end{itemize}	

\section{Sequential ordering.} 

Let $Q_i,~i=1,\ldots,n$, be statements of a formal system. Consider the ordered sequence of statements 
\ben\label{1.00}
[Q_1~Q_2~\ldots~Q_n]
\een
The list of statements is to be read in the sequential order from left to right and the procedure halts after the reading of the last statement $Q_n$.
In a machine environment the $Q$'s may be instructions such as type checking operations or assignments.
The reading of the list (\ref{1.00}) may halt prematurely if the machine encounters a statement that results in an execution error.   

The sequential order of statements in the list (\ref{1.00}) is a key property of programming languages.
At first glance the sequential order is dictated by the following properties.

\begin{itemize} 

\item The order in which each statement appears in the list is important although some interchange is possible only under special conditions.

\item Each statement in the list may have some kind of dependence on one or more statements that precede it but no statement can have a dependence on a statement that follows it.

\end{itemize}
To understand how these two properties are linked we will need to associate with each statement a list of variables that act as either input or output parameters of that statement.
We may also consider instructions that essentially split a program into a number of parallel sequentially ordered streams.
This process is related to logical statements involving disjunctions as will be discussed further in a later chapter.
For now it will suffice to consider programs defined by a single sequential stream as described above.

Programs can be constructed by either imperative or functional programming languages.
Programs constructed by an imperative language will not be discussed in great detail other than to acknowledge that they will form the collection of atoms of functional programs from which larger functional programs can be constructed.
Inference methods will be conducted in the setting of functional programs.
It will be seen that after introducing a set of rules, the process of program construction and validation can be directly linked to inference methods.
While this will open up an opportunity to study a much wider area of applications, we will remain focused on issues related to machine arithmetic.
Our interest in this will be the properties of allowable machine computations that will eventually lead us to a formal method for valid constructions of programs that model real world dynamical systems.

\chapter{Program Structure.}\label{cps}

\section{Types.}

We shall deal with objects and types, where each object has a type\index{type}.
In a machine environment the objects are strings and different string structures are identified by their type.

\textbf{Properties of types.}

\begin{itemize}

\item Types will be denoted by the symbols $\mb{A},\mb{B},\mb{C},\ldots$.
Excluded are the symbols $\mb{R},~\mb{N}$ and $\mb{Z}$, respectively, that will be used when referring to the conventional mathematical abstractions of the sets of the real numbers, the natural numbers and the integers, respectively. 

\item Object $a$ has type $\mb{A}$ is denoted by $a:\mb{A}$.

\item An object may also be dependent on another object.
We write $a(n)$ to mean that the object $a$ depends on the parameter or object $n$.
Sometimes we may index a collection of objects using subscripts, e.g. $a_1,\ldots,a_n$ denotes a collection of objects that may have different types.

\item Types may be subtypes of types\index{subtype}.
$\mb{A}$ is a subtype of $\mb{B}$ is denoted by
\be
\mb{A}<:\mb{B}
\ee
Subtypes have the property that if $a:\mb{A}$ and $\mb{A}<:\mb{B}$ then $a:\mb{B}$.
It follows that if $\mb{A} <: \mb{B}$ and $\mb{B} <: \mb{C}$ then $\mb{A} <: \mb{C}$.

\item Types may also be dependent on objects.
We write $\mb{A}(a_1,\ldots,a_n)$ to mean that type $\mb{A}$ depends on the parameters or objects $a_1,\ldots,a_n$.
Parameter dependent types are subtypes of their generic type, i.e.
\be
\mb{A}(a_1,\ldots,a_n)<:\mb{A}
\ee

\end{itemize}

Note that if $\mb{A} <: \mb{B}$ and $\mb{A}(a_1,\ldots,a_n)$ and $\mb{B}(b_1,\ldots,b_m)$ are parameter dependent subtypes it does not necessarily follow that $\mb{A}(a_1,\ldots,a_n)$ is a subtype of $\mb{B}(b_1,\ldots,b_m)$.

\section{Alphabet and strings.}

Here we shall work in a machine environment based on a real world deterministic computer.
A real world deterministic computer is characterized by the properties of finite information storage along with a collection of well defined operations. 
At any time the machine can exist in any one of a finite number of configuration states.
A program is a finite sequentially ordered list of instructions such that each instruction attempts to map the current configuration state to a new configuration state.
The context in which we will choose to work can be defined more explicitly by the following machine specific parameters.
\be
\begin{array}{ll}
\mathcal{K} & \text{number of characters in the alphabet.} \\
\mathcal{L} & \text{maximum number of characters in any string.} \\
\mathcal{M} & \text{maximum number of elements of a list stored as an array.} \\
\end{array}
\ee
We will be focused on computational processes that are entirely confined within a machine environment $\mathfrak{M}$ under the specified constraints $\mathcal{K},\mathcal{L}$ and $\mathcal{M}$.
Where it is necessary to stress this context we will write
\be
\mathfrak{M}(\mathcal{K},\mathcal{L},\mathcal{M})
\ee

We start by defining the alphabet as a collection of symbols or characters
\be
s(1),\ldots,s(\mathcal{K})
\ee
The alphabet that we will work with consists of the following characters.

\begin{itemize}

\item Letters.
\be
\bal
a~b~\ldots~z \\
\eal
\ee

\item Digits.
\be
1~2~3~4~5~6~7~8~9~0
\ee

\item Special characters.
\be
-~+~[~]~|
\ee

\end{itemize}

\textbf{Strings.}\index{string}

\begin{itemize}

\item A string of the alphabet is a sequence of characters
\be
s(i_1) s(i_2) \ldots s(i_j), \qquad 1 \leq i_1,\ldots,i_j \leq \mathcal{K},~1 \leq j \leq \mathcal{L}
\ee

\item A string is given a type denoted by $\mb{S}$.

\item There are two main subtypes of strings.
\be
\begin{array}{ll}
\mb{C}<:\mb{S} & \text{alphanumeric strings comprised of any combination of} \\
& \text{letters and digits with the first character always being} \\
& \text{a letter.} \\
 \mb{I}<:\mb{S} & \text{signed integers comprised of digits preceded by a sign $\pm$.} \\
\end{array}
\ee

\end{itemize}

Individual strings are separated by a space.
Sometimes we will allow a space to be included in an individual string.
In such cases the space will be regarded as a special character.
A space used in this way will often be employed instead of a comma such as, for example, a separator of elements of lists.

\vspace{5mm}

\textbf{Alphanumeric strings.}
Alphanumeric strings\index{alphanumeric string} are assigned the type $\mb{C}$ and are often used to represent names of programs and variable names of elements of input/output (I/O)\index{I/O lists} lists of programs.
Variable names of the elements of I/O lists of programs serve as place holders for assigned values that are defined as specific subtypes within the program.
We write $a:\mb{C}$ to stress that $a$ is a dummy variable that represents an alphanumeric string.
Upon entry to a program we may also write $a:\mb{A}$ to denote that the alphanumeric string represented by the dummy variable $a$ has been assigned a value of type $\mb{A}$.
The assigned value can be any string of a specific subtype.

\textbf{Equality\index{equality}.}
There is an important distinction that needs to be made with the notion of equality.

\begin{itemize}

\item If $a$ and $b$ are dummy variables representing two strings we write $a=b$ to denote that the two strings are identical.
The sense in which equality is being used here will always be assumed unless otherwise stated.

\item We may also write $a=b$ to mean that the assigned value of the alphanumeric string represented by the dummy variable $a$ is identical to the assigned value of the alphanumeric string represented by the dummy variable $b$.
The sense in which equality is used here will always be stated to avoid confusion.

\end{itemize}

\vspace{5mm}
\noindent
\textbf{Machine numbers.}
An object of type $\mb{I}$ is a string that can be assigned any one of the integer values
\be
0, \pm 1,\ldots, \pm N ,
\ee
where $N$ is the maximum positive integer and is a specific machine parameter.
We shall make extensive use of the following subtypes of $\mb{I}$.
\be
\begin{array}{ll}
\mb{I}_0 & \text{$a:\mb{I}_0$ denotes $a:\mb{I}$ and $0 \leq a \leq N$} \\
\mb{I}_+ & \text{$a:\mb{I}_+$ denotes $a:\mb{I}$ and $0 < a \leq N$} \\  
\end{array}
\ee
We adopt the usual convention of dropping the prefix $+$ sign when dealing with positive integers.

One of our objectives is to describe the program VPC as a tool for analysis and verification of numerical computation. 
For the purpose of demonstration only we will restrict much of the outline to machine integer arithmetic.
It will be seen later that most of the results using machine integers can also be applied to fixed precision rational numbers.
It should be kept in mind that VPC has a much wider area of application that includes floating point arithmetic.
The reasons for excluding floating point arithmetic is based on the anticipated paradigm shift in computer modeling as discussed in the introduction of Chapter 1.

\section{Lists.}\label{sl}

Throughout we shall work with lists\index{list} rather than sets.
Many properties of lists, such as list intersections and sublists, will have strong similarities with those associated with sets.
As such much of the notation used in set theory will be adopted for lists.
Since we are working in an environment $\mathfrak{M}(\mathcal{K},\mathcal{L},\mathcal{M})$ all lists will be of finite length.

\textbf{Type.}
\be
\begin{array}{ll}
\mb{L} & \text{generic type list of unspecified length.} \\
\mb{L} (n) & \text{type list with $n:\mb{I}_0$ elements, $\mb{L} (n) <: \mb{L}$.} \\
\end{array}
\ee

\textbf{Properties of lists.}

\begin{itemize}

\item Elements of lists are strings. A list $a:\mb{L}(n)$, has the representation
$a=[a_1~a_2~\ldots~a_n]$, where $a_i : \mb{S}$, $i=1,\ldots,n$, $n:\mb{I}_0$.
\emph{We use a space instead of a comma to separate elements of lists}.
The short hand notations
\be
~[a_1~a_2~\ldots~a_n]=[a_1~\ldots~a_n]=[a_i]_{i=1}^n
\ee
will often be used.
The object $n:\mb{I}_0$ is referred to as the length of the list $a$.
The notation $a_i \in a$ means that $a_i$ is an element of the list $a$.

\item An empty list $a:\mb{L}(0)$ is denoted by $a=[~]$.
If, under the list representation $a=[a_i]_{i=1}^n$, we have $n=0$ then it is understood that $a$ is the empty list. 

\item For a list $a$ of unit length we will sometimes write $a$ and $[a]$ to mean the same thing, i.e.
\be
a=[a], \quad a:\mb{L}(1)
\ee

\item In any list $a=[a_1~\ldots~a_n]$, all elements are strings, i.e. of type $\mb{S}$, but they need not all be assigned values of the same subtype.

\item Any element, $a_i \in a$, of a list $a=[a_1~\ldots~a_n]$ could itself be a list.
While all lists will be stored as arrays, we may sometimes treat a list as a single string using the hierarchy of subtypes $\mb{L}<:\mb{S}$.

\item \textbf{List equality.} If $a:\mb{L}(n)$ and $b:\mb{L}(n)$, $n:\mb{I}_0$, and $b_i=a_i$, $i=1,\ldots,n$, we write $a=b$.
We use equality in both senses of identity of strings and the values assigned to the strings.
Throughout, unless otherwise stated, equality will be assumed to be in the sense of the former, i.e. in the sense of the identity of strings.
Whenever the equality is used in the sense of assigned values it will be stated as such.

\end{itemize}

\textbf{List operations.}

\begin{itemize}

\item \textbf{Empty list extraction.} Suppose that $a=[a_i]_{i=1}^n:\mb{L}(n)$, $n:\mb{I}_+$, contains an element $a_k \in a$ that is an empty list, i.e. $a_k=[~]$.
We may extract the empty list element and write  
\be
a=[a_1~\ldots~a_{k-1}~a_{k+1}~\ldots~a_n]
\ee
After empty list extraction we can automatically redefine $a:\mb{L}(n-1)$. 

\item \textbf{List concatenation}. If $a=[a_1~\ldots~a_m]$ and $b=[b_1~\ldots~b_n]$ are two lists then the concatenation\index{concatenation} of $a$ and $b$ yields the list $c:\mb{L}(m+n)$ given by
\be
c=[a~b]= [[a_1~\ldots~a_m]~[b_1~\ldots~b_n ]] = [a_1~\ldots~a_m~b_1~\ldots~b_n]
\ee 
The internal square brackets that act as delimiters for the lists $a$ and $b$ may be removed.

\item \textbf{List partitions.} If $a^{(1)},\ldots,a^{(n)}$ are lists with representations $a^{(i)}=[a_j^{(i)}]_{j=1}^{m_i}$, $m_i:\mb{I}_+$, $i=1,\ldots,n$, then their list concatenation, $z$, is given by
\be
\bal
z = & [a^{(i)}]_{i=1}^n \\
= & [[a_j^{(i)}]_{j=1}^{m_i}]_{i=1}^n \\
= & [ [a_1^{(1)}~ \ldots ~a_{m_1}^{(1)}]~ \ldots ~[a_1^{(n)}~ \ldots ,a_{m_n}^{(n)}] ] \\
= & [ a_1^{(1)}~ \ldots ~a_{m_1}^{(1)}~ \ldots ~a_1^{(n)}~ \ldots ~a_{m_n}^{(n)} ] \\
\eal
\ee 
In the standard use of the symbol $\in$ we may write $a^{(i)}_j \in z$ to mean that $a^{(i)}_j$ is an individual element of $z$ as represented in the expanded form of the last identity of the above concatenation.
Sometimes we will write $a^{(i)} \in z$ to mean that $a^{(i)}$ is an element of the list of lists $z=[a^{(i)}]_{i=1}^n$.
When dealing with lists of lists it will be necessary to first state the sense in which elements are defined before any list operations can be performed.

\item \textbf{List intersection.} If $a=[a_1~\ldots~a_m]$ and $b=[b_1~\ldots~b_n]$ then the list intersection of $a$ and $b$ yields a new list $c:\mb{L}(k)$, $k \leq \min(m,n)$, where $c=[c_1~\ldots~c_k]$ contains all of the elements that are common to both $a$ and $b$.
We write
\be
c=a \cap b
\ee
to mean that $c$ is the list intersection of $a$ and $b$.
Whenever a list intersection is constructed the sequential order of the elements of $c$ are in the same hierarchy of sequential order as they appear in $a$.

\item \textbf{Removal of repeated elements of a list.} If $a=[a_1~\ldots~a_m]$ has repeated elements we can construct a new list $b=[a_{i_1}~\ldots~a_{i_n}]$, $i_1 <~i_2 < \ldots <~i_n < m$, by removing repeated elements as follows.
Reading the list $a$ from left to right, whenever an element is encountered that coincides with a preceding element of $a$ then that element is extracted.
In other words, each element of $b$ contains all non-repeated elements of $a$ and the first occurrence of a repeated element of the list $a$, as read from left to right, maintaining the order in which they appear in $a$. 
We write
\be
b \simeq a
\ee
to mean that $b$ is obtained by extracting repeated elements of $a$ by this procedure.

\item \textbf{List subtraction.} Suppose that $a:\mb{L}(m)$ and $b:\mb{L}(n)$, $n \leq m$, such that all elements of $b$ are contained in the list $a$.
We can construct a new list $c$ obtained by extracting from $a$ those elements found in $b$.
The new list maintains the sequential order found in $a$, i.e. $c=[a_{i_1}~\ldots~a_{i_k}]$, $i_1 <~i_2 \ldots <~i_k$, $k \leq m-n$, where $a_{i_1}~\ldots~a_{i_k}$ are all of the elements of $a$ not found in $b$.
We write
\be
c=a \setminus b
\ee
to denote the new list constructed in this way.

\item \textbf{Element substitution.}\index{element substitution}
For a list $a=[a_i]_{i=1}^m$ we write $a(a_i \to b)$ to denote substitution of the element $a_i \in a$ with $b$, i.e.
\be
a(a_i \to b) = [a_1~\ldots~a_{i-1}~b~a_{i+1}~\ldots~a_m]
\ee

\end{itemize}

\textbf{Sublists.}\index{sublist}
Because of its importance, the notion of a sublist affords a more formal definition. 

\begin{definition}(Sublist.)\label{sublists}
A list $b:\mb{L}$ is a sublist of list $a:\mb{L}$ if every element of $b$ is an element of $a$, i.e. if $x \in b$ then $x \in a$.
We write $b \subseteqq a$ to mean that $b$ is a sublist of $a$.
There are two cases that need to be distinguished.

\begin{itemize}

\item If $b \subseteqq a$ and there exist at least one element of $a$ that is not an element of $b$ then we say that $b$ is a strict sublist of $a$.
We write $b \varsubsetneqq a$ to stress that $b$ is a strict sublist of $a$.

\item If $b \subseteqq a$ and $a \subseteqq b$ we say that the two lists are equivalent and write $a \equiv b$.

\end{itemize}
The empty list, $[~]$, is regarded as a sublist of all lists.

\end{definition}

\vspace{5mm}

\textbf{Notes.}

\begin{itemize}

\item If $q:\mb{L}(m)$ is a sublist of $p:\mb{L}(n)$ it does not necessarily follow that $m \leq n$. Consider the case $q=[a~b~b~a]$ and $p=[a~b~c]$.
In this example $q$ is a strict sublist of $p$ yet its length, $m=4$, is greater than the length, $n=3$, of $p$. 

\item Similarly, two equivalent lists need not have the same length.
For example $p=[b~b~c~a]$ and $q=[a~b~c]$.
Here $q \subseteqq p$ and $p \subseteqq q$, hence $q \equiv p$.

\end{itemize}

\section{Programs.}\label{programs}

Programs are made up of strings or lists of strings with a well defined structure and are assigned the type denoted by $\mb{P}$.
Program names are assigned the type $\mb{P}_{name}$ and are specific subtypes of alphanumeric strings, i.e. $\mb{P}_{name} <: \mb{C}$.
Programs are defined inductively as follows.

\begin{definition}\label{prog}(Program.)
A program has the representation
\be
\mathfrak{p}~x~y
\ee
with the allocation of types of its component parts given by
\be
\begin{array}{ll}
\mathfrak{p}~x~y:\mb{P} & \text{program} \\
\mathfrak{p}:\mb{P}_{name} & \text{program name} \\
x:\mb{L} & \text{input list} \\
y:\mb{L} & \text{output list} \\
\end{array}
\ee

The program name $\mathfrak{p}$ and the lists $x$ and $y$ are separated by a space.
A program $\mathfrak{p}~x~y$ satisfies all of the following conditions.

\begin{itemize}

\item Elements of the I/O lists, $x$ and $y$, of a program are alphanumeric variable names (type $\mb{C}$) that serve as placeholders for assigned values.
The type of the assigned value of every element of the I/O lists is checked within the program.

\item The variable names of the elements of the output list $y$ are distinct.

\item No element of the input list, $x$, can have a variable name that coincides with a variable name of an element of the output list, $y$.

\end{itemize}

A program $\mathfrak{p}~x~y:\mb{P}$ can be represented by a list 
\be
\bal
\mathfrak{p}~x~y = & [\mathfrak{p}^{(i)}~x^{(i)}~y^{(i)}]_{i=1}^n \\
= & [\mathfrak{p}^{(1)}~x^{(1)}~y^{(1)}~\mathfrak{p}^{(2)}~x^{(2)}~y^{(2)}~\ldots~\mathfrak{p}^{(n)}~x^{(n)}~y^{(n)}] \\
\eal
\ee
for some $n:\mb{I}_+$, where $\mathfrak{p}~x~y$ is referred to as the main program and each $\mathfrak{p}^{(i)}~x^{(i)}~y^{(i)}:\mb{P}$, $i=1,\ldots,n$, is referred to as a subprogram of $\mathfrak{p}~x~y$.
Here, $x^{(i)}$ and $y^{(i)}$, $i=1,\ldots,n$, are lists
\be
x^{(i)}=[x_j^{(i)}]_{j=1}^{l_i},~~y^{(i)}=[y_j^{(i)}]_{j=1}^{m_i} , \quad l_i,m_i:\mb{I}_0,~i=1,\ldots,n
\ee
and are related to the I/O lists\index{I/O lists} $x$ and $y$ of the main program $\mathfrak{p}~x~y$ by
\be
y = [y^{(i)}]_{i=1}^n
\ee
\be
x = \bar{x} \setminus (\bar{x} \cap y)
\ee
where $\bar{x}$ is a list concatenation of the subprogram input lists with repeated variable names removed, i.e.
\be
\bar{x} \simeq [x^{(i)}]_{i=1}^n
\ee
A program list satisfies all of the following conditions.

\begin{itemize}

\item The variable names of the elements of the subprogram output lists are distinct, i.e. 
\be
y^{(j)} \cap y^{(i)} = [~] , \quad i,j=1,\ldots,n ,~j \neq i
\ee

\item For each $i=1,\ldots,n$, the variable names of the elements of the list $x^{(i)}$ must not coincide with a variable name of the elements of the lists $y^{(k)},~k=i,\ldots,n$, i.e.
\be
x^{(i)} \cap [y^{(k)}]_{k=i}^n = [~], \quad i=1,\ldots,n 
\ee

\item $\mathfrak{p} \neq \mathfrak{p}^{(i)}$, $i=1,\ldots,n$, for $n>1$.

\end{itemize}
The empty program is denoted by $[~]$\index{empty program}.

\end{definition}

\vspace{5mm}

\textbf{I/O value assignments.}\index{value assignment} The elements of the I/O lists of a program, $\mathfrak{p}~x~y$, are alphanumeric strings that serve as placeholders for assigned values.
The action of assigning a value to an alphanumeric string that represents an element of an I/O list involves the allocation of an address that links the alphanumeric string to a value and its type that is stored in memory.
These values can be integers, fixed precision rational numbers or other strings of a well defined subtype. 
An element of an I/O list may also be assigned a value that comes in the form of a list or an array where each element of the list or array has a prescribed value consistent with its subtype.   

\textbf{Programs as vertical lists.}
A program with the list representation
\be
\mathfrak{p}~x~y=[\mathfrak{p}^{(i)}~x^{(i)}~y^{(i)}]_{i=1}^n
\ee
can also be written as a vertical list
\be
\mathfrak{p}~x~y = \left \{ 
\begin{array}{l}
\mathfrak{p}^{(1)}~x^{(1)}~y^{(1)} \\
\hspace{10mm} \vdots \\
\mathfrak{p}^{(n)}~x^{(n)}~y^{(n)} \\ 
\end{array}
\right .
\ee
We shall regard the above vertical and horizontal lists of a program $\mathfrak{p}~x~y$ to be just different representations of the same program.
While all lists are stored as arrays, when $n$ is sufficiently small, programs will be written as horizontal lists and will often be regarded as representing a sentence in the form of a single string.
When $n$ is large it is more convenient to display them as vertical lists.

A list $\mathfrak{p}~x~y = [\mathfrak{p}^{(i)}~x^{(i)}~y^{(i)}]_{i=1}^n$ should be thought of as being a core program embedded in a larger program that can be represented by the vertical list
\ben \label{core}
\left \{ 
\begin{array}{l}
read~x~[~] \\
\mathfrak{p}~x~y \\
print~y~[~] \\
\end{array}
\right .
\een
where, as indicated, the programs $read~x~[~]$ and $print~y~[~]$ have empty output lists.
 
By accessing an input data file, the subprogram $read~x~[~]$ assigns to each element of the input list $x$ a value and a type consistent with the entry type checking of the main program $\mathfrak{p}~x~y$.
The assigned values and types of the elements of the output list, $y$, are generated through the dual actions of value and type assignments contained within the subprograms of $\mathfrak{p}~x~y$.  
After executing $\mathfrak{p}~x~y$ the subprogram $print~y~[~]$ prints the value assigned output list, $y$, to a file and/or screen.
If an execution error is encountered in the program $\mathfrak{p}~x~y$ the execution is halted and an error message is printed to a file and/or screen.
For the purposes of analysis the core program $\mathfrak{p}~x~y$ will always be considered in isolation with the understanding that the value and type assignments of the elements of the input list $x$ have been prescribed by an initializing program $read~x~[~]$.

\textbf{Elements of program lists.}\index{program list elements}
A program list
\be
\bal
\mathfrak{p}~x~y = & [\mathfrak{p}^{(i)}~x^{(i)}~y^{(i)}]_{i=1}^n \\
= & [\mathfrak{p}^{(1)}~x^{(1)}~y^{(1)}~\mathfrak{p}^{(2)}~x^{(2)}~y^{(2)}~\ldots~\mathfrak{p}^{(n)}~x^{(n)}~y^{(n)}] \\
\eal
\ee
is a list of ordered triplets
\be
(program~name)~[input~list]~[output~list]
\ee
The symbol $\in$ is used to denote that a single element is contained in a list and the symbol $\subseteqq$ is used to denote that a sublist of a collection of elements is contained in a list.
We will always define individual elements of a program list to be the ordered triplets $\mathfrak{p}^{(i)}~x^{(i)}~y^{(i)}$, $i=1,\ldots,n$, that represent the subprograms of $\mathfrak{p}~x~y$.
We write
\be
\mathfrak{p}^{(i)}~x^{(i)}~y^{(i)} \in \mathfrak{p}~x~y
\ee
to mean that the triplet $\mathfrak{p}^{(i)}~x^{(i)}~y^{(i)}$ is an individual element of the program list represented by the main program $\mathfrak{p}~x~y$.
The notion of a sublist of a program is defined in terms of the elements of a program list in this sense.

When reading a program list from left to right the machine will recognize each subprogram, defined by the triplet $\mathfrak{p}^{(i)}~x^{(i)}~y^{(i)}$, as an individual string so that the internal spaces that separate the program name and the I/O lists will be regarded as special characters of that string.
Each triplet has a well defined structure so that the machine will have no trouble in distinguishing the internal spaces of each triplet from the spaces that separate the subprograms in a program list.    

\textbf{I/O dependency condition.}\index{I/O dependency condition}
From Definition \ref{prog} the variable names of the elements of the concatenation of the output lists of the subprograms of a program list $\mathfrak{p}~x~y = [\mathfrak{p}^{(i)}~x^{(i)}~y^{(i)}]_{i=1}^n$ must be distinct.
The input list $x^{(i)}$ of each subprogram $\mathfrak{p}^{(i)}~x^{(i)}~y^{(i)}$ may contain variable names of elements of the output lists of subprograms that precede it in a program list but cannot contain a variable name that coincides with any element of the output lists $y^{(k)}$, $k \geq i$.
This disallows reassigning values to a variable name as is common practice in imperative programming.
We shall often refer to this property as the \emph{I/O dependency condition}.
The I/O dependency condition plays a crucial role on how program lists can be manipulated.
It is possible to reorder the subprograms in a program list provided that the I/O dependency condition is not violated.

\textbf{Constants.}\index{constants}
We need to make a distinction between common variables and constants.
For each type there may exist special objects of that type that are of particular interest because they may appear as fixed assigned values in an input list of a program.
For example, programs associated with arithmetic operations involving type integers, $\mb{I}$, will recognize as value assigned inputs three constants, $-1,0,1$.
Later we will consider higher order programs, where programs themselves serve as inputs.
For higher order programs we may regard the empty program list, $[~]$, as a constant for an assigned value input of type $\mb{P}$.

\vspace{5mm}

\textbf{Notes.}

\begin{itemize}

\item A program list has the representation $\mathfrak{p}~x~y = [\mathfrak{p}^{(i)}~x^{(i)}~y^{(i)}]_{i=1}^n$ for some $n:\mb{I}_0$.
For $n=1$ we simply drop the list representation and write $\mathfrak{p}~x~y$.
The case $n=0$ means that the program $\mathfrak{p}~x~y$ is the empty program list, denoted by $[~]$. 
The execution of the program $\mathfrak{p}~x~y$, for a given value assigned input list, $x$, is completed when all of the subprograms $\mathfrak{p}^{(i)}~x^{(i)}~y^{(i)}$, $i=1,\ldots,n$, have been executed in the sequential order from left to right in the program list.

\item If the empty program is encountered in the execution of a program list then the program does not halt and execution proceeds to the next subprogram of the list.
Subprograms of a program list that are empty programs can be immediately removed by the process of an empty list extraction.

\item Given a program $\mathfrak{p}~x~y = [\mathfrak{p}^{(i)}~x^{(i)}~y^{(i)}]_{i=1}^n$, the output $y$ of a main program is a concatenation of the output lists of its subprograms.
The input list of any subprogram can contain elements that are also elements of output lists of the subprograms that precede it in the program list.
Some of these internal outputs are used only for intermediate calculations and are not in themselves of any particular interest to the application for which the main program is designed.
It is common practice in imperative programming to regard parameters used for intermediate calculations as free parameters that are discarded upon the execution of the subprogram.  
For functional programming we shall not make use of the notion of free parameters and include in the output list of the main program all of the elements of the output lists of its subprograms.
This may lead to an accumulation of a large number of variables that need to be stored in memory but there are advantages to this approach.
Firstly, the removal of the notion of free parameters will avoid the need to introduce some cumbersome details in the definitions of the rules of program construction.   
Secondly, if the main program, $\mathfrak{p}~x~y$, is later embedded as a subprogram into another program list the elements of $y$ that may have otherwise been discarded as free parameters can sometimes be reused as input by a subprogram in the new list.

\item  Haskell is a common functional programming language that is employed in the computer sciences.
It exploits the lambda calculus formulation and can be employed as a proof checker using the inference rules of propositional and first order logic. 
A good introductory coverage of Haskell can be found in \cite {hall}. 

Our functional programs based on Definition \ref{prog} are structurally quite different from those of lambda calculus.
Other objects also defined in this book will differ from those presented to students in the computer sciences. 
For example in our definition of lists we separate elements of a list by a space rather than a comma.
We also allow value assignments of elements of the same list to have different types.
For a student in the computer sciences this has a closer resemblance to a tuple rather than a list.

\item At this point, a reader who has a background in the computer sciences may regard the I/O dependency condition as an unnecessary complication that is bypassed in the lambda calculus.
While more will be said on this in a later chapter, it can only be recommended here that the reader persevere.
An attempt will be made to demonstrate that the I/O dependency condition is quite manageable and that our language will contain some useful features for the purposes that it is has been designed.

\end{itemize}

\section{Computability.}\label{sc}\index{computability}

\textbf{Execution error.}
Within all programs type checking is performed on the assigned values of all elements of its I/O lists.
Execution errors\index{execution error} are predominantly based on type violations\index{type violation}.
A program will halt with an execution error if during its execution there is a type violation of any assigned value of the elements of its I/O lists.
The execution of a program is completed when all subprograms of the program list have been successfully executed in the sequential order from left to right.

In the next section we will introduce atomic programs that also check for the satisfaction of a relation between a pair of elements of its input list.
In such cases an execution error will also include the case where the relation is not satisfied.
When referring to type violation errors we will include the case of a failure to satisfy a prescribed relation. 

In a later chapter we will introduce program disjunctions.
Disjunctions essentially split the execution of a single program list into a number of parallel program lists.
These parallel program lists can be associated with operands of the disjunction.
If at least one of the operand programs of the disjunction does not contain a type violation then all type violations that exist in the other operand programs are overridden and the main program will not halt with an execution error.

A formal definition of an execution error will be postponed until we have introduced disjunctions.
For the moment it will suffice to regard an execution error to be solely associated with the encounter of a type violation in a single program list.   

\textbf{Computable programs.}
Our main objective is to construct computer models that can be validated by establishing computability.
By this we mean that a program will eventually halt without encountering an execution error and return a value assigned output.

In general there may exist programs for which we will be unable to rigorously establish computability or non-computability for that matter.
In the computer sciences undecidability is highlighted by the the halting problem, although this is discussed in the context of abstract computers.
In a similar way undecidability also arises in mathematics where it is often regarded as troublesome and an unwelcome reminder of a limitation in mathematics.
In a more general context of the scientific method, undecidability is an accepted concession where the best that can be hoped for is a process of continual revision from which will emerge theories with expanded scope of applicability.
We shall address this in more detail in the final chapter of this book.
With these issues in mind the following definition of computability will be sufficient for our purposes.

\begin{definition}\label{comp}(Computability.)
A program $\mathfrak{p}~x~y:\mb{P}$ is said to be computable, with respect to a value assigned input list, $x$, if upon execution it eventually halts without encountering an execution error.
A computable program returns the value assigned output, $y$, where $y$ may be the empty list.
We write $\mathfrak{p}~x~y:\mb{P}_{comp}$ to mean that the program $\mathfrak{p}~x~y$ is computable in the sense that it is computable for at least one value assigned input list $x$.
We write $\mathfrak{p}~x~y:\mb{P}_{comp}(x)$ to stress that $\mathfrak{p}~x~y$ is computable for a particular prescribed value assigned input list $x$.
\end{definition}

\textbf{Computability map.}
Let $\mathfrak{p}~x~y:\mb{P}$ and consider the pair $(\mathfrak{p},x)$, where $x$ is to be regarded as a prescribed value assigned input list of $\mathfrak{p}~x~y$.
We define the \emph{computability map}\index{computability map} by
\ben\label{cm}
\mathfrak{C}(\mathfrak{p},x) = \left \{
\begin{array}{ll}
	1, & \mathfrak{p}~x~y:\mb{P}_{comp}(x) \\
	0, & \text{otherwise} \\
\end{array}
\right .
\een
A program $\mathfrak{p}~x~y$ will be computable, i.e. $\mathfrak{C}(\mathfrak{p},x)=1$, for a given value assigned input list, $x$, if upon execution it halts without encountering an execution error in which case it returns the value assigned output, $y$, where $y$ may be an empty list.
If upon execution the program does encounter an execution error then $\mathfrak{C}(\mathfrak{p},x)=0$ and the program halts with an execution error message.

The computability map\index{computability map} is analogous to assigning truth values to the well formed formulas of classical logic.
There is, however, an important distinction that needs to be made here in that the computability map is not subject to an interpretation and is determined by executing the program for a value assigned input list of a program.
In this sense the computability map can be thought of as an empirical process.

The empirical process of the computability map will only be useful in a practical sense if a program will halt in a reasonable time.
What can be regarded as a reasonable time will be subject to an arbitrary choice of the user.
Sometimes the notion of polynomial time is adopted for this purpose.
In any case, for a program that can be observed to halt, the computability map will establish with certainty the computability or non-computability of that program with respect to a given value assigned input list.  

Most of our effort will focus on establishing the computability of programs by way of inference so the computability map will not be directly employed.
Nevertheless, it is useful to regard the computability map to be implicitly active throughout as an underlying action that defines the property of computability of a program.    
On the other hand, the empirical procedure associated with the computability map will have a more direct role to play in the final chapter of this book.

\textbf{Assignment map.}\index{assignment map}
Programs can have an empty output list.
We have already encountered two such programs such as the $read$ and $print$ programs in (\ref{core}). 
Apart from the $read$ and $print$ programs, programs with an empty output list are often associated with the sole task of checking the types of the value assignments of the elements of their input list.

A program, $\mathfrak{p}~x~y$, with a nonempty output list, $y$, will be referred to as a value assignment program.
Assignment programs are often associated with arithmetic calculations but may also involve algorithms that cannot be concisely expressed in the conventional mathematical notation.
On occasion we will refer to an assignment map $f:x \mapsto y$ as being associated with an assignment program $\mathfrak{p}~x~y$. 
The notion of an assignment map\index{assignment map} is often used when discussing the internal algorithm that is employed to assign values to the elements of the output list of the program from the assigned values of the elements of its input list.  
The internal algorithm of an atomic program can be thought of as a sequence of instructions, including arithmetic computations, written in some imperative language.
When the assignment map $f:x \mapsto y$ can be expressed in a concise conventional mathematical form we will use the notation of functions and write $y:=f(x)$.
While the association of programs with maps will be useful it should not be taken too formally since our approach will be mainly syntactic based on the manipulation of strings.

Throughout we are avoiding any reference to sets and will work with lists.
An assignment map $f$ will often be represented by the notation
\ben\label{map01}
f:\mb{A} \to \mb{B}
\een
with the meaning that $f$ maps objects of type $\mb{A}$ to objects of type $\mb{B}$.
Since we are working on a platform $\mathfrak{M}(\mathcal{K},\mathcal{L},\mathcal{M})$, the types $\mb{A}$ and $\mb{B}$ will always contain a finite number of objects.
It will be convenient to regard the collection of objects of types $\mb{A}$ and $\mb{B}$ in a map representation (\ref{map01}) to be lists, albeit with very large lengths.
For example we may write $f:\mb{I} \to \mb{I}$, where $f$ maps objects of type $\mb{I}$ to objects of type $\mb{I}$.
Here we may regard the collection of objects of type $\mb{I}$ to be represented by the list $[i]_{i=-N}^N$.
  
When used in this context we will need to distinguish collections of objects of types as lists in some abstract sense since, for practical reasons, it is inefficient to store them in the memory of a machine.
This will allow us to make use of all the list properties and in particular the notion of sublists as representing sub-domains and ranges of maps.

\textbf{False programs.}\index{false program}
It is possible for an object to have all of the structural properties of a program but will never be computable.
Such objects will be called false programs. 

\begin{definition}(False program.)
A program $\mathfrak{p}~x~y:\mb{P}$ is said to be a false program if there does not exist a value assigned input list, $x$, such that $\mathfrak{p}~x~y$ is computable.
A false program $\mathfrak{p}~x~y$ is assigned the type $\mathfrak{p}~x~y:\mb{P}_{false}$ with the hierarchy of subtypes $\mb{P}_{false}<:\mb{P}$.
\end{definition}

In the definition of a false program it is stated that $\mb{P}_{false}<:\mb{P}$.
This means that for an object to have the type assignment $\mb{P}_{false}$ it must first have the structure of a program under Definition \ref{prog}.
The statement that a program will always halt as a result of an error in syntax is not considered meaningful in this context since such an object cannot be assigned the type $\mb{P}$.

\section{Atomic programs.}\index{atomic program}
Functional programs are built up from lists of atomic programs.
It will be understood that atomic programs are constructed from some imperative language.
Except for some special cases, the imperative program list of atomic programs will not be presented, only the functionality of the atomic programs will be defined.

\begin{definition}(Atomic program.)\label{atomic}
An atomic program is a subtype of program type, $\mb{P}_{atom}<:\mb{P}$.
An atomic program $\mathfrak{p}~x~y:\mb{P}_{atom}$ must include type checking for the assigned values of every element of its I/O lists.
If for any value assigned element of the I/O lists, $x$ and $y$, there is a type violation the program halts prematurely as a type violation error.
Otherwise the atomic program returns the assigned valued output $y$, where $y$ may be the empty list.
Atomic programs may call other atomic programs but each atomic program introduces a new functionality.
\end{definition}

Atomic programs can be partitioned into the three subtypes of type checking, value/type assignment and type assignment.

\begin{definition}(Type checking programs.)\index{type checking program}
A type checking program $\mathfrak{p}~x~y$ is an atomic program with the following properties.

\begin{itemize}

\item The output list $y$ is the empty list so that type checking programs have the representation $\mathfrak{p}~x~[~]$.

\item The type of the assigned values of every element of the input list is checked upon entry.

\item If a type violation is encountered the program halts prematurely with a type violation error.

\end{itemize}

A type checking program is assigned the type $\mathfrak{p}~x~y:\mb{P}_{type}$, where $\mb{P}_{type} <: \mb{P}_{atom}$.

\end{definition}

\begin{definition}(Value assignment programs.)\index{value assignment program}
	Value assignment programs are atomic programs that combine all of the actions of entry type checking, value assignment and type assignment. 
	A value assignment program $\mathfrak{p}~x~y$ has the following properties.
	
	\begin{itemize}
		
		\item The type of the assigned values of every element of the input list, $x$, is checked upon entry.
		
		\item If there is a type violation of at least one element of the input list the program halts prematurely with a type violation error.
		
		\item If, upon entry, there are no type violations, a value assignment program then attempts to assign a value to each element of the output list through the action of an assignment map.
		
		\item If there is a type violation of an assigned value of an element of the output list the program halts prematurely with a type violation error.
		
		\item If there are no type violations each element of the output list, $y$, is simultaneously assigned a value and a type consistent with the value assignment.
		
	\end{itemize}
	
	A value assignment program is assigned the type $\mathfrak{p}~x~y:\mb{P}_{assign}$, where $\mb{P}_{assign} <: \mb{P}_{atom}$.
	
\end{definition}

There will be situations where objects of some specified type will be assigned a new type.
These newly assigned types will often be referred to as abstract types\index{abstract type}. 

\begin{definition}(Type assignment programs.)\index{type assignment program}
A type assignment program $\mathfrak{p}~x~y$ is an atomic program with the following properties.

\begin{itemize}

\item The output list $y$ is the empty list so that type assignment programs have the representation $\mathfrak{p}~x~[~]$.

\item The type of the assigned values of every element of the input list is checked upon entry.
The check is performed on the type already assigned to the variable upon entry and not the type that is to be assigned.

\item If there is a type violation the program halts prematurely with a type violation error.

\item If, upon entry, there are no type violations, a type assignment program then assigns the new type to the elements of the input list that are the target of that type assignment program.   

\item Once the target variables are assigned a new type they are internally stored in memory so that if the target variables are encountered as input of a following subprogram of a program list they are recognized as that assigned type. 
 
\end{itemize}
We write $b::\mb{B}$ to denote that object $b$ is assigned the type $\mb{B}$.

A type assignment program has the type $\mathfrak{p}~x~y:\mb{P}_{tassign}$, where $\mb{P}_{tassign} <: \mb{P}_{atom}$.

\end{definition}

\textbf{Notes.}

\begin{itemize}

\item Because the internal algorithm of an atomic program is not accessible in explicit form it will be necessary to supply a collection of rules or axioms that describe its algorithm.
It is through these axioms that VPC will be able to recognize the properties of an atomic program during proof construction. 

\item All functional programs will be constructed from atomic programs through the construction rules to be presented in the following chapters.
Hence all programs will contain the action of type checking for the assigned values of all elements of their I/O lists.

\item Type checking within a program is an action that checks the type of the assigned value of a given variable.
Type checking may also include the checking of some relation between its input variables.
For example, a type checking for valued assigned variables that are integers, say $a:\mb{I}$ and $b:\mb{I}$, may include a check for value assigned equality, $a=b$, or value assigned inequality, $a<b$.
In other words a type violation error will include failure of at least one of the actions of type checking, $a:\mb{I}$ and $b:\mb{I}$, and the value assigned equality or inequality.

\item Since atomic programs are constructed from an imperative programming language the notion of free parameters is difficult to avoid.
This is because the internal algorithms of the atomic programs are likely to employ a number of parameters in the process of computing the output parameters from a list of input parameters.
These internally defined parameters can be regarded as free parameters that are released from memory storage upon execution and do not appear in the output list of the atomic program.
Thus in our construction of programs by way of functional programming we will not see these internal free parameters.

\item Due to the I/O dependency condition there is no general rule that allows the repetition of subprograms of a program list $[\mathfrak{p}^{(i)}~x^{(i)}~y^{(i)}]_{i=1}^n$.
However, repetition of subprograms with an empty list output is allowed and will not effect the computability of the program list.
For computational efficiency such repeated subprograms are redundant and should be avoided. 

\end{itemize}

\chapter{Construction Rules.}\label{ccr}

\section{Introduction.}

Our objective is to construct a formal system from which we can determine the computability of programs, particularly programs that are designed to solve finite dynamical systems.
Adopting a language based on functional programs under the constraints of a machine environment $\mathfrak{M}(\mathcal{K},\mathcal{L},\mathcal{M})$ introduces some features that will require a departure from conventional languages employed in formal systems of proof theory.
Here we will lay down a collection of program construction rules that better reflect the operational constraints of our language based on functional programs on a working platform $\mathfrak{M}(\mathcal{K},\mathcal{L},\mathcal{M})$.

To this end we will largely deal with objects that are recognized by the machine from their string structure.
These include subtypes of strings such as scalars and lists of the machine integers, fixed precision rational numbers and programs.
As a result we will be dealing with a low level language that will lack the expressiveness found in many formal systems of proof theory.
However, we will demand that this language possess the power of analysis at a sufficiently high level for the purposes that it has been designed.
We will also demand that the language be soundly rooted as a primitive on top of which theories requiring higher levels of abstractions can be built.

The construction rules\index{construction rules} form the basis of the program VPC and can be regarded as the general inference rules that are applied to specific applications.
An application, $S$, sometimes referred to as a theory, comes with its own collection of atomic programs and axioms that serve as initializing input data to VPC.
Programs of $S$ are constructed inductively from these atomic programs as program lists.
Later we will include program constructions that are based on disjunctions.
Computability of the programs of $S$ is defined in terms of the value assignments of program input lists and is governed by the axioms associated with the application $S$ in conjunction with the construction rules.

\section{Higher order programs.}\label{hop}

Programs are strings, or lists of strings, with a well defined structure and may serve as assigned values of elements of an I/O list of a program.
A program will be said to be a higher order program if the assigned values of the elements of its I/O lists are of type $\mb{P}$.
Higher order programs\index{higher order program} essentially recognize programs as strings, or lists of strings, with a particular structure, namely that outlined in Definition \ref{prog}.

We can partition type $\mb{P}$ objects into the subtypes $\mb{P}_n$, $n:\mb{I}_0$.
A type $\mb{P}_n$ object, $n \geq 1$, is an $n$-th order program whose I/O lists contain elements that can be assigned values that are type $\mb{P}_{n-1}$ objects.

The elements of the I/O lists of zeroth order programs, type $\mb{P}_0$ objects, are assigned values such as scalars or lists of integers, fixed precision rationals and strings of a well defined structure, excluding programs.
The elements of the I/O lists of first order programs, type $\mb{P}_1$ objects, are assigned values of strings of a specific subtype, namely zeroth order programs.
\emph{First order programs do not recognize the value assignments of the I/O lists of the zeroth order programs.}

The construction rules that follow will be expressed as first order programs.
Since much of what follows will not involve objects of type $\mb{P}_n$, $n \geq 2$, we will simply refer to type $\mb{P}_1$ objects as higher order programs.       

\textbf{Notation.} From this point on throughout this book we will often use the shorthand notation of representing programs with lower case letters so that, for example, by
\be
a:\mb{P}
\ee
it is understood that the dummy variable $a$ is assigned the value of a string of subtype $\mb{P}$.
We may also regard $a$ as being assigned the value of type $\mb{P}$ explicitly given by the program
\be
a=\mathfrak{p}_a~x_a~y_a
\ee
so that we may refer to the program name of $a$ as $\mathfrak{p}_a$, the input list of $a$ as $x_a$ and the output list of $a$ as $y_a$.
From the perspective of machine hardware the value assignment $a::\mb{P}$ involves the allocation of an address that attaches the dummy variable $a:\mb{C}$ to a program $\mathfrak{p}_a~x_a~y_a : \mb{P}$ that is stored in memory.

We may also write $a=[a_i]_{i=1}^n$ to mean that $a:\mb{P}$ is a program list and each $a_i:\mb{P}$ is a subprogram of the main program $a$. 
Of particular interest will be sublists of programs, i.e. a program $b=[b_i]_{i=1}^m$ will be a sublist of the program $a=[a_i]_{i=1}^n$, written $b \subseteqq a$, if every subprogram, $b_i \in b$, is a subprogram of $a$.  

\textbf{Higher order atomic programs.} Higher order programs will be constructed from the atomic programs whose names are given in the tables below.
 
\vspace{5mm}

\begin{tabular}{|c|c|}
\hline
Atomic program names & Atomic program type \\
\hline
$typep,~eqv,~eqio,~sub,~ext,~false$ & $\mb{P}_{type}$ \\
\hline
$aext,~afalse$ & $\mb{P}_{tassign}$ \\
\hline
$conc,~disj$ & $\mb{P}_{assign}$ \\
\hline
\end{tabular}

\vspace{5mm}
We will also make use of the following special non-atomic higher order program. 

\vspace{5mm}

\begin{tabular}{|c|c|}
\hline
Special non-atomic program name & Structure \\
\hline
$epd$ & program list \\
\hline
\end{tabular} 

\vspace{5mm}

The following is a description of the higher order atomic programs that will be used to define the construction rules.
Some of the descriptions will involve definitions that will appear later on in the text.
It is important to note that higher order programs are themselves type $\mb{P}$ objects and the assigned values of the elements of their I/O lists are type $\mb{P}$ objects specific to a theory or application, $S$.
Each application $S$ will come with its own atomic programs and axioms.

\parskip0pt

\vspace{5mm}
\textbf{Check type program.}

\textbf{\textit{Syntax.}} $typep~[p]~[~]$.

\textbf{\textit{Program Type.}} $\mb{P}_{type}$.

\textbf{\textit{Type checks.}} $p:\mb{P}$.

\textbf{\textit{Description.}} $typep$ checks that the variable $p$ has been assigned a value of type $\mb{P}$, i.e. $typep$ recognizes $p$ as a string or a list of strings and checks that it does not violate any of the structural conditions stated in Definition \ref{prog}.
$typep$ halts with an execution error if there is a type violation.

\vspace{5mm}
\textbf{Check program equivalence.}

\textbf{\textit{Syntax.}} $eqv~[p~q]~[~]$.

\textbf{\textit{Program Type.}} $\mb{P}_{type}$.

\textbf{\textit{Type checks.}} $p:\mb{P}$, $q:\mb{P}$, $p \equiv q$.

\textbf{\textit{Description.}} $eqv$ first checks that the variables $p$ and $q$ have been assigned values of type $\mb{P}$.
It then checks that $p$ and $q$ are program equivalent, i.e. $p \equiv q$.
$eqv$ will examine the program structures of $p$ and $q$ as outlined in Definition \ref{equiv0}.
If the programs contain disjunctions the extended definition of program equivalence, as itemized in Definition \ref{equiv}, is used.
$eqv$ halts with an execution error if there is a type violation.
Type violation includes the case where $p$ and $q$ fail to be program equivalent.

\vspace{5mm}
\textbf{Check I/O equivalence.}

\textbf{\textit{Syntax.}} $eqio~[p~q]~[~]$.

\textbf{\textit{Program Type.}} $\mb{P}_{type}$.

\textbf{\textit{Type checks.}} $p:\mb{P}$, $q:\mb{P}$, $p \thicksim q$.

\textbf{\textit{Description.}} $eqio$ first checks that the variables $p$ and $q$ have been assigned values of type $\mb{P}$.
It then checks that $p$ is I/O equivalent to $q$, i.e. $p \thicksim q$.
$eqio$ will examine the program structures of $p$ and $q$ and identify whether the properties leading to I/O equivalence, as itemized in Definition \ref{equivio}, are satisfied.
$eqio$ halts with an execution error if there is a type violation.
Type violation includes the case where $p$ fails to be I/O equivalent to $q$. 

\vspace{5mm}
\textbf{Check program sublist.}

\textbf{\textit{Syntax.}} $sub~[q~p]~[~]$.

\textbf{\textit{Program Type.}} $\mb{P}_{type}$.

\textbf{\textit{Type checks.}} $p:\mb{P}$, $q:\mb{P}$, $q \subseteqq p$.

\textbf{\textit{Description.}} $sub$ first checks that the variables $p$ and $q$ have been assigned values of type $\mb{P}$ and then checks that $q$ is a program sublist of $p$, i.e. $q \subseteqq p$.
$sub$ halts with an execution error if there is a type violation.
Type violation includes the case where $q$ is not a sublist of $p$.

\vspace{5mm}
\textbf{Check type extension.}

\textbf{\textit{Syntax.}} $ext~[p~c]~[~]$.

\textbf{\textit{Program Type.}} $\mb{P}_{type}$.

\textbf{\textit{Type checks.}} $p:\mb{P}$, $c:\mb{P}_{ext}(p)$.

\textbf{\textit{Description.}} $ext$ checks that the variable $p$ has been assigned a value of type $\mb{P}$ and the variable $c$ has been assigned a value of type $\mb{P}_{ext}(p)$.
The program $c$ associated with a program $[p~c]$ of an application specific axiom or theorem that is stored in the file \emph{axiom.dat} is assigned the type $\mb{P}_{ext}(p)$ by default.
Otherwise a program can only acquire the type $\mb{P}_{ext}(p)$ through the type assignment program $aext$.
$ext$ halts with an execution error if there is a type violation.

\vspace{5mm}
\textbf{False program type check.}

\textbf{\textit{Syntax:}} $false~[p]~[~]$.

\textbf{\textit{Program Type:}} $\mb{P}_{type}$.

\textbf{\textit{Type checks:}} $p:\mb{P}_{false}$.

\textbf{\textit{Description:}} $false$ checks that the variable $p$ has been assigned a value of type $\mb{P}_{false}$.
Programs of axioms of falsity that are stored in the file \emph{axiom.dat} are assigned the type $\mb{P}_{false}$ by default. 
Otherwise a program can only acquire the type $\mb{P}_{false}$ through the type assignment program $afalse$.
$false$ halts with an execution error if there is a type violation.

\vspace{5mm}
\textbf{Extension type assignment.}

\textbf{\textit{Syntax.}} $aext~[p~c]~[~]$.

\textbf{\textit{Program Type.}} $\mb{P}_{tassign}$.

\textbf{\textit{Type checks.}} $p:\mb{P}$, $c:\mb{P}$.

\textbf{\textit{Type assignment.}} $c::\mb{P}_{ext}(p)$.

\textbf{\textit{Description.}} $aext$ first checks that the variables $p$ and $c$ have been assigned values of type $\mb{P}$.
If there are no type violations $aext$ then makes the assignment of subtype $c::\mb{P}_{ext}(p)$.
$aext$ halts with an execution error if there is a type violation.

\vspace{5mm}
\textbf{False program type assignment.}

\textbf{\textit{Syntax:}} $afalse~[p]~[~]$.

\textbf{\textit{Program Type:}} $\mb{P}_{tassign}$.

\textbf{\textit{Type checks:}} $p:\mb{P}$.

\textbf{\textit{Type assignment.}} $p::\mb{P}_{false}$.

\textbf{\textit{Description:}} $afalse$ first checks that the variable $p$ has been assigned a value of type $\mb{P}$.
If there is no type violation $afalse$ then makes the assignment of subtype $p::\mb{P}_{false}$.
$afalse$ halts with an execution error if there is a type violation, i.e. $p$ is not of type $\mb{P}$.

\vspace{5mm}
\textbf{Program list concatenation.}

\textbf{\textit{Syntax.}} $conc~[p~q]~[r]$.

\textbf{\textit{Program Type.}} $\mb{P}_{assign}$.

\textbf{\textit{Type checks.}} $p:\mb{P}$, $q:\mb{P}$.

\textbf{\textit{Assignment map.}} $r:=[p~q]$.

\textbf{\textit{Type assignment.}} $r::\mb{P}$.

\textbf{\textit{Description.}} $conc$ first checks that the variables $p$ and $q$ have been assigned values of type $\mb{P}$.
If successful $conc$ then attempts to assign to $r$ the program concatenation of $p$ and $q$, i.e. $r:=[p~q]$.
This may fail if $y_q \cap [x_p~y_p] \neq [~]$.
A successful value assignment is accompanied by the type assignment $r::\mb{P}$.
$conc$ halts with an execution error if there is a type violation.

\vspace{5mm}
\textbf{Program disjunction.}

\textbf{\textit{Syntax:}} $disj~[p~q]~[r]$.

\textbf{\textit{Program Type:}} $\mb{P}_{assign}$.

\textbf{\textit{Type checks:}} $p:\mb{P}$, $q:\mb{P}$.

\textbf{\textit{Assignment map.}} $r:=p~|~q$.

\textbf{\textit{Type assignment.}} $r::\mb{P}$.

\textbf{\textit{Description:}} $disj$ first checks that the variables $p$ and $q$ have been assigned values of type $\mb{P}$.
If successful $disj$ then attempts to assign to $r$ the disjunction of the programs $p$ and $q$, i.e. $r:=p~|~q$ subject to the structural properties given in Definition \ref{disj}.
A successful value assignment is accompanied by the type assignment $r::\mb{P}$.
$disj$ halts with an execution error if there is a type violation.

\parskip10pt

\textbf{Notes.}

\begin{itemize}

\item In this book we will only make use of zeroth and first order programs.
However, our formal system can be extended to include statements of higher order constructs that make use of objects of type $\mb{P}_n$, $n \geq 2$.
Here we will focus on the most basic foundations of our formal system and leave the details of these higher order constructs for future development.  

\end{itemize}

\section{Program extensions.}

The main idea behind our formal system is to construct computable programs as extensions of programs that are known to be computable.
The following definition formalizes this idea. 

\begin{definition}\label{ce}(Program extension.)\index{program extension}
A program $c:\mb{P}$ is called an \textbf{extension} of the program $p:\mb{P}$ and assigned the subtype $c:\mb{P}_{ext}(p)$ if the following conditions are satisfied.

\begin{enumerate}

\item The input list $x_c$ of the program $c$ cannot introduce new variable names other than constants, i.e. if $u \subseteqq x_c$, where $u$ is a list of all constants that appear in $x_c$, then $x_c \setminus u \subseteqq [x_p~y_p]$, where $x_p$ and $y_p$, respectively, are the input and output lists, respectively, of the program $p$.  
	
\item If $p$ is computable with respect to a valued assigned input then the program $s=[p~c]$ is also computable for the same value assigned input, i.e. if $p:\mb{P}_{comp}(x_p)$ for a value assigned input list $x_p$ of $p$ then $s:\mb{P}_{comp}(x_s)$, where $x_s$ is the value assigned input list of $s$ that acquires its value through the identities $x_s = \bar{x}_s \setminus (x_c \cap y_p)$ and
$\bar{x}_s \simeq [x_p ~x_c]$.

\end{enumerate}
We write $c:\mb{P}_{ext}(p)$ to stress that $c$ is an extension associated with $p$.
The program $s=[p~c]$, such that $c:\mb{P}_{ext}(p)$, is simply referred to as an \textbf{extended program}.\index{extended program}
The hierarchy of subtypes is $\mb{P}_{ext}(p)<:\mb{P}$.
\end{definition}

\vspace{5mm}

\begin{definition}\label{ice}(Irreducible program extension.)\index{irreducible program extension}
A program $c:\mb{P}$ is called an \textbf{irreducible extension} of the program $p:\mb{P}$ and assigned the subtype $c:\mb{P}_{iext}(p)$ if the following conditions are satisfied.

\begin{enumerate}

\item $c:\mb{P}_{ext}(p)$.

\item The program $p$ is irreducible in the following sense. There does not exist a program $r=[q~c]$ such that $q \varsubsetneqq p$ ($q$ is a strict sublist of $p$) and $c:\mb{P}_{ext}(q)$.

\end{enumerate}
The program $s=[p~c]$, such that $c:\mb{P}_{iext}(p)$, is said to be an \textbf{irreducible extended program}.\index{irreducible extended program}
The programs $p$ and $c$, respectively, are said to be the premise and conclusion, respectively, of the irreducible extended program $s$.
The hierarchy of subtypes is $\mb{P}_{iext}(p)<:\mb{P}_{ext}(p)$.
\end{definition}

\vspace{5mm}

\begin{definition}(Extended program derivation.)\index{extended program derivation}
An extended program derivation $s$ with respect to the program $[q~c]:\mb{P}$ is an assignment $s:=[p~c]$ subject to the conditions $q \subseteqq p$, $c:\mb{P}_{ext}(q)$ and $s:\mb{P}$.
It is constructed from the higher order program $epd~[q~p~c]~[s]$ defined by
\be
edp~[q~p~c]~[s] = [sub~[q~p]~[~]~ext~[q~c]~[~]~conc~[p~c]~[s]]
\ee
The program $edp~[q~p~c]~[s]$ is called the \textbf{extended program derivation}.
\end{definition}

\vspace{5mm}

\textbf{Notes.}

\begin{itemize}

\item An extension of $p$ is associated with a program concatenation $s=[p~c]$ for some $c:\mb{P}_{ext}(p)$.
The program $p$ may be a program list $p=[p_i]_{i=1}^n$.
The program $c$ is usually an atomic program.
The program $c$ may be a program list and/or disjunction but must be defined as a special non-atomic program and appear as the main program in the concatenation $[p~c]$.

\item It is important to note that an extension $c:\mb{P}_{ext}(p)$ does not imply that the program $[p~c]$ is computable.
Definition \ref{ce} merely states that if for an assigned value input, $x_p$, the program $p$ is computable then so is the program $[p~c]$ for the same value assigned input.
This observation will be crucial when dealing with false programs as will be discussed later.

\item The last statement, $conc~[p~c]~[s]$, in the definition of an extended program derivation $edp~[q~p~c]~[s]$ is not necessarily computable if the first two statements are computable.
A necessary condition for $epd~[q~p~c]~[s]$ to be computable is that $[p~c]:\mb{P}$.
The program concatenation $[p~c]$ may fail if for instance $y_c \cap [x_p~y_p] \neq [~]$.
When attempting to construct $[p~c]$ from an extended program derivation we will always be able to choose the variable names of the output list of the program $c$ such that they do not conflict with the variable names of the I/O lists of the program $p$ in the above sense.

\end{itemize}

\section{The program extension rule.}\label{stper}

Construction rules are presented as irreducible extended higher order programs.
The internal square brackets acts as a delimiter of the premise program from the conclusion program.
The standard list concatenation for programs apply so that the internal brackets can be removed.
When the premise program contains only a single statement the internal square brackets are omitted.
When applying the construction rules, the conclusion program, $c:\mb{P}_{iext}(p)$, of an irreducible extended program must appear as the last statement in the concatenation $[p~c]$.
After a construction has been completed in this way the reordering of the program $c$ in the expanded program list $[p~c]$ is allowed provided that the I/O dependency condition is not violated. 
We start with the main inference rule called the program extension rule\index{program extension rule}.

\emph{Program extension rule.}

\textbf{per}\label{per}
\be
\bal
&  [epd~[q~p~c]~[s]~aext~[p~c]~[~]] \\
\eal
\ee

The program extension rule, per, states that if $s:=[p~c]$ is an extended program derivation with respect to the extension $c:\mb{P}_{ext}(q)$ then it follows that $c$ is also an extension of $p$ and hence $s$ is an extended program of $p$.
The conclusion program of the program extension rule is a type assignment $c::\mb{P}_{ext}(p)$.

The following construction rule states that once assigned, the property of an extension is retained.
In other words, once a program has been assigned the type $\mb{P}_{ext}(p)$, for some $p:\mb{P}$, it is stored in memory as such so that it is recognized as that type whenever it is accessed by any following subprogram of a higher order program list. 

\emph{Retention of subtype assignment.} 
\begin{crule}\label{cr01}
	\be
	\bal
	&  [aext~[p~c]~[~]~ext~[p~c]~[~]] \\
	\eal
	\ee
\end{crule}

By definition, for each program $c$ that is an extension of a program $p$ there corresponds an extended program $s=[p~c]$.
The following rule constructs the program $s$ given $c:\mb{P}_{ext}(p)$.

\emph{Extended program construction.}
\begin{crule}\label{cr02}
\be
[ext~[p~c]~[~]~conc~[p~c]~[s]]
\ee
\end{crule}

The formal system based on the program extension rule, per, along with cr 1-cr 2 and the additional construction rules that will follow, will be referred to as PECR\index{PECR} (Program Extension Construction Rules).
The formal system PECR can be regarded as the rules of inference that are designed to be applied on a working platform $\mathfrak{M}(\mathcal{K},\mathcal{L},\mathcal{M})$ and forms the basis of the program VPC.

\textbf{Derivations and proofs.}
A program $[p_i]_{i=1}^m$ is called a derivation if it is constructed by a sequence of extended program derivations.
Let $s_k=[p_i]_{i=1}^k$ and consider the following iteration.

\begin{itemize}

\item The program $s_n=[p_i]_{i=1}^n$, for some $n<m$, serves as a list of premises of the derivation.

\item For each iteration $i=n+1,\ldots,m$, the statement $p_i$, in an extended program derivation $epd~[q_i~s_{i-1}~p_i]~[s_i]$, is introduced from some known extension $p_i:\mb{P}_{ext}(q_i)$ such that $q_i \subseteqq s_{i-1}$.

\end{itemize}

\noindent
A derivation may be called a proof if its final statement is of particular interest in relation to its premise program.
An irreducible extended program that is extracted from a proof is called a \emph{theorem}\index{theorem}.
An irreducible extended program for which no derivation is known is called an \emph{axiom}\index{axiom}.

Since we are working in an environment $\mathfrak{M}(\mathcal{K},\mathcal{L},\mathcal{M})$ there is a need to adopt some convention that reflects this constraint.
It will always be understood that axioms/theorems can only be expressible as horizontal lists, where the horizontal list is to be regarded as a single string that represents a sentence and has at most $\mathcal{L}$ characters.
Proofs will be expressed as vertical lists stored as arrays.
An irreducible extended program can only be considered to be a theorem on $\mathfrak{M}(\mathcal{K},\mathcal{L},\mathcal{M})$ if its proof has a list length at most $\mathcal{M}$.
An irreducible extended program that has no proof on $\mathfrak{M}(\mathcal{K},\mathcal{L},\mathcal{M})$ but has a proof on a larger machine $\mathfrak{M}(\mathcal{K},\mathcal{L},\mathcal{M}^\prime)$, where $\mathcal{M}^\prime > \mathcal{M}$, can only be regarded as a potential axiom on $\mathfrak{M}(\mathcal{K},\mathcal{L},\mathcal{M})$.

\textbf{Notes.}

\begin{itemize}
	
	\item We will often deal with objects that are subtypes of strings, $\mb{S}$, such as scalars and lists of the machine integers, fixed precision rational numbers and programs. 
	All of these objects have a well defined string structure as specified by their definitions and are recognized by the machine.
	
	The program $aext~[p~c]~[~]$ makes the type assignment $c::\mb{P}_{ext}(p)$.
	A machine can readily verify Condition 1 of Definition \ref{ce} by its string structure.
	However, from the perspective of a practical computation, the machine has no general way of recognizing that a concatenation of programs has the property associated with computability as outlined in Condition 2 of Definition \ref{ce}.
	Consequently, a machine can only interpret objects of type $\mb{P}_{ext}$ through the properties embedded in the construction rules.
	In this sense $\mb{P}_{ext}$ can be referred to as an abstract type\index{abstract type}.  
	 	
\end{itemize}

\section{Program and I/O equivalence.}
Program equivalence\index{program equivalence} refers to programs that may appear to have a different structure but are functionally identical.
Program equivalence will be defined in terms of sublists and are associated with program lists whose subprograms appear in a different sequential order.
This definition will be extended later to include disjunctions.

\begin{definition}\label{equiv0}(Program equivalence.)
Two programs $p:\mb{P}$ and $q:\mb{P}$ that are program lists are said to be program equivalent provided that $p \subseteqq q$ and $q \subseteqq p$.
Program equivalence is denoted by $p \equiv q$ and satisfies the properties of reflexivity, symmetry and transitivity.

\end{definition}

The second important kind of equivalence refers to programs where the variable names of the elements of their I/O lists differ but can be associated with some degree of functionality.

\begin{definition}\label{equivio}(I/O equivalence.)\index{I/O equivalence}
Consider two programs with the list representations $[\mathfrak{p}^{(i)}~x^{(i)}~y^{(i)}]_{i=1}^n$ and $[\mathfrak{p}^{(i)}~\bar{x}^{(i)}~\bar{y}^{(i)}]_{i=1}^n$.
Let
\be
x^{(i)}=[x_j^{(i)}]_{j=1}^{l_i},~~~\bar{x}^{(i)}=[\bar{x}_j^{(i)} ]_{j=1}^{l_i}, \qquad l_i:\mb{I}_+,~i=1,\ldots,n
\ee
\be
y^{(i)}=[y_j^{(i)}]_{j=1}^{m_i},~~~\bar{y}^{(i)}=[\bar{y}_j^{(i)} ]_{j=1}^{m_i}, \qquad m_i:\mb{I}_0,~i=1,\ldots,n
\ee
The program $[\mathfrak{p}^{(i)}~\bar{x}^{(i)}~\bar{y}^{(i)}]_{i=1}^n$ is I/O equivalent to the program $[\mathfrak{p}^{(i)}~x^{(i)}~y^{(i)}]_{i=1}^n$ provided that all of the following conditions are satisfied.

\begin{itemize}

\item If $x_k^{(i)} = x_l^{(j)}$ then $\bar{x}_k^{(i)} = \bar{x}_l^{(j)}$, $1 \leq k \leq l_i$, $1 \leq l \leq l_j$, $1 \leq j \leq i$, $1 \leq i \leq n$.

\item If $x_k^{(i)} = y_l^{(j)}$ then $\bar{x}_k^{(i)} = \bar{y}_l^{(j)}$, $1 \leq k \leq l_i$, $1 \leq l \leq m_j$, $1 \leq j \leq i-1$, $2 \leq i \leq n$.
\item If $x_k^{(i)}$ is a constant then $\bar{x}_k^{(i)}$ is the same constant, $1 \leq k \leq l_i$, $1 \leq i \leq n$.

\end{itemize}
$[\mathfrak{p}^{(i)}~\bar{x}^{(i)}~\bar{y}^{(i)}]_{i=1}^n$ is I/O equivalent to the program $[\mathfrak{p}^{(i)}~x^{(i)}~y^{(i)}]_{i=1}^n$ is denoted by
\be
[\mathfrak{p}^{(i)}~\bar{x}^{(i)}~\bar{y}^{(i)}]_{i=1}^n \thicksim [\mathfrak{p}^{(i)}~x^{(i)}~y^{(i)}]_{i=1}^n
\ee
I/O equivalence\index{I/O equivalence} does not satisfy the property of symmetry.

\end{definition}

\textbf{Notes.}

\begin{itemize}

\item In the definition of program equivalence the program lists of $p$ and $q$ will usually be permutations of each other.
This includes the case where $p$ and $q$ are identical programs.  
However, there is the possibility that the lengths of the program lists of $p$ and $q$ are not the same (see the notes of Section 2.3).
This is because repetitions of subprograms of a program list are allowed for subprograms with an empty list output.
Although such repetitions are allowed they introduce redundancies and should be avoided.
   
\end{itemize}

\section{Additional construction rules.}

In applications, proofs are largely constructed by the recursive application of the program extension rule through the extended program derivation.
In VPC there are internal procedures that employ some additional rules that are listed below.
Most of these rules follow from the definitions.
In the next chapter we will also include rules associated with false programs and disjunctions. 

Some of the construction rules listed here are in the form of existence axioms.
Others are given in two parts with an existence axiom followed by an equivalence relation.
The empty list program\index{empty program} is denoted by
\be
ep=[~]
\ee
and can be regarded as a constant for type $\mb{P}$ objects. 

\textbf{Extensions.}
The following rule involves the acquisition of the property of an extension through program equivalence.
\begin{crule}
\be
[[ext~[p~c]~[~]~eqv~[p~q]~[~]]~aext~[q~c]~[~]]
\ee
\end{crule}
\begin{crule}
\be
[[ext~[p~c]~[~]~eqv~[c~d]~[~]]~aext~[p~d]~[~]]
\ee
\end{crule}

\textbf{I/O equivalence.}

\begin{crule}
\be
\bal
&  [[ext~[q~c]~[~]~conc~[q~c]~[r]~conc~[p~d]~[s]~eqio~[p~q]~[~]~eqio~[s~r]~[~]]~aext~[p~d]~[~]] \\
\eal
\ee
\end{crule}

\textbf{Program equivalence.}

\begin{crule}
\be
[typep~[p]~[~]~eqv~[p~p]~[~]]
\ee
\end{crule}

\begin{crule}
\be
[eqv~[p~q]~[~]~eqv~[q~p]~[~]]
\ee
\end{crule}

\textbf{Sublists.}

\begin{crule}
\be
[[sub~[p~q]~[~]~sub~[q~p]~[~]]~eqv~[p~q]~[~]]
\ee
\end{crule}

\begin{crule}
\be
[typep~[p]~[~]~sub~[p~p]~[~]]
\ee
\end{crule}

\begin{crule}
\be
[[sub~[q~p]~[~]~sub~[p~r]~[~]]~sub~[q~r]~[~]]
\ee
\end{crule}

\begin{crule}
\be
[conc~[p~q]~[s]~sub~[p~s]~[~]]
\ee
\end{crule}
\begin{crule}
\be
[conc~[p~q]~[s]~sub~[q~s]~[~]]
\ee
\end{crule}


\begin{crule}
\be
[[conc~[p~q]~[r]~sub~[p~s]~[~]~sub~[q~s]~[~]]~sub~[r~s]~[~]]
\ee
\end{crule}

\textbf{Program concatenation.}

\begin{crule}
\be
[[ext~[p~a]~[~]~ext~[p~b]~[~]~conc~[a~b]~[s]]~aext~[p~s]~[~]]
\ee
\end{crule}

\emph{Concatenation with the empty program (right).}
\begin{crule}
\be
[typep~[p]~[~]~conc~[p~ep]~[s]]
\ee
\end{crule}
\begin{crule}
\be
[conc~[p~ep]~[s]~eqv~[s~p]~[~]]
\ee
\end{crule}
\emph{Concatenation with the empty program (left).}
\begin{crule}
\be
[typep~[p]~[~]~conc~[ep~p]~[s]]
\ee
\end{crule}
\begin{crule}
\be
[conc~[ep~p]~[s]~eqv~[s~p]~[~]]
\ee
\end{crule}

\textbf{Application axioms.}
We can include the following rules that take on a form that can be applied as general application axioms.
They are labeled differently from the above construction rules.

\textbf{I/O type $\mb{P}$}.
The first of these application axioms are I/O type axioms.
They are labeled by the letters aio and reflect the property that within all programs the type of the assigned values of all elements of the I/O lists are checked.
The following axiom reflects this property for higher order programs. 

\textbf{aio}
\be
 [\mathfrak{p}~x~y~typep~[a]~[~]],  \qquad a \in [x~y]
\ee

\textbf{Substitution rule.}
The substitution rule will be applied as an axiom to higher order atomic programs $\mathfrak{p}~x~y$ such that
\be
\mathfrak{p} \in [eqv~sub~conc~disj]
\ee
The program $disj$ constructs a program disjunction that will be defined in the next chapter.

The first part of the substitution rule is an existence axiom.

\textbf{sr 1}
\be
\bal
& [[\mathfrak{p}~x~y~eqv~[x_i~a]~[~]]~\mathfrak{p}~\bar{x}~\bar{y}], \qquad x_i \in x,~\bar{x} = x(x_i \to a), \\
\eal
\ee
where $x=[x_i]_{i=1}^n$, for some $n:\mb{I}_+$.
The output lists $y$ and $\bar{y}$ may be empty lists.

The second part of the substitution rule is applicable when $y$ and $\bar{y}$ are not empty lists.
To present the axiom in a general form we write
\be
y=[y_j]_{j=1}^m,~~~\bar{y}=[\bar{y}_j]_{j=1}^m, \qquad m:\mb{I}_+
\ee
For any substitution $x(x_i \to a)$ VPC will generate the following axioms for $j=1,\ldots,m$.

\textbf{sr 2}
\be
\bal
&  [[\mathfrak{p}~x~y~eqv~[x_i~a]~[~]~\mathfrak{p}~\bar{x}~\bar{y}]~eqv~[\bar{y}_j~y_j]~[~]], \\
& \hspace{60mm} x_i \in x,~\bar{x} = x(x_i \to a), \\
\eal
\ee

It is important to note that some atomic programs can be shown to satisfy the substitution rule from other axioms.
Strictly speaking, the substitution rule should not be regarded as an axiom for such programs.

\textbf{Notes.}

\begin{itemize}

\item Program equivalence satisfies the property of transitivity
\be
[[eqv~[p~q]~[~]~eqv~[q~r]~[~]]~eqv~[p~r]~[~]]
\ee
This is not included as an axiom since it follows from the substitution rule.

\end{itemize}

\section{Options file.}\label{sof}

In VPC, the program $c$ of an extended program derivation $epd~[q~p~c]~[s]$ is regarded as an extension if the program $[q~c]$ is program and I/O equivalent to an axiom or theorem that is stored in the file \emph{axiom.dat}.
In other words derivations in VPC are constructed only with respect to irreducible extended programs $[q~c]$, i.e. where $c$ is an irreducible extension $c:\mb{P}_{iext}(q)$.
The program extension rule, per, is more flexible and requires that $c:\mb{P}_{ext}(q)$.
Generality of the application of the program extension rule under this process is not lost.
This is because, given an extended program derivation $epd~[q~p~c]~[s]$, if $c:\mb{P}_{ext}(q)$ is not an irreducible extension then there must exist a program $\bar{q} \varsubsetneqq q$ ($\bar{q}$ is a strict sublist of $q$) such that $c:\mb{P}_{ext}(\bar{q})$ and $r=[q~c]$ was obtained from the extended program derivation $epd~[\bar{q}~q~c]~[r]$.
Since $\bar{q} \varsubsetneqq q \subseteqq p$ we can also construct $epd~[\bar{q}~p~c]~[t]$ and $t$ is identical to $s$.
We can now apply the same argument to $\bar{q}$ and so on until we are left with a derivation of $[p~c]$ with respect to a program that is an irreducible extension.

During proof construction, VPC accesses a file \emph{axiom.dat} that initially stores all of the axioms of the application corresponding to the specific theory under investigation.
As proofs are completed the theorems extracted from them are also stored in \emph{axiom.dat}.
The program $[q~c]$ of an extended program derivation $epd~[q~p~c]~[s]$ is identified as an extended program if it can be matched to an axiom/theorem stored in the file \emph{axiom.dat}.
The matching procedure relies on program and I/O equivalence.
In this way each axiom/theorem stored in the file \emph{axiom.dat} acts as a template from which programs of an application can be identified as extensions.
The matching procedure can be defined more precisely as follows.

Let $[q^\prime~c^\prime]$ be an axiom/theorem stored in the file \emph{axiom.dat}.
A program $[q~c]$ is identified as an extended program, i.e. $c:\mb{P}_{ext}(q)$, if there exists programs $q^{\prime \prime}$ and $c^{\prime \prime}$ such that
\begin{enumerate}

\item $q^{\prime \prime} \equiv q$ and $c^{\prime \prime} \equiv c$, i.e. $q^{\prime \prime}$ and $c^{\prime \prime}$, respectively, are program equivalent to $q$ and $c$, respectively.

\item $q^{\prime \prime} \thicksim q^\prime$ and $[q^{\prime \prime}~c^{\prime \prime}] \thicksim [q^\prime~c^\prime]$, i.e. $q^{\prime \prime}$ is I/O equivalent to $q^\prime$ and $[q^{\prime \prime}~c^{\prime \prime}]$ is I/O equivalent to $[q^\prime~c^\prime]$.

\end{enumerate}

The conclusion program $c$ is usually an atomic program so the condition $c^{\prime \prime} \equiv c$ is only relevant when $c^{\prime \prime}$ is defined as a special non-atomic program (see Section \ref{snap}).   

A program $c$ associated with the extended program $[q~c]$ that is stored as an axiom/theorem in the file \emph{axiom.dat} acquires the type $\mb{P}_{ext}(q)$ by default.
Otherwise a program $c$ can only acquire the type $\mb{P}_{ext}(q)$ through the type assignment program $aext$.

At each step of a proof construction VPC determines the conclusion programs of all possible extended program derivations that can be obtained from the main program with respect to the axioms and theorems that are currently stored in the file \emph{axiom.dat}.
These extended program derivations are listed in an options file, \emph{options.dat}, from which the user may select a desired conclusion program to generate a new statement in the main program list of the proof.
The process is repeated until the proof is completed.

If at any point of a derivation of a proof construction the user inserts a statement that is not currently stored as an option in the options file then VPC will halt with an execution error message.

Each option in the options file includes the axiom/theorem label and the associated labels of the subprograms that make up the sublist of the current proof program that can be matched to the premise program of the axiom/theorem stored in \emph{axiom.dat}.
Crucial to this search and matching procedure is program and I/O equivalence.
The procedure is one of extracting all sublists of the current program that can be matched to the premise programs of the axioms/theorems stored in \emph{axiom.dat}.

Extractions of sublists from the current proof program based on a raw search of all possible permutations followed by an I/O equivalence matching algorithm can be computationally expensive.
VPC employs special techniques that speed up this process by detecting and eliminating unsuccessful matches before a complete sublist extraction and I/O equivalence check is performed.
This significantly reduces the computations making the enumeration of all possible extended program derivations quite manageable.

\section{Connection List.}\label{connection}\index{connection list}
For each subprogram of a program list $\mathfrak{p}~x~y=[\mathfrak{p}^{(i)}~x^{(i)}~y^{(i)}]_{i=1}^n$ that is obtained from an extended program derivation is constructed a connection list that records the origin of that subprogram during the program's construction.
To construct a connection list it is necessary to provide a label for each axiom and theorem of the specific application under consideration.

\begin{definition}(Connection list.)\label{clist}
	For each subprogram $\mathfrak{p}^{(i)}~x^{(i)}~y^{(i)}$ of the program list $\mathfrak{p}~x~y=[\mathfrak{p}^{(i)}~x^{(i)}~y^{(i)}]_{i=1}^n$ that is obtained from an extended program derivation is generated a list that contains the label of the axiom/theorem and premises used to obtain that subprogram.
	For each such subprogram, $\mathfrak{p}^{(i)}~x^{(i)}~y^{(i)}$, the connection list is of the form
	\be
	a(i)~[l(i,1)~\ldots~l(i,k(i))]
	\ee
	where $1 \leq l(i,1),~\ldots~,l(i,k(i)) \leq i-1$ are the labels of the subprograms that make up the sublist of $\mathfrak{p}~x~y$ that is program and I/O equivalent to the premise program of the axiom/theorem, labeled $a(i)$, that is used to conclude $\mathfrak{p}^{(i)}~x^{(i)}~y^{(i)}$.
	Here $k(i)$ is the length of the premise program list of the axiom/theorem $a(i)$.
	
\end{definition}

Consider the proof program $[p~q]$, where $p=[p_i]_{i=1}^n$ is the list of premises of the proof and $q=[q_i]_{i=1}^m$ are the statements obtained by a sequence of extended program derivations.
In VPC, proof programs are output as a vertical list with three columns.
The first column contains the statement label, the second column contains the statement itself and the third column contains the connection list.
Statements that are premises of the main proof program do not have a connection list.
The general output layout can be illustrated as follows. 
\be
\begin{array}{lll}
	Label  & Statement & Connection~list \\
	1      & p_1       & \\
	\vdots & \vdots    & \\
	n      & p_n       &  \\
	n+1      & q_1     & a(1)~[l(1,1)~\ldots~l(1,k(1))]  \\
	\vdots & \vdots    & \vdots \\
	n+m     & q_m      & a(m)~[l(m,1)~\ldots~l(m,k(m))]  \\
\end{array}
\ee
Here $k(j)$ is the length of the premise program of the axiom/theorem $a(j)$, $j=1,\ldots,m$, and $1 \leq l(j,1),\ldots,l(j,k(j)) \leq n+j-1$, are statement labels of the sublist of the program $[p~[q_i]_{i=1}^{j-1}]$ that is program and I/O equivalent to the premise program of the axiom/theorem $a(j)$, $j=1,\ldots,m$, stored in \emph{axiom.dat}. 

\textbf{Extraction of theorems from proofs.}
We now describe an algorithm that extracts a theorem from the proof $[p~q]$ described above.
Upon completion of a proof the final statement $q_m$ is the conclusion program of the theorem $[p~q_m]$.
Consider the lists of labels
\be
c(i) = [ l(i,j) ]_{j=1}^{k(i)} , \qquad i=1,\ldots,m
\ee
obtained from the above connection lists by removing the axiom/theorem label $a(i)$. 
We construct, by iteration, a sequence of lists
\be
d(\nu) = [ \lambda(\nu,j) ]_{j=1}^{\mu(\nu)} , \qquad \nu=1,2,\ldots
\ee
where each $\lambda(\nu,j)$ is a label associated with some statement of the program $[p~q]$.
We may rewrite $d(\nu)$ as a partition
\be
d(\nu) =  [~[ \lambda_p (\nu,j) ]_{j=1}^{\mu_p(\nu)}~ [ \lambda_q(\nu,j) ]_{j=1}^{\mu_q(\nu)}~], \qquad \nu=1,2,\ldots
\ee
where $\lambda_p(\nu,j)$ are labels associated with the premise program $p$ and $\lambda_q(\nu,j)$ are labels associated with statements of $q$ obtained from extended program derivations.

For $\nu=1$ we set
\be
d(1) = c(m)
\ee
so that
\be
[ \lambda(1,j) ]_{j=1}^{\mu(1)} = [ l(m,j) ]_{j=1}^{k(m)}
\ee
and we have $\mu(1)=k(m)$ and $\lambda(1,j) = l(m,j)$, $j=1,\ldots,k(m)$.

Each list $d(\nu)$, $\nu=2,3,\dots$, consists of all labels $\lambda(\nu-1,j)$ of $d(\nu-1)$ that are associated with the premise program $p$ and the list of labels $c(\lambda(\nu-1,j))$ for labels $\lambda(\nu-1,j)$ of $d(\nu-1)$ that are associated with statements of $q$, i.e.
\be
\bal
d(\nu) = & [ \lambda(\nu,j) ]_{j=1}^{\mu(\nu)} \\
= &  [~[ \lambda_p (\nu-1,j) ]_{j=1}^{\mu_p(\nu-1)}~[ c(\lambda_q(\nu-1,j)) ]_{j=1}^{\mu_q(\nu-1)}~], \qquad \nu=2,3,\ldots \\
\eal
\ee
The iteration is continued until we obtain a final list $d(\kappa)$, for some $\kappa:\mb{I}_+$, such that all of the labels of the statements contained in $q$ have been eliminated leaving only labels of the premise program $p$, i.e. $\mu_q(\kappa)=0$.
To simplify the process we may eliminate repeated labels from each list $d(\nu-1)$ before proceeding to the construction of the new list $d(\nu)$.

If by this procedure there are labels of statements in the program list $p$ that do not appear in the final list $d(\kappa)$ then those statements are redundant as premises leading to the conclusion $q_m$.
In such a case $[p~q_m]$ will not be an irreducible extended program and hence will not be a theorem.
The proof can be reconstructed by eliminating the redundant premises.   

\chapter{Disjunctions and False Programs}\label{cdafp}

\section{Axioms/theorems of falsity}

While a program $p:\mb{P}_{false}$ will halt with an execution error for any value assigned input it does not necessarily follow that the extended program derivation $epd~[q~p~c]~[s]$ will also halt with an execution error.
The reason for this is that $epd~[q~p~c]~[s]$ is a higher order program so that type checking is based on program structure.
The program $epd~[q~p~c]~[s]$ does not recognize value assignments of the I/O lists of the programs $q,p,c$ and $s$.
Hence there is nothing stopping us from allowing the program $p$ of an extended program derivation $epd~[q~p~c]~[s]$ to be of type $\mb{P}_{false}$.
If $epd~[q~p~c]~[s]$ does not halt with an execution error then the derived object $s$ will be a program but will also be of subtype $\mb{P}_{false}$.
In this section we will demonstrate how extended program derivations can be used to identify false programs.

The statement $p:\mb{P}_{false}$ may be represented by the higher order construct
\ben\label{false}
false~[p]~[~]
\een
If $p$ is irreducible in the sense that there does not exist a program $q \varsubsetneqq p$ ($q$ is a strict sublist of $p$) such that $q:\mb{P}_{false}$ then the statement (\ref{false}) represents an axiom or theorem of falsity.
Axioms and theorems of falsity are higher order constructs of irreducible extended programs with an empty list premise.
The programs that appear as input to axioms of falsity, are application specific.
As such they are user supplied and can be regarded as constants for type $\mb{P}$ objects associated with the application.  

To the construction rules we introduce the additional rules of falsity.

\emph{Falsity rule.}

\begin{flse}
\be
\bal
&  [[sub~[q~p]~[~]~false~[q]~[~]]~afalse~[p]~[~]] \\
\eal
\ee
\end{flse}

\emph{Retention of subtype assignment.}
\begin{flse}
\be
\bal
&  [afalse~[p]~[~]~false~[p]~[~]] \\
\eal
\ee
\end{flse}

\emph{Equivalence of false programs.}
\begin{flse}
\be
[[false~[p]~[~]~eqv~[p~q]~[~]]~afalse~[q]~[~]]
\ee
\end{flse}

Consider a premise program $p$, where $p=[p_i]_{i=1}^n$.
Since we have allowed the premise program $p$ to be of type $\mb{P}_{false}$ we may iteratively apply the program $epd$ to generate the program $[p~q]$, where $q=[q_i]_{i=1}^m$ are the statements obtained by $m$ extended program derivations.
If $p:\mb{P}_{false}$ the iteration should continue until a sublist of $[p~q]$ coincides with a false program defined by an axiom or theorem of falsity stored in \emph{axiom.dat}.
When this occurs we abandon the extended program derivation format and apply the sublist falsity rule.

In VPC the output of the proof program will look like the following vertical list.
\be
\begin{array}{lll}
Label  & Statement & Connection~list \\
1      & p_1       & \\
\vdots & \vdots    & \\
n      & p_n       &  \\
n+1      & q_1     & a(1)~[l(1,1)~\ldots~l(1,k(1))]  \\
\vdots & \vdots    & \vdots \\
n+m     & q_m      & a(m)~[l(m,1)~\ldots~l(m,k(m))]  \\
n+m+1     & :false  & a(m+1)~[l(m+1,1)~\ldots~l(m+1,k(m+1))]  \\
\end{array}
\ee
Here $k(j)$ is the length of the premise program of the axiom/theorem $a(j)$, $j=1,\ldots,m+1$, and $1 \leq l(j,1),\ldots,l(j,k(j)) \leq n+j-1$ are statement labels of the sublist of the program $[p~[q_i]_{i=1}^{j-1}]$ that is program and I/O equivalent to the premise program of the axiom/theorem $a(j)$, $j=1,\ldots,m+1$, stored in \emph{axiom.dat}.

The first $n+m$ lines are in the standard derived proof format.
The addition of the final statement, $:false$, means that the standard proof format is to be abandoned and the vertical list is to be read as the statement
\be
[p~q]:\mb{P}_{false}
\ee
We may extract from this statement a theorem of falsity of the form (\ref{false}), provided that $p$ is minimal in the sense that there are no strict sublists of $p$ that are of type $\mb{P}_{false}$.

\textbf{Notes.}

\begin{itemize}

\item It should be noted that for any $p:\mb{P}_{false}$ the higher order program $false~[p]~[~]$ is computable while the program $p$ is not.

\item There is an important consequence of allowing the program $p$ of an extended program derivation $epd~[q~p~c]~[s]$ to be of type $\mb{P}_{false}$.
If we accept the program extension rule, per, without exception we must conclude that there exist extensions $c:\mb{P}_{ext}(p)$ such that $p:\mb{P}_{false}$.
Careful reading of the definition for a program extension, Definition \ref{ce}, does not disallow such a possibility.
The definition only states that if the premise $p$ is computable for a given value assigned input then it is guaranteed that $[p~c]$ is computable for the same value assigned input.

\item Given that the premise program of an extended program derivation could be of type $\mb{P}_{false}$ one should avoid terminating an iteration of derivations before a conclusion leads to a statement of falsity.
More will be said on this in a later chapter.

\item As already discussed earlier, objects that are subtypes of strings such as scalars and lists of the machine integers, fixed precision rational numbers and programs are recognized by the machine from their string structure.
This is not the case with objects of type $\mb{P}_{ext}$, where the machine can only acquire an interpretation of such objects through the properties embedded in the constructions rules. 

Type $\mb{P}_{false}$ can also be put into the same class of abstract types such as $\mb{P}_{ext}$.
Axioms of falsity are assigned the type $\mb{P}_{false}$ by default.
Otherwise an object can only acquire the type $\mb{P}_{false}$ by inference using the type assignment program $afalse$. 

\end{itemize}

\section{Disjunctions.}\label{sdisj}\index{disjunction}

In conventional theories of logic, disjunctions have an important role to play in the expressiveness and manipulation of formal statements.
Program disjunctions have a more basic role in that they effectively split a program into several parallel programs, where each parallel program is associated with an operand of the disjunction contained within the main program.
Once a disjunction splitting has been completed, extended program derivations can be performed independently on each operand program.

\begin{definition}\label{disj}(Disjunction.)
A program $\mathfrak{p}~x~y:\mb{P}$ is a disjunction if it has the form
\be
\mathfrak{p}~x~y=\mathfrak{p}^{(1)}~x^{(1)}~y~|~\mathfrak{p}^{(2)}~x^{(2)}~y
\ee
where $\mathfrak{p}^{(i)}~x^{(i)}~y:\mb{P},~i=1,2$, are the operands of $\mathfrak{p}~x~y$.
The input list, x, of the main program $\mathfrak{p}~x~y$ is defined in terms of the input lists of its operand programs by adopting the convention
\be
x = \bar{x} \setminus (\bar{x} \cap y), \qquad \bar{x} \simeq [x^{(1)}~x^{(2)}]
\ee 
The disjunction $\mathfrak{p}~x~y$ will be computable for the value assigned input list, $x$, if at least one of its operand programs, $\mathfrak{p}^{(i)}~x^{(i)}~y,~i=1,2$, is computable.
Otherwise $\mathfrak{p}~x~y$ will halt with a disjunction violation error.
Disjunctions that are computable in this sense are said to override type violation errors.

\end{definition}

\textbf{Program equivalence.}
We now extend the definition of program equivalence to include disjunctions.

\begin{definition}\label{equiv}(Program equivalence.)\index{program equivalence}
Two programs $u:\mb{P}$ and $v:\mb{P}$ are said to be program equivalent if any of the following conditions are satisfied.
\begin{itemize}

\item $u \subseteqq v$ and $v \subseteqq u$.

\item $u=a|b$, $v=b|a$.

\item $u=[p~a|b~q]$, $v=[p~a~q]~|~[p~b~q]$.

\end{itemize}
where $p$ or $q$ may be the empty list program.
Program equivalence is denoted by $u \equiv v$ and satisfies the properties of reflexivity, symmetry and transitivity.

\end{definition}

\textbf{Operand programs.}
A program $s$ containing a disjunction can be expressed in the general form $s=[p~a|b~q]$, where $a,b:\mb{P}$, and $p$ and/or $q$ may be the empty list program.
We note that $a|b$ is not a program list but rather an element of the program list $[p~a|b~q]$. 
The program $s$ can be split into the programs $[p~a~q]$ and $[p~b~q]$, by the two step procedure
\be
[p~a|b~q] \to  [[p~a]|[p~b]~q] \to [p~a~q]~|~[p~b~q]
\ee
or
\be
[p~a|b~q] \to  [p~[a~q]|[b~q]] \to [p~a~q]~|~[p~b~q]
\ee
In Section \ref{sspotcr} it will be shown that these constructions can be derived from the left and right disjunction distributive rules to be presented in the next section.
 
The programs $[p~a~q]$ and $[p~b~q]$ will be referred to as the operand programs of $s$ based on the disjunction $a|b$.
Extended program derivations can be performed independently on each operand program.
When independent derivations of the operand programs yield a common conclusion, say $c$, then the common conclusion can be contracted back onto the main program $s$ to produce an extended program $[s~c],~c:\mb{P}_{ext}(s)$.
The disjunction contraction rule demonstrates how this is done.
There are two additional disjunction contraction rules that involve type $\mb{P}_{false}$ operand programs.
 
\textbf{Disjunction contraction rules.}
To the existing construction rules we introduce the additional rule for programs containing disjunctions.

\emph{Disjunction contraction rule.}
\begin{dsj}\label{dsj01}
\be
\bal
& [[ext~[a~c]~[~]~ext~[b~c]~[~]~disj~[a~b]~[d]]~aext~[d~c]~[~]] \\
\eal
\ee
\end{dsj}

The following contraction rules involve false programs.
They are not stated as higher order axioms because they can be derived as theorems (see Section \ref{sspotcr}).

\emph{Disjunction contraction rule 2.}
\be
\bal
&  [[false~[a]~[~]~ext~[b~c]~[~]~disj~[a~b]~[d]]~aext~[d~c]~[~]] \\
\eal
\ee

\emph{Disjunction contraction rule 3.}
\be
\bal
& [[false~[a]~[~]~false~[b]~[~]~disj~[a~b]~[d]]~afalse~[d]~[~]] \\
\eal
\ee

\textbf{Execution errors.}
So far we have associated execution errors with type violations of a single program list.
We now give an extended definition of an execution error that includes programs containing disjunctions.

\begin{definition}(Execution error.)
A program $s:\mb{P}$ will halt with an execution error if any of the following conditions occur.

\begin{itemize}

\item $s$ does not contain a disjunction and there is a type violation of at least one assigned value of the elements of its I/O lists.
(Type violations may also include the failure of the satisfaction of a relation between a pair of input elements.)

\item $s$ contains a disjunction of the form $s=[p~a_1 | \cdots | a_n~q]$, where the programs $p$, $q$ and $a_i,~i=1,\ldots,n$, do not contain any disjunctions, and there is a type violation of at least one assigned value of the elements of the I/O lists of every operand program $[p~a_i~q],~i=1,\ldots,n$.
If there are no type violations of any assigned value of the elements of the I/O lists in at least one operand program then the program $s$ will not halt with an execution error.
 
\end{itemize}

\end{definition}

\textbf{Disjunction splitting.}
Let $p=[p_i]_{i=1}^m$, $q=[q_i]_{i=1}^n$ and consider the program $s=[p~d~q]$, where $d=a|b$. 
Based on the disjunction $d$, the main program $[p~d~q]$ can be split into the two operand programs $u=[p~a~q]$ and $v=[p~b~q]$.
Suppose that we have independently applied extended program derivations to each of the operand programs $u$ and $v$ to obtain a common conclusion $c$.
We may then contract the common conclusion, $c$, back onto the main program by applying the disjunction contraction rule. 
The procedure is depicted in the following table.

\be
\begin{array}{llllllll}
label  & [s~c]   &      & label   & [u~c]   & & label   & [v~c]   \\
       &         &      &         &         & &         &         \\
1      & p_1     &      & 1       & p_1     & & 1       & p_1     \\
\vdots & \vdots  &      & \vdots  & \vdots  & & \vdots  & \vdots  \\
m      & p_m     &      & m       & p_m     & & m       & p_m     \\
m+1    & d~*     & \lra & m+1     & a       & & m+1     & b       \\
m+2    & q_1     &      & m+2     & q_1     & & m+2     & q_1     \\      
\vdots & \vdots  &      & \vdots  & \vdots  & & \vdots  & \vdots  \\      
m+n+1  & q_n     &      & m+n+1   & q_n     & & m+n+1   & q_n     \\
m+n+2  & c       & \lla & m+n+2   & c       & & m+n+2   & c       \\
\end{array}
\ee
In the table, the connection lists associated with each statement have been omitted due to space restrictions. 

The asterisk next to the statement $d$ indicates that the disjunction splitting is based on the operands of that statement.
The right arrow, $\lra$, indicates that the user has requested that the main program be split into two operand programs at line $m+n+1$ of the main program based on the operands of the disjunction $d=a|b$.
The left arrow, $\lla$, indicates that the common conclusion, $c$, of the two operand programs, $u$ and $v$, is to be contracted back onto the main program at the line labeled $n+m+2$ of the main program by applying the disjunction contraction rule.

Suppose that under an extended program derivation the conclusion $c$ of $u$ and $v$, respectively, were derived from axioms/theorems labeled $a$ and $a^\prime$, respectively.
Let $j$ and $j^\prime$, respectively, be the lengths of the premise program lists of the axioms/theorems $a$ and $a^\prime$, respectively.
Upon output the statement $c$ of the main program will have an attached composite connection list
\be
 a~[l_1~\ldots~l_{j}]~a^\prime~[l_1^\prime~\ldots~l_{j^\prime}^\prime]
\ee
where $a~[l_1~\ldots~l_j]$, $1 \leq l_1,\ldots,l_j \leq m+n+1$, is the connection list appearing at line labeled $m+n+1$ of the operand program $[u~c]$ and $a^\prime~[l_1^\prime~\ldots~l_{j^\prime}^\prime ]$, $1 \leq l_1^\prime,\ldots,l_{j^\prime}^\prime \leq m+n+1$, is the connection list appearing at line labeled $m+n+1$ of the operand program $[v~c]$.

A contraction of the conclusion $c$ to the main program can also occur if one of the operand programs leads to a conclusion $c$ and the other a conclusion $: false$.
This case follows from the disjunction contraction rule 2.
If both operand programs are type $\mb{P}_{false}$ then by the disjunction contraction rule 3 the main program will be assigned the type $\mb{P}_{false}$.

In mainstream mathematics derivations associated with each operand are often conducted as separate cases within a single proof.
The reason for this is that many of these derivations are not of sufficient interest to be considered as separate theorems.
When using VPC, derivations of proofs associated with each operand program must be conducted outside of the main proof program containing the disjunction.
The theorems extracted from the separate operand program derivations are to be stored in \emph{axiom.dat}. 
The derivation of the proof associated with the main program containing the disjunction can then access the theorems associated with each operand program through the disjunction contraction rules.  

Extracting and storing theorems associated with each operand program may lead to an accumulation of theorems in \emph{axiom.dat} that are often trivial and not of particular interest in themselves.
However, this should not be a problem for storage and retrieval purposes.
In VPC one may choose to label these theorems as lemmas to weaken their status.
There is sometimes an advantage in storing these individual operand cases as separate lemmas outside of the main proof because it is not uncommon that they are reused in other proofs.

\textbf{Redundancy in disjunction splitting.}
Suppose that under disjunction splitting, $v$ is a false program and $u$ is a computable program.
It is possible that we may find an extended program derivation leading to the conclusion of $v$ that coincides with the conclusion of $u$ before detecting the falsity of $v$.
We may then proceed to contract the common conclusion, say $c$, to the main proof program containing the disjunction to obtain $[[p~d~q]~c]$.
We may suspect that this will lead to an error in our derivation of the main proof.
This will not be the case since, under the disjunction contraction rule 2, this would be the identical conclusion that would have been made if we had detected that $v:\mb{P}_{false}$.

\textbf{Notes.}

\begin{itemize}

\item There remains the possibility that all operand programs are of type $\mb{P}_{false}$ and that derivations associated with both operand programs have been terminated prematurely with a common derived conclusion, say $c$.
We may then proceed to contract this common conclusion back onto the main program to obtain $[[p~d~q]~c]$.
By the disjunction contraction rule 3 the derivations associated with each operand program should have been continued until one arrives at a common conclusion $false$ so that the program $[p~d~q]$ is identified as type $\mb{P}_{false}$.
This type of occurrence is related to the situation described in a note of the previous section and will be discussed further in a later chapter.

\item We have restricted the definition of a disjunction to one where the lengths of the output lists of all operands are identical.
This restriction will avoid complications in some application axioms.
There are practical applications where one may require that the lengths of the output lists of the operands of a disjunction differ.
In such cases one can simply introduce dummy output variables in the output list of the operand with the output list of shortest length.
These dummy variables can then be set to some arbitrary values within the operand.

\end{itemize}

\section{Additional disjunction rules.}

Some of the following construction rules are in the form of existence axioms.
Others are given in two parts with an existence axiom followed by an equivalence relation.
The disjunction distributivity rules are split into left and right, each involving two independent existence axioms followed by an equivalence axiom.
As before, the empty program is denoted by $ep=[~]$ and can be regarded as a constant for type $\mb{P}$ objects\index{empty program} associated with the application.

\textbf{Disjunctions.}

\emph{Disjunction Commutativity.}
\begin{dsj}\label{dsj02}
\be
[disj~[a~b]~[s]~disj~[b~a]~[r]]
\ee
\end{dsj}
\begin{dsj}\label{dsj03}
\be
\bal
&  [[disj~[a~b]~[s]~disj~[b~a]~[r]]~eqv~[r~s]~[~]] \\
\eal
\ee
\end{dsj}

\emph{Disjunction right distributivity.}
\begin{dsj}\label{dsj04}
\be
\bal
&  [[conc~[p~a]~[r]~conc~[p~b]~[s]~disj~[a~b]~[d]]~disj~[r~s]~[v]] \\
\eal
\ee
\end{dsj}
\begin{dsj}\label{dsj05}
\be
\bal
&  [[conc~[p~a]~[r]~conc~[p~b]~[s]~disj~[a~b]~[d]]~conc~[p~d]~[u]] \\
\eal
\ee
\end{dsj}
\begin{dsj}\label{dsj06}
\be
\bal
&  [[conc~[p~a]~[r]~conc~[p~b]~[s]~disj~[a~b]~[d]~conc~[p~d]~[u] \\
& \hspace{5mm} disj~[r~s]~[v]]~eqv~[u~v]~[~]] \\
\eal
\ee
\end{dsj}

\emph{Disjunction left distributivity.}
\begin{dsj}\label{dsj07}
\be
\bal
&  [[conc~[a~p]~[r]~conc~[b~p]~[s]~disj~[a~b]~[d]]~disj~[r~s]~[v]] \\
\eal
\ee
\end{dsj}
\begin{dsj}\label{dsj08}
\be
\bal
&  [[conc~[a~p]~[r]~conc~[b~p]~[s]~disj~[a~b]~[d]]~conc~[d~p]~[u]] \\
\eal
\ee
\end{dsj}
\begin{dsj}\label{dsj09}
\be
\bal
&  [[conc~[a~p]~[r]~conc~[b~p]~[s]~disj~[a~b]~[d]~conc~[d~p]~[u] \\
& \hspace{5mm} disj~[r~s]~[v]]~eqv~[u~v]~[~]] \\
\eal
\ee
\end{dsj}

\emph{False operand program.}
\begin{dsj}\label{dsj10}
\be
\bal
&  [[disj~[a~b]~[p]~false~[b]~[~]]~eqv~[p~a]~[~]] \\
\eal
\ee
\end{dsj}

\textbf{Notes.}

\begin{itemize}

\item The rules per, cr 1-18, flse 1-3 and dsj 1-10 along with I/O type axioms and the substitution rule, are presented as irreducible extended higher order programs.
They can be regarded as the axioms of a theory for the construction of programs as proofs in the context of the formal system PECR upon which VPC is based.
Later we will employ VPC as a self referencing tool to investigate certain properties of the construction rules themselves.

\item A collection of constants that serve as input to the higher order programs associated with the construction rules are type $\mb{P}$ objects that must be defined with respect to the application theory, $S$, to which the construction rules are being applied.
As a consequence the collection of constants called by higher order programs of the construction rules may differ among applications.
The empty list program, $ep$, is defined as a constant of type $\mb{P}$ objects and will be common to all applications.
Other type $\mb{P}$ objects that are constants include type $\mb{P}_{false}$ objects that are associated with axioms of falsity specific to the application $S$.

\end{itemize}

\chapter{Common applications axioms.}\label{appl}

\section{Introduction.}

The construction rules of PECR form the structural foundations of VPC.
They are general inference rules that should be distinguished from user supplied axioms for specific applications.
Each application corresponds to a theory, $S$, that is defined by a collection of atomic programs and axioms that are specific to the theory.
The user inserts these axioms in the file \emph{axiom.dat}.
As such the collection of axioms of $S$, that are listed in the file \emph{axiom.dat}, serve as input data to the program VPC.
In a later chapter we will use VPC as a self referencing tool to investigate certain properties of the construction rules themselves.
In this case the construction rules are inserted in the file \emph{axiom.dat} as axioms.

The axioms that serve as input data to VPC will differ for each application.
However, there are certain axioms that will have a common structure in all applications.
Of these are four classes of axioms, (1) I/O type axioms, (2) special non-atomic program axioms, (3) the substitution rule and (4) application specific axioms of falsity.
These common axioms are not treated in the same way as the other construction rules and will be discussed in this chapter.
Embedded in VPC are routines that deal with these special axioms.
The user is required to supply additional initial data to instruct VPC how to apply these axioms for the specific application under consideration.   

In this book we will consider the following four applications.

\begin{itemize}

\item Arithmetic over $\mb{I}$ that examines the properties of machine arithmetic under the elementary operations of addition, subtraction, multiplication and division.
The values assigned to elements of the I/O lists of integer functional programs are of type $\mb{I}$.  

\item Discrete interval arithmetic over $\mb{I}$ that examines the properties of machine arithmetic under the elementary operations of discrete interval addition, subtraction and multiplication (interval division over $\mb{I}$ is not well defined).
The values assigned to elements of the I/O lists of discrete interval functional programs are of a mixed type and include intervals, assigned type $\mb{B}$, and scalars of type $\mb{I}$.  

\item Theory of programs as proofs in the context of our formal system PECR.
The values assigned to elements of I/O lists of higher order functional programs are of type $\mb{P}$.  

\item Array arithmetic.
An array is an object of type $\mb{A}$ and the values assigned to the elements of an array are type $\mb{I}$ objects.
We will also include atomic integer array functional programs that are of a mixed type that allow elements of their input lists to have assigned values of both type $\mb{I}$ and $\mb{A}$.
These programs will be associated with scalar multiplication of arrays, where the scalars are type $\mb{I}$. 

\end{itemize}

\section{I/O type axioms.}

An important property of all programs is that the type of the assigned values of all elements of the I/O lists are checked within the program.
We have already encountered I/O type axioms for higher order programs.
This rule will appear in similar form for each application.

Atomic programs of type $\mb{P}_{type}$ that have the sole task of checking the type of the assigned value of a single input variable will always be assigned the name $type$ followed by some distinguishing lower case letters and/or numbers.

The four kinds of checking programs that are associated with the main applications presented in the previous section are
\be
\bal
typei~[a]~[~], & \qquad a:\mb{I} \\
typedi~[a]~[~], & \qquad a:\mb{B} \\
typea~[a]~[~], & \qquad a:\mb{A} \\
typep~[a]~[~], & \qquad a:\mb{P} \\
\eal
\ee
In array type checking programs, $typea$, the dimensions of the array are assigned prior to the entry to the program and identified internally by the program.
Discrete intervals are two element lists that contain the lower and upper bounds of the interval.
In a similar way to arrays, the interval bounds do not always appear in the I/O lists and are determined internally by the atomic program.

I/O type axioms give a conclusion of type for an assigned value of an element of an I/O list of a program.
For any program $\mathfrak{p}~x~y$ there is no restriction that all of the value assignments of elements of its I/O lists are of the same type.
As such the type checking program in the conclusion must be type related to the element of the program $\mathfrak{p}~x~y$ that is being singled out.  

Let
\ben \label{3.100}
\mathfrak{t}~[a]~[~] = \left \{ \begin{array}{ll}
typei~[a]~[~], & a:\mb{I} \\
typedi~[a]~[~], & a:\mb{B} \\
typea~[a]~[~], & a:\mb{A} \\
typep~[a]~[~], & a:\mb{P} \\
\end{array}
\right .
\een
The identity (\ref{3.100}) is based on conventional notation and is not meant represent a program list.

When using VPC, all program names and the type of each element of their associated I/O lists are specified by the user in an initializing setup file.
I/O type axioms are labeled by the letters aio.
They take the general form

\textbf{aio}
\be
[ \mathfrak{p}~x~y~\mathfrak{t}~[a]~[~]],  \qquad a \in [x~y]
\ee
In expanded form
\be
\begin{array}{lll}
& [ \mathfrak{p}~x~y~typei~[a]~[~]], & a \in [x~y],~a:\mb{I} \\
& [ \mathfrak{p}~x~y~typedi~[a]~[~]], & a \in [x~y],~a:\mb{B} \\
& [ \mathfrak{p}~x~y~typea~[a]~[~]], & a \in [x~y],~a:\mb{A} \\
& [ \mathfrak{p}~x~y~typep~[a]~[~]], & a \in [x~y],~a:\mb{P} \\
\end{array}
\ee

\section{Equality/equivalence type checking.}

We have already encountered the atomic equivalence program $eqv~[p~q]~[~]$, where $p$ and $q$ are variable names that have been assigned the values of programs.
Upon entry $eqv~[p~q]~[~]$ checks that $p$ and $q$ have been assigned values of type $\mb{P}$ and then checks that they are equivalent, i.e. $p \equiv q$.
$eqv~[p~q]~[~]$ satisfies the properties of symmetry, reflexivity and transitivity.
We note that by definition, program equivalence includes the case where $p$ and $q$ are identical programs.
For other applications there are equality checking programs that are similar in function.

Atomic programs of type $\mb{P}_{type}$ that have the additional task of checking the equivalence or equality of assigned values of pairs of input variables will always be assigned the name $eq$ followed by some distinguishing lower case letters and/or numbers.

The following equivalence/equality checking programs associated with the main applications are
\be
\bal
eqi~[a~b]~[~], & \qquad a,b:\mb{I} \\
eqdi~[a~b]~[~], & \qquad a,b:\mb{B} \\
eqa~[a~b]~[~], & \qquad a,b:\mb{A} \\
eqv~[a~b]~[~], & \qquad a,b:\mb{P} \\
\eal
\ee
For arithmetic over $\mb{I}$ we have the atomic integer equality program $eqi~[a~b]~[~]$.
Upon entry $eqi~[a~b]~[~]$ checks that $a$ and $b$ have been assigned values of type $\mb{I}$ and then checks that they are equal, i.e. $a=b$, where equality is used in the sense of assigned values.
Like the equivalence program for higher order programs, $eqi~[a~b]~[~]$ satisfies the properties of symmetry, reflexivity and transitivity.

For discrete interval arithmetic we have the atomic equality program $eqdi~[a~b]~[~]$.
Upon entry $eqdi~[a~b]~[~]$ checks that $a$ and $b$ have been assigned values of type $\mb{B}$, and then checks that they are equal, i.e. the value assignments of the bounds of both intervals are equal.
$eqdi~[a~b]~[~]$ also satisfies the properties of symmetry, reflexivity and transitivity.

For array arithmetic we have the atomic array equality program $eqa~[a~b]~[~]$.
Upon entry $eqa~[a~b]~[~]$ checks that $a$ and $b$ have been assigned values of type $\mb{A}(l)$, for some index list $l=[l_1~\ldots~l_p]$, $p:\mb{I}_+$, and then checks that they are equal, i.e. each corresponding element of the arrays $a$ and $b$ have been assigned the same values.
$eqa~[a~b]~[~]$ also satisfies the properties of symmetry, reflexivity and transitivity.
In array equality checking programs the dimensions of the arrays are identified internally and are not returned as output.

\section{Special non-atomic programs.}\label{snap}

Special non-atomic programs\index{special non-atomic program} come in the form of program lists or disjunctions. 
A program list can be represented by
\be
\mathfrak{p}~x~y=[\mathfrak{p}^{(i)}~x^{(i)}~y^{(i)}]_{i=1}^n
\ee
Here $\mathfrak{p}~x~y$ is the main program of the list and is often introduced for notational use only.
Sometimes we need to give the above program list a special status by prescribing $\mathfrak{p}~x~y$ as a special non-atomic program.
Similarly, we can also represent a disjunction as
\be
\mathfrak{p}~x~y= \mathfrak{p}^{(1)}~x^{(1)}~y~|~\mathfrak{p}^{(2)}~x^{(2)}~y
\ee
where sometimes $\mathfrak{p}~x~y$ is also given a special status by prescribing $\mathfrak{p}~x~y$ as a special non-atomic program.
Special non-atomic programs are set in an initializing file that acts as input to the program VPC.
The operand programs $\mathfrak{p}^{(i)}~x^{(i)}~y,~i=1,2$, may be atomic programs.
Otherwise they must be defined as special non-atomic programs and may take the form of programs lists or disjunctions.

For each special non-atomic program, VPC generates the axioms presented below as part of the initializing process.
These axioms are labeled with the letters spl for program lists and spd for program disjunctions.

Let
\ben \label{3.200}
\mathfrak{e}~[a~b]~[~] = \left \{ \begin{array}{ll}
eqi~[a~b]~[~], & a,b:\mb{I} \\
eqdi~[a~b]~[~], & a,b:\mb{B} \\
eqa~[a~b]~[~], & a,b:\mb{A} \\
eqv~[a~b]~[~], & a,b:\mb{P} \\
\end{array}
\right .
\een
The identity (\ref{3.200}) is based on conventional notation and is not meant represent a program list.

\textbf{Program lists.}
Here we consider the program list represented by
\be
\mathfrak{p}~x~y=[\mathfrak{p}^{(i)}~x^{(i)}~y^{(i)}]_{i=1}^n
\ee
where $\mathfrak{p}~x~y$ has been prescribed as a special non-atomic program.
Following the convention of Definition \ref{prog}, $x^{(i)}$ and $y^{(i)}$, $i=1,\ldots,n$, are lists
\be
x^{(i)}=[x_j^{(i)}]_{j=1}^{l_i},~~y^{(i)}=[y_j^{(i)}]_{j=1}^{m_i} , \quad l_i,m_i:\mb{I}_0,~i=1,\ldots,n
\ee
and are related to the I/O lists $x$ and $y$ of the main program $\mathfrak{p}~x~y$ by
\be
\bal
y = & [y^{(i)}]_{i=1}^n \\
x = & \bar{x} \setminus (\bar{x} \cap y) \\
\eal
\ee
where $\bar{x}$ is a concatenation of the subprogram input lists with repeated variable names removed, i.e.
\be
\bar{x} \simeq [x^{(i)}]_{i=1}^n
\ee
The I/O lists of the subprograms $\mathfrak{p}^{(i)}~x^{(i)}~y^{(i)}$ may be of a mixed type, i.e. the assigned values of the elements of their I/O lists need not be all of the same type. 
We have the following pair of axioms, the first part is an existence axiom and the second part establishes equivalence or equality.

The following existence axioms are generated for each $i=1,\ldots,n$.

\textbf{spl 1}
\be
[ \mathfrak{p}~x~y~\mathfrak{p}^{(i)}~x^{(i)}~\bar{y}^{(i)}]
\ee
For the case where $\bar{y}^{(i)}$ is not the empty list we write
\be
\bar{y}=[\bar{y}^{(i)}]_{i=1}^n , \quad \bar{y}^{(i)} = [\bar{y}_j^{(i)}]_{j=1}^{m_i} , \qquad m_i:\mb{I}_+,~i=1,\ldots,n
\ee
and we have the associated axioms of equivalence/equality that are generated for each $i=1,\ldots,n$ and $j=1,\ldots,m_i$.

\textbf{spl 2}
\be
[[ \mathfrak{p}~x~y~\mathfrak{p}^{(i)}~x^{(i)}~\bar{y}^{(i)}]~\mathfrak{e}~[\bar{y}_j^{(i)}~y_j^{(i)}]~[~]]
\ee

Conversely, if all of the elements of a special non-atomic program are contained in a larger program list then the special non-atomic program can be retrieved.

\textbf{spl 3}
\be
[[ \mathfrak{p}^{(i)}~x^{(i)}~y^{(i)}]_{i=1}^n~\mathfrak{p}~x~\bar{y}]
\ee
For the case where $\bar{y}$ is not an empty list the following axioms are generated for each $i=1,\ldots,n$ and $j=1,\ldots,m_i$.

\textbf{spl 4}
\be
[[[ \mathfrak{p}^{(i^\prime)}~x^{(i^\prime)}~y^{(i^\prime)}]_{i^\prime=1}^n~\mathfrak{p}~x~\bar{y}]~\mathfrak{e}~[\bar{y}_j^{(i)}~y_j^{(i)}]~[~]]
\ee

\textbf{Disjunctions.}
Special non-atomic programs can also come in the form of disjunctions.
Consider the disjunction
\be
\mathfrak{p}~x~y= \mathfrak{p}^{(1)}~x^{(1)}~y~|~\mathfrak{p}^{(2)}~x^{(2)}~y
\ee
where $\mathfrak{p}~x~y$ has been prescribed as a special non-atomic program and $\mathfrak{p}^{(i)}~x^{(i)}~y$, $i=1,2$, are either atomic programs or have been prescribed as special non-atomic programs.
This means that any one of the operand programs $\mathfrak{p}^{(i)}~x^{(i)}~y$, $i=1,2$, may itself be a disjunction.
Following the convention of Definition \ref{disj}, the output list, $x$, of the main program $\mathfrak{p}~x~y$ is given by
\be
\bal
x = & \bar{x} \setminus (\bar{x} \cap y), \quad \bar{x} \simeq [x^{(i)}]_{i=1}^2 \\
\eal
\ee

A proof may individually generate an operand of the disjunction of the special non-atomic program.
The following axioms are generated for each $i=1,2$.

\textbf{spd 1}
\be
\bal
&  [\mathfrak{p}^{(i)}~x^{(i)}~y~\mathfrak{p}~x~\bar{y}], \qquad x \setminus u \subseteqq x^{(i)} \\
\eal
\ee
where $u \subseteqq x$ is a list of all constants that appear in $x$.

For the case where $y$ is not an empty list we write
\be
y=[y_j]_{j=1}^m , \quad \bar{y} = [\bar{y}_j]_{j=1}^m , \qquad m:\mb{I}_+
\ee
and we have the associated axioms of equivalence/equality that are generated for each $i=1,2$ and $j=1,\ldots,m$.

\textbf{spd 2}
\be
\bal
& [[ \mathfrak{p}^{(i)}~x^{(i)}~y~\mathfrak{p}~x~\bar{y}]~\mathfrak{e}~[\bar{y}_j~y_j]~[~]], \qquad x \setminus u \subseteqq x^{(i)} \\
\eal
\ee

These axioms are a special form of disjunction introduction and will be discussed further in a later chapter.
They can only be applied under the conditions that the output lists of the operands of the disjunction have the same length and $x \setminus u \subseteqq x^{(i)}$, where $u \subseteqq x$ is a list of all constants that appear in $x$. 
For this reason there is no general construction rule for disjunction introduction in PECR.

\vspace{5mm}

\textbf{Notes.}

\begin{itemize}

\item Special non-atomic programs have a higher status than other programs constructed from atomic programs and can appear in the axioms that define the application.
Sometimes it will be convenient to define some programs, that can otherwise be constructed as special non-atomic programs, as atomic programs.
This is often done to suppress auxiliary I/O parameters, such as prescribed input constants and variables associated with intermediate calculations.
Only the primary variables are present in the I/O lists of the atomic program with the understanding that the prescribed constant parameters employed in the calculation of the primary output variables are reassigned their values within the program each time it is called.
The constant parameters along with the variables associated with the intermediate calculations are discarded from memory immediately following the execution of the atomic program.

Such programs are not atomic in the stricter sense of Definition \ref{atomic} and should be regarded as \emph{pseudo-atomic programs}\index{pseudo-atomic program}.
Since the internal algorithm of the pseudo-atomic program is not accessible during a proof construction there is a need to supply a collection of axioms that express the essential properties and functionality of the program. 
If the same program was constructed as a special non-atomic program, no such axioms would be necessary.
This is because the algorithm defining its functionality would be accessible during a proof construction via the above axioms.
In this way there is a trade off between reducing the lengths of I/O lists and the introduction of additional axioms.   

\end{itemize}

\section{Substitution rule.}

We have already encountered the substitution rule for higher order programs.
For other applications similar axioms exist.
The substitution rule will come in two parts, an existence axiom followed by an equivalence or equality axiom.

Let $\mathfrak{e}~[a~b]~[~]$ be defined by (\ref{3.200}).
The first part of the substitution rule is an existence axiom.

\textbf{sr 1}
\be
 [[ \mathfrak{p}~x~y~\mathfrak{e}~[x_i~a]~[~]]~\mathfrak{p}~\bar{x}~\bar{y}], \qquad x_i \in x,~\bar{x} = x(x_i \to a) 
\ee
where $x=[x_i]_{i=1}^l$, for some $l:\mb{I}_+$.
The output lists $y$ and $\bar{y}$ may be empty lists.

The second part of the substitution rule is applicable when $y$ and $\bar{y}$ are not empty lists.
To present the axiom in a more general form we write
\be
y=[y_j]_{j=1}^m,~~~\bar{y}=[\bar{y}_j]_{j=1}^m, \qquad m:\mb{I}_+
\ee
For any substitution $x(x_i \to a)$, VPC will generate the following axioms for $j=1,\ldots,m$.

\textbf{sr 2}
\be
\bal
& [[ \mathfrak{p}~x~y~\mathfrak{e}~[x_i,a]~[~]~\mathfrak{p}~\bar{x}~\bar{y}]~\mathfrak{e}~[\bar{y}_j~y_j]~[~]], \qquad y_j \in y,~\bar{y}_j \in \bar{y}, \\
& \hspace{75mm} x_i \in x,~\bar{x} = x(x_i \to a) \\
\eal
\ee

It is important to note that some atomic programs can be shown to satisfy the substitution rule from other axioms.
The substitution rule should not be regarded as an axiom for such programs.
When setting up an application for VPC the user is required to supply the names of the programs for which the substitution rule is to be applied as an axiom.

\section{Application specific axioms of falsity.}

For each application, $S$, there will be false programs that the user is required to supply in the file \emph{axiom.dat} when initializing axioms of falsity.
These programs acquire the type $\mb{P}_{false}$ by default and form the seeds from which theorems of falsity are generated for the theory $S$.
All other false programs acquire the type $\mb{P}_{false}$ through the type assignment program $afalse$ by way of inference.

Application specific axioms of falsity are higher order constructs.
They can be represented by
\ben \label{4.6.100}
false~[u]~[~]
\een
where $u$ is a program that is explicitly defined in terms of the atomic programs associated with the application $S$.
One can think of (\ref{4.6.100}) as being equivalent to the higher order axiom $[p~c]$, where the premise $p$ is the empty (first order) program and the conclusion program $c=false~[u]~[~]$.
The prescribed program $u$ can be regarded as a constant for type $\mb{P}$ objects associated with the theory $S$. 

\section{I/O data files of VPC.}

As mentioned earlier, the actual VPC code will not be presented in this book.
However, an attempt is made throughout to describe the functionality of the program in sufficient detail so that the reader will be able to construct their own version if they desire.
The functionality of VPC is also described in a way that is independent of any higher order programming language.
The current version is written in Fortran but this reflects the author's familiarity with the language based on a background in scientific computing rather than a decision based on the most efficient higher order language.
It is intended that an updated version of VPC, along with a users guide, will be made available elsewhere at a later stage.    
Nevertheless, it will be useful to describe the various I/O data files currently employed in the execution of VPC because they can have an important role in describing the way in which the construction rules are employed.

The I/O data files for the program VPC can be grouped into initializing data files, output data files and runtime output files.
There are four initializing input files, \emph{axiom.dat}, \emph{setup.dat}, \emph{list.dat} and \emph{disj.dat}.
There is one output file, \emph{theorem.dat}, that lists the theorem and its proof upon completion of a derivation.
The file \emph{options.dat} provides runtime output that can be accessed by the user for guidance in a proof construction.

At each step of a proof construction, VPC writes to the screen the lists of the premises followed by the current statements of the proof.
Before a new statement is appended to the current proof the user may consult the file \emph{options.dat} (see below).

\emph{axiom.dat.} This file stores the collection of axioms that define the application or theory, $S$, being investigated.
As theorems are extracted from proofs they are appended to the list of axioms.
Each axiom and theorem has a unique label.

\emph{setup.dat.} This file contains the list of names of all atomic and special non-atomic programs that are associated with the application, $S$, and the constants that serve as input to these programs.
It also specifies the lengths of the I/O lists of each program and the type of the value assignments of each individual element of these lists. (Here the types of the I/O elements are prescribed only. Value assignments of the elements of the I/O lists are not prescribed in proof constructions).
Included in this file are the names of the atomic programs for which the substitution rule is to be applied as an axiom and the appropriate equality/equivalent program to be used for each element substitution.

\emph{list.dat.} This file contains all of the special non-atomic programs in the form of program lists that are associated with the application, $S$.
It provides the name of each special non-atomic program and the ordered list of its subprograms.

\emph{disj.dat.} This file lists all of the initial disjunction programs that are associated with the application, $S$.
These are defined as special non-atomic programs.
It provides the names of the disjunction programs and the lists of their operand programs.       

\emph{options.dat.}    
At each step of a proof, VPC determines all possible extensions that can be derived from the main program with respect to the axioms and theorems that are currently stored in the file \emph{axiom.dat}.
These extensions are listed in the file \emph{options.dat} from which the user may select a desired conclusion program to generate a new statement in the proof.
The process is repeated until the proof is completed.
If at any point of a derivation the user inserts a new statement of the proof that is not currently stored as an option in the options file then VPC will halt with an execution error message.
Each option in the options file includes the axiom/theorem label and the associated labels of the subprograms that make up the sublist of the current proof program that are program and I/O equivalent to the premise program of an axiom or previously derived theorem that is stored in \emph{axiom.dat}.

\emph{theorems.dat.} Upon the successful completion of a proof and the associated theorem extraction, the theorem and its proof are listed.
Examples of the format of the output are given in Sections \ref{saioi}-\ref{savoi}, \ref{sspotcr}, \ref{sbia}-\ref{slap} and  \ref{mtxthm}.
VPC also checks the proof and premises for redundant statements.
If redundancies are found they are listed at the end of the proof.
The user may re-derive the proof with the redundant statements removed. 

\chapter{Arithmetic over $\mb{I}$.}\label{int}

\section{Introduction.}\label{ring}

Consider a typical computer model that at its core can be essentially represented as a dynamical system through the difference equation
\ben\label{mot01}
v^{(t)} = f(v^{(t-1)}), \quad t=1,\ldots,n
\een
where $v$ can be a scalar/vector variable and $f$ a scalar/vector function.
The sequence $v^{(t)}$, $t=1,\ldots,n$, is generated from (\ref{mot01}) after prescribing the initial scalar/vector $v^{(0)}$.
In the current dominant paradigm, the objective is to construct an approximating assignment function, $f$, that somehow captures all of the properties predicted by some continuum theory that has been put forward as defining the laws that govern the real world system being modeled.
It is often the case that using various abstractions of continuous mathematics the expected behavior of the exact solutions of the continuum theory is well analyzed before embarking upon the generation of the approximate solution.

In the first chapter we discussed some of the pathologies that exist in the current paradigm of constructing computer models that attempt to approximate theories based on continuous mathematics.
It should therefore not come as a surprise that the assignment function $f$ of the discrete system (\ref{mot01}) can often generate solutions that are at odds with the expected behavior of solutions of the exact continuum theory.
For this reason it seems worthwhile to explore possible alternatives.

One alternative is to regard the discrete system upon which the computer model is based as the language that describes the laws that govern the real world application.
We can do this if we start with the hypothesis that all of the information needed to define objects of the real world, at all scales, can be represented by finite state vectors.
Dynamical systems will now be defined in terms of information flow that involves finite state arithmetic and hence will be compatible with the operational parameters of the machine on which a solution is to be generated.
This suggests that the scale of the resolution and the machine operational parameters be an essential component of the description of the real world application.
In other words the language that we use to describe the real world application is one based on discrete and fixed precision arithmetic.

Under our working hypothesis the machine environment $\mathfrak{M}(\mathcal{K},\mathcal{L},\mathcal{M})$ is reflective of the underlying structure of the physical world.
It is then not unreasonable to expect that some of the well accepted conventional laws of the physical world will emerge as macroscopic scale manifestations of the elementary laws that govern the allowable computations on our working platform $\mathfrak{M}(\mathcal{K},\mathcal{L},\mathcal{M})$.
To initiate this endeavor we must begin with the most elementary laws associated with basic machine arithmetic.

The objective here is to construct an axiomatic system for the elementary operations of integer arithmetic that reflects the practical implementation of maps on configuration states in a machine environment.
To this end we work with objects of type $\mb{I}$ that can be assigned any one of the integer values 
\be
0, \pm 1,\ldots, \pm N,
\ee
where $N$ is the maximum positive integer and is a machine dependent parameter.
One important feature of our formal system is that we replace the notion of sets with lists.
Nevertheless, our axiomatic system will be guided by the traditional axioms of commutative rings but with important departures.
It is useful to remind ourselves of these axioms.

\textbf{Commutative rings.}\index{commutative rings}
A commutative ring $\{ \mathcal{R},+,* \}$ is a set $\mathcal{R}$ with two binary operations $+$ and $*$ subject to the following axioms.

\begin{itemize}

\item $\{ \mathcal{R},+,* \}$ is closed under the operation $+$, i.e. if $a$ and $b$ are elements of $\mathcal{R}$ then $a+b$ is also an element of $\mathcal{R}$.

\item The operation $+$ is commutative, i.e. if $a$ and $b$ are elements of $\mathcal{R}$ then $a+b=b+a$.

\item The operation $+$ is associative, i.e. if $a,b$ and $c$ are elements of $\mathcal{R}$ then $a+(b+c)=(a+b)+c$.

\item For any element $a$ of $\mathcal{R}$ there is a unique element of $\mathcal{R}$, denoted by $0$, called the zero element such that $a+0=a$.

\item For any element $a$ of $\mathcal{R}$ there is a unique element of $\mathcal{R}$, denoted by $-a$, called the additive inverse of $a$ such that $a+(-a)=0$.

\item $\{ \mathcal{R},+,* \}$ is closed under the operation $*$, i.e. if $a$ and $b$ are elements of $\mathcal{R}$ then $a*b$ is also an element of $\mathcal{R}$.

\item The operation $*$ is commutative, i.e. if $a$ and $b$ are elements of $\mathcal{R}$ then $a*b=b*a$.

\item The operation $*$ is associative, i.e. if $a,b$ and $c$ are elements of $\mathcal{R}$ then $a*(b*c)=(a*b)*c$.

\item The operation $*$ is distributive over the operation $+$, i.e. if $a,b$ and $c$ are elements of $\mathcal{R}$ then $a*(b+c)=a*b+a*c$.

\item For any element $a$ of $\mathcal{R}$ there is a unique element of $\mathcal{R}$, denoted by $1$, called the multiplicative identity such that $1*a=a$.

\item $0 \neq 1$.

\end{itemize}

The integers $\mb{Z}$ defined by the numbers $0,\pm 1, \pm 2, \ldots$ with the usual operations of addition and multiplication is an example of a commutative ring.
 
Let $\{ \mathcal{R},+,* \}$ be a commutative ring.
An element $a$ of $\mathcal{R}$ has a multiplicative inverse $b$ contained in $\mathcal{R}$ if and only if $b*a=1$.
In such a case we write $b=a^{-1}$.

An ordered set is a set $\mathcal{R}$, together with a relation $<$ such that

\begin{itemize}

\item For any elements $x,y$ of $\mathcal{R}$, exactly one of $x<y$, $x=y$, $x>y$ holds. 

\item For any elements $x,y,z$ of $\mathcal{R}$, if $x<y$ and $y<z$ then $x<z$

\end{itemize}

A ring $\{ \mathcal{R},+,* \}$ is said to be an ordered ring if $\mathcal{R}$ is an ordered set such that 

\begin{itemize}

\item For any elements $x,y,z$ of $\mathcal{R}$, if $x<y$ then $x+z<y+z$.

\item For any elements $x,y$ of $\mathcal{R}$, if $x>0$ and $y>0$ then $x*y>0$.

\end{itemize}

\section{Atomic programs for arithmetic over $\mb{I}$.}\label{apfaoi}

Derivations of the standard identities of arithmetic over fields and commutative rings are often presented to students as an introductory course to analysis.
A major difficulty when working over $\mb{I}$ is the absence of closure of the operations of addition and multiplication.
Although the derivations of the basic identities of arithmetic are elementary, it will be necessary to restate the axioms of arithmetic in the context of a machine environment $\mathfrak{M}(\mathcal{K},\mathcal{L},\mathcal{M})$.
Here we shall take a constructive approach by introducing rules that address the operations of machine arithmetic that lend themselves to a more practical guide towards establishing computability.
The results that will be presented in the following sections of this chapter will serve as a first step towards an analysis of arrays over $\mb{I}$ that will be postponed for a later chapter.

For arithmetic over $\mb{I}$ we employ the following atomic integer programs.

\vspace{5mm}

\begin{tabular}{|c|c|}
\hline
Atomic program names & Atomic program type \\
\hline
$typei,~lt,~eqi$ & $\mb{P}_{type}$ \\
\hline
$id,~add,~mult,~div$ & $\mb{P}_{assign}$ \\
\hline
\end{tabular}

\vspace{5mm}
We will also make use of the following special non-atomic integer programs. 
\vspace{5mm}

\begin{tabular}{|c|c|}
\hline
Special non-atomic program names & Structure \\
\hline
$neq,~le,~abs,~min,~max$ & disjunction \\
\hline
$abs1,~abs2,~min1,~min2,~max1,~max2$ & list \\
\hline
\end{tabular} 

\vspace{5mm}

A description of the atomic programs that will be used for arithmetic over $\mb{I}$ are given in the list that follows.

\parskip0pt

\vspace{5mm}
\textbf{Check type integer.}

\textbf{\textit{Syntax.}} $typei~[a]~[~]$.

\textbf{\textit{Program Type.}} $\mb{P}_{type}$.

\textbf{\textit{Type checks.}} $a:\mb{I}$.

\textbf{\textit{Description.}} $typei$ checks that the value assigned to the variable $a$ is type $\mb{I}$.
$typei$ halts with an execution error if there is a type violation.

\vspace{5mm}
\textbf{Less than.}

\textbf{\textit{Syntax.}} $lt~[a~b]~[~]$.

\textbf{\textit{Program Type.}} $\mb{P}_{type}$.

\textbf{\textit{Type checks.}} $a:\mb{I}$, $b:\mb{I}$, $a<b$.

\textbf{\textit{Description.}} 
$lt$ first checks that the values assigned to the variables $a$ and $b$ are type $\mb{I}$.
It then checks that $a<b$.
Here the inequality is in the sense of assigned values.
$lt$ halts with an execution error if there is a type violation.
Type violation includes the case where $a<b$ is not satisfied.

\vspace{5mm}
\textbf{Numerical equality.}

\textbf{\textit{Syntax.}} $eqi~[a~b]~[~]$.

\textbf{\textit{Program Type.}} $\mb{P}_{type}$.

\textbf{\textit{Type checks.}} $a:\mb{I}$, $b:\mb{I}$, $a=b$.

\textbf{\textit{Description.}} 
$eqi$ first checks that the values assigned to the variables $a$ and $b$ are type $\mb{I}$.
It then checks that $a=b$.
Here equality is in the sense of assigned values.
$eqi$ halts with an execution error if there is a type violation.
Type violation includes the case where the value assigned equality $a=b$ is not satisfied.

\vspace{5mm}
\textbf{Identity assignment.}

\textbf{\textit{Syntax.}} $id~[a]~[b]$.

\textbf{\textit{Program Type.}} $\mb{P}_{assign}$.

\textbf{\textit{Type checks.}} $a:\mb{I}$.

\textbf{\textit{Assignment map.}} $b:=a$.

\textbf{\textit{Type assignment.}} $b::\mb{I}$.

\textbf{\textit{Description.}}
$id$ first checks that the value assigned to the variable $a$ is type $\mb{I}$.
It then assigns to $b$ the the value assigned to $a$, i.e. $b:=a$.
The value assignment is accompanied by the type assignment $b::\mb{I}$.
$id$ returns the value $b$ as output provided that there are no type violations.
Otherwise it halts with an execution error.

\vspace{5mm}
\textbf{Addition.}

\textbf{\textit{Syntax.}} $add~[a~b]~[c]$.

\textbf{\textit{Program Type.}} $\mb{P}_{assign}$.

\textbf{\textit{Type checks.}} $a:\mb{I}$, $b:\mb{I}$.

\textbf{\textit{Assignment map.}} $c:=a+b$.

\textbf{\textit{Type assignment.}} $c::\mb{I}$.

\textbf{\textit{Description.}}
$add$ first checks that the values assigned to the variables $a$ and $b$ are type $\mb{I}$.
It then attempts to assign to $c$ the sum of $a$ and $b$, i.e. $c:=a+b$.
This may fail if the sum $a+b$ is not contained within $\mb{I}$.
A successful value assignment is accompanied by the type assignment $c::\mb{I}$.
$add$ returns the value $c$ as output provided that there are no type violations.
Otherwise it halts with an execution error.

\vspace{5mm}
\textbf{Multiplication.}

\textbf{\textit{Syntax.}} $mult~[a~b]~[c]$.

\textbf{\textit{Program Type.}} $\mb{P}_{assign}$.

\textbf{\textit{Type checks.}} $a:\mb{I}$, $b:\mb{I}$.

\textbf{\textit{Assignment map.}} $c:=a*b$.

\textbf{\textit{Type assignment.}} $c::\mb{I}$.

\textbf{\textit{Description.}}
$mult$ first checks that the values assigned to the variables $a$ and $b$ are type $\mb{I}$.
It then attempts to assign to $c$ the multiplication of $a$ and $b$, i.e. $c:=a*b$.
This may fail if $a*b$ is not contained within $\mb{I}$.
A successful value assignment is accompanied by the type assignment $c::\mb{I}$.
$mult$ returns the value $c$ as output provided that there are no type violations.
Otherwise it halts with an execution error.

\vspace{5mm}
\textbf{Division.}

\textbf{\textit{Syntax.}} $div~[a~b]~[c]$.

\textbf{\textit{Program Type.}} $\mb{P}_{assign}$.

\textbf{\textit{Type checks.}} $a:\mb{I}$, $b:\mb{I}$.

\textbf{\textit{Assignment map.}} $c:=a/b$.

\textbf{\textit{Type assignment.}} $c::\mb{I}$.

\textbf{\textit{Description.}}
$div$ first checks that the values assigned to the variables $a$ and $b$ are type $\mb{I}$.
It then attempts to assign to $c$ the value of $a$ divided by $b$, i.e. $c:=a/b$.
This may fail if $b=0$ or if $a$ is not an integer multiple of $b$.
A successful value assignment is accompanied by the type assignment $c::\mb{I}$.
$div$ returns the value $c$ as output provided that there are no type violations.
Otherwise it halts with an execution error.

\parskip10pt

\section{Axioms of arithmetic over $\mb{I}$.}\label{saoaoi}

Axioms for arithmetic over $\mb{I}$ are labeled by the letters axi followed by a number.
To these are appended the order axioms that are labeled by ord followed by a number.
These axioms are stored in a file, \emph{axiom.dat}, that is accessed by VPC during proof construction.
Following the conventions outlined in Chapter \ref{appl}, I/O type axioms are labeled by the letters aio.
Axioms associated with special non-atomic programs that are lists are labeled with the letters spl followed by a number and for disjunctions labeled with the letters spd followed by a number (see Section \ref{snap}).
The substitution rule comes in two parts, an existence axiom, labeled sr 1, and an equality axiom, labeled sr 2.

Axioms based on I/O type axioms, the substitution rule and special non-atomic programs are automated within VPC and do not appear in the initializing data file $axiom.dat$ 

In the following axioms the internal square brackets act as delimiters for the premise program.
The standard list concatenation for programs apply so that the internal brackets can be removed.
When the premise program contains a single statement the internal square brackets are omitted.

\textbf{I/O type $\mb{I}$.}
An important property of all programs is that the type of the assigned values of all elements of the I/O lists are checked within the program.
The following axioms reflect this property for integer programs.

\textbf{aio}
\be
 [ \mathfrak{p}~x~y~typei~[c]~[~]],  \qquad c \in [x~y]
\ee

\textbf{Substitution rule.}
The substitution rule will be applied as an axiom to integer atomic programs $\mathfrak{p}~x~y$ such that
\be
\mathfrak{p} \in [eqi~add~mult~div~lt]
\ee
The first part of the substitution rule for integer programs is an existence axiom.

\textbf{sr 1}
\be
 [[ \mathfrak{p}~x~y~eqi~[x_i~a]~[~]]~\mathfrak{p}~\bar{x}~\bar{y}], \qquad x_i \in x,~\bar{x} = x(x_i \to a) 
\ee
\noindent
where $x=[x_i]_{i=1}^n$, for some $n:\mb{I}_+$.
The output lists $y$ and $\bar{y}$ may be empty lists.

The second part of the substitution rule is applicable when $y$ and $\bar{y}$ are not empty lists.
To present the axiom in a more general form we write
\be
y=[y_j]_{j=1}^m,~~~\bar{y}=[\bar{y}_j]_{j=1}^m, \qquad m:\mb{I}_+
\ee
For any substitution $x(x_i \to a)$, VPC will generate the following axioms for $j=1,\ldots,m$.

\textbf{sr 2}
\be
\bal
& [[ \mathfrak{p}~x~y~eqi~[x_i~a]~[~]~\mathfrak{p}~\bar{x}~\bar{y}]~eqi~[\bar{y}_j~y_j]~[~]], \qquad x_i \in x,~\bar{x} = x(x_i \to a) \\
\eal
\ee

As already mentioned, some atomic programs can be shown to satisfy the substitution rule from other axioms.
The substitution rule should not be regarded as an axiom for such programs.

For arithmetic on $\mb{I}$ we will define the list of constants to be
\be
~[-1~0~1]
\ee  

\textbf{Identity assignment axioms.}

\begin{axi}
\be
 [typei~[a]~[~]~id~[a]~[b]]
\ee
\end{axi}
\begin{axi}
\be
 [id~[a]~[b]~eqi~[b~a]~[~]]
\ee
\end{axi}

\textbf{Equality axioms.}

\emph{Reflexivity.}
\begin{axi}
\be
 [typei~[a]~[~]~eqi~[a~a]~[~]]
\ee
\end{axi}

\emph{Symmetry.}
\begin{axi}
\be
 [eqi~[a~b]~[~]~eqi~[b~a]~[~]]
\ee
\end{axi}

The equality program satisfies the property of transitivity
\be
 [[eqi~[a~b]~[~]~eqi~[b~c]~[~]]~eqi~[a~c]~[~]]
\ee
This is not included as an axiom because it follows from the substitution rule.

\textbf{Axioms of addition and multiplication.}

\emph{Commutativity of addition.}
\begin{axi}
\be
 [add~[a~b]~[c]~add~[b~a]~[d]]
\ee
\end{axi}
\begin{axi}
\be
 [[add~[a~b]~[c]~add~[b~a]~[d]]~eqi~[d~c]~[~]]
\ee
\end{axi}

\emph{Associativity of addition.}
\begin{axi}
\be
\bal
&  [[add~[a~b]~[d]~add~[d~c]~[x]~add~[b~c]~[e]]~add~[a~e]~[y]]
\eal
\ee
\end{axi}
\begin{axi}
\be
\bal
&  [[add~[a~b]~[d]~add~[d~c]~[x]~add~[b~c]~[e]~add~[a~e]~[y]]~eqi~[y~x]~[~]] \\
\eal
\ee
\end{axi}

\emph{Addition by zero.}
\begin{axi}
\be
 [typei~[a]~[~]~add~[a~0]~[b]]
\ee
\end{axi}
\begin{axi}
\be
 [add~[a~0]~[b]~eqi~[b~a]~[~]]
\ee
\end{axi}

\emph{Additive inverse.}
\begin{axi}
\be
 [typei~[a]~[~]~mult~[-1~a]~[b]]
\ee
\end{axi}
\begin{axi}
\be
 [mult~[-1~a]~[b]~add~[a~b]~[d]]
\ee
\end{axi}
\begin{axi}
\be
 [[mult~[-1~a]~[b]~add~[a~b]~[d]]~eqi~[d~0]~[~]]
\ee
\end{axi}

\emph{Commutativity of multiplication.}
\begin{axi}
\be
 [mult~[a~b]~[c]~mult~[b~a]~[d]]
\ee
\end{axi}
\begin{axi}
\be
 [[mult~[a~b]~[c]~mult~[b~a]~[d]]~eqi~[d~c]~[~]]
\ee
\end{axi}

\emph{Associativity of multiplication.}
\begin{axi}
\be
\bal
&  [[mult~[a~b]~[d]~mult~[d~c]~[x]~mult~[b~c]~[e]]~mult~[a~e]~[y]] \\
\eal
\ee
\end{axi}
\begin{axi}
\be
\bal
&  [[mult~[a~b]~[d]~mult~[d~c]~[x]~mult~[b~c]~[e]~mult~[a~e]~[y]]~eqi~[y~x]~[~]]
\eal
\ee
\end{axi}

\emph{Multiplication by unity.}
\begin{axi}
\be
 [typei~[a]~[~]~mult~[1~a]~[b]]
\ee
\end{axi}
\begin{axi}
\be
 [mult~[1~a]~[b]~eqi~[b~a]~[~]]
\ee
\end{axi}

\noindent
\emph{Distributive law.}

\begin{axi}
\be
\bal
&  [[add~[b~c]~[d]~mult~[a~d]~[x]~mult~[a~b]~[u]~mult~[a~c]~[v]]~add~[u~v]~[y]] \\
\eal
\ee
\end{axi}
\begin{axi}
\be
\bal
&  [[mult~[a~b]~[u]~mult~[a~c]~[v]~add~[u~v]~[y]~add~[b~c]~[d]]~mult~[a~d]~[x]]
\eal
\ee
\end{axi}
\begin{axi}
\be
\bal
&  [[add~[b~c]~[d]~mult~[a~d]~[x]~mult~[a~b]~[u]~mult~[a~c]~[v]~add~[u~v]~[y]] \\
& \hspace{5mm} eqi~[y~x]~[~]] \\
\eal
\ee
\end{axi}

\textbf{Divisor.}

\begin{axi}
\be
 [[neq~[a~0]~[~]~mult~[a~b]~[c]]~div~[c~a]~[d]]
\ee
\end{axi}
\begin{axi}
\be
 [[mult~[a~b]~[c]~div~[c~a]~[d]]~eqi~[d~b]~[~]]
\ee
\end{axi}

\textbf{Order axioms.}

\begin{ord}
\be
\bal
&  [[lt~[a~b]~[~]~add~[a~c]~[x]~add~[b~c]~[y]]~lt~[x~y]~[~]]
\eal
\ee
\end{ord}
\begin{ord}
	\be
	\bal
	& [[lt~[a~b]~[~]~lt~[c~d]~[~]~add~[a~c]~[x]~add~[b~d]~[y]]~lt~[x~y]~[~]] \\
	\eal
	\ee
\end{ord}

\begin{ord}
\be
\bal
&  [[lt~[a~b]~[~]~lt~[0~c]~[~]~mult~[a~c]~[x]~mult~[b~c]~[y]]~lt~[x~y]~[~]]
\eal
\ee
\end{ord}
\begin{ord}
\be
\bal
&  [[lt~[a~b]~[~]~lt~[c~0]~[~]~mult~[a~c]~[x]~mult~[b~c]~[y]]~lt~[y~x]~[~]] \\
\eal
\ee
\end{ord}

\emph{Transitivity of inequality.}
\begin{ord}\label{toi}
\be
 [[lt~[a~b]~[~]~lt~[b~c]~[~]]~lt~[a~c]~[~]]
\ee
\end{ord}

To the order axioms we include the following axiom that has an empty list premise 
\begin{ord}
\be
lt~[0~1]~[~]
\ee
\end{ord}

\textbf{Axiom of falsity (higher order type checking axiom).}
Axioms of falsity are higher order constructs.
We depart from the convention slightly of expressing the axiom in the form of a concatenation of a premise and conclusion by simply assigning a type $p:\mb{P}_{false}$.
For arithmetic over $\mb{I}$ we include the following axiom of falsity.
\begin{ord}
\be
lt~[a~a]~[~]~:~false
\ee
\end{ord}

One can think of this as being equivalent to the higher order axiom with an empty premise list
\be
false~[p]~[~]
\ee
where the assigned value of $p$ is an object of type $\mb{P}$ and is given explicitly by $p=lt~[a~a]~[~]$.
The object $p=lt~[a~a]~[~]$ can also be regarded as a constant of type $\mb{P}$ associated with the application of integer arithmetic over $\mb{I}$.

\textbf{Notes.}

\begin{itemize}

\item The above axioms for addition and multiplication differ from the axioms of fields and commutative rings because we do not have closure, i.e. $add~[a~b]~[c]$ and $mult~[a~b]~[c]$ do not necessarily follow from $typei~[a]~[~]$ and $typei~[b]~[~]$.
Because of this the above axioms for addition and multiplication are split into one or more existence parts followed by an identity axiom.
Any occurrence of statements involving $add$ and $mult$ in a program list must either have been inferred from the axioms or have simply been inserted as premises.

\item We have departed slightly from the convention of representing all elements of program I/O lists by alphanumeric variable names by allowing some elements to be represented by the constants $-1,0,1$.
To strictly adhere to the convention we could introduce special alphanumeric names for these constants.
For convenience we allow these constants to appear in numerical form in the input lists as exceptions. 

\item For any integer $a:\mb{I}$ there is no attempt made to abstract its additive identity $a+0$, its multiplicative identity $1*a$ and its additive inverse $-1*a$.
Here, $-1,0$ and $1$ are immediately recognized as type $\mb{I}$ objects.

\item 
One advantage of first order logic is its power of expression involving quantifiers.
The absence of quantifiers in PECR is not a restriction if one learns to read the axioms in the right way.
For example axi 9 states that given $a:\mb{I}$ there exists $b:\mb{I}$ such that $b=a+0$.
Axiom axi 10 simply adds that $b=a$.
If $\mb{Z}$ is the universal set, we may write the first order statement $\forall a(a+0=a)$, but when working over $\mb{I}$ this is inadequate because we also have to include the conditional statement for the existence of the sum $a+0$.
While one may prefer the more concise expressiveness of first order logic one needs to keep in mind that PECR demands that a statement be expanded into a form that contains all of the conditional statements.

Axioms axi 18-axi 19 are similar statements and are associated with the multiplicative identity. 
Axioms axi 11-axi 13 require a little more effort to read.
Axiom axi 10 states that for any $a:\mb{I}$ the additive inverse $b=-1*a$ exists.
Axiom axi 11 states that given the additive inverse of $a$ the sum $a+(-1*a)$ also exists and axi 12 states that this sum is identically zero.

\item Associativity of addition requires an existence axiom, axi 7, followed by an identity axiom, axi 8.
The existence part is necessary because $y=a+(b+c)$ does not necessarily follow from $x=(a+b)+c$.
As an example set $a=-N,~b=N,~c=1$.
We have $d=a+b=0:\mb{I}$ and hence $x=(-N+N)+1=1:\mb{I}$ but $e=b+c=N+1$, and hence $y=a+(b+c)$, is not of type $\mb{I}$. 
In order that $y=a+(b+c):\mb{I}$ we must include in the premise the conditional statement that $e=b+c:\mb{I}$.

\item Similarly, associativity of multiplication requires an existence axiom, axi 16, followed by an identity axiom, axi 17.
The existence part is necessary because $y=a*(b*c)$ does not necessarily follow from $x=(a*b)*c$.
As an example set $a=0,~b=N,~c=2$.
We have $d=a*b=0:\mb{I}$ and hence $x=(0)*2=0:\mb{I}$ but $e=b*c=N*2$ which is not of type $\mb{I}$. 
In order that $y=a*(b*c):\mb{I}$ we must include in the premise the conditional statement that $e=b*c:\mb{I}$.

\item The axiom of distributivity has two independent existence parts, axi 20 and axi 21, followed by an identity axiom, axi 22.
The existence part, axi 20, is necessary because $y=a*b+a*c$ does not necessarily follow from $x=a*(b+c)$.
As an example set $a=N,~b=N,~c=-N$.
We have $d=b+c=0:\mb{I}$ and hence $x=N*(-N+N)=0:\mb{I}$ but neither $u=a*b$ and $v=a*c$, and hence $y=u+v$, are of type $\mb{I}$. 
In order that $y=a*b+a*c:\mb{I}$ we must include in the premise the conditional statement that $u$ and $v$ are of type $\mb{I}$.

\item Similarly, the existence axiom, axi 21, is necessary because $x=a*(b+c)$ does not necessarily follow from $y=a*b+a*c$.
As an example set $a=0,~b=N,~c=N$.
We have $u=a*b=0:\mb{I}$ and $v=a*c=0:\mb{I}$ and hence their sum $y=0:\mb{I}$.
But $d=b+c=N+N$ is not of type $\mb{I}$.
In order that $x=a*(b+c):\mb{I}$ we must include in the premise the conditional statement that $d=b+c$ is of type $\mb{I}$.

\item In the premise of ord 1 we must include the conditional conditions that the sums $a+c$ and $b+c$ exist on $\mb{I}$.
Similarly, in the premise of axiom ord 2 we must also include the conditional statements that the sums $a+c$ and $b+d$ exist on $\mb{I}$.  
If we look for a proof analogous to that found in the theory of ordered rings we may suspect that ord 2 should follow as a theorem from the ord 1.
This is not the case because we cannot assume the existence of the intermediate sum $b+c$ on $\mb{I}$.
For this reason ord 2 is included as an axiom.

\item Similarly, in an ordered ring, statements similar to ord 3 and ord 4 are theorems that can be obtained from the single axiom, if $x>0$ and $y>0$ then $x*y>0$.
For arithmetic over $\mb{I}$ this ordered ring axiom cannot be used since we cannot assume that $x*y:\mb{I}$ necessarily follows from $x,y:\mb{I}$. 

\end{itemize}

\section{Special non-atomic programs.}

For arithmetic over $\mb{I}$ we will make use of the special non-atomic programs $neq,~le,~abs,~min$ and $max$.
They are defined as disjunctions.

\textbf{Not equal.}
\be
neq~[a~b]~[~] = lt~[a~b]~[~]~|~lt~[b~a]~[~]
\ee

\vspace{5mm}
\noindent
\textbf{Less than or equal.}
\be
le~[a~b]~[~] = lt~[a~b]~[~]~|~eqi~[a~b]~[~]
\ee

\vspace{5mm}
\noindent
\textbf{Absolute value.}

The program $abs~[a~0~-1]~[b]$ is a disjunction of two special non-atomic programs $abs1~[a~0~-1]~[b]$ and $abs2~[0~a]~[b]$. 

\be
\bal
& abs~[a~0~-1]~[b] = abs1~[a~0~-1]~[b]~|~abs2~[0~a]~[b] \\
\eal
\ee
where
\be
\bal
& abs1~[a~0~-1]~[b] = [lt~[a~0]~[~]~mult~[-1~a]~[b]] \\
\eal
\ee
and
\be
\bal
& abs2~[0~a]~[b] = [le~[0~a]~[~]~id~[a]~[b]] \\
\eal
\ee

The program $abs~[a~0~-1]~[b]$ makes the assignment $b:=|a|$.
The constants -1 and 0 appear in the input list of $abs$ only to adhere to the convention used to define the input list of the main program in the formal definition of program disjunctions, Definition \ref{disj}.
$abs$ has no useful meaning here if these constants are replaced by arbitrary variable names.

\vspace{5mm}
\noindent
\textbf{Maximum/minimum value.}

The minimum value of two variables is defined by the disjunction
\be
\bal
& min~[a~b]~[c] = min1~[a~b]~[c]~|~min2~[b~a]~[c] \\
\eal
\ee
where
\be
\bal
& min1~[a~b]~[c] = [le~[a~b]~[~]~id~[a]~[c]] \\
\eal
\ee
and
\be
\bal
& min2~[b~a]~[c] = [lt~[b~a]~[~]~id~[b]~[c]] \\
\eal
\ee

Similarly, the maximum values of two variables is defined by 
\be
\bal
& max~[a~b]~[c] = max1~[a~b]~[c]~|~max2~[b~a]~[c] \\
\eal
\ee
where
\be
\bal
& max1~[a~b]~[c] = [le~[a~b]~[~]~id~[b]~[c]] \\
\eal
\ee
and
\be
\bal
& max2~[b~a]~[c] = [lt~[b~a]~[~]~id~[a]~[c]] \\
\eal
\ee

\section{Algebraic identities over $\mb{I}$.}\label{saioi}

Derivations of proofs in arithmetic over $\mb{I}$ can sometimes be much lengthier than their counterparts in field and ring theory.
The main difficulty arises from the absence of closure of addition and multiplication.
As a consequence many proofs are actually dedicated to the establishment of existence.
While many theorems look similar to those found in the theory of fields and commutative rings one should closely examine the conditional statements that appear in the premises to fully appreciate the restrictions under which the theorems hold. 

At each step of a proof construction, VPC accesses the data file \emph{axiom.dat} that initially stores all of the axioms of the theory under consideration.
In this case they are axioms axi 1-24 and ord 1-7. 
As proofs are completed the theorems extracted from them are automatically appended to the file \emph{axiom.dat}.
All axioms and theorems that are stored in \emph{axiom.dat} are provided with a label.
Theorems are labeled by thm followed by a number.
Theorems that are of less interest in themselves but are derived for the purposes of use in other proofs are referred to as lemmas and labeled lem followed by a number.
Lemmas are often used when considering the separate cases of theorems containing disjunctions.

I/O type axioms are labeled aio and the substitution rule is labeled sr followed by a number.
These have a common structure for all applications and are treated differently in VPC from the user supplied axioms in \emph{axiom.dat}.

The proofs presented below were generated interactively.
At each step of a proof, VPC determines all possible extensions that can be derived from the current proof program.
These are listed in an options file that the user can consult to select a desired conclusion program.
Each option includes the axiom/theorem label and the associated connection list.
The user then selects the desired option (conclusion program) to generate a new statement in the program list.
The process is repeated until the proof is completed.
Crucial to the matching procedure of sublists of the proof program with premise programs of the axioms/theorems stored in \emph{axiom.dat} are program and I/O equivalence.

Theorems are presented as horizontal lists while proof programs are presented as vertical lists. 
The first entry of each line of a proof is the program label (equivalent to the program list element number) followed by the statement.
Following the statement is the connection list.
The connection list is preceded by the axiom/theorem label and contains the labels associated with the premises used to generate the current statement from an extended program derivation.
The absence of a connection list means that the statement is a premise of the proof program.
When a proof is completed VPC will extract and store the theorem after it checks for redundant premise statements and redundant steps in the proof.
If redundancies are detected VPC halts with an output that lists the redundant statements.   

Many theorems that are presented below come in pairs, the first part establishing existence and the second part establishing an identity.
The proofs are presented for demonstration purposes only and are not meant to represent the most efficient proof of the given theorem.
We start with some algebraic identities.
The listings are imported directly from the output file \emph{theorem.dat} generated by VPC.

Theorems thm 1 and thm 2 highlight the difficulties associated with arithmetic over $\mb{I}$.
In the theory of fields and commutative rings the identity $a=c-b$ follows trivially from the identity $c=a+b$.   
For arithmetic over $\mb{I}$ more work is required.
In theorem thm 1 the existence of $c-b$ over $\mb{I}$ is established from the premise that $c=a+b$ exists over $\mb{I}$.
Theorem thm 2 establishes the identity $a=c-b$.
\begin{lstlisting}
Theorem thm 1.
[[add [a b] [c] mult [-1 b] [d]] add [c d] [m]]

Proof.
  1 add [a b] [c]
  2 mult [-1 b] [d]
  3 add [b d] [e]            axi 12 [2]
  4 add [d b] [f]            axi 5 [3]
  5 add [b a] [g]            axi 5 [1]
  6 eqi [e 0] [ ]            axi 13 [2 3]
  7 eqi [0 e] [ ]            axi 4 [6]
  8 eqi [e f] [ ]            axi 6 [4 3]
  9 eqi [0 f] [ ]            sr 1 [7 8]
 10 eqi [g c] [ ]            axi 6 [1 5]
 11 typei [a] [ ]            aio [1]
 12 add [a 0] [h]            axi 9 [11]
 13 add [0 a] [i]            axi 5 [12]
 14 add [f a] [j]            sr 1 [13 9]
 15 add [d g] [k]            axi 7 [4 14 5]
 16 add [d c] [l]            sr 1 [15 10]
 17 add [c d] [m]            axi 5 [16]

Theorem thm 2.
[[add [a b] [c] mult [-1 b] [d] add [c d] [m]] eqi [m a] [ ]]

Proof.
  1 add [a b] [c]
  2 mult [-1 b] [d]
  3 add [c d] [m]
  4 add [b d] [e]            axi 12 [2]
  5 add [d b] [f]            axi 5 [4]
  6 add [b a] [g]            axi 5 [1]
  7 eqi [e 0] [ ]            axi 13 [2 4]
  8 eqi [0 e] [ ]            axi 4 [7]
  9 eqi [e f] [ ]            axi 6 [5 4]
 10 eqi [0 f] [ ]            sr 1 [8 9]
 11 eqi [g c] [ ]            axi 6 [1 6]
 12 typei [a] [ ]            aio [1]
 13 add [a 0] [h]            axi 9 [12]
 14 add [0 a] [i]            axi 5 [13]
 15 add [f a] [j]            sr 1 [14 10]
 16 add [d g] [k]            axi 7 [5 15 6]
 17 add [d c] [l]            axi 5 [3]
 18 eqi [m l] [ ]            axi 6 [17 3]
 19 eqi [l k] [ ]            sr 2 [16 11 17]
 20 eqi [k j] [ ]            axi 8 [5 15 6 16]
 21 eqi [l j] [ ]            sr 1 [19 20]
 22 eqi [m j] [ ]            sr 1 [18 21]
 23 eqi [j i] [ ]            sr 2 [14 10 15]
 24 eqi [m i] [ ]            sr 1 [22 23]
 25 eqi [i h] [ ]            axi 6 [13 14]
 26 eqi [m h] [ ]            sr 1 [24 25]
 27 eqi [h a] [ ]            axi 10 [13]
 28 eqi [m a] [ ]            sr 1 [26 27]
\end{lstlisting}
Theorem thm 3 shows that if the sums $a+b$ and $a+d$ exist over $\mb{I}$ and are equal then $b=d$.
\begin{lstlisting}
Theorem thm 3.
[[add [a b] [c] add [a d] [e] eqi [c e] [ ]] eqi [b d] [ ]]

Proof.
  1 add [a b] [c]
  2 add [a d] [e]
  3 eqi [c e] [ ]
  4 add [b a] [f]            axi 5 [1]
  5 add [d a] [g]            axi 5 [2]
  6 eqi [f c] [ ]            axi 6 [1 4]
  7 eqi [g e] [ ]            axi 6 [2 5]
  8 typei [a] [ ]            aio [1]
  9 mult [-1 a] [h]          axi 11 [8]
 10 add [f h] [i]            thm 1 [4 9]
 11 add [g h] [j]            thm 1 [5 9]
 12 eqi [i b] [ ]            thm 2 [4 9 10]
 13 eqi [j d] [ ]            thm 2 [5 9 11]
 14 add [c h] [k]            sr 1 [10 6]
 15 add [e h] [l]            sr 1 [11 7]
 16 eqi [k i] [ ]            sr 2 [10 6 14]
 17 eqi [l j] [ ]            sr 2 [11 7 15]
 18 eqi [l k] [ ]            sr 2 [14 3 15]
 19 eqi [k b] [ ]            sr 1 [16 12]
 20 eqi [l b] [ ]            sr 1 [18 19]
 21 eqi [b l] [ ]            axi 4 [20]
 22 eqi [l d] [ ]            sr 1 [17 13]
 23 eqi [b d] [ ]            sr 1 [21 22]
\end{lstlisting}
Theorem thm 4 is the multiplication version of thm 3.
It shows that if $a*b$ and $a*d$ exist over $\mb{I}$ and are equal and $a \neq 0$ then $b=d$.
\begin{lstlisting}
Theorem thm 4.
[[mult [a b] [c] mult [a d] [e] eqi [c e] [ ] neq [a 0] [ ]]
 eqi [b d] [ ]]

Proof.
  1 mult [a b] [c]
  2 mult [a d] [e]
  3 eqi [c e] [ ]
  4 neq [a 0] [ ]
  5 div [c a] [f]            axi 23 [4 1]
  6 div [e a] [g]            axi 23 [4 2]
  7 eqi [f b] [ ]            axi 24 [1 5]
  8 eqi [g d] [ ]            axi 24 [2 6]
  9 eqi [g f] [ ]            sr 2 [5 3 6]
 10 eqi [g b] [ ]            sr 1 [9 7]
 11 eqi [b d] [ ]            sr 1 [8 10]
\end{lstlisting}
Theorems thm 5 and thm 6 provide another example that highlights the difficulties associated with arithmetic over $\mb{I}$ where existence is not immediate.
In the theory of fields and commutative rings the existence of $0*a$ follows immediately from the closure of multiplication.
The proof of theorem thm 5 is rather a lengthy derivation dedicated just to the establishment that $0*a$ exists over $\mb{I}$. 
This is followed by thm 6 that establishes the equality $0*a=0$.
\begin{lstlisting}
Theorem thm 5.
[[typei [a] [ ]] mult [0 a] [o]]

Proof.
  1 typei [a] [ ]
  2 mult [1 a] [b]           axi 18 [1]
  3 eqi [b a] [ ]            axi 19 [2]
  4 mult [a 1] [c]           axi 14 [2]
  5 eqi [c b] [ ]            axi 15 [2 4]
  6 eqi [c a] [ ]            sr 1 [5 3]
  7 mult [-1 a] [d]          axi 11 [1]
  8 add [a d] [e]            axi 12 [7]
  9 mult [a -1] [f]          axi 14 [7]
 10 eqi [f d] [ ]            axi 15 [7 9]
 11 typei [1] [ ]            aio [2]
 12 mult [-1 1] [g]          axi 11 [11]
 13 add [1 g] [h]            axi 12 [12]
 14 eqi [h 0] [ ]            axi 13 [12 13]
 15 mult [1 -1] [i]          axi 14 [12]
 16 eqi [i -1] [ ]           axi 19 [15]
 17 eqi [g i] [ ]            axi 15 [15 12]
 18 eqi [g -1] [ ]           sr 1 [17 16]
 19 add [1 -1] [j]           sr 1 [13 18]
 20 eqi [j h] [ ]            sr 2 [13 18 19]
 21 eqi [j 0] [ ]            sr 1 [20 14]
 22 eqi [a c] [ ]            axi 4 [6]
 23 eqi [d f] [ ]            axi 4 [10]
 24 add [a f] [k]            sr 1 [8 23]
 25 add [c f] [l]            sr 1 [24 22]
 26 mult [a j] [m]           axi 21 [4 9 25 19]
 27 mult [j a] [n]           axi 14 [26]
 28 mult [0 a] [o]           sr 1 [27 21]

Theorem thm 6.
[[mult [0 a] [b]] eqi [b 0] [ ]]

Proof.
  1 mult [0 a] [b]
  2 typei [a] [ ]            aio [1]
  3 mult [1 a] [c]           axi 18 [2]
  4 eqi [c a] [ ]            axi 19 [3]
  5 mult [a 1] [d]           axi 14 [3]
  6 eqi [d c] [ ]            axi 15 [3 5]
  7 mult [-1 a] [e]          axi 11 [2]
  8 mult [a -1] [f]          axi 14 [7]
  9 typei [-1] [ ]           aio [7]
 10 mult [1 -1] [g]          axi 18 [9]
 11 eqi [g -1] [ ]           axi 19 [10]
 12 mult [-1 1] [h]          axi 14 [10]
 13 eqi [h g] [ ]            axi 15 [10 12]
 14 eqi [h -1] [ ]           sr 1 [13 11]
 15 add [1 h] [i]            axi 12 [12]
 16 eqi [i 0] [ ]            axi 13 [12 15]
 17 add [1 -1] [j]           sr 1 [15 14]
 18 eqi [j i] [ ]            sr 2 [15 14 17]
 19 eqi [j 0] [ ]            sr 1 [18 16]
 20 eqi [0 j] [ ]            axi 4 [19]
 21 mult [j a] [k]           sr 1 [1 20]
 22 eqi [k b] [ ]            sr 2 [1 20 21]
 23 mult [a j] [l]           axi 14 [21]
 24 eqi [l k] [ ]            axi 15 [21 23]
 25 eqi [l b] [ ]            sr 1 [24 22]
 26 add [d f] [m]            axi 20 [17 23 5 8]
 27 eqi [m l] [ ]            axi 22 [17 23 5 8 26]
 28 eqi [f e] [ ]            axi 15 [7 8]
 29 add [d e] [n]            sr 1 [26 28]
 30 eqi [n m] [ ]            sr 2 [26 28 29]
 31 eqi [m b] [ ]            sr 1 [27 25]
 32 eqi [n b] [ ]            sr 1 [30 31]
 33 add [c e] [o]            sr 1 [29 6]
 34 eqi [o n] [ ]            sr 2 [29 6 33]
 35 eqi [n o] [ ]            axi 4 [34]
 36 add [a e] [p]            axi 12 [7]
 37 eqi [p o] [ ]            sr 2 [33 4 36]
 38 eqi [o p] [ ]            axi 4 [37]
 39 eqi [p 0] [ ]            axi 13 [7 36]
 40 eqi [o 0] [ ]            sr 1 [38 39]
 41 eqi [n 0] [ ]            sr 1 [35 40]
 42 eqi [b 0] [ ]            sr 1 [41 32]
\end{lstlisting}
Theorem thm 7 shows that $-(-a)=a$.
Note that it follows from the axioms that the additive inverse of an object of type $\mb{I}$ always exists.
Hence a necessary and sufficient condition for the computability of the premise of thm 7 is that $a:\mb{I}$.
Given that upon entry $mult$ checks the type of the value assignments of its input lists, the computability of premise program is guaranteed if $a:\mb{I}$.
\begin{lstlisting}
Theorem thm 7.
[[mult [-1 a] [b] mult [-1 b] [c]] eqi [c a] [ ]]

Proof.
  1 mult [-1 a] [b]
  2 mult [-1 b] [c]
  3 add [a b] [d]            axi 12 [1]
  4 add [b c] [e]            axi 12 [2]
  5 eqi [d 0] [ ]            axi 13 [1 3]
  6 eqi [e 0] [ ]            axi 13 [2 4]
  7 eqi [0 e] [ ]            axi 4 [6]
  8 eqi [d e] [ ]            sr 1 [5 7]
  9 add [b a] [f]            axi 5 [3]
 10 eqi [f d] [ ]            axi 6 [3 9]
 11 eqi [f e] [ ]            sr 1 [10 8]
 12 eqi [a c] [ ]            thm 3 [9 4 11]
 13 eqi [c a] [ ]            axi 4 [12]
\end{lstlisting}
Theorems thm 8 and thm 9 show that if $a*b$ exists over $\mb{I}$ then $a*(-b)$ also exists over $\mb{I}$ and is equal to the additive inverse of $a*b$, i.e. $a*(-b)=-(a*b)$.
\begin{lstlisting}
Theorem thm 8.
[[mult [a b] [c] mult [-1 b] [d]] mult [a d] [i]]

Proof.
  1 mult [a b] [c]
  2 mult [-1 b] [d]
  3 typei [c] [ ]            aio [1]
  4 mult [-1 c] [e]          axi 11 [3]
  5 mult [b -1] [f]          axi 14 [2]
  6 mult [c -1] [g]          axi 14 [4]
  7 eqi [f d] [ ]            axi 15 [2 5]
  8 mult [a f] [h]           axi 16 [1 6 5]
  9 mult [a d] [i]           sr 1 [8 7]

Theorem thm 9.
[[mult [a b] [c] mult [-1 b] [d] mult [a d] [i] mult [-1 c] [e]]
 eqi [i e] [ ]]

Proof.
  1 mult [a b] [c]
  2 mult [-1 b] [d]
  3 mult [a d] [i]
  4 mult [-1 c] [e]
  5 mult [b -1] [f]          axi 14 [2]
  6 mult [c -1] [g]          axi 14 [4]
  7 eqi [f d] [ ]            axi 15 [2 5]
  8 eqi [g e] [ ]            axi 15 [4 6]
  9 mult [a f] [h]           axi 16 [1 6 5]
 10 eqi [h g] [ ]            axi 17 [1 6 5 9]
 11 eqi [i h] [ ]            sr 2 [9 7 3]
 12 eqi [h e] [ ]            sr 1 [10 8]
 13 eqi [i e] [ ]            sr 1 [11 12]
\end{lstlisting}
Theorems thm 10 and thm 11 show that if $a*b$ exists over $\mb{I}$ then $(-a)*b$ also exists over $\mb{I}$ and is equal to the additive inverse of $a*b$, i.e. $(-a)*b=-(a*b)$.
\begin{lstlisting}
Theorem thm 10.
[[mult [a b] [c] mult [-1 a] [d]] mult [d b] [g]]

Proof.
  1 mult [a b] [c]
  2 mult [-1 a] [d]
  3 mult [b a] [e]           axi 14 [1]
  4 mult [b d] [f]           thm 8 [3 2]
  5 mult [d b] [g]           axi 14 [4]

Theorem thm 11.
[[mult [a b] [c] mult [-1 a] [d] mult [d b] [g] mult [-1 c] [h]]
 eqi [g h] [ ]]

Proof.
  1 mult [a b] [c]
  2 mult [-1 a] [d]
  3 mult [d b] [g]
  4 mult [-1 c] [h]
  5 eqi [h g] [ ]            axi 17 [2 3 1 4]
  6 eqi [g h] [ ]            axi 4 [5]
\end{lstlisting}
Theorems thm 12 and thm 13 show that if $a*b$ exists over $\mb{I}$ then $(-a)*(-b)$ also exists over $\mb{I}$ and is equal to $a*b$, i.e. $(-a)*(-b)=a*b$.
\begin{lstlisting}
Theorem thm 12.
[[mult [a b] [c] mult [-1 a] [d] mult [-1 b] [e]] mult [d e] [g]]

Proof.
  1 mult [a b] [c]
  2 mult [-1 a] [d]
  3 mult [-1 b] [e]
  4 mult [a e] [f]           thm 8 [1 3]
  5 mult [d e] [g]           thm 10 [4 2]

Theorem thm 13.
[[mult [a b] [c] mult [-1 a] [d] mult [-1 b] [e] mult [d e] [f]]
 eqi [f c] [ ]]

Proof.
  1 mult [a b] [c]
  2 mult [-1 a] [d]
  3 mult [-1 b] [e]
  4 mult [d e] [f]
  5 typei [c] [ ]            aio [1]
  6 mult [-1 c] [g]          axi 11 [5]
  7 typei [g] [ ]            aio [6]
  8 mult [-1 g] [h]          axi 11 [7]
  9 eqi [h c] [ ]            thm 7 [6 8]
 10 mult [a e] [i]           thm 8 [1 3]
 11 eqi [i g] [ ]            thm 9 [1 3 10 6]
 12 mult [-1 i] [j]          axi 16 [2 4 10]
 13 eqi [j f] [ ]            axi 17 [2 4 10 12]
 14 eqi [h j] [ ]            sr 2 [12 11 8]
 15 eqi [h f] [ ]            sr 1 [14 13]
 16 eqi [f c] [ ]            sr 1 [9 15]
\end{lstlisting}

\section{Inequalities.}

The inequalities derived here are fairly straight forward.
The final derivation involves an application of the disjunction contraction rule.
With the use of disjunction splitting the proofs associated with the separate operand programs precede the proof of the main program containing the disjunction.
They correspond to the separate cases that are accessed by the proof of the main program containing the disjunction and are stored as lemmas.
Lemmas are labeled by lem followed by a number.
A statement containing two connection lists indicates that disjunction splitting has been applied at the preceding line of the proof and that the statement itself is obtained from a contraction of the conclusions of the operand programs.
A statement followed by an asterisk $*$ indicates that disjunction splitting has been applied to the operands of that statement.

Theorem thm 14 shows that if $a<0$ then $-a>0$ and theorem thm 15 shows that if $a>0$ then $-a<0$.
\begin{lstlisting}
Theorem thm 14.
[[lt [0 a] [ ] mult [-1 a] [b]] lt [b 0] [ ]]

Proof.
  1 lt [0 a] [ ]
  2 mult [-1 a] [b]
  3 add [a b] [c]            axi 12 [2]
  4 eqi [c 0] [ ]            axi 13 [2 3]
  5 typei [b] [ ]            aio [2]
  6 add [b 0] [d]            axi 9 [5]
  7 eqi [d b] [ ]            axi 10 [6]
  8 add [0 b] [e]            axi 5 [6]
  9 eqi [e d] [ ]            axi 6 [6 8]
 10 eqi [e b] [ ]            sr 1 [9 7]
 11 lt [e c] [ ]             ord 1 [1 8 3]
 12 lt [b c] [ ]             sr 1 [11 10]
 13 lt [b 0] [ ]             sr 1 [12 4]

Theorem thm 15.
[[lt [a 0] [ ] mult [-1 a] [b]] lt [0 b] [ ]]

Proof.
  1 lt [a 0] [ ]
  2 mult [-1 a] [b]
  3 add [a b] [c]            axi 12 [2]
  4 eqi [c 0] [ ]            axi 13 [2 3]
  5 typei [b] [ ]            aio [2]
  6 add [b 0] [d]            axi 9 [5]
  7 eqi [d b] [ ]            axi 10 [6]
  8 add [0 b] [e]            axi 5 [6]
  9 eqi [e d] [ ]            axi 6 [6 8]
 10 eqi [e b] [ ]            sr 1 [9 7]
 11 lt [c e] [ ]             ord 1 [1 3 8]
 12 lt [c b] [ ]             sr 1 [11 10]
 13 lt [0 b] [ ]             sr 1 [12 4]
\end{lstlisting}

To the above theorems we include the following result that will also be needed in later derivations.
Like axiom ord 6, theorem thm 16 has an empty list premise.
\begin{lstlisting}
Theorem thm 16.
[lt [-1 0] [ ]]

Proof.
  1 lt [0 1] [ ]             ord 6
  2 typei [1] [ ]            aio [1]
  3 mult [-1 1] [a]          axi 11 [2]
  4 mult [1 -1] [b]          axi 14 [3]
  5 eqi [b -1] [ ]           axi 19 [4]
  6 eqi [a b] [ ]            axi 15 [4 3]
  7 eqi [a -1] [ ]           sr 1 [6 5]
  8 lt [a 0] [ ]             thm 14 [1 3]
  9 lt [-1 0] [ ]            sr 1 [8 7]
\end{lstlisting}
As a first application of the disjunction contraction rule we establish that if $a \neq 0$ and $a^2:\mb{I}$ then $a^2>0$.
Theorem thm 17 is preceded by lemmas lem 1 and lem 2 that are associated with derivations based upon the operand programs that result from the disjunction splitting in theorem thm 17.
\begin{lstlisting}
Lemma lem 1.
[[lt [0 a] [ ] mult [a a] [b]] lt [0 b] [ ]]

Proof.
  1 lt [0 a] [ ]
  2 mult [a a] [b]
  3 typei [a] [ ]            aio [1]
  4 mult [0 a] [c]           thm 5 [3]
  5 eqi [c 0] [ ]            thm 6 [4]
  6 lt [c b] [ ]             ord 3 [1 1 4 2]
  7 lt [0 b] [ ]             sr 1 [6 5]

Lemma lem 2.
[[lt [a 0] [ ] mult [a a] [b]] lt [0 b] [ ]]

Proof.
  1 lt [a 0] [ ]
  2 mult [a a] [b]
  3 typei [a] [ ]            aio [1]
  4 mult [0 a] [c]           thm 5 [3]
  5 eqi [c 0] [ ]            thm 6 [4]
  6 lt [c b] [ ]             ord 4 [1 1 2 4]
  7 lt [0 b] [ ]             sr 1 [6 5]
\end{lstlisting}
We now apply the disjunction contraction rule.
\begin{lstlisting}
Theorem thm 17.
[[neq [a 0] [ ] mult [a a] [b]] lt [0 b] [ ]]

Proof.
  1 neq [a 0] [ ] *
  2 mult [a a] [b]
  3 lt [0 b] [ ]             lem 2 [1 2] lem 1 [1 2]
\end{lstlisting}
VPC splits the premise of theorem thm 17 into the two operand programs
\be
~[lt~[0~a]~[~]~mult~[a~a]~[b]]
\ee
and
\be
~[lt~[a~0]~[~]~mult~[a~a]~[b]]
\ee
A search is conducted for premises of the axioms/theorems stored in the file \emph{axiom.dat} that can be matched to sublists of each operand program and their conclusions stored in memory.
It then searches through the two collections of conclusions associated with each operand program and extracts those conclusions that are common to both.

\section{Semi-inequalities over $\mb{I}$.}

Before moving onto absolute values we need to generalize some of the inequalities just derived by replacing the strict inequality $<$ with the semi-inequality $\leq$.

The following two theorems involve mixed inequalities.
\begin{lstlisting}
Theorem thm 18.
[[lt [a b] [ ] le [b c] [ ]] lt [a c] [ ]]

Proof.
  1 lt [a b] [ ]
  2 le [b c] [ ] *
  3 lt [a c] [ ]             ord 5 [1 2] sr 1 [1 2]

Lemma lem 3.
[[eqi [a b] [ ] lt [b c] [ ]] lt [a c] [ ]]

Proof.
  1 eqi [a b] [ ]
  2 lt [b c] [ ]
  3 eqi [b a] [ ]            axi 4 [1]
  4 lt [a c] [ ]             sr 1 [2 3]

Theorem thm 19.
[[le [a b] [ ] lt [b c] [ ]] lt [a c] [ ]]

Proof.
  1 le [a b] [ ] *
  2 lt [b c] [ ]
  3 lt [a c] [ ]             ord 5 [1 2] lem 3 [1 2]
\end{lstlisting}

Theorem thm 20 shows that the semi-inequality satisfies the substitution rule for the second variable and theorem thm 21 shows that the semi-inequality satisfies the substitution rule for the first variable.
\begin{lstlisting}
Lemma lem 4.
[[eqi [a b] [ ] eqi [b c] [ ]] le [a c] [ ]]

Proof.
  1 eqi [a b] [ ]
  2 eqi [b c] [ ]
  3 eqi [a c] [ ]            sr 1 [1 2]
  4 le [a c] [ ]             spd 1 [3]

Lemma lem 5.
[[lt [a b] [ ] eqi [b c] [ ]] le [a c] [ ]]

Proof.
  1 lt [a b] [ ]
  2 eqi [b c] [ ]
  3 lt [a c] [ ]             sr 1 [1 2]
  4 le [a c] [ ]             spd 1 [3]

Theorem thm 20.
[[le [a b] [ ] eqi [b c] [ ]] le [a c] [ ]]

Proof.
  1 le [a b] [ ] *
  2 eqi [b c] [ ]
  3 le [a c] [ ]             lem 5 [1 2] lem 4 [1 2]

Lemma lem 6.
[[eqi [a b] [ ] eqi [a c] [ ]] le [c b] [ ]]

Proof.
  1 eqi [a b] [ ]
  2 eqi [a c] [ ]
  3 eqi [c b] [ ]            sr 1 [1 2]
  4 le [c b] [ ]             spd 1 [3]

Lemma lem 7.
[[lt [a b] [ ] eqi [a c] [ ]] le [c b] [ ]]

Proof.
  1 lt [a b] [ ]
  2 eqi [a c] [ ]
  3 lt [c b] [ ]             sr 1 [1 2]
  4 le [c b] [ ]             spd 1 [3]

Theorem thm 21.
[[le [a b] [ ] eqi [a c] [ ]] le [c b] [ ]]

Proof.
  1 le [a b] [ ] *
  2 eqi [a c] [ ]
  3 le [c b] [ ]             lem 7 [1 2] lem 6 [1 2]
\end{lstlisting}

Theorem thm 22 generalizes the order axiom of transitivity, ord 5, and can be translated to the statement that if $a \leq b$ and $b \leq c$ then $a \leq c$.
\begin{lstlisting}
Lemma lem 8.
[[le [a b] [ ] lt [b c] [ ]] le [a c] [ ]]

Proof.
  1 le [a b] [ ]
  2 lt [b c] [ ]
  3 lt [a c] [ ]             thm 19 [1 2]
  4 le [a c] [ ]             spd 1 [3]

Theorem thm 22.
[[le [a b] [ ] le [b c] [ ]] le [a c] [ ]]

Proof.
  1 le [a b] [ ]
  2 le [b c] [ ] *
  3 le [a c] [ ]             lem 8 [1 2] thm 20 [1 2]
\end{lstlisting}

Theorem thm 23 generalizes theorem thm 14 and can be translated to the statement that if $c \geq 0$ then $-c \leq 0$.
\begin{lstlisting}
Lemma lem 9.
[[lt [0 c] [ ] mult [-1 c] [d]] le [d 0] [ ]]

Proof.
  1 lt [0 c] [ ]
  2 mult [-1 c] [d]
  3 lt [d 0] [ ]             thm 14 [1 2]
  4 le [d 0] [ ]             spd 1 [3]

Lemma lem 10.
[[eqi [0 c] [ ] mult [-1 c] [d]] le [d 0] [ ]]

Proof.
  1 eqi [0 c] [ ]
  2 mult [-1 c] [d]
  3 eqi [c 0] [ ]            axi 4 [1]
  4 mult [-1 0] [a]          sr 1 [2 3]
  5 mult [0 -1] [b]          axi 14 [4]
  6 eqi [b a] [ ]            axi 15 [4 5]
  7 eqi [a b] [ ]            axi 4 [6]
  8 eqi [b 0] [ ]            thm 6 [5]
  9 eqi [a 0] [ ]            sr 1 [7 8]
 10 eqi [d a] [ ]            sr 2 [4 1 2]
 11 eqi [d 0] [ ]            sr 1 [10 9]
 12 le [d 0] [ ]             spd 1 [11]

Theorem thm 23.
[[le [0 c] [ ] mult [-1 c] [d]] le [d 0] [ ]]

Proof.
  1 le [0 c] [ ] *
  2 mult [-1 c] [d]
  3 le [d 0] [ ]             lem 9 [1 2] lem 10 [1 2]
\end{lstlisting}
Theorem thm 24 generalizes axiom ord 1 and can be translated to the statement that if $a \leq b$ and the sums $a+c$ and $b+c$ exist over $\mb{I}$ then $a+c \leq b+c$.
\begin{lstlisting}
Lemma lem 11.
[[lt [a b] [ ] add [a c] [x] add [b c] [y]] le [x y] [ ]]

Proof.
  1 lt [a b] [ ]
  2 add [a c] [x]
  3 add [b c] [y]
  4 lt [x y] [ ]             ord 1 [1 2 3]
  5 le [x y] [ ]             spd 1 [4]

Lemma lem 12.
[[eqi [a b] [ ] add [a c] [x] add [b c] [y]] le [x y] [ ]]

Proof.
  1 eqi [a b] [ ]
  2 add [a c] [x]
  3 add [b c] [y]
  4 eqi [y x] [ ]            sr 2 [2 1 3]
  5 eqi [x y] [ ]            axi 4 [4]
  6 le [x y] [ ]             spd 1 [5]

Theorem thm 24.
[[le [a b] [ ] add [a c] [x] add [b c] [y]] le [x y] [ ]]

Proof.
  1 le [a b] [ ] *
  2 add [a c] [x]
  3 add [b c] [y]
  4 le [x y] [ ]             lem 11 [1 2 3] lem 12 [1 2 3]
\end{lstlisting}
Theorem thm 25 generalizes axiom ord 2 and can be translated to the statement that if $a \leq b$ and $c \leq d$ and the sums $a+c$ and $b+d$ exist over $\mb{I}$ then $a+c \leq b+d$.
\begin{lstlisting}
Lemma lem 13.
[[lt [a b] [ ] eqi [c d] [ ] add [a c] [x] add [b d] [y]]
 le [x y] [ ]]

Proof.
  1 lt [a b] [ ]
  2 eqi [c d] [ ]
  3 add [a c] [x]
  4 add [b d] [y]
  5 eqi [d c] [ ]            axi 4 [2]
  6 add [b c] [e]            sr 1 [4 5]
  7 eqi [e y] [ ]            sr 2 [4 5 6]
  8 le [x e] [ ]             lem 11 [1 3 6]
  9 le [x y] [ ]             thm 20 [8 7]

Lemma lem 14.
[[lt [a b] [ ] lt [c d] [ ] add [a c] [x] add [b d] [y]]
 le [x y] [ ]]

Proof.
  1 lt [a b] [ ]
  2 lt [c d] [ ]
  3 add [a c] [x]
  4 add [b d] [y]
  5 lt [x y] [ ]             ord 2 [1 2 3 4]
  6 le [x y] [ ]             spd 1 [5]

Lemma lem 15.
[[lt [a b] [ ] le [c d] [ ] add [a c] [x] add [b d] [y]]
 le [x y] [ ]]

Proof.
  1 lt [a b] [ ]
  2 le [c d] [ ] *
  3 add [a c] [x]
  4 add [b d] [y]
  5 le [x y] [ ]             lem 14 [1 2 3 4] lem 13 [1 2 3 4]

Lemma lem 16.
[[eqi [a b] [ ] le [c d] [ ] add [a c] [x] add [b d] [y]]
 le [x y] [ ]]

Proof.
  1 eqi [a b] [ ]
  2 le [c d] [ ]
  3 add [a c] [x]
  4 add [b d] [y]
  5 add [c a] [e]            axi 5 [3]
  6 add [c b] [f]            sr 1 [5 1]
  7 add [d b] [g]            axi 5 [4]
  8 le [f g] [ ]             thm 24 [2 6 7]
  9 eqi [g y] [ ]            axi 6 [4 7]
 10 eqi [f e] [ ]            sr 2 [5 1 6]
 11 eqi [e x] [ ]            axi 6 [3 5]
 12 eqi [f x] [ ]            sr 1 [10 11]
 13 le [f y] [ ]             thm 20 [8 9]
 14 le [x y] [ ]             thm 21 [13 12]

Theorem thm 25.
[[le [a b] [ ] le [c d] [ ] add [a c] [x] add [b d] [y]]
 le [x y] [ ]]

Proof.
  1 le [a b] [ ] *
  2 le [c d] [ ]
  3 add [a c] [x]
  4 add [b d] [y]
  5 le [x y] [ ]             lem 15 [1 2 3 4] lem 16 [1 2 3 4]
\end{lstlisting}

\section{Absolute values over $\mb{I}$.}\label{savoi}
\noindent
We start by showing that $|a| \geq 0$.
\begin{lstlisting}
Lemma lem 17.
[[abs1 [a 0 -1] [b]] le [0 b] [ ]]

Proof.
  1 abs1 [a 0 -1] [b]
  2 lt [a 0] [ ]             spl 1 [1]
  3 mult [-1 a] [c]          spl 1 [1]
  4 lt [0 c] [ ]             thm 15 [2 3]
  5 eqi [c b] [ ]            spl 2 [1 3]
  6 lt [0 b] [ ]             sr 1 [4 5]
  7 le [0 b] [ ]             spd 1 [6]

Lemma lem 18.
[[abs2 [0 a] [b]] le [0 b] [ ]]

Proof.
  1 abs2 [0 a] [b]
  2 le [0 a] [ ]             spl 1 [1]
  3 id [a] [c]               spl 1 [1]
  4 eqi [c a] [ ]            axi 2 [3]
  5 eqi [c b] [ ]            spl 2 [1 3]
  6 eqi [b a] [ ]            sr 1 [4 5]
  7 eqi [a b] [ ]            axi 4 [6]
  8 le [0 b] [ ]             thm 20 [2 7]

Theorem thm 26.
[[abs [a 0 -1] [b]] le [0 b] [ ]]

Proof.
  1 abs [a 0 -1] [b] *
  2 le [0 b] [ ]             lem 17 [1] lem 18 [1]
\end{lstlisting}
The next two theorems are examples where a disjunction splitting and contraction involves detecting a false program in one of the operand program proofs.
Theorem thm 27 is equivalent to the statement that if $|a|=0$ then $a=0$.
It is preceded by two lemmas, lem 19 and lem 20, that are associated with the two operand programs that result from disjunction splitting in thm 27.
The premise of the first lemma, lem 19, is type $\mb{P}_{false}$.
Appealing to the disjunction contraction rule 2, the conclusion of the second lemma, lem 20, is contracted back onto the main proof of thm 27.
\begin{lstlisting}
Lemma lem 19.
[[abs1 [a 0 -1] [b] eqi [b 0] [ ]] :false]

Proof.
  1 abs1 [a 0 -1] [b]
  2 eqi [b 0] [ ]
  3 lt [a 0] [ ]             spl 1 [1]
  4 mult [-1 a] [c]          spl 1 [1]
  5 eqi [c b] [ ]            spl 2 [1 4]
  6 lt [0 c] [ ]             thm 15 [3 4]
  7 lt [0 b] [ ]             sr 1 [6 5]
  8 lt [0 0] [ ]             sr 1 [7 2]
  9 :false                   ord 7 [8]

Lemma lem 20.
[[abs2 [0 a] [b] eqi [b 0] [ ]] eqi [a 0] [ ]]

Proof.
  1 abs2 [0 a] [b]
  2 eqi [b 0] [ ]
  3 id [a] [c]               spl 1 [1]
  4 eqi [c b] [ ]            spl 2 [1 3]
  5 eqi [c a] [ ]            axi 2 [3]
  6 eqi [a b] [ ]            sr 1 [4 5]
  7 eqi [a 0] [ ]            sr 1 [6 2]

Theorem thm 27.
[[abs [a 0 -1] [b] eqi [b 0] [ ]] eqi [a 0] [ ]]

Proof.
  1 abs [a 0 -1] [b] *
  2 eqi [b 0] [ ]
  3 eqi [a 0] [ ]            lem 19 [1 2] lem 20 [1 2]
\end{lstlisting}
\noindent
Theorem thm 28 is the converse of thm 27 and is equivalent to the statement that if $a=0$ then $|a|=0$.
\begin{lstlisting}
Lemma lem 21.
[[abs1 [a 0 -1] [b] eqi [a 0] [ ]] :false]

Proof.
  1 abs1 [a 0 -1] [b]
  2 eqi [a 0] [ ]
  3 lt [a 0] [ ]             spl 1 [1]
  4 lt [0 0] [ ]             sr 1 [3 2]
  5 :false                   ord 7 [4]

Lemma lem 22.
[[abs2 [0 a] [b] eqi [a 0] [ ]] eqi [b 0] [ ]]

Proof.
  1 abs2 [0 a] [b]
  2 eqi [a 0] [ ]
  3 id [a] [c]               spl 1 [1]
  4 eqi [c b] [ ]            spl 2 [1 3]
  5 eqi [b c] [ ]            axi 4 [4]
  6 eqi [c a] [ ]            axi 2 [3]
  7 eqi [b a] [ ]            sr 1 [5 6]
  8 eqi [b 0] [ ]            sr 1 [7 2]

Theorem thm 28.
[[abs [a 0 -1] [b] eqi [a 0] [ ]] eqi [b 0] [ ]]

Proof.
  1 abs [a 0 -1] [b] *
  2 eqi [a 0] [ ]
  3 eqi [b 0] [ ]            lem 21 [1 2] lem 22 [1 2]
\end{lstlisting}
\noindent
We now prove that if $|a| \leq c$ then $-c \leq a \leq c$.
Theorems thm 29 and thm 30, respectively, split this into the two parts leading to the conclusions $a \leq c$ and $-c \leq a$, respectively.
\begin{lstlisting}
Lemma lem 23.
[[abs1 [a 0 -1] [b] le [b c] [ ]] le [a c] [ ]]

Proof.
  1 abs1 [a 0 -1] [b]
  2 le [b c] [ ]
  3 lt [a 0] [ ]             spl 1 [1]
  4 mult [-1 a] [d]          spl 1 [1]
  5 eqi [d b] [ ]            spl 2 [1 4]
  6 lt [0 d] [ ]             thm 15 [3 4]
  7 lt [0 b] [ ]             sr 1 [6 5]
  8 lt [a b] [ ]             ord 5 [3 7]
  9 lt [a c] [ ]             thm 18 [8 2]
 10 le [a c] [ ]             spd 1 [9]

Lemma lem 24.
[[abs2 [0 a] [b] le [b c] [ ]] le [a c] [ ]]

Proof.
  1 abs2 [0 a] [b]
  2 le [b c] [ ]
  3 id [a] [d]               spl 1 [1]
  4 eqi [d b] [ ]            spl 2 [1 3]
  5 eqi [b d] [ ]            axi 4 [4]
  6 eqi [d a] [ ]            axi 2 [3]
  7 eqi [b a] [ ]            sr 1 [5 6]
  8 le [a c] [ ]             thm 21 [2 7]

Theorem thm 29.
[[abs [a 0 -1] [b] le [b c] [ ]] le [a c] [ ]]

Proof.
  1 abs [a 0 -1] [b] *
  2 le [b c] [ ]
  3 le [a c] [ ]             lem 23 [1 2] lem 24 [1 2]

Lemma lem 25.
[[abs1 [a 0 -1] [b] lt [b c] [ ] mult [-1 c] [d]] le [d a] [ ]]

Proof.
  1 abs1 [a 0 -1] [b]
  2 lt [b c] [ ]
  3 mult [-1 c] [d]
  4 mult [-1 a] [e]          spl 1 [1]
  5 eqi [e b] [ ]            spl 2 [1 4]
  6 lt [e c] [ ]             lem 3 [5 2]
  7 typei [e] [ ]            aio [4]
  8 mult [-1 e] [f]          axi 11 [7]
  9 mult [e -1] [g]          axi 14 [8]
 10 mult [c -1] [h]          axi 14 [3]
 11 lt [-1 0] [ ]            thm 16
 12 lt [h g] [ ]             ord 4 [6 11 9 10]
 13 eqi [g f] [ ]            axi 15 [8 9]
 14 eqi [f a] [ ]            thm 7 [4 8]
 15 eqi [g a] [ ]            sr 1 [13 14]
 16 lt [h a] [ ]             sr 1 [12 15]
 17 eqi [h d] [ ]            axi 15 [3 10]
 18 lt [d a] [ ]             sr 1 [16 17]
 19 le [d a] [ ]             spd 1 [18]

Lemma lem 26.
[[abs1 [a 0 -1] [b] eqi [b c] [ ] mult [-1 c] [d]] le [d a] [ ]]

Proof.
  1 abs1 [a 0 -1] [b]
  2 eqi [b c] [ ]
  3 mult [-1 c] [d]
  4 mult [-1 a] [e]          spl 1 [1]
  5 eqi [e b] [ ]            spl 2 [1 4]
  6 eqi [b e] [ ]            axi 4 [5]
  7 eqi [c e] [ ]            sr 1 [6 2]
  8 eqi [e c] [ ]            axi 4 [7]
  9 mult [-1 e] [f]          sr 1 [3 7]
 10 eqi [f a] [ ]            thm 7 [4 9]
 11 eqi [d f] [ ]            sr 2 [9 8 3]
 12 eqi [d a] [ ]            sr 1 [11 10]
 13 le [d a] [ ]             spd 1 [12]

Lemma lem 27.
[[abs1 [a 0 -1] [b] le [b c] [ ] mult [-1 c] [d]] le [d a] [ ]]

Proof.
  1 abs1 [a 0 -1] [b]
  2 le [b c] [ ] *
  3 mult [-1 c] [d]
  4 le [d a] [ ]             lem 25 [1 2 3] lem 26 [1 2 3]

Lemma lem 28.
[[abs2 [0 a] [b] le [b c] [ ] mult [-1 c] [d]] le [d a] [ ]]

Proof.
  1 abs2 [0 a] [b]
  2 le [b c] [ ]
  3 mult [-1 c] [d]
  4 le [0 a] [ ]             spl 1 [1]
  5 le [a c] [ ]             lem 24 [1 2]
  6 le [0 c] [ ]             thm 22 [4 5]
  7 le [d 0] [ ]             thm 23 [6 3]
  8 le [d a] [ ]             thm 22 [7 4]

Theorem thm 30.
[[abs [a 0 -1] [b] le [b c] [ ] mult [-1 c] [d]] le [d a] [ ]]

Proof.
  1 abs [a 0 -1] [b] *
  2 le [b c] [ ]
  3 mult [-1 c] [d]
  4 le [d a] [ ]             lem 27 [1 2 3] lem 28 [1 2 3]
\end{lstlisting}
Theorem thm 31 proves the converse statement that if $-c \leq a \leq c$ then $|a| \leq c$.
\begin{lstlisting}
Lemma lem 29.
[[mult [-1 c] [d] eqi [d a] [ ] abs1 [a 0 -1] [b]] le [b c] [ ]]

Proof.
  1 mult [-1 c] [d]
  2 eqi [d a] [ ]
  3 abs1 [a 0 -1] [b]
  4 mult [-1 a] [e]          spl 1 [3]
  5 eqi [e b] [ ]            spl 2 [3 4]
  6 typei [d] [ ]            aio [1]
  7 mult [-1 d] [f]          axi 11 [6]
  8 eqi [f c] [ ]            thm 7 [1 7]
  9 eqi [e f] [ ]            sr 2 [7 2 4]
 10 eqi [e c] [ ]            sr 1 [9 8]
 11 eqi [b c] [ ]            sr 1 [10 5]
 12 le [b c] [ ]             spd 1 [11]

Lemma lem 30.
[[mult [-1 c] [d] lt [d a] [ ] abs1 [a 0 -1] [b]] le [b c] [ ]]

Proof.
  1 mult [-1 c] [d]
  2 lt [d a] [ ]
  3 abs1 [a 0 -1] [b]
  4 mult [-1 a] [e]          spl 1 [3]
  5 mult [a -1] [f]          axi 14 [4]
  6 typei [d] [ ]            aio [1]
  7 mult [-1 d] [g]          axi 11 [6]
  8 mult [d -1] [h]          axi 14 [7]
  9 lt [-1 0] [ ]            thm 16
 10 lt [f h] [ ]             ord 4 [2 9 8 5]
 11 eqi [f e] [ ]            axi 15 [4 5]
 12 eqi [e b] [ ]            spl 2 [3 4]
 13 eqi [f b] [ ]            sr 1 [11 12]
 14 lt [b h] [ ]             sr 1 [10 13]
 15 eqi [h g] [ ]            axi 15 [7 8]
 16 eqi [g c] [ ]            thm 7 [1 7]
 17 eqi [h c] [ ]            sr 1 [15 16]
 18 lt [b c] [ ]             sr 1 [14 17]
 19 le [b c] [ ]             spd 1 [18]

Lemma lem 31.
[[mult [-1 c] [d] le [d a] [ ] abs1 [a 0 -1] [b]] le [b c] [ ]]

Proof.
  1 mult [-1 c] [d]
  2 le [d a] [ ] *
  3 abs1 [a 0 -1] [b]
  4 le [b c] [ ]             lem 30 [1 2 3] lem 29 [1 2 3]

Lemma lem 32.
[[le [a c] [ ] abs2 [0 a] [b]] le [b c] [ ]]

Proof.
  1 le [a c] [ ]
  2 abs2 [0 a] [b]
  3 id [a] [d]               spl 1 [2]
  4 eqi [d a] [ ]            axi 2 [3]
  5 eqi [d b] [ ]            spl 2 [2 3]
  6 eqi [b a] [ ]            sr 1 [4 5]
  7 eqi [a b] [ ]            axi 4 [6]
  8 le [b c] [ ]             thm 21 [1 7]

Theorem thm 31.
[[mult [-1 c] [d] le [a c] [ ] le [d a] [ ] abs [a 0 -1] [b]]
 le [b c] [ ]]

Proof.
  1 mult [-1 c] [d]
  2 le [a c] [ ]
  3 le [d a] [ ]
  4 abs [a 0 -1] [b] *
  5 le [b c] [ ]             lem 31 [1 3 4] lem 32 [2 4]
\end{lstlisting}
When combined, theorems thm 32 and thm 33, state that $-|a| \leq a \leq |a|$.
No disjunction splitting is required in the proofs. 
\begin{lstlisting}
Theorem thm 32.
[[abs [a 0 -1] [b]] le [a b] [ ]]

Proof.
  1 abs [a 0 -1] [b]
  2 typei [b] [ ]            aio [1]
  3 eqi [b b] [ ]            axi 3 [2]
  4 le [b b] [ ]             spd 1 [3]
  5 le [a b] [ ]             thm 29 [1 4]

Theorem thm 33.
[[abs [a 0 -1] [b] mult [-1 b] [c]] le [c a] [ ]]

Proof.
  1 abs [a 0 -1] [b]
  2 mult [-1 b] [c]
  3 typei [b] [ ]            aio [1]
  4 eqi [b b] [ ]            axi 3 [3]
  5 le [b b] [ ]             spd 1 [4]
  6 le [c a] [ ]             thm 30 [1 5 2]
\end{lstlisting}
Theorem thm 34 states that if $|x|+|y|$ and $|x+y|$ exist over $\mb{I}$ then $|x+y| \leq |x|+|y|$.
No disjunction splitting is required.
\begin{lstlisting}
Theorem thm 34.
[[abs [x 0 -1] [u] abs [y 0 -1] [v] add [u v] [w] add [x y] [z]
 abs [z 0 -1] [p]] le [p w] [ ]]

Proof.
  1 abs [x 0 -1] [u]
  2 abs [y 0 -1] [v]
  3 add [u v] [w]
  4 add [x y] [z]
  5 abs [z 0 -1] [p]
  6 le [x u] [ ]             thm 32 [1]
  7 le [y v] [ ]             thm 32 [2]
  8 le [z w] [ ]             thm 25 [6 7 4 3]
  9 typei [u] [ ]            aio [1]
 10 mult [-1 u] [a]          axi 11 [9]
 11 typei [v] [ ]            aio [2]
 12 mult [-1 v] [b]          axi 11 [11]
 13 le [a x] [ ]             thm 33 [1 10]
 14 le [b y] [ ]             thm 33 [2 12]
 15 typei [w] [ ]            aio [3]
 16 mult [-1 w] [c]          axi 11 [15]
 17 add [a b] [d]            axi 20 [3 16 10 12]
 18 eqi [d c] [ ]            axi 22 [3 16 10 12 17]
 19 le [d z] [ ]             thm 25 [13 14 17 4]
 20 le [c z] [ ]             thm 21 [19 18]
 21 le [p w] [ ]             thm 31 [16 8 20 5]
\end{lstlisting}

\textbf{Notes.}

\begin{itemize}

\item As with many of the derivations given above, theorem thm 34, is weaker than its counterpart in the theory of fields and commutative rings.
This is because for arithmetic over $\mb{I}$ the existence of $|x|+|y|$ is not guaranteed given the existence of $x+y$.
In the premise of thm 35 we must also include the conditional statement that $|x|+|y|$ exist over $\mb{I}$.   

\item There are a few derivations of standard identities for absolute values that have been omitted.
We leave as an exercise to the reader to establish the following.
(i) $|-a|=|a|$, (ii) If $a*b:\mb{I}$ and $|a|*|b|:\mb{I}$ then $|a*b|=|a|*|b|$, (iii) If $a^2:\mb{I}$ then $|a|^2=a^2$.
These have been omitted because their proofs can be rather lengthy due to the need to apply a few more applications of disjunction splitting.
Otherwise they are fairly straight forward.  

\end{itemize}

\section{Arithmetic over $\mb{J}$.}

We could also work with a finite collection of rationals $\mb{J} = \epsilon \mb{I}$, where $0<\epsilon << 1$ and $\epsilon$ is also a machine specific parameter.
An object of type $\mb{J}$ can take on any one of the assigned values
\be
0, \pm \epsilon, \pm 2 \epsilon, \ldots , \pm N \epsilon
\ee
The finite collection of rationals $\mb{J}$ has a fixed resolution size so that arithmetic over $\mb{J}$, as defined here, differs from floating point arithmetic.
Because of this all of the results of arithmetic over $\mb{I}$ of the previous sections can be directly applied to $\mb{J}$.

We use the same atomic programs with the important modification that all type checks within the atomic programs that are associated with $\mb{I}$ are replaced by $\mb{J}$.
We accept the same axioms axi 1-24 and ord 1-7, along with the aio axioms, the substitution rule and the special non-atomic program axioms by replacing all references to type $\mb{I}$ objects by type $\mb{J}$ objects.

From the ordered ring axioms it can be shown that if $0<x<y$ then $0 < \frac{1}{y} < \frac{1}{x}$.
The standard proof follows by first establishing that if $0<x$ then $0 < \frac{1}{x}$.
For an ordered ring we can derive the result that for any nonzero element $x$, $x^2>0$.
Hence we have $(\frac{1}{x})^2 >0$ and using the second axiom of an ordered ring to obtain $x (\frac{1}{x})^2 >0$ and the desired result follows.

Adapting theorem thm 18 to $\mb{J}$ we have that if $a:\mb{J}$, $a \neq 0$ and $a^2:\mb{J}$ then $a^2>0$.
But we do not have $a^2:\mb{J}$ necessarily follows from $a:\mb{J}$.
For this reason the standard proof that starts with the result $(\frac{1}{x})^2 >0$ cannot be used.

In the absence of a known proof, for arithmetic over $\mb{J}$ we include the additional order axioms
\begin{ord}
\be
 [[lt~[0~a]~[~]~div~[1~a]~[x]]~lt~[0~x]~[~]]
\ee
\end{ord}

\begin{ord}
\be
 [[lt~[0~a]~[~]~lt~[a~b]~[~]~div~[1~a]~[x]~div~[1~b]~[y]]~lt~[y~x]~[~]]
\ee
\end{ord}
Note that $div~[1~a]~[x]$ does not necessarily follow from $a:\mb{J}$ and $a \neq 0$.
Because of this the premises in the above axioms are conditional on the computability of the statements $div~[1~a]~[x]$ and $div~[1~b]~[y]$.

The previous sections largely addressed the prevention of arithmetic operations that lead to overflows.
Special care needs to be exercised when dealing with objects of type $\mb{J}$ in that we are now faced with possible underflows as well as overflows.
This is because the absolute values of objects of $\mb{J}$ are not only bounded above by $\epsilon N$ but also have a finite resolution $\epsilon$ that provides a lower bound on operations of multiplication of nonzero elements of $\mb{J}$.

Working over $\mb{J}$ is often desirable because dynamical systems over $\mb{I}$ can generate integers that become exceedingly large.
However, it is often the case that when constructing a model over $\mb{J}$ we have actually applied some scaling law to a dynamical system that has originally been posed over $\mb{I}$.
The scaling is introduced not only to avoid dealing with large integers but also to generalize the integer based model.
Thus, when attempting to establish the computability of a model based over $\mb{J}$ it is usually safer to return to the original formulation and carry out the analysis on the associated model based over $\mb{I}$.

\chapter{Finite Dynamical Systems.}\label{cfds}

\section{Introduction.}

We are interested in computer models of dynamical systems.
In particular we shall focus exclusively on dynamical systems as they are implemented on a deterministic machine with finite memory.
Under this regime we will always deal with dynamical systems that are defined by the map
\be
\Phi : \mb{T} \times \mb{M} \to \mb{M}
\ee
where objects of $\mb{T}$ are integers or fixed precision rational numbers associated with discrete time and $\mb{M}$ is a finite state space.
In our context we can associate objects of $\mb{T}$ as type $\mb{I}$ (or $\mb{J}$) and objects of the state space, $\mb{M}$, to be discrete vectors (or arrays) whose elements are of the same type $\mb{I}$ (or $\mb{J}$) (see Section \ref{sdl}).
In the simplest case $\mb{M}$ is a one-dimensional state space containing objects that are scalars of type $\mb{I}$ (or $\mb{J}$).

It will often be convenient to set the ordered list of objects of type $\mb{T}$ as a sublist of the nonnegative integers $\mb{I}_0$.
If $t:\mb{I}_0$ is the time parameter and $v$ is the initial state, then the map $\Phi$ satisfies the properties
\be
\bal
\Phi(0,v) = & v \\
\Phi(t_2,\Phi(t_1,v)) = & \Phi(t_1+t_2,v), \qquad t_1,~t_2,~t_1+t_2:\mb{T}
\eal
\ee

We will often refer to systems that are directly implemented on a deterministic machine with finite memory as \emph{finite dynamical systems}\index{finite dynamical system}.
This may differ slightly from definitions of finite dynamical systems found in contemporary literature.   
Sometimes we will use the expression \emph{fully discrete system}\index{fully discrete system} to stress that $\Phi$ is a map over the finite lists of objects that are associated with the types $\mb{T}$ and $\mb{M}$.
Dynamical systems are usually defined in a more general sense that include the following.

\emph{Real dynamical system:}\index{real dynamical system} Here $\mb{T}$ is an open interval of the set of real numbers $\mb{R}$ and $\mb{M}$ is a manifold locally diffeomorphic to a Banach space.
For the case $\mb{T}=\mb{R}$ the system is called global.
If $\Phi$ is continuously differentiable then the system is said to be a differentiable dynamical system.

\emph{Discrete dynamical system:}\index{discrete dynamical system} As with a real dynamical system, $\mb{M}$ is a manifold locally diffeomorphic to a Banach space but $\mb{T}$ is a set of integers.

\emph{Cellular automata:}\index{cellular automata} Our definition of a dynamical system closely resembles cellular automata.
More generally cellular automata are characterized by $\mb{T}$ as a lattice of integers or can be a higher dimensional integer lattice.
$\mb{M}$ is a finite integer lattice in one or more dimensions.

A survey of the current literature indicates that most of the analysis on nonlinear systems is carried out in the context of real and discrete dynamical systems.
There is no single global method of analysis and the choice of the theoretical tools that are used depend on the specific properties of the system under investigation.

The behavior of the solutions in the vicinity of fixed points\index{fixed points} are of particular interest.
Fixed points can act as local attractors where a trajectory of $v$ can enter a basin of attraction about the fixed point and remain within that region.
Once captured within a basin of attraction the trajectory of $v$ need not converge to the fixed point.
If the trajectory does converge to the fixed point the attractor is said to be locally asymptotically stable.
If the fixed point convergence is independent of the initial condition then the fixed point is said to be globally asymptotically stable.

While these continuous based methods provide useful insights into the properties of complex solutions generated by nonlinear systems they may not be compatible with tests of computability in our formal system.
This means that we need to find other methods of analysis that target the specific issues that arise when working on $\mathfrak{M}(\mathcal{K},\mathcal{L},\mathcal{M})$.

In this chapter we will consider the simplest case where the state space $\mb{M}$ is one-dimensional, i.e. $\mb{M} = \mb{I}$.
Extensions to higher dimensions will be discussed in more detail in a later chapter. 

A finite dynamical system based on a one-dimensional state space can be represented by the difference equation 
\ben\label{int010}
\bal
v^{(t)} = f(v^{(t-1)}), \quad t=1,\ldots,n
\eal
\een
where $f:\mb{I} \to \mb{I}$ and $v^{(t)}:\mb{I}$, $t=0,1,\dots,n$, for some $n:\mb{I}_+$, is a sequence that defines the evolution of the variable $v$ in discrete time $t$.
The sequence $v^{(t)}:\mb{I}$, $t=1,\dots,n$, is generated from (\ref{int010}) by first prescribing an initial condition $v^{(0)}:\mb{I}$. 
Here we shall work over $\mb{I}$ but the results should also be applicable over $\mb{J}$, i.e $f: \mb{J} \to \mb{J}$.

The difference equation (\ref{int010}) suggest that the assignment map $f:\mb{I} \to \mb{I}$ can be expressed concisely as a function in terms of conventional mathematical notation.
This will not always be the case and we should regard the assignment map $f:\mb{I} \to \mb{I}$ to be constructed in the more general context of an algorithm.
When dealing with application specific assignment programs associated with some assignment map $f:\mb{I} \to \mb{I}$ we can expect that the input list will often include the primary input variable, $v$, along with some additional constant parameters that are employed in the algorithm that lead to the evaluation of the primary output variable, $w$.
If the assignment program is constructed as a non-atomic program we also need to include in the output list the variables associated with the intermediate calculations.

As an example, consider the primary variables $v,w:\mb{I}$ of the assignment map $f:v \mapsto w$ evaluated by the quadratic equation
\ben\label{slap10}
w := a*v*v + b*v + c
\een
where $a,b,c:\mb{I}$ are prescribed constant coefficients.

The map (\ref{slap10}) can be constructed from the non-atomic program list
\ben\label{slap20}
f~[a~v~b~c]~[d~e~g~h~w] = \left \{
\begin{array}{l}
~mult~[a~v]~[d] \\
~mult~[d~v]~[e] \\
~mult~[b~v]~[g] \\
~add~[e~g]~[h] \\
~add~[h~c]~[w] \\
\end{array}
\right .
\een
Following the structural rules for program lists as outlined in Definition \ref{prog}, under the representation $f~x~y$ the assignment program (\ref{slap20}) has the input list $x=[a~v~b~c]$ and the output list $y=[d~e~g~h~w]$.
Here $v$ is the primary input variable and $a,~b$ and $c$ are constant parameters whose values have been assigned prior to entry to the program $f$.
The output list contains the primary output variable $w$ along with $d,~e,~g$ and $h$ that are variables associated with the intermediate calculations leading to the evaluation of $w$.

We will often use the shorthand notation for an application specific assignment program
\be
f~\la v \ra~\la w \ra = f~x~y
\ee
The input list $x = \la v \ra$ indicates that $v \in x$ is the primary input variable and the remaining constant parameter elements of $x$ are not shown.
Similarly, the output list $y=\la w \ra$ indicates that $w \in y$ is the primary output variable and the variables associated with the intermediate calculations are not shown.

When using this shorthand notation, if a program of the form $f~\la v \ra~\la w \ra$ appears in a program list it is understood that the introduction of new variable names in the subprograms that follow it do not coincide with any variable names not shown in the I/O lists $\la v \ra$ and $\la w \ra$.

Alternatively, we can define $f~[v]~[w]$ as a atomic program so that the values of the constant parameters are set within the program and, along with the intermediate variables, are discarded immediately after its execution.
Here $f~[v]~[w]$ is not atomic in the stricter sense of Definition \ref{atomic} and can be referred to as a pseudo-atomic program\index{pseudo-atomic program}.
This, to some extent, alleviates the need to carry on too many variables but there are disadvantages.
Firstly, the properties of the internal core algorithm of the program $f~[v]~[w]$ can only be revealed by supplying a collection of application specific axioms.
Such axioms are not needed through an explicit program list such as (\ref{slap20}). 
The second disadvantage is of a practical computational nature and arises when the program $f~[v]~[w]$ is called many times when it forms the core of an iteration.
In such a case there will be a loss of computational efficiency due to the repeated resetting of the constant parameters every time the program $f$ is called.

\section{Atomic iteration programs.}

We associate the assignment map $f:\mb{I} \to \mb{I}$ with a program $f~\la v \ra~\la w \ra$ and replace (\ref{int010}) by the program list
\ben \label{int020}
[f~\la v^{(t-1)} \ra~\la v^{(t)} \ra]_{t=1}^n
\een
with the understanding that the values of the initial condition $v^{(0)}$ and the iteration count, $n$, are assigned by the $read$ program in the general program structure (\ref{core}).

In application the input parameter $n:\mb{I}_+$ of (\ref{int020}) is typically very large and it is inconvenient to store the entire sequence of the output lists $\la v^{(t)} \ra$, $t=1,\ldots,n$.
One can construct an iteration assignment program\index{iteration program}
\ben \label{int040}
itf~\la v~n \ra~\la w \ra
\een
where $v$ is the initial condition (equivalent to $v^{(0)}$ in (\ref{int020})) and $w$ is the value assigned output obtained after $n:\mb{I}_0$ iterations of $f$ (equivalent to $v^{(n)}$ in (\ref{int020})).

We shall regard $itf~\la v~n \ra~\la w \ra$ as an atomic program that is constructed by some imperative language using an iteration loop as follows.  
\ben \label{int050}
itf~\la v~n \ra~\la w \ra = \left \{
\begin{array}{l}
	n:\mb{I},~v:\mb{I},~w:\mb{I} \\
	\ldots \\
	t:\mb{I},~z:\mb{I} \\
	le~[0~n]~[~] \\
	w:=v \\
	do~t=1,n \\
	~~~z:=w \\
	~~~f~\la z \ra~\la w \ra \\
	end~do \\
\end{array}
\right .
\een
Under the representation $itf~x~y = itf~\la v~n \ra~\la w \ra$, the second line indicated by the dots, $\ldots$, is meant to represent the type checks of the parameters of $x \setminus [v~n]$ and $y \setminus w$.
These are the same parameter names not shown in the I/O lists of $f~\la z \ra~\la w \ra$.   

Note that (\ref{int050}) represents a program constructed from an imperative language and each preceding solution of the iteration is discarded through the reassignment $z:=w$.
Here the variables $t:\mb{I}$ and $z:\mb{I}$ are defined internally and are released from memory storage once the program has been executed.
The $do$-loop is not activated when $n=0$, in which case the value assignment $w:=v$ is returned as output.
If the user prescribes $n$ as a negative integer the atomic program $le~[0~n]~[~]$, and hence $itf~\la v~n \ra~\la w \ra$, will halt with an execution error.

We observe that the iteration assignment program $itf~\la v~n \ra~\la w \ra$ can be associated with the function $\Phi(n,v)$ of our definition of a finite dynamical system given in the first section of this chapter.
Sometimes one may be interested in storing intermediate steps.
In such a case we introduce the desired intermediate state variables $v^{(l)}$, $l=1,\ldots,k$, for some $k:\mb{I}_+$, associated with the prescribed iteration numbers $n^{(l)}:\mb{I}_+$, $l=1,\ldots,k-1$ and construct the program
\be
[itf~\la v^{(l-1)}~n^{(l)} \ra~\la v^{(l)} \ra ]_{l=1}^k
\ee
where each $v^{(l)}$ is evaluated after $n^{(l)}$ iterations from the starting value $v^{(l-1)}$.
Here $v^{(0)}$ is the prescribed initial state.
To avoid introducing too many variables we shall work with the iteration program (\ref{int050})

As already mentioned, in contemporary analysis of real and discrete dynamical systems, the behavior of the solutions in the vicinity of fixed points is of particular interest.
The behavior of solutions about fixed points of fully discrete systems are less well known.
While our primary concern here is to establish computability of applications based on some assignment program $f~\la v \ra~\la w \ra$ we also desire tools that will allow us to investigate solution behaviors, often involving fixed points.
A fixed point of an assignment program $f~\la v \ra~\la w \ra$ can be defined as $v^*:\mb{I}$ such that the program
\ben \label{int030}
[f~\la v^* \ra~\la w \ra~eqi~[w~v^*]~[~]]
\een 
is computable.

\section{Discrete intervals.}\label{si}\index{discrete interval}

If it is feasible to obtain bounds on the primary output variable of the program $f~\la v \ra~\la w \ra$ over a discrete sub-domain then computability may be established on that sub-domain.
Discrete sub-domains can be constructed using discrete intervals\index{interval}. 
We introduce the notion of a discrete interval over $\mb{I}$ but all of the results presented below will also hold for discrete intervals over $\mb{J}$.

A discrete interval over $\mb{I}$ can be represented by the two element list
\be
[a~b], \qquad a,b:\mb{I},~a \leq b
\ee
where $a$ and $b$, respectively, are the lower and upper bounds, respectively, of the interval.
Note that $[a~b]$ is a two element list that represents the larger list $[a~\ldots~b]$ that contains all elements of $\mb{I}$ between and including $a$ and $b$.
So that the machine can recognize its distinction from a standard list, a two element list $p=[a~b]$ that represents an interval over $\mb{I}$ will be assigned the type $p:\mb{B}$.
We say that $v:\mb{I}$ is an element of the interval $[a~b]:\mb{B}$ to mean that $v \in [a~\ldots~b]$.
We also say that the interval represented by $[a~b]:\mb{B}$ is an \emph{interval enclosure}\index{interval enclosure} of the interval represented by $[c~d]:\mb{B}$ to mean that $[c~\ldots~d] \subseteqq [a~\dots~b]$.
For this to hold we simply require that $a \leq c$ and $d \leq b$.
An interval enclosure $[a~b]:\mb{B}$ of $[c~d]:\mb{B}$ is denoted by $[c~d] \subseteq [a~b]$, where the symbol $\subseteq$ is to be distinguished from the symbol $\subseteqq$ that is used for sublists of standard lists.
A single point interval $[a~a]$ contains only the single element $a:\mb{I}$.
The empty interval is denoted by
\be
ei=[~]
\ee
and can be regarded as a constant for type $\mb{B}$ objects.

Let $p:\mb{B}$ be the interval that represents the standard list $[v_1 \dots v_n]$, for some $n:\mb{I}_+$.
The map $f: \mb{I} \to \mb{I}$ under the restriction to $p$ generates the standard list $q=[w_1 \ldots w_n]$, where each $w_i$ is the evaluation of $f~\la v_i \ra~\la w_i \ra$, $i=1,\ldots,n$.
We can define the interval that represents the tightest bound of $f: \mb{I} \to \mb{I}$ under the restriction to $p$ by
\be
R(f,p) = [\max q~\min q]
\ee   
where $\min q$ and $\max q$, respectively, are the lower and upper bounds, respectively, of the list $q$. 

Let $c:\mb{I}$ and $d:\mb{I}$, respectively, be any lower and upper bound, respectively, of $f:\mb{I} \to \mb{I}$ over the interval $p=[a~b]$.
We write
\ben \label{int100}
B(f,p)=[c~d], \quad c,d:\mb{I}
\een 
to represent an interval that bounds $f:\mb{I} \to \mb{I}$ over the interval represented by $p$.
Here, $c$ and $d$, respectively, need not be the greatest lower bound and least upper bound, respectively, of $f$ over $p$.
In general we have
\ben \label{int101}
R(f,p) \subseteq B(f,p)
\een 
The aim is to find $p:\mb{B}$ and a suitable interval $q=B(f,p)$ such that $q$ is a sufficiently tight enclosure of $R(f,p)$.

If the assignment map $f:\mb{I} \to \mb{I}$ can be expressed as simple operations of arithmetic the following rules for interval addition, subtraction and multiplication can be useful.\index{interval arithmetic}
\ben\label{iarith01}
\bal
& [a~b]+[c~d] = [a+c~b+d] \\
& [a~b]-[c~d] = [a-d~b-c] \\
& [a~b]*[c~d] = [e~f] \\
& e= \min~[a*c~a*d~b*c~b*d] \\
& f= \max~[a*c~a*d~b*c~b*d] \\
\eal
\een
Since we are working over $\mb{I}$, the above rules are conditional on the existence of the sums and multiplications of the interval bounds.

For discrete intervals the operation of interval division is not well defined. 
It is sometimes convenient to define the multiplication of an interval with a scalar constant $c$ given by
\ben \label{iarith02}
c*[a~b] = \left \{
\begin{array}{ll}
~[c*a~c*b], & c \geq 0 \\
~[c*b~c*a], & c < 0 \\
\end{array}
\right .
\een

Interval arithmetic is not distributive and satisfies the weaker rule
\be
r*(p+q) \subseteq r*p + r*q, \quad p,q,r:\mb{B}
\ee
An enclosure of the union of two intervals can be defined by
\ben\label{iarith04}
\bal
& [a~b] \cup [c~d] \subseteq [e~f] , \quad e= \min [a~c],~f= \max [b~d] \\
\eal
\een

It is often the case that for a given interval $p=[a~b]:\mb{B}$ the construction of $B(f,p)$ will be a large over estimate of $R(f,p)$.
One can construct a tighter enclosure by splitting the single interval $[a~b]$ into smaller intervals
\ben\label{iarith05}
\bal
& p = \cup_{i=1}^m p^{(i)}, \quad p^{(i)}=[a^{(i)}~a^{(i+1)}],~i=1,\ldots,m-1 \\
\eal
\een
where
\ben\label{iarith06}
\bal
& a^{(1)}=a,~a^{(m)}=b \\
\eal
\een
If we can make use of the above rules of discrete interval arithmetic we have in general
\ben\label{iarith07}
\bal
R(f,p) \subseteq \cup_{i=1}^m B(f,p^{(i)}) \subseteq B(f,p) \\
\eal
\een

\section{Atomic programs.} \label{sintap}

To implement the ideas of the previous section we present an application for the computability of finite dynamical systems on the working platform $\mathfrak{M}(\mathcal{K},\mathcal{L},\mathcal{M})$.
For brevity we present only a condensed version that may be considered to contain some core features from which a more comprehensive theory can be built. 
In such an application we will need to include all of the atomic programs and axioms of arithmetic over $\mb{I}$.
To these are appended additional axioms that specifically address the computability of finite dynamical systems.

The following is a description of the atomic programs for this application.

\vspace{5mm}

\begin{tabular}{|c|c|}
	\hline
	Atomic program names & Atomic program type \\
	\hline
	$typedi,~eqdi,~intelt,~intenc$ & $\mb{P}_{type}$ \\
	\hline
	$int,~iddi,~adddi,~multdi,~smultdi,~cup$ & $\mb{P}_{assign}$ \\
	\hline
\end{tabular}

There are three special non-atomic programs that will be defined later.

\begin{tabular}{|c|c|}
\hline
Special non-atomic program names & Structure \\
\hline
$min4,~max4,~mlt4$ & program list \\
\hline
\end{tabular} 

\vspace{5mm}

\parskip0pt

\vspace{5mm}
\textbf{Check type interval list.}

\textbf{\textit{Syntax.}} $typedi~[p]~[~]$.

\textbf{\textit{Program Type.}} $\mb{P}_{type}$.

\textbf{\textit{Type checks.}} $p:\mb{B}$.

\textbf{\textit{Description.}} $typedi$ checks that the value assignment of the variable $p$ is of type $\mb{B}$.
Here $p$ represents a discrete interval and is given by the two element list
\be
p=[a~b]
\ee
for some $a,b:\mb{I}$ such that $a \leq b$ are the lower and upper bounds of the interval.
The values of the bounds of the interval $p$ are assigned prior to entry to the program $typedi$ and are determined internally by $typedi$.
If the interval $p$ has been set to the empty interval, $ei=[~]$, prior to entry to $typedi$ no attempt is made to determine its bounds. 
$typedi$ halts with an execution error if there is a type violation.

\vspace{5mm}
\textbf{Interval equality.}

\textbf{\textit{Syntax:}} $eqdi~[p~q]~[~]$.

\textbf{\textit{Program Type:}} $\mb{P}_{type}$.

\textbf{\textit{Type checks:}} $p,q:\mb{B}$ and $p=q$.

\textbf{\textit{Description:}} $eqdi$ first checks that the value assignments of the variables $p$ and $q$ are of type $\mb{B}$.
It then checks the value assignment equality $p=q$.
If $p=[a~b]$ and $q=[c~d]$, for some $a,b,c,d:\mb{I}$, then for $p=q$ to hold $eqdi$ just checks the value equalities $a=c$ and $b=d$.
The values of the bounds of the intervals $p$ and $q$ are assigned prior to entry to the program $eqdi$ and are determined internally by $eqdi$.
For any interval that has been set to the empty interval, $ei=[~]$, prior to entry to $eqdi$ no attempt is made to determine its bounds.
$eqdi$ halts with an execution error if there is a type violation.
This includes the case where the interval equality does not hold.

\vspace{5mm}
\textbf{Check element of an interval.}

\textbf{\textit{Syntax:}} $intelt~[v~p]~[~]$.

\textbf{\textit{Program Type:}} $\mb{P}_{type}$.

\textbf{\textit{Type checks:}} $v:\mb{I}$, $p:\mb{B}$ and $p \neq ei$.

\textbf{\textit{Description:}} $intelt$ first checks that the value assignment of the variable $v$ is of type $\mb{I}$ and the value assignment of the variable $p$ is of type $\mb{B}$ such that $p \neq ei$.
It then checks that $v$ is an element contained in the interval list represented by $p$.
If $p=[a~b]$, for some $a,b:\mb{I}$, then for $v$ to be an element of $p$ $intelt$ just checks that $a \leq v \leq b$.
The values of the bounds of the interval $p$ are assigned prior to entry to the program $intelt$ and are determined internally by $intelt$.
$intelt$ halts with an execution error if there is a type violation.
This includes the case where $v$ is not an element contained in the interval $p$ or $p$ has been set to the empty interval prior to entry to $intelt$.

\vspace{5mm}
\textbf{Check interval enclosure.}

\textbf{\textit{Syntax:}} $intenc~[q~p]~[~]$

\textbf{\textit{Program Type:}} $\mb{P}_{type}$.

\textbf{\textit{Type checks:}} $p,q:\mb{B}$, $q \subseteq p$ and $p,q \neq ei$.

\textbf{\textit{Description:}} $intenc$ first checks that the value assignments of the variables $p$ and $q$ are of type $\mb{B}$ such that $q \subseteq p$ and $p,q \neq ei$.
If $p=[a~b]$ and $q=[c~d]$, for some $a,b,c,d:\mb{I}$, then for $p$ to be an interval enclosure of $q$ $intenc$ just checks that $a \le c$ and $d \le b$.
The values of the bounds of the intervals $p$ and $q$ are assigned prior to entry to the program $intenc$ and are determined internally by $intenc$.
$intenc$ halts with an execution error if there is a type violation. 
This includes the case where $p$ is not an interval enclosure of $q$ or any one of the intervals $p$ and $q$ has been set to the empty interval, $ei=[~]$, prior to entry to $intenc$.

\vspace{5mm}
\textbf{Construct an interval.}

\textbf{\textit{Syntax:}} $int~[a~b]~[p]$.

\textbf{\textit{Program Type:}} $\mb{P}_{assign}$.

\textbf{\textit{Type checks:}} $a:\mb{I}$, $b:\mb{I}$ and $a \leq b$.

\textbf{\textit{Assignment map:}} $p:=[a~b]$.

\textbf{\textit{Type assignment.}} $p::\mb{B}$.

\textbf{\textit{Description:}} $int$ first checks that the variables $a$ and $b$ have been assigned values of type $\mb{I}$, such that $a \leq b$.
If there are no type violations $int$ constructs the two element list
\be
p:=[a~b]
\ee
This list construction is accompanied by the type assignment $p::\mb{B}$ to indicate that $p$ is an interval that represents the standard list $[a~\ldots~b]$.
Here $a:\mb{I}$ and $b:\mb{I}$, respectively, are the lower and upper bounds, respectively, of the interval $p$.
$int$ halts with an execution error if there is a type violation.
A type violation includes the case $a>b$.

\vspace{5mm}
\textbf{Interval identity assignment.}

\textbf{\textit{Syntax:}} $iddi~[p]~[q]$.

\textbf{\textit{Program Type:}} $\mb{P}_{assign}$.

\textbf{\textit{Type checks:}} $p:\mb{B}$.

\textbf{\textit{Assignment map:}} $q:=p$.

\textbf{\textit{Type assignment.}} $q::\mb{B}$.

\textbf{\textit{Description:}} $iddi$ first checks that the value assignment of the variable $p$ is of type $\mb{B}$.
If there is no type violation $iddi$ makes the interval identity assignment $q:=p$.
This is accompanied by the type assignment $q::\mb{B}$.
The values of the bounds of the interval $p$ are assigned prior to entry to the program $iddi$ and are determined internally by $iddi$.
If the interval $p$ has been set to the empty interval, $ei=[~]$, no attempt is made to determine its bounds and the assignment $q:=ei$ is made.
$iddi$ halts with an execution error if there is a type violation.

\vspace{5mm}
\textbf{Interval addition.}

\textbf{\textit{Syntax:}} $adddi~[p~q]~[r]$.

\textbf{\textit{Program Type:}} $\mb{P}_{assign}$.

\textbf{\textit{Type checks:}} $p,q:\mb{B}$, $p,q \neq ei$.

\textbf{\textit{Assignment map:}} $r:=p+q$.

\textbf{\textit{Type assignment.}} $r::\mb{B}$.

\textbf{\textit{Description:}} $adddi$ first checks that the value assignments of the variables $p$ and $q$ are of type $\mb{B}$ such that $p,q \neq ei$.
If $p=[a~b]$ and $q=[c~d]$, for some $a,b,c,d:\mb{I}$, then $adddi$ attempts to construct the two element list
\be
\bal
r:= & [a~b]+[c~d] = [a+c~b+d] \\
\eal
\ee
This may fail if any one of the sums $a+c$ and $b+d$ does not exist.
A successful value assignment is accompanied by the type assignment $r::\mb{B}$.
The values of the bounds of the intervals $p$ and $q$ are assigned prior to entry to the program $adddi$ and are determined internally by $adddi$.
$adddi$ halts with an execution error if there is a type violation.

\vspace{5mm}
\textbf{Interval multiplication.}

\textbf{\textit{Syntax:}} $multdi~[p~q]~[r]$.

\textbf{\textit{Program Type:}} $\mb{P}_{assign}$.

\textbf{\textit{Type checks:}} $p,q:\mb{B}$, $p,q \neq ei$.

\textbf{\textit{Assignment map:}} $r:=p*q$.

\textbf{\textit{Type assignment.}} $r::\mb{B}$.

\textbf{\textit{Description:}} $multdi$ first checks that the value assignments of the variables $p$ and $q$ are of type $\mb{B}$ such that $p,q \neq ei$.
If $p=[a~b]$ and $q=[c~d]$, for some $a,b,c,d:\mb{I}$, then $multdi$ attempts to construct the two element list
\be
\bal
& r := [a~b]*[c~d] = [e~f] \\
& e= \min [a*c~a*d~b*c~b*d] \\
& f= \max [a*c~a*d~b*c~b*d] \\
\eal
\ee
This may fail if any one of the scalar multiplications $a*c,~a*d,~b*c$ and $b*d$ does not exist.
A successful value assignment is accompanied by the type assignment $r::\mb{B}$.
The values of the bounds of the intervals $p$ and $q$ are assigned prior to entry to the program $multdi$ and are determined internally by $multdi$.
$multdi$ halts with an execution error if there is a type violation.

\vspace{5mm}
\textbf{Interval scalar multiplication.}

\textbf{\textit{Syntax:}} $smultdi~[c~p]~[r]$.

\textbf{\textit{Program Type:}} $\mb{P}_{assign}$.

\textbf{\textit{Type checks:}} $c:\mb{I}$, $p:\mb{B}$, $p \ne ei$.

\textbf{\textit{Assignment map:}} $r:=c*p$.

\textbf{\textit{Type assignment.}} $r::\mb{B}$.

\textbf{\textit{Description:}} $smultdi$ first checks that the value assignment of the variable $c$ is of type $\mb{I}$ and  the value assignment of the variable $p$ is of type $\mb{B}$ such that $p \ne ei$.
If $p=[a~b]$, for some $a,b:\mb{I}$, then $smultdi$ attempts to construct the two element list
\be
r := c*[a~b] = \left \{
\begin{array}{ll}
~[c*a~c*b] & c \geq 0 \\
~[c*b~c*a] & c < 0 \\
\end{array}
\right .
\ee
This may fail if any one of the scalar multiplications $c*a$ and $c*b$ does not exist.
A successful value assignment is accompanied by the type assignment $r::\mb{B}$.
The values of the bounds of the interval $p$ are assigned prior to entry to the program $smultdi$ and are determined internally by $smultdi$.
$smultdi$ halts with an execution error if there is a type violation.

\vspace{5mm}
\textbf{Enclosure of the union of intervals.}

\textbf{\textit{Syntax:}} $cup~[p~q]~[r]$.

\textbf{\textit{Program Type:}} $\mb{P}_{assign}$.

\textbf{\textit{Type checks:}} $p,q:\mb{B}$.

\textbf{\textit{Assignment map:}} $p \cup q \subseteq r$.

\textbf{\textit{Type assignment.}} $r::\mb{B}$.

\textbf{\textit{Description:}} $cup$ first checks that the value assignments of the variables $p$ and $q$ are of type $\mb{B}$.
If $p=[a~b]$ and $q=[c~d]$, for some $a,b,c,d:\mb{I}$, then $cup$ constructs the two element list
\be
\bal
r:= & [e~f], \quad e=\min [a~c],~f=\max [b~d] \\
\eal
\ee
The value assignment is accompanied by the type assignment $r::\mb{B}$.
The values of the bounds of the intervals $p$ and $q$ are assigned prior to entry to the program $cup$ and are determined internally by $cup$.
For any interval that has been set to the empty interval, $ei=[~]$, prior to entry to $cup$ no attempt is made to determine its bounds.
Interval union with the empty interval follows the rule $p \cup ei=p$.
$cup$ halts with an execution error if there is a type violation.

\parskip10pt

\vspace{5mm}

We will also make use of the following application specific atomic program

\begin{tabular}{|c|c|}
	\hline
	Atomic program name & Atomic program type \\
	\hline
	$itf$ & $\mb{P}_{assign}$ \\
	\hline
\end{tabular}

and special non-atomic programs

\vspace{5mm}

\begin{tabular}{|c|c|}
\hline
Special non-atomic program names & Structure \\
\hline
$f,~intf$ & program list and/or disjunction \\
\hline
\end{tabular}

\parskip0pt

\vspace{5mm}
\textbf{Assignment program (application specific non-atomic program).}

\textbf{\textit{Syntax:}} $f~\la v \ra~\la w \ra$.

\textbf{\textit{Program Type:}} $\mb{P}_{assign}$.

\textbf{\textit{Type checks:}} $v:\mb{I}$.

\textbf{\textit{Assignment map:}} $f:\mb{I} \to \mb{I}$, application specific.

\textbf{\textit{Type assignment.}} $w::\mb{I}$.

\textbf{\textit{Description:}} $f$ first checks that the variable $v$ has been assigned the type $\mb{I}$.
The types of the values assigned to all other input parameters not shown in the input list $\la v \ra$ are also checked.
Type checks are performed through the subprograms that make up the list and/or operands of the disjunctions of $f~\la v \ra~\la w \ra$. 
If there are no entry type violations $f~\la v \ra~\la w \ra$ then attempts to assign a value to $w:\mb{I}$ through its associated assignment map $f:\mb{I} \to \mb{I}$.
A successful value assignment is accompanied by the type assignment $w::\mb{I}$ along with type assignments of all variables associated with intermediate calculations not shown in the output list $\la w \ra$.
The program $f~\la v \ra~\la w \ra$ is an application specific program constructed by the user.
$f$ halts with an execution error if there is a type violation.

\vspace{5mm}
\textbf{Iteration assignment program (application specific atomic program).}

\textbf{\textit{Syntax:}} $itf~\la v~n \ra~\la w \ra$.

\textbf{\textit{Program Type:}} $\mb{P}_{assign}$.

\textbf{\textit{Type checks:}} $v:\mb{I}$, $n:\mb{I}_0$.

\textbf{\textit{Assignment map:}} See algorithm (\ref{int050}).

\textbf{\textit{Type assignment.}} $w::\mb{I}$.

\textbf{\textit{Description:}} $itf$ first checks that the variable $n$ has been assigned a value of type $\mb{I}_0$ and the variable $v$ has been assigned a value of type $\mb{I}$.
Type checks of the values assigned to all other input parameters not shown in the input list $\la v~n \ra$ are also checked.
$itf$ then attempts to assign a value to $w$ through the iteration program (\ref{int050}) for the associated assignment program $f~\la v^\prime \ra~\la w^\prime \ra$.
A successful value assignment is accompanied by the type assignment $w::\mb{I}$ along with type assignments of the values assigned to all variables associated with intermediate calculations not shown in the output list $\la w \ra$.
The program $itf$ is application specific and depends on the user supplied assignment program $f~\la v^\prime \ra~\la w^\prime \ra$.
$itf$ halts with an execution error if there is a type violation.

\vspace{5mm}
\textbf{Construct an interval bound of $f$ over $p$ (application specific non-atomic program).}

\textbf{\textit{Syntax:}} $intf~\la p \ra~\la q \ra$.

\textbf{\textit{Program Type:}} $\mb{P}_{assign}$.

\textbf{\textit{Type checks:}} $p:\mb{B}$.

\textbf{\textit{Assignment map:}} $q=\cup_{i=1}^m B(f,p^{(i)})$, $p=\cup_{i=1}^m p^{(i)}$ for some $m:\mb{I}_+$.

\textbf{\textit{Type assignment.}} $q::\mb{B}$.

\textbf{\textit{Description:}} $intf$ first checks that the value assignment of the variable $p$ is of type $\mb{B}$.
Type checks of the values assigned to all other input parameters not shown in the input list $\la p \ra$ are also checked.
Type checks are performed through the subprograms that make up the list and/or operands of the disjunctions of $intf~\la p \ra~\la q \ra$.
The algorithm of the program $intf~\la p \ra~\la q \ra$ will be constructed by the user such that it computes a sufficiently tight interval enclosure of $R(f,p)$.
Hence the construction of the program $intf~\la p \ra~\la q \ra$ will depend on the properties of $f~\la u^\prime \ra~\la v^\prime \ra$.
If the single interval $p$ is adequate then $intf$ constructs the interval list, $q:\mb{B}$, given by
\be
q=B(f,p)
\ee
Otherwise $intf$ will seek a suitable partition
\be
p=\cup_{i=1}^m p^{(i)}
\ee
for some $m:\mb{I}_+$ and constructs the interval $q:\mb{B}$ given by
\be
q=\cup_{i=1}^m B(f,p^{(i)})
\ee
A successful value assignment is accompanied by the type assignment $q::\mb{B}$ along with type assignments of the values assigned to all variables associated with intermediate calculations not shown in the output list $\la q \ra$.
The intervals $B(f,p^{(i)})$ can be obtained by the applications of (\ref{iarith01})-(\ref{iarith02}) on the arithmetic operations contained in $f~\la v^\prime \ra~\la w^\prime \ra$.
In such a case we are guaranteed that $q$ will be a discrete interval enclosure of $R(f,p)$.
The values of the bounds of the interval $p$ are assigned prior to entry to the program $intf$ and are determined internally by $intf$.
$intf$ halts with an execution error if there is a type violation.

\parskip10pt

\vspace{5mm}

\textbf{Notes.}

\begin{itemize}

\item An enclosure of the union of more than two intervals can be obtained by successive application of the binary operation of the program $cup$.

\item An atomic program for interval subtraction is not included since it can be constructed by the rules of interval addition and scalar multiplication.

\end{itemize}

\section{Properties of discrete intervals.}\label{spoi1}

Most of the atomic programs presented in the previous section can be constructed as special non-atomic programs using the atomic scalar programs of Section \ref{apfaoi}.
Such programs are not atomic in the the stricter sense of Definition \ref{atomic} and can be regarded as pseudo-atomic. 
Since we have defined them as atomic programs we need to supply the following axioms so that VPC can recognize their internal algorithms.

\textbf{Interval construction.}

\begin{axint} \label{axdi01}
	\be
	\bal
	& [le~[a~b]~[~]~int~[a~b]~[p]]
	\eal
	\ee
\end{axint}

\begin{axint}
	\be
	\bal
	& [int~[a~b]~[p]~le~[a~b]~[~]]
	\eal
	\ee
\end{axint}

\begin{axint}
	\be
	\bal
	& [[int~[a~b]~[p]~int~[c~d]~[q]~eqdi~[q~p]~[~]]~eqi~[c~a]~[~]]
	\eal
	\ee
\end{axint}
\begin{axint}
	\be
	\bal
	& [[int~[a~b]~[p]~int~[c~d]~[q]~eqdi~[q~p]~[~]]~eqi~[d~b]~[~]]
	\eal
	\ee
\end{axint}

\begin{axint}
	\be
	\bal
	& [[int~[a~b]~[p]~int~[c~d]~[q]~eqi~[c~a]~[~]~eqi~[d~b]~[~]]~eqdi~[q~p]~[~]]
	\eal
	\ee
\end{axint}

\textbf{Interval identity assignment.}
\begin{axint}
	\be
	\bal
	& [iddi~[p]~[q]~eqdi~[q~p]~[~]]
	\eal
	\ee
\end{axint}

\textbf{Elements of intervals.}
\begin{axint}
	\be
	\bal
	& [[int~[a~b]~[p]~intelt~[v~p]~[~]]~le~[a~v]~[~]]
	\eal
	\ee
\end{axint}
\begin{axint}
	\be
	\bal
	& [[int~[a~b]~[p]~intelt~[v~p]~[~]]~le~[v~b]~[~]]
	\eal
	\ee
\end{axint}

\begin{axint}
	\be
	\bal
	& [[int~[a~b]~[p]~le~[a~v]~[~]~le~[v~b]~[~]]~intelt~[v~p]~[~]] \\
	\eal
	\ee
\end{axint}

\textbf{Interval enclosures.}

\begin{axint}
	\be
	\bal
	& [[int~[a~b]~[p]~int~[c~d]~[q]~intenc~[q~p]~[~]]~le~[a~c]~[~]]
	\eal
	\ee
\end{axint}
\begin{axint}
	\be
	\bal
	& [[int~[a~b]~[p]~int~[c~d]~[q]~intenc~[q~p]~[~]]~le~[d~b]~[~]]
	\eal
	\ee
\end{axint}

\begin{axint}
	\be
	\bal
	& [[int~[a~b]~[p]~int~[c~d]~[q]~le~[a~c]~[~]~le~[d~b]~[~]]~intenc~[q~p]~[~]]
	\eal
	\ee
\end{axint}

\textbf{Union of intervals.}

\begin{axint}
	\be
	\bal
	& [[typedi~[p]~[~]~typedi~[q]~[~]]~cup~[p~q]~[r]]
	\eal
	\ee
\end{axint}
\begin{axint} \label{cup02}
	\be
	\bal
	& [[cup~[p~q]~[r]~cup~[q~p]~[s]]~eqdi~[s~r]~[~]]
	\eal
	\ee
\end{axint}
\begin{axint} \label{cup03}
	\be
	\bal
	& [[int~[a~b]~[p]~int~[c~d]~[q]~min~[a~c]~[e]~max~[b~d]~[f]~int~[e~f]~[s] \\
	&~~~cup~[p~q]~[r]]~eqdi~[r~s]~[~]] \\
	\eal
	\ee
\end{axint}

\begin{axint}
	\be
	\bal
	& [cup~[p~ei]~[r]~eqdi~[r~p]~[~]]
	\eal
	\ee
\end{axint}

\textbf{Notes.}

\begin{itemize}

\item For non-empty intervals the commutativity rule for the union of intervals, axdi \ref{cup02}, can be derived from axdi \ref{cup03}.
Axiom axdi \ref{cup02} is included to incorporate the commutativity of unions involving the empty interval.

\end{itemize}

\section{Discrete interval arithmetic.}\label{spoi2}

The axioms associated with the operations of discrete interval arithmetic will be denoted by axdia followed by a number.
In order to shorten the lengths of these axioms we introduce the following special non-atomic programs.

\textbf{Special non-atomic programs.}

\be
\bal
min4~[e~f~g~h]~[i~j~k] = & [min~[e~f]~[i]~min~[i~g]~[j]~min~[j~h]~[k]] \\
\eal
\ee
\be
\bal
max4~[e~f~g~h]~[i~j~k] = & [max~[e~f]~[i]~max~[i~g]~[j]~max~[j~h]~[k]] \\
\eal
\ee

\be
mlt4~[a~c~b~d]~[e~f~g~h] = [mult~[a~c]~[e]~mult~[b~c]~[f]~mult~[a~d]~[g]~mult~[b~d]~[h]]
\ee

\textbf{Interval addition.}

\begin{axdia}
	\be
	\bal
	& [[int~[a~b]~[p]~int~[c~d]~[q]~adddi~[p~q]~[r]]~add~[a~c]~[e]]
	\eal
	\ee
\end{axdia}
\begin{axdia}
	\be
	\bal
	& [[int~[a~b]~[p]~int~[c~d]~[q]~adddi~[p~q]~[r]]~add~[b~d]~[f]]
	\eal
	\ee
\end{axdia}
\begin{axdia}
	\be
	\bal
	& [[int~[a~b]~[p]~int~[c~d]~[q]~add~[a~c]~[e]~add~[b~d]~[f]]~adddi~[p~q]~[r]]
	\eal
	\ee
\end{axdia}

\begin{axdia}
	\be
	\bal
	& [[int~[a~b]~[p]~int~[c~d]~[q]~add~[a~c]~[e]~add~[b~d]~[f]~adddi~[p~q]~[r] \\
	&~~~int~[e~f]~[s]]~eqdi~[s~r]~[~]]
	\eal
	\ee
\end{axdia}

\textbf{Interval multiplication.}

\begin{axdia}
	\be
	\bal
	& [[int~[a~b]~[p]~int~[c~d]~[q]~multdi~[p~q]~[r]]~mlt4~[a~b~c~d]~[e~f~g~h]]
	\eal
	\ee
\end{axdia}
\begin{axdia}
	\be
	\bal
	& [[int~[a~b]~[p]~int~[c~d]~[q]~mlt4~[a~c~b~d]~[e~f~g~h]]~multdi~[p~q]~[r]]
	\eal
	\ee
\end{axdia}

\begin{axdia}
	\be
	\bal
    & [[int~[a~b]~[p]~int~[c~d]~[q]~multdi~[p~q]~[r]~mlt4~[a~c~b~d]~[e~f~g~h] \\
    &~~~min4~[e~f~g~h]~[i1~j1~k]~max4~[e~f~g~h]~[i2~j2~l] \\
    &~~~int~[k~l]~[s]]~eqdi~[s~r]~[~]] \\
	\eal
	\ee
\end{axdia}

\textbf{Scalar interval multiplication.}

\begin{axdia}
	\be
	\bal
	& [[int~[a~b]~[p]~smultdi~[c~p]~[r]]~mult~[c~a]~[e]]
	\eal
	\ee
\end{axdia}
\begin{axdia}
	\be
	\bal
	& [[int~[a~b]~[p]~smultdi~[c~p]~[r]]~mult~[c~b]~[f]]
	\eal
	\ee
\end{axdia}
\begin{axdia}
	\be
	\bal
	& [[int~[a~b]~[p]~mult~[c~a]~[e]~mult~[c~b]~[f]]~smultdi~[c~p]~[r]]
	\eal
	\ee
\end{axdia}

\begin{axdia}
	\be
	\bal
    & [[int~[a~b]~[p]~smultdi~[c~p]~[r]~mult~[c~a]~[e]~mult~[c~b]~[f]~le~[0~c]~[~] \\
    &~~~int~[e~f]~[s]]~eqdi~[s~r]~[~]] \\
	\eal
	\ee
\end{axdia}
\begin{axdia}
	\be
	\bal
    & [[int~[a~b]~[p]~smultdi~[c~p]~[r]~mult~[c~a]~[e]~mult~[c~b]~[f]~lt~[c~0]~[~] \\
    &~~~int~[f~e]~[s]]~eqdi~[s~r]~[~]] \\
	\eal
	\ee
\end{axdia}

\section{Axioms of computability.}\label{saoc}

Here we present the axioms of computability for application specific dynamical systems.
These axioms are labeled by axoc followed by a number.

Suppose that
\be\label{int300}
f~\la v \ra~\la w \ra
\ee
is the assignment program with the associated assignment map $f:\mb{I} \to \mb{I}$.
By construction the iteration assignment program $itf~\la v~n \ra~\la w \ra$, defined by (\ref{int050}), satisfies the axiom
\begin{axoc} \label{axoc01}
	\be
	\bal
	& [itf~\la v~0 \ra~\la w \ra~eqi~[w~v]~[~]]
	\eal
	\ee
\end{axoc}
This axiom reflects the property that the $do$-loop in (\ref{int050}) is not activated when $n=0$.
The iteration assignment program $itf$ also obeys the semi-group properties
\begin{axoc}  \label{axoc02}
	\be
	\bal
	& [[itf~\la v~1 \ra~\la w \ra~f~\la v \ra~\la z \ra]~eqi~[z~w]~[~]]
	\eal
	\ee
\end{axoc}
\begin{axoc} \label{axoc03}
	\be
	\bal
	& [[itf~\la v~n \ra~\la s \ra~itf~\la s~m \ra~\la w \ra~add~[n~m]~[l]~itf~\la v~l \ra~\la z \ra]~eqi~[z~w]~[~]] \\
	\eal
	\ee
\end{axoc}

Rather than investigate the detailed properties of solutions, our primary concern is that of computability.
More precisely, we wish to establish that given an initial condition, $v$, contained in some interval, $p$, the iteration assignment program, $itf~\la v~n \ra~\la w \ra$, will be computable for any $n:\mb{I}_0$.

We construct an interval represented by the two element list
\be
p=[a~b]
\ee
for some $a,b:\mb{I}$ such that $a \leq b$.
The program
\ben\label{int320}
intf~\la p \ra~\la q \ra
\een
is a user supplied program that attempts to construct the interval, $q:\mb{B}$, such that $q$ is a sufficiently tight interval enclosure of $R(f,p)$.

If the assignment program $f~\la v \ra~\la w \ra$ can be expressed in terms of basic operations of arithmetic we can construct the program $intf~\la p \ra~\la q \ra$ such that it applies the rules of interval arithmetic on the operations of scalar arithmetic at the core of $f~\la v \ra~\la w \ra$.
In such a case we can accept as an axiom
\begin{axoc} \label{axoc04}
	\be
	\bal
	~[[intf~\la p \ra~\la q \ra~intelt~[v~p]~[~]~f~\la v \ra~\la w \ra]~intelt~[w~q]~[~]]
	\eal
	\ee
\end{axoc}
This axiom states that if $p:\mb{B}$ represents an interval, $v:\mb{I}$ is an element contained in $p$ and $w:\mb{I}$ is obtained from the evaluation $f~\la v \ra~\la w \ra$ then $w$ is an element contained in $q:\mb{B}$.
This is equivalent to the statement that $q$ will be an interval enclosure of $R(f,p)$.
The aim is to find suitable intervals $p$ and $q$ such that $q$ will be a sufficiently tight interval enclosure of $R(f,p)$.

If in addition we can construct $p$ and $q$ such that $p$ is an interval enclosure of $q$ then we can apply the following axiom of computability.
\begin{axoc} \label{axoc05}
	\be
	\bal
	& [[intf~\la p \ra~\la q \ra~intenc~[q~p]~[~]~intelt~[v~p]~[~]~le~[0~n]~[~]]~itf~\la v~n \ra~\la w \ra] \\
	\eal
	\ee
\end{axoc}
Axiom axoc \ref{axoc05} states that given $p:\mb{B}$ such that $p$ is an interval enclosure of $q:\mb{B}$, i.e. $R(f,p) \subseteq q \subseteq p$, then for any element $v:\mb{I}$ contained in $p$ the iteration program $itf~\la v~n \ra~\la w \ra$ will be computable for any $n:\mb{I}_0$.
The main task is to find appropriate intervals $p:\mb{B}$ and $q:\mb{B}$ such that $p$ encloses $q$.
Once we have $q \subseteq p$ we can then apply the axiom of computability axoc \ref{axoc05} and we are done.

\vspace{5mm}

\textbf{Notes.}
\begin{itemize}

\item If the program $intf~\la p \ra~\la q \ra$ employs the rules of interval arithmetic to construct $q$ from $p$ then axiom axoc \ref{axoc04} can be accepted as given.
For applications where methods other than interval arithmetic are employed in $intf~\la p \ra~\la q \ra$ there will be a need ensure that the algorithms used will always guarantee that $R(f,p) \subseteq q$.
In such a case axoc \ref{axoc04} is no longer an axiom and must be derived as a theorem subject to the properties of the methods used to construct $q$ from $p$. 

\item In conventional mathematics we often desire a result that proves that a predicate $P(n)$ is true for all positive integers, $n$, of $\mb{N}$.
In axiom axoc \ref{axoc05} we can only make the statement that $itf~\la v~n \ra~\la w \ra$ will be computable for any $n:\mb{I}_0$ because we are working with integer values of $\mb{I}$.

\end{itemize}

\section{The tent map.}\index{tent map}

As an example consider the dynamical system that can be represented by the difference equation
\ben\label{tent010}
\bal
v^{(t)} = f(v^{(t-1)}), \quad t=1,\ldots,n
\eal
\een
where $f:\mb{I} \to \mb{I}$.
We will examine the evolution of the state variable $v:\mb{I}$ for the discrete tent map that can be defined explicitly as
\ben\label{tent020}
f(v) = \left \{
\begin{array}{ll}
	2*v , & 0 \leq v \leq a \\
	2*(b-v) , & a < v \leq b \\
	\text{undefined} , & \text{otherwise} \\
\end{array}
\right .
\een
where
\be
b=2*a
\ee
for some $a:\mb{I}_+$.
The internal algorithm of the associated program $f~\la v \ra~\la w \ra$ will be constructed such that it halts with an execution error if $v < 0$ or $v > b$ is inserted as input data.

\begin{figure}[!h]
	\begin{center}
		\includegraphics[width=6cm]{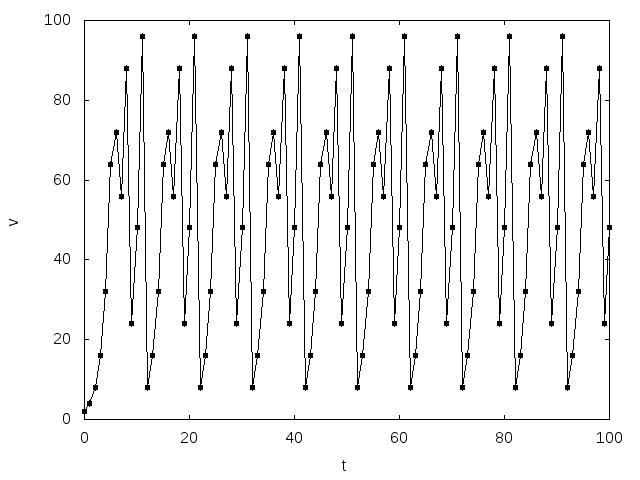}
	\end{center}
	\caption{\textit{The evolution of the fully discrete tent map over the discrete time interval $0 \leq t \leq 100$ with initial condition $v^{(0)} = 2$ for the case $a = 50$.}}
	\label{fig_tent0}
\end{figure}

Figure \ref{fig_tent0} shows the evolution of the finite dynamical system (\ref{tent010})-(\ref{tent020}) for the case $a=50$ with the initial condition $v^{(0)}=2$.
It is seen that the solution increases rapidly from its initial condition, begins to oscillate and then quickly settles into a cycle.

We are generating fully discrete solutions of the tent map so that the lines joining the points in the figures are included only as a visual aid and do not represent part of the solution.
This should be distinguished from traditional studies of the tent map where $f$ is treated as a continuous real valued function.

\begin{figure}[!h]
	\begin{center}
		\includegraphics[width=6cm]{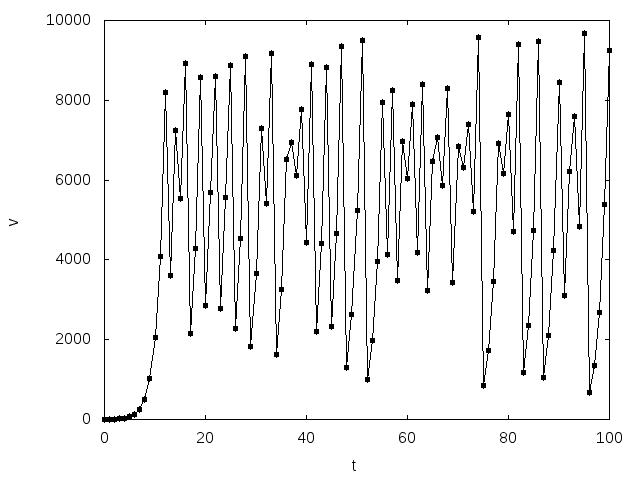}
		\includegraphics[width=6cm]{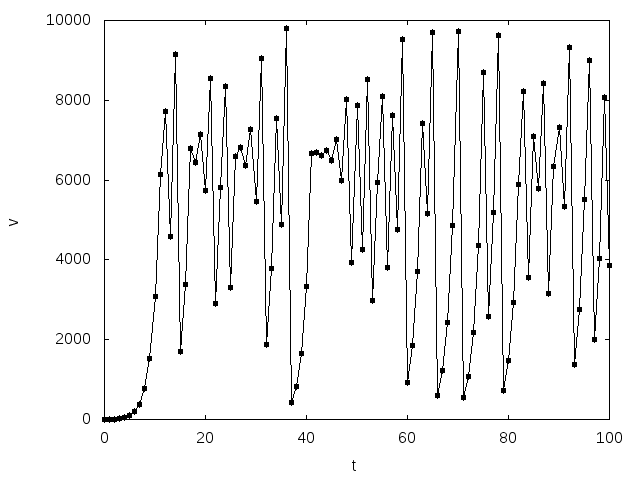}
	\end{center}
	\caption{\textit{The evolution of the fully discrete tent map over the discrete time interval $0 \leq t \leq 100$ for the case $a = 5,000$ with two initial conditions, $v^{(0)} = 2$ (left) and $v^{(0)} = 3$ (right).}}
	\label{fig_tent1}
\end{figure}

The solution behavior becomes more complicated when we increase the degrees of freedom of the state variable $v$. 
Figure \ref{fig_tent1} shows the evolution of the finite dynamical system (\ref{tent010})-(\ref{tent020}) for the case $a=5,000$ with the initial conditions $v^{(0)}=2$ and $v^{(0)}=3$.
Both solutions exhibit the same qualitative behavior.
The trajectories increase rapidly from their initial condition and begin to oscillate with no cycle being evident in the time period $0 \leq t \leq 100$.
It is visually evident in Figure \ref{fig_tent1} that there is a significant quantitative difference between the two solutions.   
This indicates that the fully discrete tent map is, in a discrete sense, sensitive to initial conditions.

A striking feature of the fully discrete tent map is the relatively large differences in the magnitudes of the solution between consecutive iterations.
There are many fully discrete maps that display this feature.
Although such maps may have some academic interest there exists a class of fully discrete systems that have very different properties and are of greater interest in real world applications.
In a later chapter we will examine systems on multidimensional lattices that are constrained by the law of conservation of information.
Under increasing grid refinement the generated solutions of such systems do not oscillate in such a wild manner and tend to take on the appearance of continuous maps.

\textbf{Notes.}

\begin{itemize}

\item In continuous mathematics the notion of sensitivity to initial conditions has been explored in some depth.
The sensitivity to initial conditions of the example depicted in Figure \ref{fig_tent1} is one based on discrete values so that direct comparisons with conventional theories may not be appropriate.
The implications of sensitivity to initial conditions for fully discrete systems needs to be examined further.

\end{itemize}

\section{Partitioned maps.}

It is easy to see that the tent map is computable over the interval $p:\mb{B}$ given by
\be
p=[0~b]
\ee
The tightest interval enclosure of the assignment map $f:\mb{I} \to \mb{I}$ over $p$ will be bounded by $p$, i.e. $R(f,p) \subseteq p$.

The program associated with the map $f:\mb{I} \to \mb{I}$ given by (\ref{tent020}) can be defined explicitly as a disjunction.
The program $intf~\la p \ra~\la q \ra$ can be constructed such that it first performs the partition
\be
[0~b] = [0~a] \cup [a~b]
\ee
followed by the construction
\be
q = B(f,[0~a]) \cup B(f,[a~b])
\ee
where each interval partition, $B(f, \cdot )$, is evaluated using the rules of interval arithmetic, (\ref{iarith01})-(\ref{iarith02}).
Since the construction is based on interval arithmetic, axiom axoc \ref{axoc04} tells us that for any interval $p$ such that $intf~\la p \ra~\la q \ra$ is computable, $q$ will be an interval enclosure of $R(f,p)$.
Since $R(f,p) \subseteq q=p \subseteq p$ we can apply axiom axoc \ref{axoc05} to establish that the program $itf~\la u~n \ra~\la v \ra$ will be computable for any element $v$ contained in $p$ and $n:\mb{I}_0$.

While the tent map, (\ref{tent020}), has a simple structure it is not a linear map.
Because it involves a disjunction it can be somewhat tedious to work with as a program.
The tent map can be generalized to any map that is linear on interval partitions, $p^{(i)}=[p_l^{(i)}~p_u^{(i)}],~i=1,\ldots,m$, with the representation
\ben
f(v) = \left \{
\begin{array}{ll}
	a^{(i)}*v+b^{(i)} , & p_l^{(i)} \leq v \leq p_u^{(i)},~i=1,\ldots,m \\
	\text{undefined} , & \text{otherwise} \\
\end{array}
\right .
\een
where $a^{(i)},b^{(i)}:\mb{I}$ are prescribed constant coefficients associated with the interval $p^{(i)}$.
The domain of determinacy, $p$, can be enclosed by the program
\be
~[iddi~[ei]~[r^{(0)}]~[cup~[p^{(i)}~r^{(i-1)}~]~[r^{(i)}]]_{i=1}^m~iddi~[r^{(m)}]~[p]]
\ee
It follows that the bounds for the interval $p$ are given by $p=r^{(m)}=[p_l^{(1)}~p_u^{(m)}]$.

Given an element $v:\mb{I}$ in any subinterval of a partition $p^{(i)},~i=1,\dots,m$, of $p$, the core evaluations of $f~\la v \ra~\la w \ra$ are based on the program
\be
evf~[a^{(i)}~v~b^{(i)}]~[z~w] = \left \{
\begin{array}{l}
mult~[a^{(i)}~v]~[z] \\
add~[z~b^{(i)}]~[w] \\
\end{array}
\right .
\ee
The program $f~\la v \ra~\la w \ra$ is constructed such that for a given input $v$ it first identifies the partition $p^{(i)}$ that contains $v$, assigns the appropriate values to the coefficients $a^{(i)}$ and $b^{(i)}$ for that partition and then performs the evaluation $evf~[a^{(i)}~v~b^{(i)}]~[z~w]$.

The core evaluations of $intf~\la p \ra~\la q \ra$ over any partition, $p^{(i)},~i=1,\dots,m$, of $p$, are obtained by translating the program operations of scalar arithmetic of $evf$ into operations of interval arithmetic through the program 
\be
evintf~[a^{(i)}~p^{(i)}~b^{(i)}]~[z~u~q] = \left \{
\begin{array}{l}
smultdi~[a^{(i)}~p^{(i)}]~[z] \\
int~[b^{(i)}~b^{(i)}]~[u] \\
adddi~[z~u]~[q] \\
\end{array}
\right .
\ee

The computability of $intf~\la p \ra~\la q \ra$ will depend on the computability of the interval evaluations $evintf~[a^{(i)}~p^{(i)}~b^{(i)}]~[z^{(i)}~u^{(i)}~q^{(i)}]$ for each partition $p^{(i)}$.
This reduces to the task of establishing the existence of the scalar additions and multiplications of the bounding points of the intervals involved in the operations of interval arithmetic employed in $evintf$.
Although this is a straight forward application of the theorems of the previous chapter it can be a rather lengthy and tedious process even for the case $m=2$.

\section{Multi-dimensional intervals.}

While methods of discrete intervals can be laborious for the class of assignment programs described in the previous section there are important applications where the methods of discrete intervals are better suited.
The methods of discrete intervals can be applied to maps involving more than one primary variable where the associated assignment programs can be represented by
\be
f~\la v_1 \ldots v_m \ra~\la w_1 \ldots w_m \ra
\ee
When $m$ is large it is better to work with arrays and extend the notion of discrete intervals to multi-dimensional discrete boxes.
Axioms axdi 1-16, axdia 1-12 and axoc 1-5 will still apply with minimal modifications that include the replacement of intervals with boxes where the box bounds are expressed by arrays.

As will be discussed in a later chapter, of particular interest is the application of the methods of discrete intervals to finite dynamical systems on multi-dimensional lattices where the primary laws are governed by the conservation of information.
Here closure models are less tedious to work with since they can be expressed in a form that do not involve disjunctions.

\chapter{Program Constructions as Proofs.}\label{cpcap}

\section{Human verses machine proofs.}

Proofs in contemporary mathematics are constructed from a language comprised of symbols and natural language and their merits often judged by their elegance.
This style of proof construction is natural to humans and has been accepted as the standard for good reasons.
Purely symbolic proofs can be difficult to read and lack the expressiveness that humans demand to satisfy their interpretation of meaning.
Indeed, many mathematicians have a preference of reducing the amount of symbolic content in their proofs in favor of natural language.

Proofs are presented as an outline of a sequence of steps that are often bound together by trivial and tedious calculations.
The author of a proof attempts to provide the reader with an outline of the important steps leading to a conclusion by omitting the details of what may be regarded as obvious and trivial calculations.
Thus the reader is spared from the tedious details that can otherwise be a distraction from the main thrust of the proof.

For longer proofs elegance is difficult to maintain and can even be a challenge to read by experts in the particular subject area.
Indeed it is not uncommon for referees of mathematical proofs to call upon proof checking software to establish the correctness of a proof.
This raises some questions as to the extent by which contemporary proofs are rigorous constructions and not merely outlines.
There is no defining line here and no formal criteria exist to distinguish an outline of a proof from one that can be designated as rigorous. 

Machine proofs are uncompromising in rigor and demand the inclusion of even the most trivial calculations.
As a result machine proofs can be much longer than those written down by humans.
They are symbolic in structure, devoid of natural language and demand a completely different way of interpreting their meaning.

A good example of this is found in a later chapter dedicated to matrix arithmetic.
In Section \ref{mtxthm} a number of theorems for matrices are derived using VPC.
At first glance the reader may conclude that they are just trivial extensions of some of the basic results that were derived in Chapter \ref{int}.
In Section \ref{linsys} a more conventional language is used to demonstrate that the machine proofs of Section \ref{mtxthm} reveal some important properties of linear systems over $\mb{I}$ that are similar, although not equivalent, to well established properties of linear systems over fields and rings.

Developers of proof assistance software often make some effort to provide an interface that allows the user to interact with the machine in the more familiar language of contemporary mathematics.
Here we shall make very little effort in this regard.
This is a choice that is deliberate and is made to encourage the reader to acquire familiarity with machine proofs that are presented as a list of functional programs.
Although such proofs may be unsightly at first, the reader needs to be assured that with some effort and experience they will find that this style of derivation will become no less natural than the more traditional style that they have become accustomed to.

Given current trends it does appear that efforts in acquiring familiarity with machine proofs, whether they be based on functional programs or any other machine language, are not wasted.
It is not unreasonable to anticipate that machine proofs will eventually become more widely used.
This will be especially beneficial for the construction and rigorous validation of computer models based on rule based algorithms and finite state arithmetic.

\section{PECR versus conventional theories of logic.}

The motives behind the development of the conventional formal systems of proof theory were primarily aimed at solving theoretical questions in logic and the efficiency of translating them into computer programs were at best a secondary concern.
Notwithstanding this, various applications of some of these formal systems can be readily translated into computer programs and form the basis of many proof assistance software currently available.
Here we shall examine the program extension rule and the associated construction rules of our formal system, PECR, in relation to conventional methods of propositional calculus, first order logic and other formal systems of proof theory.

Proof assistance software are powerful tools for checking proofs of theorems in mathematics.
There is an important distinction to be made here with respect to the program VPC in that it is designed specifically for the purpose of checking the computability of programs in a machine environment $\mathfrak{M}(\mathcal{K},\mathcal{L},\mathcal{M})$ rather than a general theorem prover.
Nevertheless it is useful to examine the formal system PECR with respect to the conventional theories of mathematical logic.    

\textbf{Propositional calculus.} Propositional calculus\index{propositional calculus} is a formal system $\mathcal{L}(A,\Omega,Z,I)$, where $A,\Omega,Z,I$, are defined as follows.

\begin{itemize}

\item A is the set of propositional variables referred to as atomic formulas or terminal elements.

\item $\Omega$ is a finite set of logical connectives.
Typically these include $\land, \lor, \to, \leftrightarrow$ and the negation symbol $\neg$.
We may also include in the set $\Omega$ the symbols $\top,\bot$ that are associated with the value assignments of truth and falsity.

\item Z is the set of transformation or inference rules.

\item I is a finite set of axioms.

\end{itemize}

The language of $\mathcal{L}$ is a set of well-formed formulas that are constructed inductively by the following rules.
\begin{itemize}

\item Any element of A is a formula of the language of $\mathcal{L}$.

\item For any formulas $p_1,\ldots,p_n$ of the language of $\mathcal{L}$ and any transformation rule $f$ of Z, $f(p_1,\ldots,p_n)$ is also a formula of the language of $\mathcal{L}$.

\end{itemize}

In the language of PECR, well-formed formulas of propositional calculus are replaced by formal statements that are inductively constructed as program lists and/or disjunctions from atomic functional programs.
There are no connectives because the interpretation of the formal meaning of an axiom/theorem relies heavily on the sequential order of the statements.
The value assignments of true and false are simply replaced by the notion of program computability via the computability map (\ref{cm}).  

\textbf{First order logic.}
First order logic\index{first order logic} is an extension of propositional logic that allows atomic sentences to include predicates.
Predicates depend on variables that can either come in the form of free or bound variables that are defined relative to the quantifiers $\forall$ and $\exists$.  
When formal statements involve predicates special care needs to be placed on the treatment of variables.
Various rules have been devised to maintain some form of consistency of the predicate variable names under the actions of substitutions and other manipulations of predicates constrained by quantifiers.

The issue of variables and how they are represented requires a different approach when we choose to represent formal statements as functional programs under a typed system.
The main source of difficulty arises from the restrictions on how variable names of program I/O lists are chosen and the I/O dependency condition.
Therefore our assessment of the suitability of any of the standard methods of proof theory must be based upon how efficiently they can be coded up to handle the special issues that can arise when manipulating program lists.
With this in mind we need to construct a formal scheme that efficiently deals with these issues but at the same time be guided by the conventional theories of proof theory.

\textbf{The sequent calculus.}
The sequent calculus\index{sequent calculus} is based on sequents
\be
\Gamma = A_1 , \ldots , A_m
\ee 
where each $A_i$ is a formula.
We will use the following notation.
\begin{itemize}

\item Upper case letters $A,B,\ldots,Z$ represent formulas.

\item Upper case Greek letters $\Gamma,\Delta,\Lambda,\Theta$ represent sequents, that consist of a finite (possibly empty) sequence of formulas.

\end{itemize} 
The general structure of the sequent calculus is given by the expression
\ben\label{5.10}
\Gamma \vdash \Delta
\een
where the turnstile, $\vdash$, represents entailment, i.e. the sequent on the right hand side of the turnstile follows from the sequent on the left hand side.
For intuitionistic logic $\Delta$ can only contain at most a single formula.
If $\Gamma$ and $\Delta$ are the sequents
\be
\Gamma = A_1 , \ldots , A_m
\ee 
\be
\Delta = C_1, \ldots ,C_l
\ee
then the statement (\ref{5.10}) has the intuitive interpretation
\ben\label{5.20}
(A_1 \land \cdots \land A_m) \to (C_1 \lor \ldots \lor C_l)
\een

The rules of the sequent calculus are as follows.

\emph{Identity axiom.}
\ben\label{seq_id}
\begin{array}{l}
~~~~~~~ \\
\overline{A \vdash A} \\
\end{array}
\een
\emph{Cut rule.}
\ben\label{seq_cut}
\begin{array}{l}
	\Gamma \vdash \Delta,A \hspace{5mm} A,\Lambda \vdash \Theta \\
	\hline
	\hspace{5mm} \Gamma,\Lambda \vdash \Delta,\Theta \\
\end{array}
\een
\emph{Exchange.}
\ben\label{seq_exch}
\begin{array}{l}
	\Gamma,A,B,\Lambda \vdash \Delta \\
	\hline
	\Gamma,B,A,\Lambda \vdash \Delta \\
\end{array}
\hspace{20mm}
\begin{array}{l}
	\Gamma \vdash \Delta,A,B,\Lambda \\
	\hline
	\Gamma \vdash \Delta,B,A,\Lambda \\
\end{array}
\een

\emph{Weakening.}
\ben\label{seq_weak}
\begin{array}{l}
	\Gamma \vdash \Delta \\
	\hline
	\Gamma,A \vdash \Delta \\
\end{array}
\hspace{20mm}
\begin{array}{l}
	\Gamma \vdash \Delta \\
	\hline
	\Gamma \vdash \Delta,A \\
\end{array}
\een

\emph{Contraction.}
\ben\label{seq_contr}
\begin{array}{l}
	\Gamma,A,A \vdash \Delta \\
	\hline
	\Gamma,A \vdash \Delta \\
\end{array}
\hspace{20mm}
\begin{array}{l}
	\Gamma \vdash \Delta,A,A \\
	\hline
	\Gamma \vdash \Delta,A \\
\end{array}
\een

\emph{Negation.}
\ben\label{seq_neg}
\begin{array}{l}
	\Gamma \vdash \Delta,A \\
	\hline
	\neg A,\Gamma \vdash \Delta \\
\end{array}
\hspace{20mm}
\begin{array}{l}
	B,\Gamma \vdash \Delta \\
	\hline
	\Gamma \vdash \Delta, \neg B \\
\end{array}
\een

\emph{Implication.}
\ben\label{seq_imp}
\begin{array}{l}
\Gamma \vdash \Delta,A \hspace{5mm} B,\Lambda \vdash \Theta \\
\hline
\hspace{5mm} A \to B,\Gamma,\Lambda \vdash \Delta,\Theta \\
\end{array}
\hspace{20mm}
\begin{array}{l}
	A,\Gamma \vdash \Delta,B \\
	\hline
	\Gamma \vdash \Delta,A \to B
\end{array}
\een

\emph{Conjunction (left).}
\ben\label{seq_conjl}
\begin{array}{l}
	A,\Gamma \vdash \Delta \\
	\hline
	A \land B,\Gamma \vdash \Delta \\
\end{array}
\hspace{20mm}
\begin{array}{l}
	B,\Gamma \vdash \Delta \\
	\hline
	A \land B,\Gamma \vdash \Delta \\
\end{array}
\een
\emph{Conjunction (right).}
\ben\label{seq_conjr}
\begin{array}{l}
	\Gamma \vdash \Delta,A \hspace{5mm} \Gamma \vdash \Delta,B \\
	\hline
	\hspace{10mm} \Gamma \vdash \Delta,A \land B \\
\end{array}
\een

\emph{Disjunction (left).}
\ben\label{seq_disjl}
\begin{array}{l}
	A,\Gamma \vdash \Delta \hspace{5mm} B,\Gamma \vdash \Delta \\
	\hline
	\hspace{10mm} A \lor B,\Gamma \vdash \Delta \\
\end{array}
\een
\emph{Disjunction (right).}
\ben\label{seq_disjr}
\begin{array}{l}
	\Gamma \vdash \Delta,A \\
	\hline
	\Gamma \vdash \Delta,A \lor B \\
\end{array}
\hspace{20mm}
\begin{array}{l}
	\Gamma \vdash \Delta,B \\
	\hline
	\Gamma \vdash \Delta,A \lor B \\
\end{array}
\een

Rules (\ref{seq_exch})-(\ref{seq_contr}) are referred to as structural rules and rules (\ref{seq_neg})-(\ref{seq_disjr}) are referred to as inference rules.

The sequent calculus includes rules associated with the manipulation of the predicate variables and quantifiers.
They fall under the category of inference rules and are given by the following.

\emph{Quantifiers.}
\ben\label{seq_quant1}
\begin{array}{l}
	F(t),\Gamma \vdash \Delta \\
	\hline
	\forall xF(x)\Gamma \vdash \Delta \\
\end{array}
\hspace{20mm}
\begin{array}{l}
	\Gamma \vdash \Delta,F(a) \\
	\hline
	\Gamma \vdash \Delta,\forall xF(x) \\
\end{array}
\een
\ben\label{seq_quant2}
\begin{array}{l}
	F(a),\Gamma \vdash \Delta \\
	\hline
	\exists xF(x)\Gamma \vdash \Delta \\
\end{array}
\hspace{20mm}
\begin{array}{l}
	\Gamma \vdash \Delta,F(t) \\
	\hline
	\Gamma \vdash \Delta,\exists xF(x) \\
\end{array}
\een
In the above rules, $t$ is an arbitrary term.
The variable $a$ is known as an eigenvariable of the respective inference and must not occur in the lower sequents.

The rules (\ref{seq_id})-(\ref{seq_quant2}) are associated with classical logic\index{classical logic}.
For intuitionistic logic\index{intuitionistic logic} sequents appearing on the right hand side of the turnstile, $\vdash$, can contain at most a single formula.  
This means that some rules must be discarded.
These include the right exchange and right contraction rules.
The sequents $\Delta$ and $\Theta$ in the other rules either contain a single formula or are empty sequents. 

In the remaining discussion of this and the next section we will find that there is some comparison to be made with PECR and the rules of the sequent calculus.
Most of the similarities that can be identified will involve the modifications of the above sequent rules for intuitionistic logic. 

Where the similarities end are with the rules associated with predicates and quantifiers.
In PECR we deal with types and the manipulation of functional programs relies heavily on rules associated with the variable names of their I/O lists.
We can consider the rules (\ref{seq_quant1})-(\ref{seq_quant2}) as representing the most significant departure from the overall methodology used in PECR.

\textbf{The cut rule.}
In applications of the sequent calculus the cut rule (\ref{seq_cut}) is used extensively in the construction of proofs.
In PECR there is no need for a similar rule when constructing proofs.
However, when a proof is completed, theorem extraction relies on the elimination of intermediate statements of the proof.
This is achieved by the algorithm of Section \ref{connection} that makes use of connection lists.

One may be tempted to introduce an analogy to the cut rule in PECR by concluding that $aext~[p~c]~[~]$ is an extension of the program $[ext~[p~q]~[~]~ext~[q~c]~[~]]$.
This cannot hold as a general rule since the input list $x_c$ of the program $c$ may include variable names of elements of the output list $y_q$ of $q$.
This would mean that in the concatenated program $[p~c]$ the input list of the program $c$ introduces new variable names that are not contained in the I/O lists of $p$.
This is not allowed in an extension.

As an example, consider the integer program axioms axi 9-axi 10 introduced in Chapter 6.
Set $p=typei~[a]~[~]$, $q=add~[a~0]~[b]$ and $c=eqi~[b~a]~[~]$.
From axi 9 we have $q:\mb{P}_{iext}(p) <: \mb{P}_{ext}(p)$ and from axi 10 we have $c:\mb{P}_{iext}(q) <: \mb{P}_{ext}(q)$.
While $[p~c]=[typei~[a]~[~]~eqi~[b~a]~[~]]$ is a well defined program it cannot be an extended program since the input list of $c$ contains the variable $b$ that is not contained in the input list of $p$. 

\textbf{Conjunction commutativity.}
Conjunctions satisfy the property of commutativity, i.e. for any two statements $P$ and $Q$
\be
P \land Q = Q \land P
\ee
In the sequent calculus this is expressed through the exchange rule (\ref{seq_exch}).  

An extension $c:\mb{P}_{ext}(p)$ has the association with the program concatenation $[p~c]$.
The list representation $p=[p_i]_{i=1}^n$ can be thought of as a kind of conjunction of the formal statements $p_1,\ldots,p_n$ but with one major difference.
The order of elements of the program list $[p_i]_{i=1}^n$ can only be rearranged provided that the I/O dependency condition is not violated.
Thus the classical properties of commutativity of conjunctions cannot be applied as a general rule in PECR. 

\textbf{Repetition of statements.}
Propositional logic allows for repetitions to occur freely in formal statements.
For instance the statement $P \land P$ is allowed and may simply be contracted to the statement $P$. 
In the sequent calculus repetitions are also allowed and may be removed by the contraction rules (\ref{seq_contr}).

For program lists there is no general rule allowing repetitions because no two subprograms of a program list can have elements of their output lists that contain the same variable names.
Program lists can contain repeated subprograms only when those subprograms have an empty list output.

To maintain efficiency of computation it is better to avoid repetitions for subprograms that have an empty list output.
However, there are situations where this is not possible.
For example the irreducible extended program for a special non-atomic program (spd 1) of arithmetic over $\mb{I}$, 
\be
[lt~[a~b]~[~]~le~[a~b]~[~]]
\ee
has an implied repetition under the equivalent representation
\be
[lt~[a~b]~[~]~lt~[a~b]~[~]]~|~[lt~[a~b]~[~]~eqi~[a~b]~[~]]
\ee
obtained through the disjunction distributivity rules.
The first operand contains a repeated program but is computable.
The second operand is a false program.
By application of disjunction commutativity and the disjunction contraction rule 2 this statement reduces to
\be
[lt~[a~b]~[~]~lt~[a~b]~[~]]
\ee

We note that for any program $p$ with an empty list output we can define $p:\mb{P}_{ext}(p)$ and $[p~p]$ is a well defined extended program.
This has a similarity with the identity axiom (\ref{seq_id}) but is of little use in application of PECR because of its restrictive nature. 

\textbf{The program extension rule.}
Consider the classical expression of $m$ conjunctions
\ben\label{5.90a}
A = A_1 \land \cdots \land A_m
\een
The conjunction $A$ is true if and only if every statement $A_1, \ldots , A_m$ is true.
Suppose further that we also have
\be
Q \to C
\ee
$Q$ may represent a single statement or $n$ conjunctions
\ben\label{5.90b}
Q = Q_1 \land \cdots \land Q_n
\een
In classical logic the following statement is a tautology.
\ben\label{5.90}
 ( (A \land Q) \land (Q \to C)  ) \to  ( (A \land Q) \to C  )
\een
Because conjunctions satisfy the property of commutativity we may readily rearrange the conjunctions in any order.
Let $P(A,Q)$ be any sequential order of the conjunctions in $A \land Q$.
We can now generalize (\ref{5.90}) as
\ben\label{5.100}
 ( P(A,Q) \land (Q \to C)  ) \to  ( P(A,Q) \to C  )
\een
with the interpretation that if all $Q_i$, $i=1,\ldots,n$, are contained in the conjunctions of $P(A,Q)$ and $C$ follows from $Q$ then $C$ also follows from $P(A,Q)$.
With the exclusion of the commutativity of conjunctions, the similarities between the program extension rule of our formal system and the interpretation based on (\ref{5.100}) is then evident.

We may also find some analogy of the program extension rule in the sequent calculus.
Let $A$ and $Q$ be defined by the conjunctions (\ref{5.90a}) and (\ref{5.90b}).
By applying the left weakening rule (\ref{seq_weak}) we may rewrite the identity axiom (\ref{seq_id}) as
\ben\label{5.110}
\begin{array}{l}
~~~~~ \\
\overline{Q \vdash Q} \\
\overline{A,Q \vdash Q} \\
\end{array}
\een
Suppose that we have
\ben\label{5.120}
\begin{array}{l}
Q \vdash C \\
\end{array}
\een
Employing the cut rule (\ref{seq_cut}) to (\ref{5.110}) and (\ref{5.120}) we obtain
\ben\label{5.130}
\begin{array}{l}
A,Q \vdash Q \hspace{5mm} Q \vdash C \\
\hline
\hspace{10mm} A,Q \vdash C \\
\end{array}
\een

\textbf{Disjunction introduction.}
In propositional logic we have the axiom of disjunction introduction
\be
P \vdash P \lor Q
\ee
In the sequence calculus we have the right disjunction rules (\ref{seq_disjr}), where, for intuitionistic logic, we modify the right hand side of the turnstile, $\vdash$, to contain at most a single formula.
For classical logic where the sequent $\Delta$ may contain multiple statements we obtain a similar outcome, under the interpretation (\ref{5.20}), by the right weakening rule (\ref{seq_weak}).

The special non-atomic program axioms spd 1, presented in Section \ref{snap}, have some similarity with the above rules involving disjunction introduction.
However, in our formal system PECR there is no straight forward way of defining a general higher order construction rule for disjunction introduction because a program cannot be weakened by a disjunction with just any other program.
Again, this is largely due to the fact that programs are defined in terms of the variable names of their I/O lists and therefore are constrained by certain rules of composition.
For example the program $add~[a~b]~[c]$ cannot be weakened by a disjunction with the program $add~[a~c]~[d]$ because the output $c$ of the former would not be available to the latter.

\textbf{Quantifiers.}
One of the powers of first order logic is its expressiveness, particularly through the use of quantifiers.
Under the typed system of PECR there are ways to construct theories that are not restrictive due to the lack of quantifiers.
This may require some significant change from the conventional mindset of first order logic that will be acquired through more experience in the use of PECR.

In first order logic we may also make a formal statement $\forall x \exists z(P(x) \to C(x,z))$, for two predicates $P(x),C(x,z)$.
The variables $x$ and $z$ are bound by the quantifiers.
Without the quantifiers a statement of the form $P(x) \to C(x,z)$ can be problematic due to the introduction of the new variable $z$ in the conclusion statement.

Consider the extended program $[p~c],~c:\mb{P}_{ext}(p)$.
In Condition 1 of the definition of a program extension, Definition \ref{ce}, states that the variable name of any element of the input list $x_c$ of $c$ that is not a constant must either appear in the input list $x_p$ or output list $y_p$ of the program $p$, i.e. the input list $x_c$ of $c$ cannot introduce new variable names other than constants.
In fact Condition 1 of Definition \ref{ce} is a necessary condition for a program to be an extension.

Consider for example a program of the form
\be
q~[x~z]~[y~\bar{y}]=[p~x~y~c~[x~z]~\bar{y}]
\ee
where for generality $x$ and $z$ are lists.
While there may exist value assignments of $x$ and $z$ such that $q~[x~z]~[y~\bar{y}]$ is computable, the program $c~[x~z]~\bar{y}$ cannot be an extension of $p~x~y$.
This is because it is possible that $p~x~y$ is computable for a given value assignment of $x$, but at the same time there may exist an independent value assignment of $z$ such that the program $c~[x~z]~\bar{y}$ is not computable.
Thus, computability of the program $p~x~y$ under a value assigned input $x$ will not guarantee the computability of $[p~ x~y~c~[x~z]~\bar{y}]$.

\textbf{Negations.}
In conventional theories of logic negations play an important role in the expressiveness of formal statements and how they are manipulated.
In our formal system PECR there is no concise way to construct for each program a corresponding program that can serve as its negation. 
This is because programs often contain several actions that include type checking of input variables as well as possible assignments.
In this way programs often contain statements that are instructions that lead to the construction of new variables and hence cannot be directly related to predicates of conventional theories of logic.

For example the integer program $add~[a~b]~[c]$ is equivalent to the statement that if $a:\mb{I}$ and $b:\mb{I}$ then perform the assignment $c:=a+b$ provided that $a+b:\mb{I}$.
Otherwise halt with an execution error.
While the statement $a:\mb{I}$ and $b:\mb{I}$ and $b+c:\mb{I}$ can be negated there is no meaningful negation of an assignment instruction $c:=a+b$.
For this reason rules such as (\ref{seq_neg}) have no utility in PECR.

\textbf{Implication.}
Under the interpretation (\ref{5.20}) the turnstile, $\vdash$, takes on the role of an implication, $\to$.
This connection between entailment and implication is rather loose and cannot be taken too formally.
In (\ref{seq_imp}) we see that entailment and implication have distinct roles to play.

In an extended program $[p~c],~c:\mb{P}_{ext}(p)$, attempts to make the distinction that the conclusion $c$ follows from $p$ as either an entailment or an implication becomes too ambiguous to be useful.
The closest association that one can make with rules such as (\ref{seq_imp}) in PECR will likely involve higher order programs.
However, even armed with these higher order constructs seeking general rules that are close analogies of both the left and right implication rules will be subject to conditional constraints due to the I/O dependence condition. 

The definition of an extension $c:\mb{P}_{ext}(p)$ removes the need for any kind of entailment connective between the programs $p$ and $c$ in the concatenation $[p~c]$.
The sequential order that requires $c$ to follow $p$ is necessarily fixed when it is used in an extended program derivation, say $epd~[q~p~c]~[s]$.
Once the program $s=[p~c]$ has been constructed from an extended program derivation the sequential order of the subprograms of $s$ can be rearranged provided that there is no violation of the I/O dependency condition.

\textbf{Context.}
The most commonly used propositional proof systems are based on the use of modus ponens as the sole rule of inference.
Modus ponens can be represented by
\be
P,P \to Q \vdash Q
\ee
and states, from $P$ and $P \to Q$ infer $Q$.
Under a truth assignment the statement that $Q$ is true may have no meaning if all reference to its origin as a conclusion has been removed.

In an extended program $[p~c],~c:\mb{P}_{ext}(p)$, the claim that the conclusion $c$ is computable can only be made within the context of the premise $p$.
Consider for example the conclusion of the transitivity of integer value assigned equality axiom ord \ref{toi} that states
\be
 [[lt~[a~b]~[~]~lt~[b~c]~[~]]~lt~[a~c]~[~]]
\ee
In isolation the subprogram $lt~[a~c]~[~]$ is not always computable.
Its computability is guaranteed if the premise program $[lt~[a~b]~[~]~lt~[b~c]~[~]]$ is computable.
Program extensions have the desired property that they retain the context within which a conclusion statement can be said to be computable.

\textbf{Programs as structured strings.} It is tempting to associate a program $\mathfrak{p}~x~y$ with some map $f:x \mapsto y$.
While this may be useful in many cases some care needs to be taken in following this association too formally.
Firstly, we have allowed some programs to have an empty list output.
These are usually associated with programs with the sole purpose of type checking and abstract type assignment.

The construction rules of PECR rely heavily on the representation of programs as strings that obey certain structural rules and the representation of programs as maps is not fully exploited.
One formal scheme where the notion of a map is of primary importance is found in the typed lambda calculus.  

\textbf{Typed lambda calculus.}
Lambda calculus\index{lambda calculus} is largely regarded as the foundations of functional programming languages.
It is a formal system that employs a function abstraction using an elegant method for variable binding and substitution.
Lambda calculus employs the function abstraction $\lambda x . M$ with the interpretation of the map $x \mapsto M$, that maps the variable $x$ to the term $M$.
The lambda abstraction $\lambda x . M$ is also regarded as a term, where the term $M$ can be an expression involving the variable $x$ and other variables.
In the expression $\lambda x . M$ the variable $x$ is said to be bound by $\lambda$.
Input value assignments are applied through the expression $\lambda x . M~(z)$, where $z$ is a value that is assigned to the variable $x$ under the map $x \mapsto M$.

The typed lambda calculus is a formalism that uses the lambda function abstraction as well as incorporating types.
Although there are many variants, typed lambda calculi provide an important link between typed functional programming, mathematical logic and proof theory. 
The formalism is based on the typing judgments
\ben\label{5.200}
x_1:\mb{A}_1, \cdots , x_n:\mb{A}_n \vdash M:\mb{A}
\een
where $x_1,\ldots,x_n$ are variables, $\mb{A}_1,\ldots,\mb{A}_n,\mb{A}$ are types and $M$ is referred to as a pre-term.
The main typing rules are
\ben\label{5.210}
\begin{array}{l}
~~~~~~~~~~~~~~~~~~~~~~~~~~~~~~~~~~~~\\
\hline
x_1:\mb{A}_1, \cdots , x_n:\mb{A}_n \vdash x_i:\mb{A}_i \\
\end{array}
\qquad 1 \leq i \leq n
\een
\ben\label{5.220}
\begin{array}{l}
\Gamma \vdash M:\mb{A} \to \mb{B} \hspace{5mm} \Gamma \vdash N:\mb{A} \\
\hline
\hspace{15mm} \Gamma \vdash MN:\mb{B} \\
\end{array}
\een
\ben\label{5.230}
\begin{array}{l}
\Gamma,x:\mb{A} \vdash M:\mb{B} \\
\hline
\Gamma \vdash \lambda x^A.M:\mb{A} \to \mb{B}
\end{array}
\een

In our formal system PECR we have a preference of representing functional programs in the less abstract form of $\mathfrak{p}~x~y$, but the similarities with some of the construction rules of Chapter \ref{ccr} and the above typing rules should be evident.
We note that in PECR the type assignments of all initial input elements of a core program $\mathfrak{p}~x~y$ are understood to be set by an initializing program $read~x~[~]$ under the general program structure (\ref{core}).
Along with the application axioms this provides an equivalent initialization to the typing judgments (\ref{5.200}) of the typed lambda calculus.

In the lambda calculus a derived statement is a binding of previous statements by an association of a lambda function abstraction or a composition of terms as a kind of implicit function construct.
It would be of interest to find an abstraction of PECR such that it can be demonstrated to be fully contained within the typed lambda calculus, or at least a well established version of it.
Here our choice of preference is to represent proofs as a vertical list of functional programs with each statement attached to a connection list.
An important difference between PECR and the typed lambda calculus is that the manipulation rules of functional programs in PECR are largely based on the representation of functional programs as structured strings rather than maps.

There are clearly many advantages of programming languages that directly employ the lambda abstraction.
The inherent properties of the lambda calculus removes the complications that arise from the I/O dependency condition of PECR.

One might regard the I/O dependency condition of PECR as an unnecessary complication that is bypassed by the typed lambda calculus.
Whether this is a major issue is a matter of choice. 
The construction rules of PECR are designed to minimize the complications that arise from the I/O dependency condition while ensuring that no inconsistencies occur.
As a result the language of PECR still retains the property of simplicity while providing a sufficiently high level of power as a tool for analysis, at least for the primary purposes that it was designed.
Its simplicity means that it is accessible to those whose backgrounds are not rooted in the computer sciences.
As a language based on functional programs it is easily adapted to a machine environment and has adequate automated capabilities that reduce many of the laborious tasks when generating proofs.

\section{Exploring the Platonic world.}
Students of high school mathematics are largely taught to derive identities and inequalities through the actions of substitutions and elementary rules of algebraic manipulation.
With the exposure to mathematical symbolic software they soon begin to understand that there is a more fundamental process going on, namely the manipulation of symbols and strings subject to certain basic rules of syntax. 

When entering college, students of pure mathematics are confronted with an apparently new way of doing things.
The focus moves towards the construction of proofs.
Apart from the introduction of some new algebraic rules that are specific to the abstract objects that are under consideration, the most notable change in style comes in the form of employing natural language in proofs.
As discussed in the previous sections, a language based on a combination of symbols and natural language is a shorthand strategy aimed at emphasizing the main steps of the proof while leaving out details of what may otherwise be regarded as trivial but lengthy calculations.
In this way pure mathematicians can explore the properties of their abstract objects unhindered by the laborious task of verifying their proofs in a purely symbolic language.
 
In principle essentially nothing has changed because the application of elementary rules of syntax to manipulate strings is still implicitly active when dealing with the abstract mathematical objects in their proof constructions.
Rather than an emphasis on acquiring a general understanding of rigorous proof construction based on syntax, students of contemporary pure mathematics focus on exploring the properties of specific mathematical abstractions by way of elegant shorthand proofs.

These observations are crucial to an understanding of the motivation behind the language PECR.
It should be noted that there does not appear to be any reason why one cannot introduce theories as applications of PECR that employ the abstract objects of conventional mathematics.
For example, one may start with an application that is defined by the user supplied axioms of first order logic and the field axioms.
When this is done, sets of any cardinality (including infinite sets) are objects that can be assigned as abstract types.
Members of sets are defined by way of sentences that express a relationship between strings that define the variable names of the member and the set. 
Predicates of the theory, along with quantifier binding of their variables, can be constructed by atomic programs that define them as sentences in the form of structured strings.
The manipulation of such objects is entirely based on syntax through the rules supplied as axioms that define the application.
In this way we see that higher order abstractions in PECR will rely heavily on the employment of type $\mb{P}_{tassign}$ atomic programs that assign abstract types to objects.

From this perspective we can regard PECR as a primitive language upon which theories based on higher levels of abstractions can be constructed.  
Through the ability to construct abstract types we see that a machine language such as PECR allows us to explore the platonic world in much the same way as we explore the real physical world.
As such it becomes inappropriate to regard PECR as an alternative formal system and comparisons of PECR with the conventional theories of proof theory and the typed lambda calculus become less relevant.
For applications involving program computability through finite state arithmetic very little abstraction is needed.
If an analysis of a more abstract notion of computability is desired, say in the context of Turing machines, then PECR can also serve as a primitive in the sense described above.

\textbf{Law of the excluded middle.}
Mainstream mathematicians have a more relaxed attitude than their constructivist counterparts regarding the need to establish the truth of a premise before a proof is derived.
The aim of a proof in classical logic is to establish a conclusion that is understood to be true if the premise is true.
This feature of classical logic is reflected in our definition of a program extension.
However, it would not be correct to conclude that our formal system is contained fully within classical logic.
It shares features that lean towards intuitionistic logic.

Mathematicians often appeal to the law of the excluded middle by starting with a premise that they believe to be false and proceed to derive a proof that leads to a contradiction.
Leading up to the contradiction they obtain formal statements that are derived from axioms and previously derived theorems.
From the point of view of a constructivist this is unacceptable because in constructive logic it is meaningless to derive statements from a false premise.

It is important to note that in PECR derivations based on axioms and theorems of falsity are not equivalent to the classical method of proof by contradiction.
Proofs leading to a conclusion of falsity are based on a higher order type assignment and not a contradiction of the premise.
Furthermore, there is no feature built into VPC that calls upon the law of the excluded middle.
Derivations leading up to conclusions of falsity are simply aimed at detecting programs that are not computable for any value assigned input. 
Detecting false programs is of particular interest when dealing with disjunctions through an appeal to the disjunction contraction rules.
This involves a contraction back onto the main proof containing the disjunction only after proofs of all operand programs have been completed.

\textbf{Soundness, consistency and completeness.}
The strength of formal systems are measured by their satisfaction of consistency, soundness, and completeness.
The standard systems of propositional logic and sequent calculus can be shown to satisfy these properties by employing a meta-theory of logic.
Crucial to the establishment of soundness and completeness are the notions of interpretations and models. 

The main objective of the formal system PECR is to establish program computability for dynamical systems that are based on finite state arithmetic.
In a practical real world application sense, program computability is ultimately an empirical concept.
A program can be empirically tested for its computability with respect to a value assigned input by simply executing the program and observing whether it halts with an execution error or returns an output in a reasonable time.

Empirical observation plays an important role in the scientific method.
This is an iterative process of self correction where theories are strengthened or replaced by continual revision.
In the final chapter of this book we will explore these ideas in the context of our formal system, PECR.              

\textbf{Notes.}

\begin{itemize}

\item The law of the excluded middle is not assumed in applications of PECR in its most primitive form.
However, the law of the excluded middle can be implemented in the axioms of applications of PECR where higher level abstractions are employed.

\item In applications employing higher levels of abstractions we are still constrained by the machine parameters of $\mathfrak{M}(\mathcal{K},\mathcal{L},\mathcal{M})$.
This means that the restrictions of axiom/theorem and proof lengths, as outlined in Section \ref{stper}, still apply.

\item The restrictions of axiom/theorem and proof lengths, as just noted, are not as severe as they may seem at first glance.
For example, we may construct an application on PECR that explores the properties of propositional logic.
One can expect that we can translate conventional mathematical arguments into our functional programming language to construct theorems and proofs of the soundness and completeness of propositional logic where the machine parameters $\mathcal{L}$ and $\mathcal{M}$ are of moderate size and well within the capacity of a relatively small computer.

\end{itemize}

\section{Some properties of the construction rules.}\label{sspotcr}

The construction rules of PECR are presented as irreducible extended programs of higher order programs.
They can be regarded as the axioms of a theory for the construction of programs as proofs in the context of the formal system PECR on which VPC is based.
Here we will employ VPC as a self referencing tool to examine some properties of the construction rules themselves.
To this end the axioms that are supplied to the file \emph{axiom.dat} include the rules per, cr 1-18, flse 1-3 and dsj 1-10.
I/O type axioms and the substitution rule are automated within VPC.
In addition, we will need to include the special non-atomic program axioms spl 1-4 applied to the extended program derivation $epd~[q~p~c]~[s]$ (see Section \ref{snap}).

\textbf{Weakening.}
It is worthwhile to further explore some results where analogies do exist between the sequent calculus and PECR.
We restrict comparisons with the sequent calculus to intuitionistic logic where the right hand side of the turnstile, $\vdash$, contains at most a single statement.
This feature of intuitionistic logic is also reflected in PECR where the program $c:\mb{P}_{ext}(p)$ of an extended program $[p~c]$, is either an atomic program or must be defined as a special non-atomic program.

Take for instance the left weakening rule (\ref{seq_weak}).
We can derive the following rule where the program $p$ of an extended program $[p~c],~c:\mb{P}_{ext}(p)$, is weakened by the concatenation $r=[p~a]$ for some program $a$.
We want to show that $c:\mb{P}_{ext}(r)$.
Here there is a restriction in that the I/O lists of the introduced program $a$ must be compatible with the I/O lists of the programs $p$ and $c$ such that $r=[p~a]:\mb{P}$ and $s=[r~c]:\mb{P}$.
As a consequence we must include in the premise of the following theorem the conditional statements $conc~[p~a]~[r]$ and $conc~[r~c]~[s]$.  
\begin{lstlisting}
Theorem thm 1.
[[ext [p c] [ ] conc [p a] [r] conc [r c] [s]] ext [r c] [ ]]

Proof.
  1 ext [p c] [ ]
  2 conc [p a] [r]
  3 conc [r c] [s]
  4 sub [p r] [ ]            cr 11 [2]
  5 epd [p r c] [b]          spl 3 [4 1 3]
  6 aext [r c] [ ]           per [5]
  7 ext [r c] [ ]            cr 1 [6]
\end{lstlisting}

In PECR we do not have a general rule that is analogous to the left exchange rule (\ref{seq_exch}) of the sequent calculus so we need to check that weakening will also work for the concatenation $r=[a~p]$.
\begin{lstlisting}
Theorem thm 2.
[[ext [p c] [ ] conc [a p] [r] conc [r c] [s]] ext [r c] [ ]]

Proof.
  1 ext [p c] [ ]
  2 conc [a p] [r]
  3 conc [r c] [s]
  4 sub [p r] [ ]            cr 12 [2]
  5 epd [p r c] [b]          spl 3 [4 1 3]
  6 aext [r c] [ ]           per [5]
  7 ext [r c] [ ]            cr 1 [6]
\end{lstlisting}

\textbf{Conjunction introduction.}
In the sequence calculus there are two left conjunction rules, (\ref{seq_conjl}), where for intuitionistic logic $\Delta$ has at most a single formula.
Theorem thm 3 is a derivation of an analogy of the first of these rules.
In the premise of the following theorem we include two conditional statements $conc~[s~p]~[t]$ and $conc~[t~c]~[u]$ that reflect the requirement that the I/O lists of the introduced program $b$ must be compatible with the I/O lists of the programs $a$, $p$ and $c$ such that $t=[s~p]:\mb{P}$ and $u=[t~c]:\mb{P}$.
\begin{lstlisting}
Theorem thm 3.
[[conc [a p] [r] ext [r c] [ ] conc [a b] [s] conc [s p] [t]
 conc [t c] [u]] ext [t c] [ ]]

Proof.
  1 conc [a p] [r]
  2 ext [r c] [ ]
  3 conc [a b] [s]
  4 conc [s p] [t]
  5 conc [t c] [u]
  6 sub [a s] [ ]            cr 11 [3]
  7 sub [s t] [ ]            cr 11 [4]
  8 sub [a t] [ ]            cr 10 [6 7]
  9 sub [p t] [ ]            cr 12 [4]
 10 sub [r t] [ ]            cr 13 [1 8 9]
 11 epd [r t c] [d]          spl 3 [10 2 5]
 12 aext [t c] [ ]           per [11]
 13 ext [t c] [ ]            cr 1 [12]
\end{lstlisting}
Because we do not have an analogous rule for the exchange rule (\ref{seq_exch}) of the sequent calculus we need to check that conjunction introduction will also work for the concatenation $r=[p~a]$.
\begin{lstlisting}
Theorem thm 4.
[[conc [p a] [r] ext [r c] [ ] conc [a b] [s] conc [s p] [t]
 conc [t c] [u]] ext [t c] [ ]]

Proof.
  1 conc [p a] [r]
  2 ext [r c] [ ]
  3 conc [a b] [s]
  4 conc [s p] [t]
  5 conc [t c] [u]
  6 sub [a s] [ ]            cr 11 [3]
  7 sub [s t] [ ]            cr 11 [4]
  8 sub [a t] [ ]            cr 10 [6 7]
  9 sub [p t] [ ]            cr 12 [4]
 10 sub [r t] [ ]            cr 13 [1 9 8]
 11 epd [r t c] [d]          spl 3 [10 2 5]
 12 aext [t c] [ ]           per [11]
 13 ext [t c] [ ]            cr 1 [12]
\end{lstlisting}
Theorems thm 5 and thm 6, respectively, are derivations that are analogous to the second left conjunction sequent rule for $[b~p]$ and $[p~b]$, respectively.
\begin{lstlisting}
Theorem thm 5.
[[conc [b p] [r] ext [r c] [ ] conc [a b] [s] conc [s p] [t]
 conc [t c] [u]] ext [t c] [ ]]

Proof.
  1 conc [b p] [r]
  2 ext [r c] [ ]
  3 conc [a b] [s]
  4 conc [s p] [t]
  5 conc [t c] [u]
  6 sub [b s] [ ]            cr 12 [3]
  7 sub [s t] [ ]            cr 11 [4]
  8 sub [b t] [ ]            cr 10 [6 7]
  9 sub [p t] [ ]            cr 12 [4]
 10 sub [r t] [ ]            cr 13 [1 8 9]
 11 epd [r t c] [d]          spl 3 [10 2 5]
 12 aext [t c] [ ]           per [11]
 13 ext [t c] [ ]            cr 1 [12]

Theorem thm 6.
[[conc [p b] [r] ext [r c] [ ] conc [a b] [s] conc [s p] [t]
 conc [t c] [u]] ext [t c] [ ]]

Proof.
  1 conc [p b] [r]
  2 ext [r c] [ ]
  3 conc [a b] [s]
  4 conc [s p] [t]
  5 conc [t c] [u]
  6 sub [b s] [ ]            cr 12 [3]
  7 sub [s t] [ ]            cr 11 [4]
  8 sub [b t] [ ]            cr 10 [6 7]
  9 sub [p t] [ ]            cr 12 [4]
 10 sub [r t] [ ]            cr 13 [1 9 8]
 11 epd [r t c] [d]          spl 3 [10 2 5]
 12 aext [t c] [ ]           per [11]
 13 ext [t c] [ ]            cr 1 [12]
\end{lstlisting}

One should note that, although subject to conditional constraints, the rules of weakening and conjunction introduction in PECR are derivable, i.e. they are not axioms. 
It should also be observed that if we set $a$ to the empty program, i.e. $a=ep$, in theorems thm 3-thm 4 we obtain theorems thm 1-thm 2 that are similar to the left weakening rule of the sequent calculus.
Similarly, if we set $b$ to the empty program in theorems thm 5-thm 6 we also obtain theorems thm 1-thm 2.
Thus, as well as being derivable, the rules of weakening and conjunction introduction in PECR are not independent. 

In the sequent calculus there is a right conjunction rule (\ref{seq_conjr}), where, for intuitionistic logic, we modify the right hand side of the turnstile, $\vdash$, to contain at most a single formula.
In PECR this is expressed in the form of construction rule cr 14.

\textbf{The program extension rule.}
At each step of a proof construction a new statement $c$ of the proof is generated from an extended program derivation $epd~[q~p~c]~[s]$, where $p$ is the current program of the proof and $q \subseteqq p$.
The process is one of finding an axiom/theorem whose premises can be matched with some sublist $q$ of the current program $p$.
Crucial to identifying the program $[q~c]$ as an extended program, is to establish that $[q~c]$ is program and I/O equivalent to some known axiom/theorem.
Having achieved this VPC then constructs the appropriate extended program derivation $epd~[q~p~c]~[s]$.

The following theorem demonstrates the formal procedure of identifying $[q~c]$ as an extended program to finally conclude that the program $[p~c]$, constructed from the extended program derivation $epd~[q~p~c]~[s]$, is also an extended program.
To be more precise, the program $[q~c]$ is identified as an extended program, i.e. $c:\mb{P}_{ext}(q)$, if there exists programs $[q_1~c_1]$ and $[q_2~c_2]$ such that $q_1 \equiv q$, $c_1 \equiv c$ and $q_1 \thicksim q_2$, $c_1 \thicksim c_2$, where $[q_2~c_2],~c_2:\mb{P}_{iext}(q_2)$, is an axiom/theorem.
Note that for the program extension rule to hold we only require that $[q~c],~c:\mb{P}_{ext}(q)$.
As discussed in Section \ref{sof}, generality of the program extension rule is not lost by the restriction of matching $[q~c]$ to an axiom/theorem. 

\begin{lstlisting}
Theorem thm 7.
[[sub [q p] [ ] conc [p c] [s] eqv [q1 q] [ ] eqv [c1 c] [ ]
 eqio [q1 q2] [ ] conc [q1 c1] [u] conc [q2 c2] [v] eqio [u v] [ ]
 ext [q2 c2] [ ]] ext [p c] [ ]]

Proof.
  1 sub [q p] [ ]
  2 conc [p c] [s]
  3 eqv [q1 q] [ ]
  4 eqv [c1 c] [ ]
  5 eqio [q1 q2] [ ]
  6 conc [q1 c1] [u]
  7 conc [q2 c2] [v]
  8 eqio [u v] [ ]
  9 ext [q2 c2] [ ]
 10 aext [q1 c1] [ ]         cr 5 [9 7 6 5 8]
 11 ext [q1 c1] [ ]          cr 1 [10]
 12 aext [q c1] [ ]          cr 3 [11 3]
 13 ext [q c1] [ ]           cr 1 [12]
 14 aext [q c] [ ]           cr 4 [13 4]
 15 ext [q c] [ ]            cr 1 [14]
 16 epd [q p c] [a]          spl 3 [1 15 2]
 17 aext [p c] [ ]           per [16]
 18 ext [p c] [ ]            cr 1 [17]
\end{lstlisting}
At line 16 of the above proof, VPC has selected the first freely available parameter name $a$ of the extended program derivation $epd~[q~p~c]~[a]$ but it does recognize the equivalence of $a$ and $s$.
This equivalence statement is supplied as an option by VPC during the proof construction but is not included in the above proof since it is redundant to the derivation of the conclusion.

In the premise of thm 7 is included the conditional statement $conc~[p~c]~[s]$.
This is necessary to ensure that the variable names of the I/O lists of $c$ are compatible with the variable names of the I/O lists of the program $p$.
Since $q \subseteqq p$ and $c:\mb{P}_{ext}(q)$, the only incompatibility that can occur is that $y_c \cap [x_p~y_p] \neq [~]$.
During a proof construction, we are free to choose new variable names for the output list, $y_c$, of $c$ to ensure that no conflict of variable names occurs.   

\textbf{Disjunction distributive rules.}
The right and left disjunction distributive rules, dsj \ref{dsj04}-dsj \ref{dsj09}, can be combined to form the more general rules that follow.
Theorems thm 8-thm 10 are derivations for these generalized distributive rules for disjunctions.

Theorem thm 8 proves that if $d=a|b$ and $u=[p~a~q]$ and $v=[p~b~q]$ are type $\mb{P}$ then $e=u|v$ is type $\mb{P}$.
This combines dsj \ref{dsj04} and dsj \ref{dsj07} into a single rule.
\begin{lstlisting}
Theorem thm 8.
[[disj [a b] [d] conc [p a] [f] conc [f q] [u] conc [p b] [g]
 conc [g q] [v]] disj [u v] [e]]

Proof.
  1 disj [a b] [d]
  2 conc [p a] [f]
  3 conc [f q] [u]
  4 conc [p b] [g]
  5 conc [g q] [v]
  6 disj [f g] [c]           dsj 4 [2 4 1]
  7 disj [u v] [e]           dsj 7 [3 5 6]
\end{lstlisting}
Theorem thm 9 proves that if $d=a|b$, $u=[p~a~q]$, $v=[p~b~q]$ and $r=[p~d]$ are type $\mb{P}$ then $h=[r~q]=[p~d~q]$ is type $\mb{P}$.
This combines dsj \ref{dsj05} and dsj \ref{dsj08} into a single rule.
\begin{lstlisting}
Theorem thm 9.
[[disj [a b] [d] conc [p a] [f] conc [f q] [u] conc [p b] [g]
 conc [g q] [v] conc [p d] [r]] conc [r q] [h]]

Proof.
  1 disj [a b] [d]
  2 conc [p a] [f]
  3 conc [f q] [u]
  4 conc [p b] [g]
  5 conc [g q] [v]
  6 conc [p d] [r]
  7 disj [f g] [c]           dsj 4 [2 4 1]
  8 conc [c q] [e]           dsj 8 [3 5 7]
  9 eqv [r c] [ ]            dsj 6 [2 4 1 6 7]
 10 eqv [c r] [ ]            cr 7 [9]
 11 conc [r q] [h]           sr 1 [8 10]
\end{lstlisting}
Theorem thm 10 proves that if $d=a|b$ and $s=[p~d~q]$, $j=[p~a~q] | [p~b~q]$ are type $\mb{P}$ then $j \equiv s$.
This combines dsj \ref{dsj06} and dsj \ref{dsj09} into a single rule.

\begin{lstlisting}
Theorem thm 10.
[[disj [a b] [d] conc [p d] [r] conc [r q] [s] conc [p a] [f]
 conc [f q] [u] conc [p b] [g] conc [g q] [v] disj [u v] [j]]
 eqv [s j] [ ]]

Proof.
  1 disj [a b] [d]
  2 conc [p d] [r]
  3 conc [r q] [s]
  4 conc [p a] [f]
  5 conc [f q] [u]
  6 conc [p b] [g]
  7 conc [g q] [v]
  8 disj [u v] [j]
  9 disj [f g] [c]           dsj 4 [4 6 1]
 10 conc [c q] [e]           dsj 8 [5 7 9]
 11 eqv [r c] [ ]            dsj 6 [4 6 1 2 9]
 12 eqv [e s] [ ]            sr 2 [3 11 10]
 13 eqv [s e] [ ]            cr 7 [12]
 14 eqv [e j] [ ]            dsj 9 [5 7 9 10 8]
 15 eqv [s j] [ ]            sr 1 [13 14]
\end{lstlisting}

\textbf{Disjunction contraction rule.}
The contraction rule, dsj \ref{dsj01}, can be written in the more general form given by theorem thm 13, below.

Theorem thm 11 proves that if $d=a|b$, $s=[p~d~q]$, $j=[p~a~q] | [p~b~q]$ and $k=[j~c]$ are type $\mb{P}$ then $e=[s~c]$ is type $\mb{P}$.
\begin{lstlisting}
Theorem thm 11.
[[disj [a b] [d] conc [p d] [r] conc [r q] [s] conc [p a] [f]
 conc [f q] [u] conc [p b] [g] conc [g q] [v] disj [u v] [j]
 conc [j c] [k]] conc [s c] [e]]

Proof.
  1 disj [a b] [d]
  2 conc [p d] [r]
  3 conc [r q] [s]
  4 conc [p a] [f]
  5 conc [f q] [u]
  6 conc [p b] [g]
  7 conc [g q] [v]
  8 disj [u v] [j]
  9 conc [j c] [k]
 10 eqv [s j] [ ]            thm 10 [1 2 3 4 5 6 7 8]
 11 eqv [j s] [ ]            cr 7 [10]
 12 conc [s c] [e]           sr 1 [9 11]
\end{lstlisting}
Theorem thm 12 proves that if $d=a|b$, $s=[p~d~q]$, $j=[p~a~q] | [p~b~q]$, $k=[j~c]$ and $l=[s~c]$ are type $\mb{P}$ then $l \equiv k$.
\begin{lstlisting}
Theorem thm 12.
[[disj [a b] [d] conc [p d] [r] conc [r q] [s] conc [p a] [f]
 conc [f q] [u] conc [p b] [g] conc [g q] [v] disj [u v] [j]
 conc [j c] [k] conc [s c] [l]] eqv [l k] [ ]]

Proof.
  1 disj [a b] [d]
  2 conc [p d] [r]
  3 conc [r q] [s]
  4 conc [p a] [f]
  5 conc [f q] [u]
  6 conc [p b] [g]
  7 conc [g q] [v]
  8 disj [u v] [j]
  9 conc [j c] [k]
 10 conc [s c] [l]
 11 disj [f g] [e]           dsj 4 [4 6 1]
 12 conc [e q] [h]           dsj 8 [5 7 11]
 13 eqv [r e] [ ]            dsj 6 [4 6 1 2 11]
 14 eqv [h s] [ ]            sr 2 [3 13 12]
 15 eqv [h j] [ ]            dsj 9 [5 7 11 12 8]
 16 eqv [j h] [ ]            cr 7 [15]
 17 conc [h c] [i]           sr 1 [9 16]
 18 eqv [i k] [ ]            sr 2 [9 16 17]
 19 eqv [l i] [ ]            sr 2 [17 14 10]
 20 eqv [l k] [ ]            sr 1 [19 18]
\end{lstlisting}
Theorem thm 13 is the derivation of the general version of the contraction rule.
It proves that if $d=a|b$, $s=[p~d~q]$, $u=[p~a~q]$, $v=[p~b~q]$ and $j=u | v$ are type $\mb{P}$ and $c:\mb{P}_{ext}~(u)$, $c:\mb{P}_{ext}~(v)$ then the type assignment $c::\mb{P}_{ext}(s)$ is valid.
\begin{lstlisting}
Theorem thm 13.
[[disj [a b] [d] conc [p d] [r] conc [r q] [s] conc [p a] [f]
 conc [f q] [u] conc [p b] [g] conc [g q] [v] disj [u v] [j]
 ext [u c] [ ] ext [v c] [ ]] aext [s c] [ ]]

Proof.
  1 disj [a b] [d]
  2 conc [p d] [r]
  3 conc [r q] [s]
  4 conc [p a] [f]
  5 conc [f q] [u]
  6 conc [p b] [g]
  7 conc [g q] [v]
  8 disj [u v] [j]
  9 ext [u c] [ ]
 10 ext [v c] [ ]
 11 aext [j c] [ ]           dsj 1 [9 10 8]
 12 ext [j c] [ ]            cr 1 [11]
 13 eqv [s j] [ ]            thm 10 [1 2 3 4 5 6 7 8]
 14 eqv [j s] [ ]            cr 7 [13]
 15 aext [s c] [ ]           cr 3 [12 14]
\end{lstlisting}

\textbf{Disjunction contraction rule 2.}
Theorem thm 14 shows that the contraction rule 2 (Section \ref{sdisj}) is a theorem.
It states that if $d=a|b$ is type $\mb{P}$, $c:\mb{P}_{ext}(a)$ and $b:\mb{P}_{false}$ then the type assignment $c::\mb{P}_{ext}(d)$ is valid.
\begin{lstlisting}
Theorem thm 14.
[[disj [a b] [d] ext [a c] [ ] false [b] [ ]] aext [d c] [ ]]

Proof.
  1 disj [a b] [d]
  2 ext [a c] [ ]
  3 false [b] [ ]
  4 eqv [d a] [ ]            dsj 10 [1 3]
  5 eqv [a d] [ ]            cr 7 [4]
  6 aext [d c] [ ]           cr 3 [2 5]
\end{lstlisting}
Theorem thm 15 generalizes the contraction rule 2.
It states that if $d=a|b$, $s=[p~d~q]$, $u=[p~a~q]$ and $v=[p~d~q]$ are type $\mb{P}$ and $c:\mb{P}_{ext}(u)$, $v:\mb{P}_{false}$, then the type assignment $c::\mb{P}_{ext}(s)$ is valid.
\begin{lstlisting}
Theorem thm 15.
[[disj [a b] [d] conc [p d] [r] conc [r q] [s] conc [p a] [f]
 conc [f q] [u] conc [p b] [g] conc [g q] [v] ext [u c] [ ]
 false [v] [ ]] aext [s c] [ ]]

Proof.
  1 disj [a b] [d]
  2 conc [p d] [r]
  3 conc [r q] [s]
  4 conc [p a] [f]
  5 conc [f q] [u]
  6 conc [p b] [g]
  7 conc [g q] [v]
  8 ext [u c] [ ]
  9 false [v] [ ]
 10 disj [u v] [e]           thm 8 [1 4 5 6 7]
 11 eqv [s e] [ ]            thm 10 [1 2 3 4 5 6 7 10]
 12 eqv [e u] [ ]            dsj 10 [10 9]
 13 eqv [s u] [ ]            sr 1 [11 12]
 14 eqv [u s] [ ]            cr 7 [13]
 15 aext [s c] [ ]           cr 3 [8 14]
\end{lstlisting}

\textbf{Disjunction contraction rule 3.}
The following shows that the contraction rule 3 (Section \ref{sdisj}) is a theorem. 
Theorem thm 16 states that if $d=a|b$ is type $\mb{P}$ and $a,b:\mb{P}_{false}$ then the type assignment $a::\mb{P}_{false}$ is valid.
\begin{lstlisting}
Theorem thm 16.
[[disj [a b] [d] false [a] [ ] false [b] [ ]] afalse [d] [ ]]

Proof.
  1 disj [a b] [d]
  2 false [a] [ ]
  3 false [b] [ ]
  4 eqv [d a] [ ]            dsj 10 [1 3]
  5 eqv [a d] [ ]            cr 7 [4]
  6 afalse [d] [ ]           flse 3 [2 5]
\end{lstlisting}
Theorem thm 17 generalizes the contraction rule 3.
It states that if $d=a|b$ $s=[p~d~q]$, $u=[p~a~q]$ and $v=[p~d~q]$ are type $\mb{P}$ and $u,v:\mb{P}_{false}$ then the type assignment $s::\mb{P}_{false}$ is valid.
\begin{lstlisting}
Theorem thm 17.
[[disj [a b] [d] conc [p d] [r] conc [r q] [s] conc [p a] [f]
 conc [f q] [u] conc [p b] [g] conc [g q] [v] false [u] [ ]
 false [v] [ ]] afalse [s] [ ]]

Proof.
  1 disj [a b] [d]
  2 conc [p d] [r]
  3 conc [r q] [s]
  4 conc [p a] [f]
  5 conc [f q] [u]
  6 conc [p b] [g]
  7 conc [g q] [v]
  8 false [u] [ ]
  9 false [v] [ ]
 10 disj [u v] [c]           thm 8 [1 4 5 6 7]
 11 eqv [s c] [ ]            thm 10 [1 2 3 4 5 6 7 10]
 12 eqv [c s] [ ]            cr 7 [11]
 13 afalse [c] [ ]           thm 16 [10 8 9]
 14 false [c] [ ]            flse 2 [13]
 15 afalse [s] [ ]           flse 3 [14 12]
\end{lstlisting}

\chapter{Arrays.}\label{array}

\section{Dimension lists.}\label{sdl}

Here we will be interested in arrays whose elements are integers of type $\mb{I}$.
Matrices can be represented as arrays and the results derived from the properties of arrays also apply to matrices as a special case. 

An array $a$ is associated with the general type $a:\mb{A}$.
If we want to specify its dimensions we shall write $a:\mb{A}(n_1,\ldots,n_d)$ for some $d:\mb{I}_+$.
The elements of an array $a:\mb{A}(n_1,\ldots,n_d)$ are type $\mb{I}$ objects and are written as $a(x_1,\ldots,x_d)$, where $x_i: \mb{I}_+$,~$1 \leq x_i \leq n_i,~i=1,\ldots,d$.
When discussing arrays of general dimensions in a conventional mathematical language indexing can become cumbersome to write down.
We will use some shorthand notation.

First we will define the array dimension lists
\be
\begin{array}{ll}
l=[l_1~\ldots~l_p], \qquad & l_i:\mb{I}_+,~i=1,\ldots,p,~p:\mb{I}_+ \\
m=[m_1~\ldots~m_q], \qquad & m_i:\mb{I}_+,~i=1,\ldots,q,~q:\mb{I}_+ \\
n=[n_1~\ldots~n_r], \qquad & n_i:\mb{I}_+,~i=1,\ldots,r,~r:\mb{I}_+ \\
\end{array}
\ee
Array dimension lists are fixed and will be used to define dimensions of arrays.
We write
\be
\bal
\mb{A}(l) = & \mb{A}(l_1,\ldots,l_p) \\
\mb{A}(l,m) =& \mb{A}(l_1,\ldots,l_p,m_1,\ldots,m_q) \\
\mb{A}(l,m,n) = & \mb{A}(l_1,\ldots,l_p,m_1,\ldots,m_q,n_1,\ldots,n_r) \\
\eal
\ee  
Elements of arrays will be expressed as functions of the index lists
\be
\begin{array}{ll}
x=[x_1~\ldots~x_p], \qquad & x_i:\mb{I}_+,~1 \leq x_i \leq l_i,~i=1,\ldots,p \\
y=[y_1~\ldots~y_q], \qquad & y_i:\mb{I}_+,~1 \leq y_i \leq m_i,~i=1,\ldots,q \\
z=[z_1~\ldots~z_r], \qquad & z_i:\mb{I}_+,~1 \leq z_i \leq n_i,~i=1,\ldots,r \\
\end{array}
\ee
We write
\be
x \in \Omega (l)
\ee
to mean that $x$ is the index list of an array in the domain
\be
1 \leq x_i \leq l_i,~i=1,\ldots,p
\ee

The rank of an array is equal to the length of its dimension list.
Using this notation, an array $a:\mb{A}(l)$ has rank $p$ with the element representation $a(x)=a(x_1,\ldots,x_p)$, $x \in \Omega(l)$.
A scalar has zero rank.

Sometimes it will be convenient to add an additional array partition.
An array $a:\mb{A}(l,m)$ has rank $p+q$ with the element representation $a(x,y)=a(x_1,\ldots,x_p,y_1,\ldots,y_q)$, $x \in \Omega(l)$, $y \in \Omega(m)$, and an array $a:\mb{A}(l,m,n)$ has rank $p+q+r$ with the element representation  $a(x,y,z)=a(x_1,\ldots,x_p,y_1,\ldots,y_q,z_1,\ldots,z_r)$, $x \in \Omega(l)$, $y \in \Omega(m)$, $z \in \Omega(n)$.

We can now define addition of arrays as follows.
For two arrays $a,b:\mb{A}(l)$ the array addition, $a+b$, is defined by the assignment of its elements
\be
c(x):=a(x)+b(x), \qquad x \in \Omega(l)
\ee
where the array $c:\mb{A}(l)$ provided that each element sum $a(x)+b(x)$ exists.
It will also be meaningful to sum two arrays $a,b:\mb{A}(l,m)$ with the element representation
\be
c(x,y):=a(x,y)+b(x,y), \qquad x \in \Omega(l),~y \in \Omega(m)
\ee
and, similarly, the sum of two arrays $a,b:\mb{A}(l,m,n)$ with the element representation
\be
c(x,y,z):=a(x,y,z)+b(x,y,z), \qquad x \in \Omega(l),~y \in \Omega(m),~z \in \Omega(n)
\ee

If $a:\mb{A}(l)$ and $b:\mb{A}(m)$ such that $l_p=m_1=k$ then the array multiplication $ab$ yields the array $c:\mb{A}(n)$, $n=[l_1~\ldots~l_{p-1}~m_2~\ldots~m_q]$, given by 
\be
c(l_1,\ldots,l_{p-1},m_2,\ldots,m_q) := \sum_{j=1}^k a(l_1,\ldots,l_{p-1},j)*b(j,m_2,\ldots,m_q)
\ee
The ranks of the arrays $a$ and $b$, respectively, are $p$ and $q$, respectively, and the rank of $c$ is $p+q-2$.

For $r:\mb{I}$ and $a:\mb{A}(l)$ the scalar multiplication of an array, $c=r*a$, has the element representation
\be
c(x):=r*a(x), \qquad x \in \Omega(l) 
\ee
provided that each element multiplication $r*a(x)$ exists.

\textbf{Arrays as list partitions.}
Arrays are stored as lists with a specific partition.
The position of an element, $a(x)$, of an array, $a:\mb{A}(l)$, is given by the list index
\be
x_1 + (x_2-1)*l_1 + \cdots + (x_p-1)*l_{p-1}*l_{p-2}* \cdots *l_1
\ee
where
\be
s=l_1*l_2* \cdots *l_p
\ee
is the length of the list that stores the elements of the array.

\textbf{Vectors.}
A vector can simply be represented by an unpartitioned list and is equivalent to an array of rank 1.
A vector $v$ can be represented by the type $v:\mb{A}(m)$, where $m:\mb{I}_+$ is a scalar rather than a dimension list.

\textbf{Matrices.}
We can also express an array as a matrix.
A matrix can be thought of as an array of rank two and is given the type $\mb{A}(s,t)$, where $s,t:\mb{I}_+$.
If $a:\mb{A}(l,m)$, whose elements are given by $a(x,y)$, then we can construct the matrix $c:\mb{A}(s,t)$, $s,t:\mb{I}_+$, where
\be
\bal
s= & l_1*l_2* \cdots *l_p \\
t= & m_1*m_2* \cdots *m_q \\
\eal
\ee
Each element $c(i,j)$ of the matrix $c$ can be obtained from the array order index functions 
\be
\bal
i(x) = & x_1 + (x_2-1)*l_1 + \cdots + (x_p-1)*l_{p-1}*l_{p-2}* \cdots *l_1 \\
j(y) = & y_1 + (y_2-1)*m_1 + \cdots + (y_q-1)*m_{q-1}*m_{q-2}* \cdots *m_1 \\
\eal
\ee

\vspace{5mm}

\textbf{Notes.}

\begin{itemize}

\item Because our main focus here is to establish some basic properties of finite state arithmetic we have defined our arrays such that all elements of an array are of the same type $\mb{I}$ (or $\mb{J}$).
In a more general context arrays can be used to store any objects that are strings of a well defined structure.
For example programs that are constructed as lists are stored as vector arrays where each element of the array is a subprogram.
The list properties outlined in Section \ref{sl} apply to all arrays.
The atomic programs and axioms introduced in the following sections apply only to arrays whose elements are assigned numerical values.

\item The lower limits of the index lists of an array $a:\mb{A}(l)$ need not be limited to 1, as given above.
Common alternatives are
\be
\begin{array}{ll}
x=[x_1~\ldots~x_p], \qquad & x_i:\mb{I}_+,~0 \leq x_i \leq l_i,~i=1,\ldots,p \\
\end{array}
\ee
or
\be
\begin{array}{ll}
x=[x_1~\ldots~x_p], \qquad & x_i:\mb{I}_+,~-l_i \leq x_i \leq l_i,~i=1,\ldots,p \\
\end{array}
\ee
For any case where the lower limits on each index $x_i$ are 1, 0 or $-l_i$ we will still use the notation $a:\mb{A}(l)$.
We could generalize further by allowing the bounds of each index $x_i$ to be any pair $l_i^\prime, l_i : \mb{I}$ such that $l_i^\prime \leq x_i \leq l_i$.
To avoid introducing more notation we will restrict the index list bounds to the three cases just mentioned.  

\end{itemize}

\section{Array atomic programs.}

We will seek an element and dimension free formulation.
A dimension free formulation means that the dimensions of any array will not be explicitly specified in the I/O lists of array atomic programs.
Every atomic program will internally identify the dimensions of the arrays and where operations of arithmetic  are involved will check for the appropriate compatibility of the arrays under that operation.
We can regard the dimensions of the initial arrays, along with value assignments of their elements, to be prescribed by the $read~x~[~]$ program under the general program structure (\ref{core}).
New arrays are generated by atomic assignment programs that internally set the dimensions of the new array variables.
Once the dimensions of an array have been assigned they are stored in memory and accessed whenever that array is employed as an assigned value input of a program.

Here we will include scalar multiplication of arrays where the scalars are integers of type $\mb{I}$.
As such we will need to introduce atomic programs whose input lists will be of a mixed type.
Since the scalars themselves must obey the usual axioms of arithmetic over $\mb{I}$ we will need to append to the collection of array axioms the axioms of integer arithmetic over $\mb{I}$.
This means that when setting up an application involving arrays we must also include the atomic programs associated with integer arithmetic over $\mb{I}$.

The following array atomic programs are used.

\vspace{5mm}

\begin{tabular}{|c|c|}
\hline
Atomic program names & Atomic program type \\
\hline
$typea,~eqa,~dima$ & $\mb{P}_{type}$ \\
\hline
$adda,~multa,~zarr,~smult$ & $\mb{P}_{assign}$ \\
\hline
\end{tabular}

\vspace{5mm}

In the description of the array atomic programs given below the following notation for the array dimension lists will be assumed.
\be
\begin{array}{ll}
l=[l_1~\ldots~l_p], \qquad & l_i:\mb{I}_+,~i=1,\ldots,p,~p:\mb{I}_+ \\
m=[m_1~\ldots~m_q], \qquad & m_i:\mb{I}_+,~i=1,\ldots,q,~q:\mb{I}_+ \\
n=[n_1~\ldots~n_r], \qquad & n_i:\mb{I}_+,~i=1,\ldots,r,~r:\mb{I}_+ \\
\end{array}
\ee

\parskip0pt

\vspace{5mm}
\textbf{Check type array.}

\textbf{\textit{Syntax:}} $typea~[a]~[~]$.

\textbf{\textit{Program Type:}} $\mb{P}_{type}$.

\textbf{\textit{Type checks:}} $a:\mb{A}(m)$.

\textbf{\textit{Description:}} $typea$ checks that the assigned value of the variable $a$ is of type $\mb{A}(m)$ for some dimension list $m$.
The dimension list $m$ of the array $a$ is assigned prior to entry to the program $typea$ and is recognized internally by $typea$.
$typea$ halts with an execution error if there is a type violation.

\vspace{5mm}
\textbf{Check equality of array.}

\textbf{\textit{Syntax:}} $eqa~[a~b]~[~]$.

\textbf{\textit{Program Type:}} $\mb{P}_{type}$.

\textbf{\textit{Type checks:}} $a:\mb{A}(m)$, $b:\mb{A}(m)$ and $a=b$.

\textbf{\textit{Description:}} $eqa$ first checks that the assigned values of the variables $a$ and $b$ are of type $\mb{A}(m)$ for some dimension list $m$.
It then checks the assigned value equality $a=b$, i.e. the assigned values of each corresponding element of $a$ and $b$ are type $\mb{I}$ and are equal.
The dimension lists of the arrays $a$ and $b$ are assigned prior to entry to the program $eqa$ and are recognized internally by $eqa$.
$eqa$ halts with an execution error if there is a type violation.
Type violation includes the case where the value assignment array equality $a=b$ is not satisfied.

\vspace{5mm}
\textbf{Check equality of array dimensions.}

\textbf{\textit{Syntax:}} $dima~[a~b]~[~]$.

\textbf{\textit{Program Type:}} $\mb{P}_{type}$.

\textbf{\textit{Type checks:}} $a:\mb{A}(m)$, $b:\mb{A}(m)$.

\textbf{\textit{Description:}} $dima$ checks that the assigned values of the variables $a$ and $b$ are of type $\mb{A}(m)$ for some dimension list $m$.
The dimension lists of the arrays $a$ and $b$ are assigned prior to entry to the program $dima$ and are recognized internally by $dima$.
$dima$ halts with an execution error if there is a type violation.

\vspace{5mm}
\textbf{Array addition.}

\textbf{\textit{Syntax:}} $adda~[a~b]~[c]$.

\textbf{\textit{Program Type:}} $\mb{P}_{assign}$.

\textbf{\textit{Type checks:}} $a:\mb{A}(m)$, $b:\mb{A}(m)$.

\textbf{\textit{Assignment map.}} $c:=a+b$.

\textbf{\textit{Type assignment.}} $c::\mb{A}(m)$.

\textbf{\textit{Description:}} $adda$ first checks that the assigned values of the variables $a$ and $b$ are of type $\mb{A}(m)$ for some dimension list $m$.
It then attempts to assign to $c$ the array addition of $a$ and $b$, i.e. $c:=a+b$. 
A successful value assignment is accompanied by the type assignment $c::\mb{A}(m)$.
The dimension lists of the arrays $a$ and $b$ are assigned prior to entry to the program $adda$ and are recognized internally by $adda$.
$adda$ halts with an execution error if there is a type violation.

\vspace{5mm}
\textbf{Array multiplication.}

\textbf{\textit{Syntax:}} $multa~[a~b]~[c]$.

\textbf{\textit{Program Type:}} $\mb{P}_{assign}$.

\textbf{\textit{Type checks:}} $a:\mb{A}(l)$, $b:\mb{A}(m)$ such that $l_p=m_1$.

\textbf{\textit{Assignment map.}} $c:=ab$.

\textbf{\textit{Type assignment.}} $c::\mb{A}(n)$, $n=[l_1~\ldots~l_{p-1}~m_2~\ldots~m_q]$.

\textbf{\textit{Description:}} $multa$ first checks that the assigned value of the variable $a$ is of type $\mb{A}(l)$, for some dimension list $l=[l_1~\ldots~l_p]$, and the assigned value of the variable $b$ is of type $\mb{A}(m)$, for some dimension list $m=[m_1~\ldots~m_q]$ such that $l_p=m_1$.
It then attempts to assign to $c$ the array multiplication of $a$ and $b$, i.e. $c:=ab$.
If $a:\mb{A}(l)$ and $b:\mb{A}(m)$ such that $l_p=m_1=k$ then the array multiplication $ab$ yields the array $c:\mb{A}(n)$, $n=[l_1~\ldots~l_{p-1}~m_2~\ldots~m_q]$, given by 
\be
c(l_1,\ldots,l_{p-1},m_2,\ldots,m_q) := \sum_{j=1}^k a(l_1,\ldots,l_{p-1},j)*b(j,m_2,\ldots,m_q)
\ee
A successful value assignment is accompanied by the type assignment $c::\mb{A}(n)$.
The dimension lists of the arrays $a$ and $b$ are assigned prior to entry to the program $multa$ and are recognized internally by $multa$.
$multa$ halts with an execution error if there is a type violation.

\vspace{5mm}
\textbf{Construct the null array.}

\textbf{\textit{Syntax:}} $zarr~[a]~[b]$.

\textbf{\textit{Program Type:}} $\mb{P}_{assign}$.

\textbf{\textit{Type checks:}} $a:\mb{A}(m)$.

\textbf{\textit{Assignment map.}} $b(i_1,\ldots,i_q):=0$ for all $1 \leq i_j \leq m_j,~1 \leq j \leq q$.

\textbf{\textit{Type assignment.}} $b::\mb{A}(m)$.

\textbf{\textit{Description:}} $zarr$ first checks that the assigned value of the variable $a$ is of type $\mb{A}(m)$ for some dimension list $m$.
If successful it then constructs the matrix $b$  with the same dimensions of $a$ such that $b(i_1,\ldots,i_q):=0$ for $1 \leq i_j \leq m_j$, $1 \leq j \leq q$.
If $a:\mb{A}(m)$ then the existence of $b:\mb{A}(m)$ is guaranteed.
The value assignment of $b$ is accompanied by the type assignment $b::\mb{A}(m)$.
The dimension list of the array $a$ is assigned prior to entry to the program $zarr$ and is recognized internally by $zarr$.
$zarr$ halts with an execution error if there is a type violation.

\vspace{5mm}
\textbf{Scalar multiplication.}

\textbf{\textit{Syntax:}} $smult~[r~a]~[b]$.

\textbf{\textit{Program Type:}} $\mb{P}_{assign}$.

\textbf{\textit{Type checks:}} $r:\mb{I}$, $a:\mb{A}(m)$.

\textbf{\textit{Assignment map.}} $b:=r*a$.

\textbf{\textit{Type assignment.}} $b::\mb{A}(m)$.

\textbf{\textit{Description:}} $smult$ first checks that the assigned value of the variable $r$ is of type $\mb{I}$ and the assigned value of the variable $a$ is of type $\mb{A}(m)$ for some dimension list $m$.
It then attempts to assign to $b$ the scalar multiplication of $r$ and $a$, i.e. $b:=r*a$.
A successful value assignment is accompanied by the type assignment $b::\mb{A}(m)$.
The dimension list of the array $a$ is assigned prior to entry to the program $smult$ and is recognized internally by $smult$.
$smult$ halts with an execution error if there is a type violation.

\parskip10pt

\textbf{Notes.}

\begin{itemize}

\item The results presented here will also be applicable for arrays over $\mb{J}$.
We can define the elements of an object $a:\mb{A}$ to be type $\mb{J}$.
The dimensions of arrays remain objects of type $\mb{I}_+$.
All of the axioms and derivations that follow will also be valid if we regard the elements of $\mb{A}$ to be of type $\mb{J}$ instead of type $\mb{I}$.
We may also regard the scalars of scalar array multiplications to be of type $\mb{J}$.
Whenever an object is referred to as type $\mb{I}$ and that object is not a dimension of an array we simply replace $\mb{I}$ with $\mb{J}$.

\end{itemize}

\section{Axioms for array arithmetic.}

\textbf{I/O type $\mb{A}$.}
The following are the I/O type axioms for integer array atomic programs.

\textbf{aio}
\be
\bal
~[ \mathfrak{p}~x~y~typea~[a]~[~]] , & \qquad a \in [x~y],~a:\mb{A} \\
~[ \mathfrak{p}~x~y~typei~[a]~[~]] , & \qquad a \in x,~a:\mb{I} \\
\eal
\ee
The second expression is necessary for atomic programs involving scalar multiplication.
The output lists of all array atomic assignment programs will contain only a single element of type $\mb{A}$. 

\textbf{Substitution rule.}
The substitution rule will be applied as an axiom to array atomic programs $\mathfrak{p}~x~y$ such that
\be
\mathfrak{p} \in [eqa~adda~multa~zarr~smult~dima]
\ee
The first part of the substitution rule is an existence axiom.
Since we are including atomic programs for scalar multiplication we will present the two versions.

\textbf{sr 1}
\be
 [[ \mathfrak{p}~x~y~eqa~[x_i~a]~[~]]~\mathfrak{p}~\bar{x}~\bar{y}], \qquad x_i \in x,~\bar{x} = x(x_i \to a),~x_i:\mb{A} 
\ee
\be
 [[ \mathfrak{p}~x~y~eqi~[x_i~a]~[~]]~\mathfrak{p}~\bar{x}~\bar{y}], \qquad x_i \in x,~\bar{x} = x(x_i \to a),~x_i:\mb{I}
\ee
where $x=[x_i]_{i=1}^n$, for some $n:\mb{I}_+$.
The output lists $y$ and $\bar{y}$ may be empty lists.

The second part of the substitution rule is applicable when $y$ and $\bar{y}$ are not empty lists.
To present the axiom in a more general form we write
\be
y=[y_j]_{j=1}^m,~~~\bar{y}=[\bar{y}_j]_{j=1}^m, \qquad m:\mb{I}_+
\ee
For any substitution $x(x_i \to a)$, VPC will generate the following axioms for $j=1,\ldots,m$.
The two versions of the second part of the substitution rule are

\textbf{sr 2}
\be
\bal
&  [[ \mathfrak{p}~x~y~eqa~[x_i~a]~[~]~\mathfrak{p}~\bar{x}~\bar{y}]~eqa~[\bar{y}_j~y_j]~[~]], \\
& \hspace{60mm} x_i \in x,~\bar{x} = x(x_i \to a),~x_i:\mb{A} \\
\eal
\ee
\be
\bal
&  [[ \mathfrak{p}~x~y~eqi~[x_i~a]~[~]~\mathfrak{p}~\bar{x}~\bar{y}~]~eqa~[y_j~\bar{y}_j]~[~]], \\
& \hspace{60mm} x_i \in x,~\bar{x} = x(x_i \to a),~x_i:\mb{I} \\
\eal
\ee

\textbf{Equality axioms.}

\emph{Reflexivity.}
\begin{axarr}
\be
 [typea~[a]~[~]~eqa~[a~a]~[~]]
\ee
\end{axarr}
\emph{Symmetry.}
\begin{axarr}
\be
 [eqa~[a~b]~[~]~eqa~[b~a]~[~]]
\ee
\end{axarr}
Array equality satisfies the property of transitivity
\be
 [[eqa~[a~b]~[~]~eqa~[b~c]~[~]]~eqa~[a~c]~[~]]
\ee
This is not included as an axiom because it follows from the substitution rule.

\textbf{Axioms for addition and multiplication.}

\emph{Commutativity of addition.}
\begin{axarr}
\be
 [adda~[a~b]~[c]~adda~[b~a]~[d]]
\ee
\end{axarr}
\begin{axarr}
\be
 [[adda~[a~b]~[c]~adda~[b~a]~[d]]~eqa~[d~c]~[~]]
\ee
\end{axarr}

\emph{Associativity of addition.}
\begin{axarr}
\be
\bal
&  [[adda~[a~b]~[d]~adda~[d~c]~[x]~adda~[b~c]~[e]]~adda~[a~e]~[y]]
\eal
\ee
\end{axarr}
\begin{axarr}
\be
\bal
&  [[adda~[a~b]~[d]~adda~[d~c]~[x]~adda~[b~c]~[e]~adda~[a~e]~[y]]~eqa~[y~x]~[~]] \\
\eal
\ee
\end{axarr}

\emph{Addition with the null array.}
\begin{axarr}
\be
 [typea~[a]~[~]~zarr~[a]~[b]]
\ee
\end{axarr}
\begin{axarr}
\be
 [zarr~[a]~[b]~adda~[a~b]~[c]
\ee
\end{axarr}
\begin{axarr}
\be
 [[zarr~[a]~[b]~adda~[a~b]~[c]]~eqa~[c~a]~[~]]
\ee
\end{axarr}

\emph{Associativity of array multiplication.}
\begin{axarr}
\be
\bal
&  [[multa~[a~b]~[d]~multa~[b~c]~e]~multa~[d~c]~[x]]~multa~[a~e]~[y]]
\eal
\ee
\end{axarr}
\begin{axarr}
\be
\bal
&  [[multa~[a~b]~[d]~multa~[b~c]~e]~multa~[a~e]~[y]]~multa~[d~c]~[x]]
\eal
\ee
\end{axarr}
\begin{axarr}
\be
\bal
&  [[multa~[a~b]~[d]~multa~[b~c]~e]~multa~[d~c]~[x]~multa~[a~e]~[y]]~eqa~[y~x]~[~]] \\
\eal
\ee
\end{axarr}

\emph{Distributive law (left).}
\begin{axarr}
\be
\bal
&  [[adda~[b~c]~[d]~multa~[a~d]~[x]~multa~[a~b]~[u]~multa~[a~c]~[v]]~adda~[u~v]~[y]]
\eal
\ee
\end{axarr}
\begin{axarr}
\be
\bal
&  [[multa~[a~b]~[u]~multa~[a~c]~[v]~adda~[u~v]~[y]~adda~[b~c]~[d]]~multa~[a~d]~[x]]
\eal
\ee
\end{axarr}
\begin{axarr}
\be
\bal
&  [[adda~[b~c]~[d]~multa~[a~d]~[x]~multa~[a~b]~[u]~multa~[a~c]~[v]~adda~[u~v]~[y]] \\
& \hspace{5mm} eqa~[y~x]~[~]] \\
\eal
\ee
\end{axarr}

\emph{Distributive law (right).}
\begin{axarr}
\be
\bal
&  [[adda~[b~c]~[d]~multa~[d~a]~[x]~multa~[b~a]~[u]~multa~[c~a]~[v]]~adda~[u~v]~[y]]
\eal
\ee
\end{axarr}
\begin{axarr}
\be
\bal
&  [[multa~[b~a]~[u]~multa~[c~a]~[v]~adda~[u~v]~[y]~adda~[b~c]~[d]]~multa~[d~a]~[x]]
\eal
\ee
\end{axarr}
\begin{axarr}
\be
\bal
&  [[adda~[b~c]~[d]~multa~[d~a]~[x]~multa~[b~a]~[u]~multa~[c~a]~[v]~adda~[u~v]~[y]] \\
& \hspace{5mm} eqa~[y~x]~[~]] \\
\eal
\ee
\end{axarr}

\section{Scalar multiplication of arrays.}

With the introduction of atomic programs of the mixed type we define the additive inverse of an array explicitly as a scalar multiplication.
Axioms involving scalar multiplication will be labeled by smlt followed by a number.  

\emph{Additive inverse.}
\begin{smlt}
\be
 [typea~[a]~[~]~smult~[-1~a]~[b]]
\ee
\end{smlt}
\begin{smlt}
\be
 [smult~[-1~a]~[b]~adda~[a~b]~[c]]
\ee
\end{smlt}
\begin{smlt}
\be
\bal
&  [[smult~[-1~a]~[b]~adda~[a~b]~[c]~zarr~[a]~[d]]~eqa~[c~d]~[~]]
\eal
\ee
\end{smlt}

\emph{Multiplication by unity.}
\begin{smlt}
\be
 [typea~[a]~[~]~smult~[1~a]~[b]]
\ee
\end{smlt}
\begin{smlt}
\be
 [smult~[1~a]~[b]~eqa~[b~a]~[~]]
\ee
\end{smlt}

For scalar, matrix and array arithmetic in an environment $\mathfrak{M}(\mathcal{K},\mathcal{L},\mathcal{M})$ one cannot express the associativity and distributivity laws in a concise way.
This is because of the absence of closure over $\mb{I}$ of the operations of scalar addition and multiplication.
For mixed scalar/array programs the list of axioms gets even longer.
We have three associative laws, one involving a scalar multiplication and two involving an array multiplication.
There are two distributive laws, one involving an array addition and the other involving a scalar addition.

%

\emph{Associativity $(r,s:\mb{I}$, $a:\mb{A})$.}   

\begin{smlt}
\be
\bal
&  [[mult~[r~s]~[d]~smult~[s~a]~[e]~smult~[r~e]~[x]]~smult~[d~a]~[y]]
\eal
\ee
\end{smlt}
\begin{smlt}
\be
\bal
&  [[mult~[r~s]~[d]~smult~[s~a]~[e]~smult~[d~a]~[y]]~smult~[r~e]~[x]]
\eal
\ee
\end{smlt}
\begin{smlt}
\be
\bal
&  [[mult~[r~s]~[d]~smult~[s~a]~[e]~smult~[r~e]~[x]~smult~[d~a]~[y]]~eqa~[y~x]~[~]] \\
\eal
\ee
\end{smlt}

\emph{Associativity $(r:\mb{I}$, $a,b:\mb{A})$.}   
\begin{smlt}
\be
\bal
&  [[smult~[r~a]~[d]~multa~[a~b]~[e]~multa~[d~b]~[x]]~smult~[r~e]~[y]]
\eal
\ee
\end{smlt}
\begin{smlt}
\be
\bal
&  [[smult~[r~a]~[d]~multa~[a~b]~[e]~smult~[r~e]~[y]]~multa~[d~b]~[x]]
\eal
\ee
\end{smlt}
\begin{smlt}
\be
\bal
&  [[smult~[r~a]~[d]~multa~[a~b]~[e]~smult~[r~e]~[y]~multa~[d~b]~[x]]~eqa~[y~x]~[~]] \\
\eal
\ee
\end{smlt}

\emph{Associativity $(r:\mb{I}$, $a,b:\mb{A})$.}    
\begin{smlt}
\be
\bal
&  [[multa~[a~b]~[d]~smult~[r~b]~[e]~multa~[a~e]~[x]]~smult~[r~d]~[y]]
\eal
\ee
\end{smlt}
\begin{smlt}
\be
\bal
&  [[multa~[a~b]~[d]~smult~[r~b]~[e]~smult~[r~d]~[y]]~multa~[a~e]~[x]]
\eal
\ee
\end{smlt}
\begin{smlt}
\be
\bal
&  [[multa~[a~b]~[d]~smult~[r~b]~[e]~smult~[r~d]~[y]~multa~[a~e]~[x]]~eqa~[y~x]~[~]] \\
\eal
\ee
\end{smlt}

\emph{Distributivity $(r:\mb{I}$, $a,b:\mb{A})$.}     
\begin{smlt}
\be
\bal
&  [[adda~[a~b]~[c]~smult~[r~c]~[x]~smult~[r~a]~[u]~smult~[r~b]~[v]]~adda~[u~v]~[y]]
\eal
\ee
\end{smlt}
\begin{smlt}
\be
\bal
&  [[adda~[a~b]~[c]~smult~[r~a]~[u]~smult~[r~b]~[v]~adda~[u~v]~[y]]~smult~[r~c]~[x]]
\eal
\ee
\end{smlt}
\begin{smlt}
\be
\bal
&  [[adda~[a~b]~[c]~smult~[r~c]~[x]~smult~[r~a]~[u]~smult~[r~b]~[v]~adda~[u~v]~[y]] \\
& \hspace{5mm} eqa~[y~x]~[~]]
\eal
\ee
\end{smlt}

\emph{Distributivity $(r,s:\mb{I}$, $a:\mb{A})$.}     

\begin{smlt}
\be
\bal
&  [[add~[r~s]~[t]~smult~[t~a]~[x]~smult~[r~a]~[u]~smult~[s~a]~[v]]~adda~[u~v]~[y]]
\eal
\ee
\end{smlt}
\begin{smlt}
\be
\bal
&  [[add~[r~s]~[t]~smult~[r~a]~[u]~smult~[s~a]~[v]~adda~[u~v]~[y]]~smult~[t~a]~[x]]
\eal
\ee
\end{smlt}
\begin{smlt}
\be
\bal
&  [[add~[r~s]~[t]~smult~[t~a]~[x]~smult~[r~a]~[u]~smult~[s~a]~[v]~adda~[u~v]~[y]] \\
& \hspace{5mm} eqa~[y~x]~[~]]
\eal
\ee
\end{smlt}

\section{Compatibility.}

We have adopted a dimension free formulation which means that the dimensions of the arrays do not appear in the I/O lists of the array atomic programs.
Within all array atomic programs, type checking requires that the elements of all arrays be of type $\mb{I}$ along with compatibility of array dimensions.
The following compatibility rules are included as axioms.
They are labeled by dim followed by a number.  

\emph{Array dimensions.}
\begin{dm}
\be
 [dima~[a~b]~[~]~dima~[b~a]~[~]]
\ee
\end{dm}
\begin{dm}
\be
 [[dima~[a~b]~[~]~dima~[b~c]~[~]]~dima~[a~c]~[~]]
\ee
\end{dm}

\emph{Equality.}
\begin{dm}
\be
 [eqa~[a~b]~[~]~dima~[b~a]~[~]]
\ee
\end{dm}

\noindent
\emph{Null array.}
\begin{dm}
\be
 [zarr~[a]~[b]~dima~[b~a]~[~]]
\ee
\end{dm}
\begin{dm}
\be
\bal
&  [[zarr~[a]~[b]~zarr~[c]~[d]~dima~[a~c]~[~]]~eqa~[d~b]~[~]]
\eal
\ee
\end{dm}

\emph{Array addition.}
\begin{dm}
\be
 [adda~[a~b]~[c]~dima~[c~a]~[~]]
\ee
\end{dm}

\emph{Scalar Multiplication.}
\begin{dm}
\be
 [smult~[a~b]~[c]~dima~[c~b]~[~]]
\ee
\end{dm}

\section{Basic identities.}\label{sbia}

We start with a few preliminary results that will shorten proofs that follow.
Theorem thm 1 extends the compatibility axiom dim 6 for the second input element.

\begin{lstlisting}
Theorem thm 1.
[[adda [a b] [c]] dima [b a] [ ]]

Proof.
  1 adda [a b] [c]
  2 dima [c a] [ ]           dim 6 [1]
  3 adda [b a] [d]           axa 3 [1]
  4 eqa [d c] [ ]            axa 4 [1 3]
  5 dima [d b] [ ]           dim 6 [3]
  6 dima [c b] [ ]           sr 1 [5 4]
  7 dima [b c] [ ]           dim 1 [6]
  8 dima [b a] [ ]           dim 2 [7 2]
\end{lstlisting}
Theorems thm 2-thm 3 extend the axiom of associativity of array addition by making use of the commutativity of array addition.
Since we are working with arrays in an environment $\mathfrak{M}(\mathcal{K},\mathcal{L},\mathcal{M})$ we must first establish the existence of $(c+a)+b$ given the existence of $a+b$, $c+(a+b)$ and $c+a$.
Having established existence (thm 2), theorem thm 3 shows that $c+(a+b)=(c+a)+b$.
\begin{lstlisting}
Theorem thm 2.
[[adda [a b] [d] adda [c d] [e] adda [c a] [f]] adda [f b] [m]]

Proof.
  1 adda [a b] [d]
  2 adda [c d] [e]
  3 adda [c a] [f]
  4 adda [b a] [g]           axa 3 [1]
  5 eqa [d g] [ ]            axa 4 [4 1]
  6 adda [d c] [h]           axa 3 [2]
  7 adda [g c] [i]           sr 1 [6 5]
  8 adda [a c] [j]           axa 3 [3]
  9 adda [b j] [k]           axa 5 [4 7 8]
 10 adda [j b] [l]           axa 3 [9]
 11 eqa [j f] [ ]            axa 4 [3 8]
 12 adda [f b] [m]           sr 1 [10 11]

Theorem thm 3.
[[adda [a b] [d] adda [c d] [e] adda [c a] [f] adda [f b] [m]]
 eqa [m e] [ ]]

Proof.
  1 adda [a b] [d]
  2 adda [c d] [e]
  3 adda [c a] [f]
  4 adda [f b] [m]
  5 eqa [e m] [ ]            axa 6 [3 4 1 2]
  6 eqa [m e] [ ]            axa 2 [5]
\end{lstlisting}
Theorem thm 4 shows that the array sum $c+(-b)$ exists if the array sum $c=a+b$ exists.
Theorem thm 5 establishes the identity $a=c+(-b)$.  
\begin{lstlisting}
Theorem thm 4.
[[adda [a b] [c] smult [-1 b] [d]] adda [c d] [j]]

Proof.
  1 adda [a b] [c]
  2 smult [-1 b] [d]
  3 adda [b d] [e]           smlt 2 [2]
  4 typea [b] [ ]            aio [1]
  5 zarr [b] [f]             axa 7 [4]
  6 eqa [e f] [ ]            smlt 3 [2 3 5]
  7 typea [a] [ ]            aio [1]
  8 zarr [a] [g]             axa 7 [7]
  9 adda [a g] [h]           axa 8 [8]
 10 dima [b a] [ ]           thm 1 [1]
 11 eqa [g f] [ ]            dim 5 [5 8 10]
 12 eqa [f e] [ ]            axa 2 [6]
 13 eqa [g e] [ ]            sr 1 [11 12]
 14 adda [a e] [i]           sr 1 [9 13]
 15 adda [c d] [j]           thm 2 [3 14 1]

Theorem thm 5.
[[adda [a b] [c] smult [-1 b] [d] adda [c d] [j]] eqa [j a] [ ]]

Proof.
  1 adda [a b] [c]
  2 smult [-1 b] [d]
  3 adda [c d] [j]
  4 adda [b d] [e]           smlt 2 [2]
  5 typea [b] [ ]            aio [1]
  6 zarr [b] [f]             axa 7 [5]
  7 eqa [e f] [ ]            smlt 3 [2 4 6]
  8 typea [a] [ ]            aio [1]
  9 zarr [a] [g]             axa 7 [8]
 10 adda [a g] [h]           axa 8 [9]
 11 eqa [h a] [ ]            axa 9 [9 10]
 12 dima [b a] [ ]           thm 1 [1]
 13 eqa [g f] [ ]            dim 5 [6 9 12]
 14 eqa [f e] [ ]            axa 2 [7]
 15 eqa [g e] [ ]            sr 1 [13 14]
 16 adda [a e] [i]           axa 5 [1 3 4]
 17 eqa [i h] [ ]            sr 2 [10 15 16]
 18 eqa [j i] [ ]            thm 3 [4 16 1 3]
 19 eqa [i a] [ ]            sr 1 [17 11]
 20 eqa [j a] [ ]            sr 1 [18 19]
\end{lstlisting}
Theorem thm 6 establishes that if the array sums $a+b$ and $a+d$ exist and are equal then $b=d$.
\begin{lstlisting}
Theorem thm 6.
[[adda [a b] [c] adda [a d] [e] eqa [c e] [ ]] eqa [b d] [ ]]

Proof.
  1 adda [a b] [c]
  2 adda [a d] [e]
  3 eqa [c e] [ ]
  4 adda [b a] [f]           axa 3 [1]
  5 adda [d a] [g]           axa 3 [2]
  6 eqa [f c] [ ]            axa 4 [1 4]
  7 eqa [g e] [ ]            axa 4 [2 5]
  8 typea [a] [ ]            aio [1]
  9 smult [-1 a] [h]         smlt 1 [8]
 10 adda [f h] [i]           thm 4 [4 9]
 11 adda [g h] [j]           thm 4 [5 9]
 12 eqa [i b] [ ]            thm 5 [4 9 10]
 13 eqa [j d] [ ]            thm 5 [5 9 11]
 14 adda [c h] [k]           sr 1 [10 6]
 15 adda [e h] [l]           sr 1 [11 7]
 16 eqa [k i] [ ]            sr 2 [10 6 14]
 17 eqa [l j] [ ]            sr 2 [11 7 15]
 18 eqa [l k] [ ]            sr 2 [14 3 15]
 19 eqa [k b] [ ]            sr 1 [16 12]
 20 eqa [l d] [ ]            sr 1 [17 13]
 21 eqa [l b] [ ]            sr 1 [18 19]
 22 eqa [b d] [ ]            sr 1 [20 21]
\end{lstlisting}
For any array, $a$, the existence of the additive inverses $-a$ and $-(-a)$ are guaranteed.
Theorem thm 7 establishes that $-(-a)=a$.
\begin{lstlisting}
Theorem thm 7.
[[smult [-1 a] [b] smult [-1 b] [c]] eqa [c a] [ ]]

Proof.
  1 smult [-1 a] [b]
  2 smult [-1 b] [c]
  3 adda [a b] [d]           smlt 2 [1]
  4 typea [a] [ ]            aio [1]
  5 zarr [a] [e]             axa 7 [4]
  6 eqa [d e] [ ]            smlt 3 [1 3 5]
  7 adda [b c] [f]           smlt 2 [2]
  8 typea [b] [ ]            aio [1]
  9 zarr [b] [g]             axa 7 [8]
 10 eqa [f g] [ ]            smlt 3 [2 7 9]
 11 dima [b a] [ ]           dim 7 [1]
 12 eqa [e g] [ ]            dim 5 [9 5 11]
 13 eqa [g e] [ ]            axa 2 [12]
 14 adda [b a] [h]           axa 3 [3]
 15 eqa [h d] [ ]            axa 4 [3 14]
 16 eqa [h e] [ ]            sr 1 [15 6]
 17 eqa [f e] [ ]            sr 1 [10 13]
 18 eqa [e f] [ ]            axa 2 [17]
 19 eqa [h f] [ ]            sr 1 [16 18]
 20 eqa [a c] [ ]            thm 6 [14 7 19]
 21 eqa [c a] [ ]            axa 2 [20]
\end{lstlisting}
Theorem thm 8 shows that if the array multiplication $ab$ exists then the array multiplication $a(-b)$ also exists.
Theorem thm 9 then establishes that $a(-b)=-(ab)$.
\begin{lstlisting}
Theorem thm 8.
[[multa [a b] [c] smult [-1 b] [d]] multa [a d] [f]]

Proof.
  1 multa [a b] [c]
  2 smult [-1 b] [d]
  3 typea [c] [ ]            aio [1]
  4 smult [-1 c] [e]         smlt 1 [3]
  5 multa [a d] [f]          smlt 13 [1 2 4]

Theorem thm 9.
[[multa [a b] [c] smult [-1 b] [d] multa [a d] [f] smult [-1 c] [e]]
 eqa [f e] [ ]]

Proof.
  1 multa [a b] [c]
  2 smult [-1 b] [d]
  3 multa [a d] [f]
  4 smult [-1 c] [e]
  5 eqa [e f] [ ]            smlt 14 [1 2 3 4]
  6 eqa [f e] [ ]            axa 2 [5]
\end{lstlisting}
Theorem thm 10 shows that if the array multiplication $ab$ exists then the array multiplication $(-a)b$ also exists.
Theorem thm 11 then establishes that $(-a)b=-(ab)$.
\begin{lstlisting}
Theorem thm 10.
[[multa [a b] [c] smult [-1 a] [d]] multa [d b] [f]]

Proof.
  1 multa [a b] [c]
  2 smult [-1 a] [d]
  3 typea [c] [ ]            aio [1]
  4 smult [-1 c] [e]         smlt 1 [3]
  5 multa [d b] [f]          smlt 10 [2 1 4]

Theorem thm 11.
[[multa [a b] [c] smult [-1 a] [d] multa [d b] [f] smult [-1 c] [e]]
 eqa [f e] [ ]]

Proof.
  1 multa [a b] [c]
  2 smult [-1 a] [d]
  3 multa [d b] [f]
  4 smult [-1 c] [e]
  5 eqa [e f] [ ]            smlt 11 [2 1 3 4]
  6 eqa [f e] [ ]            axa 2 [5]
\end{lstlisting}
Theorem thm 12 shows that if the array multiplication $ab$ exists then the array multiplication $(-a)(-b)$ also exists.
Theorem thm 13 then establishes that $(-a)(-b)=ab$.
\begin{lstlisting}
Theorem thm 12.
[[multa [a b] [c] smult [-1 a] [d] smult [-1 b] [e]] multa [d e] [g]]

Proof.
  1 multa [a b] [c]
  2 smult [-1 a] [d]
  3 smult [-1 b] [e]
  4 multa [a e] [f]          thm 8 [1 3]
  5 multa [d e] [g]          thm 10 [4 2]

Theorem thm 13.
[[multa [a b] [c] smult [-1 a] [d] smult [-1 b] [e] multa [d e] [g]]
 eqa [g c] [ ]]

Proof.
  1 multa [a b] [c]
  2 smult [-1 a] [d]
  3 smult [-1 b] [e]
  4 multa [d e] [g]
  5 typea [c] [ ]            aio [1]
  6 smult [-1 c] [f]         smlt 1 [5]
  7 typea [f] [ ]            aio [6]
  8 smult [-1 f] [h]         smlt 1 [7]
  9 eqa [h c] [ ]            thm 7 [6 8]
 10 multa [a e] [i]          smlt 13 [1 3 6]
 11 eqa [i f] [ ]            thm 9 [1 3 10 6]
 12 smult [-1 i] [j]         smlt 9 [2 10 4]
 13 eqa [h j] [ ]            sr 2 [12 11 8]
 14 eqa [j g] [ ]            smlt 11 [2 10 4 12]
 15 eqa [h g] [ ]            sr 1 [13 14]
 16 eqa [g c] [ ]            sr 1 [9 15]
\end{lstlisting}
It does not immediately follow from the axioms that the multiplication of the scalar $0$ with any array exists. 
Theorem thm 14 proves that it does exist and theorem thm 15 establishes that it is equal to a null array.
\begin{lstlisting}
Theorem thm 14.
[[typea [a] [ ]] smult [0 a] [k]]

Proof.
  1 typea [a] [ ]
  2 smult [1 a] [b]          smlt 4 [1]
  3 eqa [b a] [ ]            smlt 5 [2]
  4 smult [-1 a] [c]         smlt 1 [1]
  5 adda [a c] [d]           smlt 2 [4]
  6 eqa [a b] [ ]            axa 2 [3]
  7 adda [b c] [e]           sr 1 [5 6]
  8 typei [1] [ ]            aio [2]
  9 mult [-1 1] [f]          axi 11 [8]
 10 add [1 f] [g]            axi 12 [9]
 11 mult [1 -1] [h]          axi 14 [9]
 12 eqi [h f] [ ]            axi 15 [9 11]
 13 eqi [h -1] [ ]           axi 19 [11]
 14 eqi [f -1] [ ]           sr 1 [13 12]
 15 eqi [g 0] [ ]            axi 13 [9 10]
 16 add [1 -1] [i]           sr 1 [10 14]
 17 eqi [i g] [ ]            sr 2 [10 14 16]
 18 eqi [i 0] [ ]            sr 1 [17 15]
 19 smult [i a] [j]          smlt 19 [16 2 4 7]
 20 smult [0 a] [k]          sr 1 [19 18]

Theorem thm 15.
[[smult [0 a] [l] zarr [a] [m]] eqa [l m] [ ]]

Proof.
  1 smult [0 a] [l]
  2 zarr [a] [m]
  3 typea [a] [ ]            aio [1]
  4 smult [1 a] [b]          smlt 4 [3]
  5 eqa [b a] [ ]            smlt 5 [4]
  6 smult [-1 a] [c]         smlt 1 [3]
  7 adda [a c] [d]           smlt 2 [6]
  8 eqa [a b] [ ]            axa 2 [5]
  9 adda [b c] [e]           sr 1 [7 8]
 10 typei [1] [ ]            aio [4]
 11 mult [-1 1] [f]          axi 11 [10]
 12 add [1 f] [g]            axi 12 [11]
 13 mult [1 -1] [h]          axi 14 [11]
 14 eqi [h f] [ ]            axi 15 [11 13]
 15 eqi [h -1] [ ]           axi 19 [13]
 16 eqi [f -1] [ ]           sr 1 [15 14]
 17 eqi [g 0] [ ]            axi 13 [11 12]
 18 add [1 -1] [i]           sr 1 [12 16]
 19 eqi [i g] [ ]            sr 2 [12 16 18]
 20 eqi [i 0] [ ]            sr 1 [19 17]
 21 smult [i a] [j]          smlt 19 [18 4 6 9]
 22 eqa [d m] [ ]            smlt 3 [6 7 2]
 23 eqa [l j] [ ]            sr 2 [21 20 1]
 24 eqa [e j] [ ]            smlt 20 [18 21 4 6 9]
 25 eqa [j e] [ ]            axa 2 [24]
 26 eqa [e d] [ ]            sr 2 [7 8 9]
 27 eqa [j d] [ ]            sr 1 [25 26]
 28 eqa [j m] [ ]            sr 1 [27 22]
 29 eqa [l m] [ ]            sr 1 [23 28]
\end{lstlisting}
Theorem thm 16 shows that if an array multiplication $ba$ exists, where $b$ is a null array, then the array multiplication of $ba$ will be a null array.
\begin{lstlisting}
Theorem thm 16.
[[multa [b a] [c] zarr [b] [d] eqa [b d] [ ] zarr [c] [e]]
 eqa [c e] [ ]]

Proof.
  1 multa [b a] [c]
  2 zarr [b] [d]
  3 eqa [b d] [ ]
  4 zarr [c] [e]
  5 typea [b] [ ]            aio [1]
  6 smult [0 b] [f]          thm 14 [5]
  7 eqa [f d] [ ]            thm 15 [6 2]
  8 eqa [d f] [ ]            axa 2 [7]
  9 eqa [b f] [ ]            sr 1 [3 8]
 10 multa [f a] [g]          sr 1 [1 9]
 11 eqa [g c] [ ]            sr 2 [1 9 10]
 12 eqa [c g] [ ]            axa 2 [11]
 13 smult [0 c] [h]          smlt 9 [6 1 10]
 14 eqa [h g] [ ]            smlt 11 [6 1 10 13]
 15 eqa [g h] [ ]            axa 2 [14]
 16 eqa [h e] [ ]            thm 15 [13 4]
 17 eqa [g e] [ ]            sr 1 [15 16]
 18 eqa [c e] [ ]            sr 1 [12 17]
\end{lstlisting}
Since there is no commutativity rule for array multiplication we also need to show that $ab$ is a null array if $b$ is a null array. 
Theorem thm 17 shows that if an array multiplication $ab$ exists, where $b$ is a null array, then the array multiplication of $ab$ will be a null array.
The proof is almost identical to that of thm 16.
\begin{lstlisting}
Theorem thm 17.
[[multa [a b] [c] zarr [b] [d] eqa [b d] [ ] zarr [c] [e]]
 eqa [c e] [ ]]

Proof.
  1 multa [a b] [c]
  2 zarr [b] [d]
  3 eqa [b d] [ ]
  4 zarr [c] [e]
  5 typea [b] [ ]            aio [1]
  6 smult [0 b] [f]          thm 14 [5]
  7 eqa [f d] [ ]            thm 15 [6 2]
  8 eqa [d f] [ ]            axa 2 [7]
  9 eqa [b f] [ ]            sr 1 [3 8]
 10 multa [a f] [g]          sr 1 [1 9]
 11 eqa [g c] [ ]            sr 2 [1 9 10]
 12 eqa [c g] [ ]            axa 2 [11]
 13 smult [0 c] [h]          smlt 12 [1 6 10]
 14 eqa [h g] [ ]            smlt 14 [1 6 10 13]
 15 eqa [g h] [ ]            axa 2 [14]
 16 eqa [h e] [ ]            thm 15 [13 4]
 17 eqa [g e] [ ]            sr 1 [15 16]
 18 eqa [c e] [ ]            sr 1 [12 17]
\end{lstlisting}

\section{Array inequalities.}\label{sai}

Inequalities for matrices and arrays usually involve positive scalars that are associated with some norm of the matrix or array.
For instance one may define the norm, $\| a \|$, to be the maximum absolute value of the elements of the array $a$.
Rather than dealing with norms we will find it useful to define array inequalities that involve a scalar inequality applied to all corresponding elements of two arrays.
This kind of inequality will be found to be useful in a later chapter when considering multidimensional interval methods.
For completion we shall include some axioms for the inequality as defined here. 
Theorems involving array inequalities will not be derived here but follow in a similar manner to those derived in Chapter \ref{int} for scalar arithmetic on $\mb{I}$.
The following array atomic program is used to define an array inequality.

\vspace{5mm}

\begin{tabular}{|c|c|}
	\hline
	Atomic program names & Atomic program type \\
	\hline
	$lta$ & $\mb{P}_{type}$ \\
	\hline
\end{tabular}

\vspace{5mm}
We will also make use of the following special non atomic array program. 
\vspace{5mm}

\begin{tabular}{|c|c|}
	\hline
	Special non atomic program name & Structure \\
	\hline
	$lea$ & disjunction \\
	\hline
\end{tabular} 

\vspace{5mm}

In the description of the array atomic programs given below the following notation for the array dimension lists will be assumed.
\be
\begin{array}{ll}
	m=[m_1~\ldots~m_q], \qquad & m_i:\mb{I}_+,~i=1,\ldots,q,~q:\mb{I}_+ \\
\end{array}
\ee

\parskip0pt

\vspace{5mm}
\textbf{Check array inequality.}

\textbf{\textit{Syntax:}} $lta~[a~b]~[~]$.

\textbf{\textit{Program Type:}} $\mb{P}_{type}$.

\textbf{\textit{Type checks:}} $a:\mb{A}(m)$, $b:\mb{A}(m)$ and $a < b$.

\textbf{\textit{Description:}} $lta$ first checks that the values assigned to the variables $a$ and $b$ are of type $\mb{A}(m)$ for some dimension list $m$.
It then checks that $a<b$ where inequality is applied to each element of the arrays, i.e. 
\be
a(x) < b(x), \qquad x \in \Omega(m)
\ee
The dimensions of the arrays $a$ and $b$ are assigned prior to entry to the program $lta$ and are determined internally by $lta$.
$lta$ halts with an execution error if there is a type violation.
This includes the case that $a < b$ is violated.

\parskip10pt

\textbf{Order Axioms for array arithmetic on $\mb{I}$.}
The order axioms for arrays are similar to those for scalars and are labeled by orda followed by a number.  

\begin{orda}
	\be
	\bal
	&  [[lta~[a~b]~[~]~adda~[a~c]~[x]~adda~[b~c]~[y]]~lta~[x~y]~[~]~]
	\eal
	\ee
\end{orda}
\begin{orda}
	\be
	\bal
	& [[lta~[a~b]~[~]~lta~[c~d]~[~]~adda~[a~c]~[x]~adda~[b~d]~[y]]~lta~[x~y]~[~]] \\
	\eal
	\ee
\end{orda}
\begin{orda}
	\be
	\bal
	&  [[lta~[a~b]~[~]~lta~[0~c]~[~]~multa~[a~c]~[x]~multa~[b~c]~[y]]~lta~[x~y]~[~]]
	\eal
	\ee
\end{orda}
\begin{orda}
	\be
	\bal
	&  [[lta~[a~b]~[~]~lta~[c~0]~[~]~multa~[a~c]~[x]~multa~[b~c]~[y]]~lta~[y~x]~[~]] \\
	\eal
	\ee
\end{orda}

\emph{Transitivity of inequality.}
\begin{orda}
	\be
	[[lta~[a~b]~[~]~lta~[b~c]~[~]]~lta~[a~c]~[~]]
	\ee
\end{orda}

\textbf{Axiom of falsity (higher order type checking axiom).}
For arithmetic on $\mb{A}$ we include the following axiom of falsity.
\begin{orda}
	\be
	lta~[a~a]~[~]~:~\mb{P}_{false}
	\ee
\end{orda}

One can think of this as being equivalent to the higher order axiom with an empty premise list
\be
false~[p]~[~]
\ee
where the assigned value of $p$ is an object of type $\mb{P}$ and is given explicitly by $p=lt~[a~a]~[~]$.
The object $p=lt~[a~a]~[~]$ can also be regarded as a constant of type $\mb{P}$ objects associated with the application of array arithmetic.

\textbf{Special non-atomic program.}

It will often be more convenient to make use of the non strict array inequality, $a \leq b$, defined as a program disjunction 
\be
lea~[a~b]~[~] = lta~[a~b]~[~]~|~eqa~[a~b]~[~]
\ee

\section{Linear assignment programs.}\label{slap}

We are interested in expressing the conventional notion of linear maps in the language of functional programs.
We shall work with the array dimension lists
\be
\begin{array}{ll}
	l=[l_1~\ldots~l_p], \qquad & l_i:\mb{I}_+,~i=1,\ldots,p,~p:\mb{I}_+ \\
	m=[m_1~\ldots~m_q], \qquad & m_i:\mb{I}_+,~i=1,\ldots,q,~q:\mb{I}_+ \\
\end{array}
\ee
Suppose that $f~[u]~[v]$ is an assignment program with an associated assignment map $f:\mb{A}(l) \to \mb{A}(m)$.
(We use the same name for the assignment program and its associated assignment map.)

An assignment program $f~[u]~[v]$ is said to be linear if all of the following, lin1a-lin1c and lin2a-lin2c, can be derived as theorems.
Since we are working under the constraints imposed by our working platform $\mathfrak{M}(\mathcal{K},\mathcal{L},\mathcal{M})$, the two collections of theorems, lin1a-lin1c and lin2a-lin2c, each start with two conditional existence statements followed by an equality statement.

The first property of linearity is expressed by the three irreducible extended programs

\textbf{lin1a}
\be
\bal
~[[f~[u]~[r]~f~[v]~[s]~adda~[u~v]~[z]~adda~[r~s]~[t]]~f~[z]~[w]] \\
\eal
\ee

\textbf{lin1b}
\be
\bal
~[[f~[u]~[r]~f~[v]~[s]~adda~[u~v]~[z]~f~[z]~[w]]~adda~[r~s]~[t]] \\
\eal
\ee

\textbf{lin1c}
\be
\bal
~[[f~[u]~[r]~f~[v]~[s]~adda~[u~v]~[z]~f~[z]~[w]~adda~[r~s]~[t]]~eqa~[w~t]~[~]] \\
\eal
\ee

If the assignment map $f:\mb{A}(l) \to \mb{A}(m)$ of $f~[u^\prime]~[v^\prime]$ can be represented as $v^\prime:=f(u^\prime)$ then lin1a is a generalization of the statement that if $f(u),~f(v),~u+v$ and $f(u)+f(v)$ exist then $f(u+v)$ exists.
lin1b is the converse of lin1a and states that if $f(u),~f(v),~u+v$ and $f(u+v)$ exist then $f(u)+f(v)$ exists.
lin1c generalizes the statement that given $f(u),~f(v),~u+v,~f(u+v)$ and $f(u)+f(v)$ it follows that $f(u+v)=f(u)+f(v)$.

The second property of linearity is expressed by the three irreducible extended programs

\textbf{lin2a}
\be
\bal
~[[f~[u]~[v]~smult~[a~v]~[r]~smult~[a~u]~[z]]~f~[z]~[s]] \\
\eal
\ee

\textbf{lin2b}
\be
\bal
~[[f~[u]~[v]~smult~[a~u]~[z]~f~[z]~[s]]~smult~[a~v]~[r]] \\
\eal
\ee

\textbf{lin2c}
\be
\bal
~[[f~[u]~[v]~smult~[a~v]~[r]~smult~[a~u]~[z]~f~[z]~[s]]~eqa~[r~s]~[~]] \\
\eal
\ee

If the assignment map $f:\mb{A}(l) \to \mb{A}(m)$ of $f~[u^\prime]~[v^\prime]$ can be represented as $v^\prime:=f(u^\prime)$ then lin2a is a generalization of the statement that if $f(u),~a*u$ and $a*f(u)$ exist then $f(a*u)$ exists, where $a:\mb{I}$.
lin2b is the converse of lin2a and states that if $f(u),~a*u$ and $f(a*u)$ exist then $a*f(u)$ exists.
lin2c generalizes the statement that given $f(u),~a*u,~a*f(u)$ and $f(a*u)$ it follows that $a*f(u)=f(a*u)$.

Suppose that $f$ is a linear assignment program, i.e. satisfies lin1a-lin1c and lin2a-lin2c.
The following theorem states that if $u$ is a null array and $v$ is the evaluation of $f~[u]~[v]$ then $v$ is also a null array.
Where the map $f:\mb{A}(l) \to \mb{A}(m)$ of $f~[u]~[v]$ can be represented by $v:=f(u)$ we have the familiar result $f(0)=0$.
\begin{lstlisting}
Theorem thm 18.
[[f [u] [v] zarr [u] [z] eqa [u z] [ ] zarr [v] [w]]
 eqa [v w] [ ]]

Proof.
  1 f [u] [v]
  2 zarr [u] [z]
  3 eqa [u z] [ ]
  4 zarr [v] [w]
  5 typea [u] [ ]            aio [1]
  6 smult [0 u] [a]          thm 14 [5]
  7 eqa [a z] [ ]            thm 15 [6 2]
  8 eqa [z a] [ ]            axa 2 [7]
  9 f [z] [b]                sr 1 [1 3]
 10 f [a] [c]                sr 1 [9 8]
 11 smult [0 v] [d]          lin2b [1 6 10]
 12 eqa [d w] [ ]            thm 15 [11 4]
 13 eqa [d c] [ ]            lin2c [1 6 10 11]
 14 eqa [c w] [ ]            sr 1 [12 13]
 15 eqa [b c] [ ]            sr 2 [10 7 9]
 16 eqa [b w] [ ]            sr 1 [15 14]
 17 eqa [b v] [ ]            sr 2 [1 3 9]
 18 eqa [v w] [ ]            sr 1 [16 17]
\end{lstlisting}

Suppose again that $f$ is a linear assignment program, i.e. satisfies lin1a-lin1c and lin2a-lin2c.
Where the map $f:\mb{A}(l) \to \mb{A}(m)$ of $f~[u^\prime]~[v^\prime]$ can be represented by $v^\prime:=f(u^\prime)$ we have the familiar result $f(a*u+b*v)=a*f(u)+b*f(v)$ for scalars $a,b:\mb{I}$.
The following theorem generalizes this result for functional assignment programs.
Since we are working in $\mathfrak{M}(\mathcal{K},\mathcal{L},\mathcal{M})$ we need to include in the premises a number of conditional statements for the existence of certain additions and scalar multiplications.
\begin{lstlisting}
Theorem thm 19.
[[f [u] [p] f [v] [q] smult [a p] [r] smult [b q] [s]
 adda [r s] [t] smult [a u] [c] smult [b v] [d] adda [c d] [e]
 f [e] [w]] eqa [w t] [ ]]

Proof.
  1 f [u] [p]
  2 f [v] [q]
  3 smult [a p] [r]
  4 smult [b q] [s]
  5 adda [r s] [t]
  6 smult [a u] [c]
  7 smult [b v] [d]
  8 adda [c d] [e]
  9 f [e] [w]
 10 f [c] [f]                lin2a [1 6 3]
 11 f [d] [g]                lin2a [2 7 4]
 12 eqa [r f] [ ]            lin2c [1 6 10 3]
 13 eqa [s g] [ ]            lin2c [2 7 11 4]
 14 adda [r g] [h]           sr 1 [5 13]
 15 eqa [h t] [ ]            sr 2 [5 13 14]
 16 adda [f g] [i]           lin1b [10 11 8 9]
 17 eqa [w i] [ ]            lin1c [10 11 8 16 9]
 18 eqa [i h] [ ]            sr 2 [14 12 16]
 19 eqa [w h] [ ]            sr 1 [17 18]
 20 eqa [w t] [ ]            sr 1 [19 15]
\end{lstlisting}
(Note that the output variable name $f$ that is introduced at line 10 should not be confused with the name of the assignment program.)

It is tempting to continue this analysis to construct a theory for linear programs that yield results that are similar to those contained in the conventional theory of linear transformations (see for example \cite{waer}).
However, in any attempt to do so, it soon becomes apparent that the constraints imposed by the working platform $\mathfrak{M}(\mathcal{K},\mathcal{L},\mathcal{M})$ results in a theory that may be too restrictive for practical use in our constructive approach.

It is important to keep in mind that lin1a-lin1c and lin2a-lin2c are application specific and will only apply if for a given assignment program $f~[u]~[v]$ they can be derived as theorems.
To this end we need to have some idea of the internal algorithm of the program $f~[u]~[v]$.
We know that any linear transformation can be expressed in the form of a linear system involving matrices.
In the following chapter we will examine some basic identities of matrices as special kinds of arrays.

\chapter{Matrices.}\label{mtx}

\section{Atomic matrix programs.}

A matrix is an array of rank two and is represented by a list partition
\be
a=[[a(i,j)]_{j=1}^n]_{i=1}^m,~~a(i,j):\mb{I},~1 \leq i \leq m,~1 \leq j \leq n,~m,n:\mb{I}_+
\ee
We can write $a:\mb{A}(l)$, where the dimension list $l=[l_1~l_2]$.
It will be more convenient to write $a:\mb{A}(m,n)$, $m,n:\mb{I}_+$.
   
Here we will present a collection of axioms that are aimed at addressing some specific properties associated with the computability of integer matrix arithmetic subject to the constraints of a machine environment $\mathfrak{M}(\mathcal{K},\mathcal{L},\mathcal{M})$.
Of particular interest are linear systems that involve multiplicative inverses of matrices.
Since matrices are also arrays we can employ all of the axioms and theorems of the previous chapter.
We introduce only a few more atomic programs that are specific to arrays of rank two.

As before we are working with an element and dimension free formulation.
A dimension free formulation means that the dimensions of any matrix will not be explicitly specified in the I/O lists of matrix atomic programs.
Every atomic program will internally identify the dimensions of the matrices and where arithmetic operations are involved will check for the appropriate compatibility of the matrices under that operation.
We can regard the dimensions of the starting matrices, along with value assignments of their elements, to be initialized by the $read~x~[~]$ program under the general program structure (\ref{core}).
New matrices are generated by atomic assignment programs that internally set the dimensions of the new matrix variables.
In other words, while any matrix $a$ is stored as a list represented by $a=[[a(i,j)]_{j=1}^n]_{i=1}^m$, its partition as an $m \times n$ array is recognized as $a:\mb{A}(m,n)$, for some $m,n:\mb{I}_+$, that has already been stored in memory when it is introduced by the initializing subprogram $read~x~[~]$ or generated through the action of some assignment program.

The following integer array atomic programs that are specific to matrices are used.

\vspace{5mm}

\begin{tabular}{|c|c|}
\hline
Atomic program names & Atomic program type \\
\hline
$invm,~sqrm$ & $\mb{P}_{type}$ \\
\hline
$lid,~rid$ & $\mb{P}_{assign}$ \\
\hline
\end{tabular}

The following gives a description of the atomic programs given in the above table.

\parskip0pt

\vspace{5mm}
\textbf{Check of inverse matrix.}

\textbf{\textit{Syntax:}} $invm~[a~b]~[~]$.

\textbf{\textit{Program Type:}} $\mb{P}_{type}$.

\textbf{\textit{Type checks:}} $a:\mb{A}(n,n)$, $b:\mb{A}(n,n)$, for some $n:\mb{I}_+$, and $b=a^{-1}$.

\textbf{\textit{Description:}} $invm$ first checks that the values assigned to the variables $a$ and $b$ are of type $\mb{A}(n,n)$ for some $n:\mb{I}_+$.
It then checks that $b$ is the inverse matrix of $a$, i.e. $ab=ba=i$, where $i:\mb{A}(n,n)$ is the identity matrix.
The dimensions of the matrices $a$ and $b$ are assigned prior to entry to the program $invm$ and are recognized internally by $invm$.
$invm$ halts with an execution error if there is a type violation.
This includes the case where $b$ is not the inverse of $a$.

\vspace{5mm}
\textbf{Check square matrix.}

\textbf{\textit{Syntax:}} $sqrm~[a]~[~]$.

\textbf{\textit{Program Type:}} $\mb{P}_{type}$.

\textbf{\textit{Type checks:}} $a:\mb{A}(n,n)$, for some $n:\mb{I}_+$.

\textbf{\textit{Description:}} $sqrm$ checks that the value assigned to the variable $a$ is of type $\mb{A}(n,n)$ for some $n:\mb{I}_+$.
The dimensions of the array $a$ are assigned prior to entry to the program $sqrm$ and are recognized internally by $sqrm$.
$sqrm$ halts with an execution error if there is a type violation.

\vspace{5mm}
\textbf{Left identity matrix construction.}

\textbf{\textit{Syntax:}} $lid~[a]~[b]$.

\textbf{\textit{Program Type:}} $\mb{P}_{assign}$.

\textbf{\textit{Type checks:}} $a:\mb{A}(l)$, for some dimension list $l=[l_1 \ldots l_p]$, $p:\mb{I}_+$.

\textbf{\textit{Assignment map.}} $b(i,j):=1$ for $i=j$ and $b(i,j):=0$ otherwise, $1 \leq i,j \leq l_1$.

\textbf{\textit{Type assignment.}} $b::\mb{A}(m,m)$, $m=l_1$.

\textbf{\textit{Description:}} $lid$ first checks that the value assigned to the variable $a$ is of type $\mb{A}(l)$ for some dimension list $l=[l_1 \ldots l_p]$, $p:\mb{I}_+$. 
It then constructs the left identity matrix $b$ of $a$, i.e. $ba=a$.
The value assignment is accompanied by the type assignment $b::\mb{A}(m,m)$, where $m=l_1$.
If $a:\mb{A}(l)$ then the type check $b:\mb{A}(m,m)$ is never violated.
The dimensions of the matrix $a$ are assigned prior to entry to the program $lid$ and are recognized internally by $lid$.
$lid$ halts with an execution error if there is a type violation.

\vspace{5mm}
\textbf{Right identity matrix construction.}

\textbf{\textit{Syntax:}} $rid~[a]~[b]$.

\textbf{\textit{Program Type:}} $\mb{P}_{assign}$.

\textbf{\textit{Type checks:}} $a:\mb{A}(l)$, for some dimension list $l=[l_1 \ldots l_p]$, $p:\mb{I}_+$.

\textbf{\textit{Assignment map.}} $b(i,j):=1$ for $i=j$ and $b(i,j):=0$ otherwise, $1 \leq i,j \leq l_p$.

\textbf{\textit{Type assignment.}} $b::\mb{A}(m,m)$, $m=l_p$.

\textbf{\textit{Description:}} $rid$ first checks that the value assigned to the variable $a$ is of type $\mb{A}(l)$ for some dimension list $l=[l_1 \ldots l_p]$, $p:\mb{I}_+$. 
It then constructs the right identity matrix $b$ of $a$, i.e. $ab=a$.
The value assignment is accompanied by the type assignment $b::\mb{A}(m,m)$, where $m=l_p$.
If $a:\mb{A}(l)$ then the type check $b:\mb{A}(m,m)$ is never violated.
The dimensions of the matrix $a$ are assigned prior to entry to the program $rid$ and are recognized internally by $rid$.
$rid$ halts with an execution error if there is a type violation.

\parskip10pt

\textbf{Notes.}

\begin{itemize}

\item The atomic programs $lid$ and $rid$ could have been included in the previous chapter for general arrays.
They are introduced here because they have greater utility when dealing with matrices.

\item The results presented here will also be applicable for matrices whose elements are type $\mb{J}$ objects.
We can define an object, $a$, of type $\mb{A}(m,n)$, $m,n:\mb{I}_+$, to be represented by the list partition $a=[[a(i,j)]_{j=1}^n]_{i=1}^m$, $a(i,j):\mb{J},~1 \leq i \leq m,~1 \leq j \leq n$.
The dimensions $m$ and $n$ remain objects of type $\mb{I}_+$.
Since we are working with a coordinate free formulation, the dimensions of each matrix will be assigned and stored in memory.
Here we only need to regard the elements of objects of type $\mb{A}$ to be of type $\mb{J}$ instead of type $\mb{I}$.
Whenever an object is referred to as type $\mb{I}$ and that object is not a dimension of a matrix we simply replace $\mb{I}$ with $\mb{J}$.

\end{itemize}

\section{Axioms for Matrices.}

We shall introduce axioms that address the specific properties of matrices and are labeled by axm followed by a number.
They are appended to the general array axioms of the previous chapter.
The substitution rule will be applied as an axiom to matrix atomic programs $\mathfrak{p}~x~y$ such that
\be
\mathfrak{p} \in [lid~rid~invm~sqrm]
\ee

\textbf{Multiplication with the left identity matrix.}
\begin{axmtx}
\be
 [typea~[a]~[~]~lid~[a]~[i]]
\ee
\end{axmtx}
\begin{axmtx}
\be
 [lid~[a]~[i]~multa~[i~a]~[b]]
\ee
\end{axmtx}
\begin{axmtx}
\be
 [[lid~[a]~[i]~multa~[i~a]~[b]]~eqa~[b~a]~[~]]
\ee
\end{axmtx}

\textbf{Multiplication with the right identity matrix.}
\begin{axmtx}
\be
 [typea~[a]~[~]~rid~[a]~[i]]
\ee
\end{axmtx}
\begin{axmtx}
\be
 [rid~[a]~[i]~multa~[a~i]~[b]]
\ee
\end{axmtx}
\begin{axmtx}
\be
 [[rid~[a]~[i]~multa~[a~i]~[b]]~eqa~[b~a]~[~]]
\ee
\end{axmtx}

\textbf{Multiplicative inverse.}

The atomic program $invm~[a~b]~[~]$ checks that $a$ and $b$ are type $\mb{A}(n,n)$, for some $n:\mb{I}_+$, and then checks that $b$ is the multiplicative inverse of $a$, i.e. $ab=ba=i$, where $i:\mb{A}(n,n)$ is the identity matrix. 
By definition of the multiplicative inverse, the program $invm~[a~b]~[~]$ yields the following axioms.

\begin{axmtx}
\be
 [invm~[a~b]~[~]~sqrm~[a]~[~]]
\ee
\end{axmtx}
\begin{axmtx}
\be
 [invm~[a~b]~[~]~sqrm~[b]~[~]]
\ee
\end{axmtx}
\begin{axmtx}
\be
\bal
&  [invm~[a~b]~[~]~multa~[a~b]~[c]] \\
\eal
\ee
\begin{axmtx}
\end{axmtx}
\be
~[[invm~[a~b]~[~]~multa~[a~b]~[c]~lid~[a]~[i]]~eqa~[c~i]~[~]] \\
\ee
\end{axmtx}
\begin{axmtx}
\be
~[invm~[a~b]~[~]~multa~[b~a]~[d]] \\
\ee
\end{axmtx}
\begin{axmtx}
\be
~[[invm~[a~b]~[~]~multa~[a~b]~[c]~multa~[b~a]~[d]]~eqa~[d~c]~[~]] \\
\ee
\end{axmtx}

We also have the converse.
\begin{axmtx}
\be
\bal
&  [[sqrm~[a]~[~]~sqrm~[b]~[~]~multa~[a~b]~[c]~multa~[b~a]~[d]~eqa~[d~c]~[~] \\
& \hspace{5mm} lid~[a]~[i]~eqa~[c~i]~[~]]~invm~[a~b]~[~]] \\
\eal
\ee
\end{axmtx}

\textbf{Notes.}

\begin{itemize}

\item In the axioms for the multiplicative inverse the left and right identity matrices for the matrix $a$ are the same, since $a$ must be a square matrix.
This is expressed through the compatibility axiom dimm \ref{sqrm2} presented in the next section.
Without loss of generality, the above axioms for the multiplicative inverse employ the left identity matrix only.

\item The program $invm~[a~b]~[~]$ is type $\mb{P}_{type}$ and does not construct the inverse of a matrix.
It simply checks that $ab=ba=i$, where $i$ is the identity matrix with the same dimensions of the square matrices $a$ and $b$.

\item In the stricter sense of Definition \ref{atomic}, $invm$ is not atomic and should be regarded as pseudo-atomic. 
This is because $invm~[a~b]~[~]$ could have been constructed as a special non-atomic program using the atomic programs of $multa$, $sqrm$, $eqa$ and $lid$.
In light of this observation, the similarities of axioms axm 7-axm 13 with the special non-atomic program axioms spl 1-sp 4 should be apparent. 

\end{itemize}

\section{Compatibility.}

Matrix dimensions are not stated explicitly but certain compatibility conditions need to be maintained.
When initiating any derivation, the matrices that appear in the premises are assumed to be of general dimensions, say $\mb{A}(m,n)$ for some $m,n:\mb{I}_+$.
We may have $m=n$ if statements involving $invm$, $lid$, $rid$ and $sqrm$ are present. 
Within all matrix atomic programs type checking requires that the elements of all matrices be of type $\mb{I}$ along with compatibility of matrix dimensions.
The following compatibility rules are included as axioms.
They are labeled by dimm followed by a number.

\emph{Matrix multiplication and identity matrices.}
\begin{dmm}
\be
\bal
&  [[lid~[a]~[b]~lid~[c]~[d]~dima~[a~c]~[~]]~eqa~[d~b]~[~]]
\eal
\ee
\end{dmm}
\begin{dmm}
\be
\bal
&  [[rid~[a]~[b]~rid~[c]~[d]~dima~[a~c]~[~]]~eqa~[d~b]~[~]]
\eal
\ee
\end{dmm}
\begin{dmm}
\be
\bal
&  [[multa~[a~b]~[c]~rid~[a]~[d]~lid~[b]~[e]]~eqa~[e~d]~[~]]
\eal
\ee
\end{dmm}

\emph{Square matrices.}
\begin{dmm}
\be
 [[sqrm~[a]~[~]~dima~[a~b]~[~]]~sqrm~[b]~[~]]
\ee
\end{dmm}
\begin{dmm}\label{sqrm2}
\be
\bal
&  [[sqrm~[a]~[~]~lid~[a]~[c]~rid~[a]~[d]]~eqa~[d~c]~[~]]
\eal
\ee
\end{dmm}
\begin{dmm}
\be
\bal
&  [[sqrm~[a]~[~]~sqrm~[b]~[~]~multa~[a~b]~[c]]~dima~[c~a]~[~]]
\eal
\ee
\end{dmm}
\begin{dmm}
\be
\bal
&  [[sqrm~[a]~[~]~sqrm~[b]~[~]~multa~[a~b]~[c]]~dima~[c~b]~[~]]
\eal
\ee
\end{dmm}

\section{Basic identities for matrices.}\label{mtxthm}

The following theorems are just a continuation of the general array theorems derived in the previous chapter.
We start with Theorem thm 20 that shows that if $b$ is the multiplicative inverse of $a$ then $a$ is the multiplicative inverse of $b$.
\begin{lstlisting}
Theorem thm 20.
[[invm [a b] [ ]] invm [b a] [ ]]

Proof.
  1 invm [a b] [ ]
  2 sqrm [a] [ ]             axm 7 [1]
  3 sqrm [b] [ ]             axm 8 [1]
  4 multa [a b] [c]          axm 9 [1]
  5 multa [b a] [d]          axm 11 [1]
  6 eqa [d c] [ ]            axm 12 [1 4 5]
  7 eqa [c d] [ ]            axa 2 [6]
  8 typea [a] [ ]            aio [1]
  9 lid [a] [e]              axm 1 [8]
 10 eqa [c e] [ ]            axm 10 [1 4 9]
 11 typea [b] [ ]            aio [1]
 12 lid [b] [f]              axm 1 [11]
 13 rid [a] [g]              axm 4 [8]
 14 eqa [d e] [ ]            sr 1 [6 10]
 15 eqa [g e] [ ]            dimm 5 [2 9 13]
 16 eqa [e g] [ ]            axa 2 [15]
 17 eqa [f g] [ ]            dimm 3 [4 13 12]
 18 eqa [g f] [ ]            axa 2 [17]
 19 eqa [d g] [ ]            sr 1 [14 16]
 20 eqa [d f] [ ]            sr 1 [19 18]
 21 invm [b a] [ ]           axm 13 [3 2 5 4 7 12 20]
\end{lstlisting}
For matrices over rings it trivially follows that if the inverse of the matrix $a$ exists and $c=ab$ then $b=a^{-1}c$. 
For matrices in an environment $\mathfrak{M}(\mathcal{K},\mathcal{L},\mathcal{M})$ more work is required.
Theorem thm 21 shows that if the inverse of the matrix $a$ and $c=ab$ exist then $a^{-1}c$ also exists.
Theorem thm 22 establishes the identity $b=a^{-1}c$.
\begin{lstlisting}
Theorem thm 21.
[[multa [a b] [c] invm [a d] [ ]] multa [d c] [l]]

Proof.
  1 multa [a b] [c]
  2 invm [a d] [ ]
  3 invm [d a] [ ]           thm 20 [2]
  4 multa [a d] [e]          axm 9 [2]
  5 multa [d a] [f]          axm 9 [3]
  6 typea [d] [ ]            aio [2]
  7 lid [d] [g]              axm 1 [6]
  8 eqa [f g] [ ]            axm 10 [3 5 7]
  9 eqa [g f] [ ]            axa 2 [8]
 10 typea [b] [ ]            aio [1]
 11 lid [b] [h]              axm 1 [10]
 12 multa [h b] [i]          axm 2 [11]
 13 typea [a] [ ]            aio [1]
 14 rid [a] [j]              axm 4 [13]
 15 eqa [h j] [ ]            dimm 3 [1 14 11]
 16 eqa [g j] [ ]            dimm 3 [4 14 7]
 17 eqa [j g] [ ]            axa 2 [16]
 18 eqa [h g] [ ]            sr 1 [15 17]
 19 eqa [h f] [ ]            sr 1 [18 9]
 20 multa [f b] [k]          sr 1 [12 19]
 21 multa [d c] [l]          axa 10 [5 1 20]

Theorem thm 22.
[[multa [a b] [c] invm [a d] [ ] multa [d c] [l]] eqa [l b] [ ]]

Proof.
  1 multa [a b] [c]
  2 invm [a d] [ ]
  3 multa [d c] [l]
  4 invm [d a] [ ]           thm 20 [2]
  5 multa [a d] [e]          axm 9 [2]
  6 multa [d a] [f]          axm 9 [4]
  7 typea [d] [ ]            aio [2]
  8 lid [d] [g]              axm 1 [7]
  9 eqa [f g] [ ]            axm 10 [4 6 8]
 10 eqa [g f] [ ]            axa 2 [9]
 11 typea [b] [ ]            aio [1]
 12 lid [b] [h]              axm 1 [11]
 13 multa [h b] [i]          axm 2 [12]
 14 typea [a] [ ]            aio [1]
 15 rid [a] [j]              axm 4 [14]
 16 eqa [h j] [ ]            dimm 3 [1 15 12]
 17 eqa [g j] [ ]            dimm 3 [5 15 8]
 18 eqa [j g] [ ]            axa 2 [17]
 19 eqa [h g] [ ]            sr 1 [16 18]
 20 eqa [h f] [ ]            sr 1 [19 10]
 21 multa [f b] [k]          axa 11 [6 1 3]
 22 eqa [i b] [ ]            axm 3 [12 13]
 23 eqa [k i] [ ]            sr 2 [13 20 21]
 24 eqa [l k] [ ]            axa 12 [6 1 21 3]
 25 eqa [l i] [ ]            sr 1 [24 23]
 26 eqa [l b] [ ]            sr 1 [25 22]
\end{lstlisting}
Theorems thm 23-thm 24 are similar to theorems thm 21-thm 22.
Because of the absence of commutativity of matrix multiplication, we need to show that if $c=ab$ and the inverse of $b$ exist then $cb^{-1}$ exists and $a=cb^{-1}$.
\begin{lstlisting}
Theorem thm 23.
[[multa [a b] [c] invm [b d] [ ]] multa [c d] [j]]

Proof.
  1 multa [a b] [c]
  2 invm [b d] [ ]
  3 multa [b d] [e]          axm 9 [2]
  4 typea [b] [ ]            aio [1]
  5 lid [b] [f]              axm 1 [4]
  6 eqa [e f] [ ]            axm 10 [2 3 5]
  7 typea [a] [ ]            aio [1]
  8 rid [a] [g]              axm 4 [7]
  9 multa [a g] [h]          axm 5 [8]
 10 eqa [f g] [ ]            dimm 3 [1 8 5]
 11 eqa [e g] [ ]            sr 1 [6 10]
 12 eqa [g e] [ ]            axa 2 [11]
 13 multa [a e] [i]          sr 1 [9 12]
 14 multa [c d] [j]          axa 11 [1 3 13]

Theorem thm 24.
[[multa [a b] [c] invm [b d] [ ] multa [c d] [j]] eqa [j a] [ ]]

Proof.
  1 multa [a b] [c]
  2 invm [b d] [ ]
  3 multa [c d] [j]
  4 multa [b d] [e]          axm 9 [2]
  5 typea [b] [ ]            aio [1]
  6 lid [b] [f]              axm 1 [5]
  7 eqa [e f] [ ]            axm 10 [2 4 6]
  8 typea [a] [ ]            aio [1]
  9 rid [a] [g]              axm 4 [8]
 10 multa [a g] [h]          axm 5 [9]
 11 eqa [f g] [ ]            dimm 3 [1 9 6]
 12 eqa [e g] [ ]            sr 1 [7 11]
 13 eqa [g e] [ ]            axa 2 [12]
 14 multa [a e] [i]          axa 10 [1 4 3]
 15 eqa [i j] [ ]            axa 12 [1 4 3 14]
 16 eqa [j i] [ ]            axa 2 [15]
 17 eqa [i h] [ ]            sr 2 [10 13 14]
 18 eqa [h a] [ ]            axm 6 [9 10]
 19 eqa [j h] [ ]            sr 1 [16 17]
 20 eqa [j a] [ ]            sr 1 [19 18]
\end{lstlisting}
Theorem thm 25 establishes the result that if $a$ is an invertible matrix and the matrix multiplications $ab$ and $ad$ exist and are equal then $b=d$.
\begin{lstlisting}
Theorem thm 25.
[[multa [a b] [c] multa [a d] [e] eqa [c e] [ ] invm [a f] [ ]]
 eqa [b d] [ ]]

Proof.
  1 multa [a b] [c]
  2 multa [a d] [e]
  3 eqa [c e] [ ]
  4 invm [a f] [ ]
  5 multa [f c] [g]          thm 21 [1 4]
  6 multa [f e] [h]          thm 21 [2 4]
  7 eqa [g b] [ ]            thm 22 [1 4 5]
  8 eqa [b g] [ ]            axa 2 [7]
  9 eqa [h g] [ ]            sr 2 [5 3 6]
 10 eqa [g h] [ ]            axa 2 [9]
 11 eqa [b h] [ ]            sr 1 [8 10]
 12 eqa [h d] [ ]            thm 22 [2 4 6]
 13 eqa [b d] [ ]            sr 1 [11 12]
\end{lstlisting}
Theorem thm 26 is similar to theorem thm 25.
Because of the absence of commutativity of matrix multiplication, we need to show that if $a$ is an invertible matrix and the matrix multiplications $ba$ and $da$ exist and are equal then $b=d$.
\begin{lstlisting}
Theorem thm 26.
[[multa [b a] [c] multa [d a] [e] eqa [c e] [ ] invm [a f] [ ]]
 eqa [b d] [ ]]

Proof.
  1 multa [b a] [c]
  2 multa [d a] [e]
  3 eqa [c e] [ ]
  4 invm [a f] [ ]
  5 multa [c f] [g]          thm 23 [1 4]
  6 multa [e f] [h]          thm 23 [2 4]
  7 eqa [g b] [ ]            thm 24 [1 4 5]
  8 eqa [b g] [ ]            axa 2 [7]
  9 eqa [h g] [ ]            sr 2 [5 3 6]
 10 eqa [g h] [ ]            axa 2 [9]
 11 eqa [b h] [ ]            sr 1 [8 10]
 12 eqa [h d] [ ]            thm 24 [2 4 6]
 13 eqa [b d] [ ]            sr 1 [11 12]
\end{lstlisting}
Theorem thm 27 establishes that for a given matrix $a$ for which the inverses $a^{-1}$ and $(a^{-1})^{-1}$ exist then $(a^{-1})^{-1}=a$.
\begin{lstlisting}
Theorem thm 27.
[[invm [a b] [ ] invm [b c] [ ]] eqa [c a] [ ]]

Proof.
  1 invm [a b] [ ]
  2 invm [b c] [ ]
  3 invm [c b] [ ]           thm 20 [2]
  4 multa [a b] [d]          axm 9 [1]
  5 multa [c b] [e]          axm 9 [3]
  6 typea [a] [ ]            aio [1]
  7 lid [a] [f]              axm 1 [6]
  8 rid [a] [g]              axm 4 [6]
  9 typea [b] [ ]            aio [1]
 10 lid [b] [h]              axm 1 [9]
 11 typea [c] [ ]            aio [2]
 12 lid [c] [i]              axm 1 [11]
 13 rid [c] [j]              axm 4 [11]
 14 sqrm [a] [ ]             axm 7 [1]
 15 sqrm [c] [ ]             axm 7 [3]
 16 eqa [d f] [ ]            axm 10 [1 4 7]
 17 eqa [g f] [ ]            dimm 5 [14 7 8]
 18 eqa [f g] [ ]            axa 2 [17]
 19 eqa [d g] [ ]            sr 1 [16 18]
 20 eqa [h g] [ ]            dimm 3 [4 8 10]
 21 eqa [g h] [ ]            axa 2 [20]
 22 eqa [d h] [ ]            sr 1 [19 21]
 23 eqa [h j] [ ]            dimm 3 [5 13 10]
 24 eqa [j i] [ ]            dimm 5 [15 12 13]
 25 eqa [h i] [ ]            sr 1 [23 24]
 26 eqa [e i] [ ]            axm 10 [3 5 12]
 27 eqa [i e] [ ]            axa 2 [26]
 28 eqa [h e] [ ]            sr 1 [25 27]
 29 eqa [d e] [ ]            sr 1 [22 28]
 30 eqa [a c] [ ]            thm 26 [4 5 29 2]
 31 eqa [c a] [ ]            axa 2 [30]
\end{lstlisting}
Theorem thm 28 establishes the uniqueness of the multiplicative inverse.
\begin{lstlisting}
Theorem thm 28.
[[invm [a b] [ ] invm [a c] [ ]] eqa [b c] [ ]]

Proof.
  1 invm [a b] [ ]
  2 invm [a c] [ ]
  3 multa [a b] [d]          axm 9 [1]
  4 multa [a c] [e]          axm 9 [2]
  5 typea [a] [ ]            aio [1]
  6 lid [a] [f]              axm 1 [5]
  7 eqa [d f] [ ]            axm 10 [1 3 6]
  8 eqa [e f] [ ]            axm 10 [2 4 6]
  9 eqa [f e] [ ]            axa 2 [8]
 10 eqa [d e] [ ]            sr 1 [7 9]
 11 eqa [b c] [ ]            thm 25 [3 4 10 1]
\end{lstlisting}
Theorem thm 29 shows that if $ab$, $a^{-1}$, $b^{-1}$ and $b^{-1} a^{-1}$ exist then $b^{-1} a^{-1}$ is the multiplicative inverse of $ab$.
\begin{lstlisting}
Theorem thm 29.
[[multa [a b] [c] invm [a d] [ ] invm [b e] [ ] multa [e d] [f]]
 invm [c f] [ ]]

Proof.
  1 multa [a b] [c]
  2 invm [a d] [ ]
  3 invm [b e] [ ]
  4 multa [e d] [f]
  5 sqrm [a] [ ]             axm 7 [2]
  6 sqrm [d] [ ]             axm 8 [2]
  7 sqrm [b] [ ]             axm 7 [3]
  8 sqrm [e] [ ]             axm 8 [3]
  9 multa [d c] [g]          thm 21 [1 2]
 10 eqa [g b] [ ]            thm 22 [1 2 9]
 11 eqa [b g] [ ]            axa 2 [10]
 12 typea [b] [ ]            aio [1]
 13 lid [b] [h]              axm 1 [12]
 14 multa [e b] [i]          axm 11 [3]
 15 multa [b e] [j]          axm 9 [3]
 16 eqa [j h] [ ]            axm 10 [3 15 13]
 17 eqa [i j] [ ]            axm 12 [3 15 14]
 18 eqa [i h] [ ]            sr 1 [17 16]
 19 eqa [h i] [ ]            axa 2 [18]
 20 multa [e g] [k]          sr 1 [14 11]
 21 eqa [k i] [ ]            sr 2 [14 11 20]
 22 eqa [i k] [ ]            axa 2 [21]
 23 multa [f c] [l]          axa 11 [4 9 20]
 24 eqa [k l] [ ]            axa 12 [4 9 23 20]
 25 multa [c e] [m]          thm 23 [1 3]
 26 eqa [m a] [ ]            thm 24 [1 3 25]
 27 eqa [a m] [ ]            axa 2 [26]
 28 typea [a] [ ]            aio [1]
 29 lid [a] [n]              axm 1 [28]
 30 multa [a d] [o]          axm 9 [2]
 31 eqa [o n] [ ]            axm 10 [2 30 29]
 32 multa [m d] [p]          sr 1 [30 27]
 33 eqa [p o] [ ]            sr 2 [30 27 32]
 34 multa [c f] [q]          axa 10 [25 4 32]
 35 eqa [q p] [ ]            axa 12 [25 4 32 34]
 36 eqa [q o] [ ]            sr 1 [35 33]
 37 eqa [q n] [ ]            sr 1 [36 31]
 38 typea [c] [ ]            aio [1]
 39 lid [c] [r]              axm 1 [38]
 40 dima [c a] [ ]           dimm 6 [5 7 1]
 41 eqa [n r] [ ]            dimm 1 [39 29 40]
 42 eqa [q r] [ ]            sr 1 [37 41]
 43 dima [c b] [ ]           dimm 7 [5 7 1]
 44 eqa [h r] [ ]            dimm 1 [39 13 43]
 45 eqa [r i] [ ]            sr 1 [19 44]
 46 eqa [r k] [ ]            sr 1 [45 22]
 47 eqa [r l] [ ]            sr 1 [46 24]
 48 eqa [q l] [ ]            sr 1 [42 47]
 49 eqa [l q] [ ]            axa 2 [48]
 50 dima [f e] [ ]           dimm 6 [8 6 4]
 51 dima [e f] [ ]           dim 1 [50]
 52 sqrm [f] [ ]             dimm 4 [8 51]
 53 sqrm [m] [ ]             sr 1 [5 27]
 54 dima [c m] [ ]           sr 1 [40 27]
 55 dima [m c] [ ]           dim 1 [54]
 56 sqrm [c] [ ]             dimm 4 [53 55]
 57 invm [c f] [ ]           axm 13 [56 52 34 23 49 39 42]
\end{lstlisting}

\textbf{Notes.}

\begin{itemize}
	
	\item As with the theorems derived for scalar arithmetic over $\mb{I}$, many of the above theorems for matrices are weaker than their counterparts found in the theory of rings. These weaknesses are evident from the conditional statements required in their premises.
	Theorem thm 29 is a good example of this.
	In the theory of rings the existence of the inverse of dimension compatible matrices $a^{-1}$ and $b^{-1}$ guarantees the existence of $b^{-1} a^{-1}$.
	This is not so for matrices on a working platform $\mathfrak{M}(\mathcal{K},\mathcal{L},\mathcal{M})$.
	Consequently we must include in the premise the conditional statement $multa~[e~d]~[f]$, where $d=a^{-1}$ and $e=b^{-1}$.
	  
\end{itemize}

\section{Linear systems of equations.}\label{linsys}

Embedded in the theorems just derived are proofs of some useful properties of linear systems of equations over $\mb{I}$ that are similar to those that are more familiar with linear systems involving matrices over fields and rings.
All we need to do is change the variable names to make them recognizable.
In the outlines given below we may replace the context $\mb{I}$ with the fixed precision rationals $\mb{J}$, i.e. the elements of matrices may be type $\mb{J}$ objects instead of type $\mb{I}$ objects.
The dimensions of the matrices over $\mb{J}$ remain type $\mb{I}_+$ objects.

Here we are interested in systems of the form $ax=b$, where $a$ is a square matrix.
Suppose that we have $b=ax$ and $a^{-1}$ exist, where $a:\mb{A}(n,n)$, $x:\mb{A}(n,k)$ and $b:\mb{A}(n,k)$, for some $n,k:\mb{I}_+$.
This defines a linear system of equations over $\mb{I}$ (or $\mb{J}$).
Given that $ax$ and $a^{-1}$ exist we have from theorem thm 21 that $a^{-1}b$ exists and theorem thm 22 shows that from $ax=b$ it follows that $x=a^{-1}b$.

Conversely, suppose that $c:\mb{A}(n,n)$, $b:\mb{A}(n,k)$, $cb$, $c^{-1}$ and $(c^{-1})^{-1}$ exist.
From theorem thm 27 we have $c=(c^{-1})^{-1}$.
Set $x=cb$, where compatibility demands that $x:\mb{A}(n,k)$.
By theorems thm 21 and thm 22 we have $c^{-1}x$ exists and $b=c^{-1}x$.
We simply label the inverse of $c$ as $a$ (by way of $invm~[c~a]~[~]$) to relabel $b=c^{-1}x$ as $b=ax$.
From theorem thm 27 it follows that $a^{-1}=(c^{-1})^{-1}=c$ and we conclude that $x=cb=a^{-1}b$ is a solution of the linear system $ax=b$.

Suppose that $az=b$ and $ax=b$ are two solutions of the same linear system.
Given that the inverse $a^{-1}$ exists it follows from theorem thm 25 that $z=x$, so we have uniqueness of the solution of $ax=b$.

In the previous chapter we established that given the existence of an array multiplication $cb$, where $b$ is a multiplicative compatible null array, the array multiplication $cb$ will also be a null array.
The above linear system $ax=b$, where $a$ is a square matrix, is treated as a special case of an array multiplication.
We have $x=a^{-1}b$ so that if $b:\mb{A}(n,k)$ is a null matrix we must also have $x:\mb{A}(n,k)$ is a null matrix.
Uniqueness states that this is the only solution for $ax=b$ such that $b:\mb{A}(n,k)$ is a null matrix.

Note that the premises of the above arguments differ from those associated with linear systems over fields and rings.
The first part, where we establish that a solution of the system $ax=b$ is given by $x=a^{-1}b$, is conditional not only on the existence of $a^{-1}$ but also the matrix multiplication $ax$.
The existence of $ax$ is not guaranteed by the existence of $a$ and $x$.
Similarly, in the converse argument we have the conditional statement that $cb$ exists because the existence of $cb$ is not guaranteed by the existence of $c$ and $b$.
This highlights the importance of reading the premises carefully to identify the restrictions of the derived theorems.
The loss of generality that occurs when attempting to translate well known results from the theory of fields and rings to a machine environment has consequences.

\chapter{Dynamical Systems on Lattices.}\label{chap_ds}

\section{Introduction.}

In this chapter we will return to the topic that was the primary motivation behind the discourse presented in this book.
The lack of a general formal method for the construction and rigorous validation of computer models is a situation that needs to be addressed.
We have attempted to begin the process towards a more formal approach to computer modeling by proposing a collection of tools and methods that attempt to establish computability.

As outlined at the beginning of the first chapter, the most popular approach involves drawing upon a number of \emph{ad hoc} procedures that attempt to establish the validity of a discrete model as an approximation of a theoretical continuum model, the latter being regarded as the defining language that represents the real world system to be modeled.
Typical examples are found in applications of the general area of hydrodynamics where the theoretical model is presented as a continuum system based on second order partial differential equations. 

The problems associated with this approach have been discussed at the beginning of the first chapter.
One may take an alternative approach by abandoning the continuum model altogether and construct a discrete computer model that is not just a tool for simulation but also represents the language that describes the real world system under investigation.
Under this regime one employs the hypothesis that all of the information needed to fully define any real world object, including its state of motion, can be represented by a finite state vector.
In this way the formal validation methods for the map from the continuum to discrete system is replaced by the single notion of computability of the computer model.  

The first task is to lay down the laws that govern the construction of any model that can be adequately employed to simulate a real world system.
As we shall see, any dynamical system that is based on a law of conservation of information will be characterized by state vectors whose elements can only take on finite integer values.
Since the information flow is confined to a finite number of states we can expect that the laws governing the computability of any finite dynamical system must have some association with the laws that govern the allowable computational operations on a deterministic machine with finite memory.
A preliminary investigation into what these laws may look like is the topic of Chapters \ref{int}, \ref{cfds}, \ref{array} and \ref{mtx}.
We start by investigating the most fundamental core structure of finite dynamical systems on multidimensional lattices.
We close with a brief discussion on how the ideas of Chapter \ref{cfds}, based on the axiom of computability for finite dynamical systems, can be extended to multi-dimensions.

\section{Dynamic networks.}

Consider a directed graph $G=(v,e)$, where $v=[ i ]_{i=1}^m$, $m:\mb{I}_+$, is a list of nodes and $e(i,j)$ are directed arcs connecting nodes $i$ and $j$.
Each directed arc $e(i,j)$ represents a conduit of information flow from node $i \in v$ to a node $j \in v$.
In this sense $e(i,j)$ and $e(j,i)$ are the same arcs but represent opposite directions of flow along that arc.
Not all arcs need be conduits of information flow.
We can regard $e$ as a map $e: v \otimes v \to [0~1]$ where $e(i,j)=1$ for arcs that are conduits of information flow and $e(i,j)=0$ for arcs across which no information can flow.

The time $t$ can take on the discrete values $0,1,\ldots,n$, for some $n:\mb{I}_+$.
Associated with each arc $e(i,j)=1$ is a discrete transit time $\tau(i,j)$ for information flow.
The transit time on the same arc but in the opposite direction to $e(i,j)$ is given by $\tau(j,i)$, where $\tau(i,j)=\tau(j,i)$.

For each node $i$ we introduce the list $[d(i,t)~u(i,j,t)]$, $d,u:\mb{I}_0$, that represents the nodal state vector, where
\ben
\begin{array}{ll}
d(i,t) & \text{ capacity at node $i \in v$ at time $t$.} \\
u(i,j,t) & \text{ amount of information leaving node $i \in v$} \\
& \text{ to node $j \in v$ at time $t$.} \\
\end{array}
\een

Suppose that the initial state $[d~u]_{t=0}$ is prescribed.
The conservation of information at each node $i$ for $t>0$ is given by
\ben\label{dn01}
\bal
&  \sum_{\substack{j=1 \\ t-\tau(j,i) \geq 0}}^m u(j,i,t-\tau(j,i)) - \sum_{j=1}^m u(i,j,t) = d(i,t), \quad t>0 \\
\eal
\een
where we set
\ben\label{dn02}
u(i,j,t)|_{e(i,j)=0}=0
\een
We require that
\ben\label{dn03}
d:\mb{I}_0,~u:\mb{I}_0
\een
We may also impose the constraints, for each arc $e(i,j)$ and time $t$,
\ben\label{dn04}
0 \leq u(i,j,t) \leq c(i,j)
\een
for some prescribed constants $c(i,j):\mb{I}$.

At each time $t$, we may solve the system (\ref{dn01})-(\ref{dn04}) as a linear programming problem by introducing a linear objective function.
A solution is sought for the optimization of the objective function.

\section{Lattice network.}\label{sln}

We now consider a similar formulation to the dynamic network just described but with a more ordered structure. 
Consider a $p$-dimensional lattice that defines a discrete space of nodes, where each node is labeled by an index list $x=[x_1~\ldots~x_p]$, where $x_i : \mb{I}_0$, $i=1,\dots,p$.
For many applications we will be interested in the dimensions $p=1,2,3$.

\begin{figure}[!h]
	
	\setlength{\unitlength}{1.5mm}
	\begin{picture}(30,30)
	\multiput(10,2)(5,0){5}{\circle*{1}}
	\multiput(10,7)(5,0){5}{\circle*{1}}
	\multiput(10,12)(5,0){5}{\circle*{1}}
	\multiput(10,17)(5,0){5}{\circle*{1}}
	\multiput(10,22)(5,0){5}{\circle*{1}}
	\put(15,12){\line(1,0){10}}
	\put(20,7){\line(0,1){10}}

	\multiput(52,2)(5,0){5}{\circle*{1}}
	\multiput(52,7)(5,0){5}{\circle*{1}}
	\multiput(52,12)(5,0){5}{\circle*{1}}
	\multiput(52,17)(5,0){5}{\circle*{1}}
	\multiput(52,22)(5,0){5}{\circle*{1}}
	\put(57,12){\line(1,0){10}}
	\put(62,7){\line(0,1){10}}
	\put(57,7){\line(1,1){10}}
    \put(57,17){\line(1,-1){10}}
	\end{picture}
	
	\caption{\textit{A 2-dimensional representation of a lattice of nodes, (left) Neumann neighborhood, (right) Moore neighborhood.
	The solid lines emanating from the central node represent connecting arcs along which information can flow to and from the central node and its adjacent nodes.}}
	\label{lattice}
		
\end{figure}
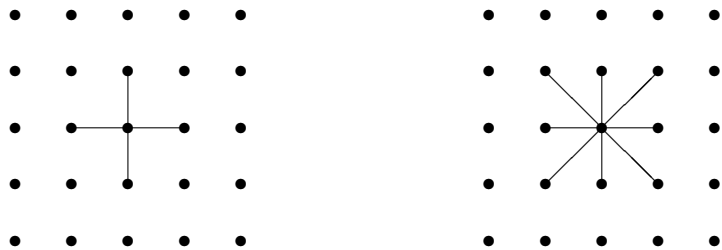

At any given time each node will contain a finite amount of information and the transfer of information can only occur across connecting arcs of adjacent nodes.
The most popular lattices are based on the Neumann\index{Neumann neighborhood} and Moore neighborhoods\index{Moore neighborhood}. 
Figure \ref{lattice} is a 2-dimensional graphic representation of these lattices.
We should keep in mind that other lattices or grid structures could also be investigated.

We define the array dimension list
\be
\begin{array}{ll}
l=[l_1~\ldots~l_p], \qquad & l_i:\mb{I}_+,~i=1,\ldots,p,~p:\mb{I}_+ \\
\end{array}
\ee
Elements of arrays will be expressed as functions of the index lists
\be
\begin{array}{ll}
x=[x_1~\ldots~x_p], \qquad & x_i:\mb{I}_+,~0 \leq x_i \leq l_i,~i=1,\ldots,p \\
y=[y_1~\ldots~y_p], \qquad & y_i:\mb{I}_+,~0 \leq y_i \leq l_i,~i=1,\ldots,p \\
\end{array}
\ee
Here $a:\mb{A}(l)=\mb{A}(l_1,\ldots,l_p)$ is an array of rank $p$ with the element representation $a(x)=a(x_1,\ldots,x_p)$, $x \in \Omega(l)$, where $\Omega(l)$ denotes the domain of the index list with the lower limit on each index $x_i$ being $0$.

Define the lists
\be
\bal
& e^{(i)}=[\delta(i,j)]_{j=1}^p , \quad i=1,\ldots,p\\
\eal
\ee
where $\delta(i,j)$ is the Kronecker delta given by
\be
\delta(i,j) = \left \{
\begin{array}{ll}
1 & i=j \\
0 & i \neq j \\
\end{array} 
\right .
\ee
If
\be
y \in \m{B}(x)
\ee
denotes nodes $y$ in the neighborhood $\m{B}(x)$ then for the Neumann neighborhood we can write
\be
\bal
& y = [x \pm e^{(i)}], \quad 1 \leq i \leq p \\
\eal
\ee
For the Moore neighborhood similar lists can be constructed.

We will examine the dynamics of the system in the discrete time $t:\mb{I}_0$ that can take on the integral values $0,1,\ldots,n$ for some $n:\mb{I}_+$.
We introduce the nodal state vector $[r(x,t)~u(x,y,t)]$ where
\be
\begin{array}{ll}
r(x,t) & \text{ amount of information at node $x$ at time $t$.} \\
u(x,y,t) & \text{ amount of information leaving node $x$ to node $y$ at time $t$.} \\
\end{array}
\ee
Suppose that the initial state $[r(x,0)~u(x,y,0)]$ is prescribed.
The conservation of information at each node, $x$, and time, $t>0$, is given by
\ben \label{6.100}
\bal
& r(x,t) = r(x,t-1) + W(x,t-1) - V(x,t), \quad t>0 \\
\eal
\een
where
\ben \label{6.101}
\bal
V(x,t) = & \sum_{y \in \m{B}(x)} u(x,y,t) \\
W(x,t) = & \sum_{y \in \m{B}(x)} u(y,x,t) \\
\eal
\een
To this we may need to impose certain constraints on boundary nodes.
These may involve noflow, inflow or outflow boundaries. 

For each time, $t$, we set
\ben \label{6.110}
u(x,y,t) = 0, \qquad y \notin \m{B}(x)
\een
This reflects the condition that information can only be transferred across connecting arcs of adjacent nodes.
In practice it will be more computationally efficient to store only those elements of $u(x,y,t)$ across which information can flow.
We define
\ben\label{6.110a}
z(x,t) = [u(x,y,t)]_{y \in \m{B}(x)}
\een
so that at each node, $x$, and time, $t$, we work with the state vector
\ben\label{6.110b}
[r(x,t)~z(x,t)]
\een
The \emph{system state vector}\index{system state vector} is defined by
\ben\label{6.110c}
[r(x,t)~z(x,t)]_{x \in \Omega(l)}
\een
and is the list concatenation of all nodal state vectors of the lattice at time, $t$.

The system governed by (\ref{6.100})-(\ref{6.110}) along with suitable boundary constraints is under determined.
We may consider two options.

\begin{itemize}

\item Closure. This involves the introduction of additional laws and possibly new state variables.

\item At each time $t$, pose the problem as a linear programming problem.
To this end one needs to construct an appropriate objective function.
We may also need to introduce additional laws and possibly new state variables.

\end{itemize}
Because of the application specific nature of these options we will not explore a complete construction of any model.
Our main objective is to demonstrate that the assigned values of the state variables of dynamical systems of the class just described will always be constrained by some finite bound.   

\section{Lattice refinement.}

To better understand how we may interpret the lattice network formulation of the previous section we will introduce a scale parameter $s$.

Consider now a sequence of $p$-dimensional lattice refinements.
We introduce the scale parameter $s:\mb{I}_0$ that can take on the integral values $0,1,\ldots,m$, for some $m:\mb{I}_+$.
The scale of finest possible resolution is represented by $s=0$.
On each scale slice $s=const$ we will examine the dynamics of the system in the discrete time $t(s):\mb{I}_0$ that can take on the integral values, $0 \leq t(s) \leq n(s)$, for some $n(s):\mb{I}_+$.

The array dimension list and the index list are scale dependent.
The array dimension list at each scale $s=0,1,\ldots,m$ is represented by
\be
\begin{array}{ll}
l(s)=[l_1(s)~\ldots~l_p(s)], \qquad & l_i(s):\mb{I}_+,~i=1,\ldots,p,~p:\mb{I}_+ \\
\end{array}
\ee
On each scale $s=0,1,\ldots,m$, the elements of arrays will be expressed as functions of the index lists
\be
\begin{array}{ll}
x(s)=[x_1(s)~\ldots~x_p(s)], \qquad & x_i(s):\mb{I}_0,~0 \leq x_i(s) \leq l_i(s),~i=1,\ldots,p \\
y(s)=[y_1(s)~\ldots~y_p(s)], \qquad & y_i(s):\mb{I}_0,~0 \leq y_i(s) \leq l_i(s),~i=1,\ldots,p \\
\end{array}
\ee
Here $a:\mb{A}(l(s))=\mb{A}(l_1(s),\ldots,l_p(s))$ is an array of rank $p$ with the element representation $a(x(s))=a(x_1(s),\ldots,x_p(s))$, $x(s) \in \Omega(l(s))$, where the lower limit on each index $x_i(s)$ is $0$. 

\begin{figure}[!h]
	
	\center
	
	\setlength{\unitlength}{1.5mm}
	\begin{picture}(30,30)
	\linethickness{0.075mm}
	\linethickness{0.35mm}
	\multiput(0,0)(10,0){4}{\line(0,1){30}}
	\multiput(0,0)(0,10){4}{\line(1,0){30}}
	\multiput(5,5)(10,0){3}{\circle{1}}
	\multiput(5,15)(10,0){3}{\circle{1}}
	\multiput(5,25)(10,0){3}{\circle{1}}
	\multiput(2.5,2.5)(5,0){6}{\circle*{1}}
	\multiput(2.5,7.5)(5,0){6}{\circle*{1}}
	\multiput(2.5,12.5)(5,0){6}{\circle*{1}}
	\multiput(2.5,17.5)(5,0){6}{\circle*{1}}
	\multiput(2.5,22.5)(5,0){6}{\circle*{1}}
	\multiput(2.5,27.5)(5,0){6}{\circle*{1}}
	\end{picture}
	
	\caption{\textit{A 2-dimensional representation of a two layered lattice refinement.
	The nodes of the finer lattice are represented by a bullet, $\bullet$.
	The nodes of the coarser lattice are offset from the nodes of the finer lattice and are represented by $\circ$.
	The grid lines depict the boundaries of cells where the information contained in the coarser nodes are the sum of information contained in the nodes contained within the cell.}}
    \label{refine}
	
\end{figure}
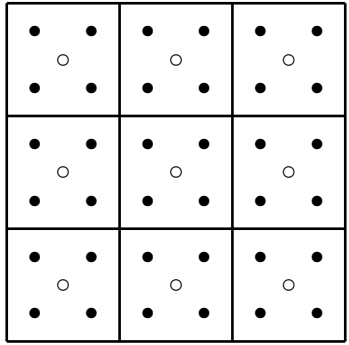

A typical scaled lattice system is associated with a doubling of the lattice grid size as the resolution is increased, i.e. as the scale parameter $s \to 0$. 
\be
\bal
x(s-1) = & 2x(s) \\
l_i(s-1)= & 2 l_i(s), \quad i=1,\ldots,p \\
\eal
\ee
The $(s-1)$-scale time is double the $s$-scale time, i.e.
\be
\bal
t(s-1) = & 2t(s) \\
n(s-1) = & 2 n(s) \\
\eal
\ee
By construction we need to assume that $n(0)$ and $l_i(0)$, $i=1,\ldots,p$, are even integers.  
For illustration purposes, Figure \ref{refine} depicts a typical two level lattice refinement on a 2-dimensional lattice network.
The extension of the two level refinement to $p$-dimensions should be apparent.

At each node, $x(s)$, of an $s$-scale lattice we introduce the state vector
\be
[r(x(s),t(s),s)~u(x(s),y(s),t(s),s)]
\ee
where
\be
\begin{array}{ll}
r(x(s),t(s),s) & \text{ amount of information at node $x(s)$} \\
& \text{ at time $t(s)$ and scale $s$.} \\
u(x(s),y(s),t(s),s) & \text{ amount of information leaving node $x(s)$} \\
& \text{ to node $y(s)$ at time $t(s)$ and scale $s$.} \\
\end{array}
\ee

On each $s$-scale lattice we assume that an initial state
\be
[r(x(s),0,s)~u(x(s),y(s),0,s)]
\ee
is prescribed.
The conservation of information at each $s$-scale node, $x(s)$, and time, $t(s)>0$, is given by
\ben \label{6.150}
\bal
& r(x(s),t(s),s) = r(x(s),t(s)-1,s) \\
& \hspace{30mm} + W(x(s),t(s)-1,s) - V(x(s),t(s),s), \quad t(s) > 0 \\
\eal
\een
where
\ben \label{6.151}
\bal
V(x(s),t(s),s) = & \sum_{y(s) \in \m{B}(x(s))} u(x(s),y(s),t(s),s) \\
W(x(s),t(s),s) = & \sum_{y(s) \in \m{B}(x(s))} u(y(s),x(s),t(s),s) \\
\eal
\een
To this we may need to impose certain constraints on boundary nodes. 

On each scale, $s$, and each time, $t(s)$, we set
\ben \label{6.160}
u(x(s),y(s),t(s),s) = 0, \qquad y(s) \notin \m{B}(x(s))
\een
This reflects the condition that information can only be transferred across connecting arcs of adjacent nodes.

\textbf{General filter.}
The $s$-scale state variables can be related to the $(s-1)$-scale state variables by the following discrete filter equations given in the general form

\ben \label{6.200}
\bal
& \sum_{\bar{x}(s)} a(x(s),\bar{x}(s)) * r(\bar{x}(s),t(s),s) = \\
& \sum_{x(s-1)} b(x(s),x(s-1))*r(x(s-1),t(s-1)/2,s-1) \\
\eal
\een
\ben \label{6.210}
\bal
& \sum_{\bar{x}(s)} \sum_{\bar{y}(s)} c(x(s),y(s),\bar{x}(s),\bar{y}(s)) * u(\bar{x}(s),\bar{y}(s),t(s),s) = \\
& \sum_{k=0}^1 \sum_{x(s-1)} \sum_{y(s-1)} d(x(s),y(s),x(s-1),y(s-1),k) * \\
& \hspace{20mm} u(x(s-1),y(s-1),t(s-1)/2 - k,s-1) \\
\eal
\een
where $a,b,c$ and $d$ are constant coefficient arrays.
The coefficient arrays $a,b,c$ and $d$ must be chosen such that consistency of scale invariance is satisfied, i.e. the conservation constraint (\ref{6.150}) is satisfied on each scale $s > 0$.\index{scale invariance}

Let $ cell(x(s)) $ be a list of $(s-1)-$scale nodes contained in the cell with centroid node labeled $x(s)$. 
Consider the choice of restricting the right hand side of (\ref{6.200}) to each cell such that
\ben \label{6.215}
b(x(s),x(s-1)) = \left \{
\begin{array}{ll}
	1, & x(s-1) \in cell(x(s)) \\ 
	0, & x(s-1) \notin cell(x(s)) \\ 
\end{array}
\right .
\een

\emph{Cell average.}
A cell average can be defined by using (\ref{6.215}) and setting
\ben \label{6.230}
a(x(s),\bar{x}(s)) = \left \{
\begin{array}{ll}
2p, & \bar{x}(s) = x(s) \\ 
0, & \bar{x}(s) \neq x(s) \\ 
\end{array}
\right .
\een

\emph{Cell sum.}
The simplest $s$ to $(s-1)$-scale relationship can be defined such that the information content of an $s$-scale node, $x(s)$, is the sum of the net information stored in the cell with centroid labeled $x(s)$.
Using (\ref{6.215}) we set
\ben \label{6.240}
a(x(s),\bar{x}(s)) = \left \{
\begin{array}{ll}
1, & \bar{x}(s) = x(s) \\ 
0, & \bar{x}(s) \neq x(s) \\ 
\end{array}
\right .
\een
Strictly speaking the cell sum is not a filter and reflects the exact relationship that connects states of different scales.

\textbf{Initial state.}
The initial state $[r(x(s),0,s)~u(x(s),y(s),0,s)]_{s=0}$ is assumed to be prescribed.
The initial states $[r(x(s),0,s)~u(x(s),y(s),0,s)]$ for $s>1$ are defined as the recursive application of the identities (\ref{6.200}) and (\ref{6.210}).

\textbf{Limiting state.} We need to examine the properties of the maximum resolution possible for any discrete system.
It is reasonable to assume that the smallest grid size that is possible is one on which each node contains only a single bit of information.
Any finer resolution would be meaningless.
Thus at a scale $s=0$ the state variables $r$ and $u$ at each node can only take on the assigned values $[0~1]$.
This reflects the features of the simplest cellular automaton.
We can expect that a cellular automata based model at scale $s=0$ will be largely governed by deterministic rules of node pair interactions.
  
Consider the cell summed lattice system based on (\ref{6.215}) and (\ref{6.240}).
The maximum assigned value of the state variable $r$ at $s=0$ is $1$.
On the coarser lattice system $s=1$ the maximum assigned value of the state variable $r$ is $2^p$ and on the lattice system $s=2$ the maximum assigned value of the state variable $r$ is $2^{2p}$.
Continuing in this way we see that the maximum assigned value of the state variable $r$ at each scale $s$ at each node is given by $2^{ps}$, i.e. the state variable $r$ at each scale $s$ can only take on the assigned values $0,1,\ldots,2^{ps}$.
Thus we have the constraint
\ben \label{6.290}
0 \leq r(x(s),t(s),s) \leq 2^{ps}
\een

We can also impose the constraint that at any time the amount of information leaving each node to an adjacent node cannot exceed the amount of information contained at that node.
We include the constraint
\ben \label{6.291}
0 \leq u(x(s),y(s),t(s),s) \leq r(x(s),t(s),s)
\een

In practice it will be necessary to formulate the dynamical model as an isolated macroscopic system, i.e. for some fixed scale $s=const >0$.
Here the understanding is that in the limit $s \to 0$ the macroscopic model will reduce to some simple deterministic rule based algorithm of node pair interactions that define the fundamental laws governing the application.
We can expect that the macroscopic model will largely be based on algorithms that will involve numerical computations that employ operations of basic arithmetic.
This will include the conservation law of information as expressed by (\ref{6.100}).

\textbf{Saturated flows.}
There is a class of flows such that each node remains saturated at all times, i.e. $r(x(s),t(s),s) = 2^{ps}$, $t(s) \geq 0$.
In such a case we have the conservation law
\be
\bal
& W(x(s),t(s)-1,s) - V(x(s),t(s),s) = 0, \quad t(s) > 0 \\
\eal
\ee

\textbf{Multicomponent flows.}
The term multicomponent flows is taken from continuum theories that describe a medium made up of more than one material species.
The individual species may be molecules, ions or elementary particles.
For a system involving $m$ species we need to introduce the state vector $[r_k~u_k]_{k=1}^m$, where each $[r_k~u_k]$ is associated with the $k$th species.
At each node we will also have $m$ conservation equations
\ben \label{6.300}
\bal
& r_k(x(s),t(s),s) = r_k(x(s),t(s)-1,s) \\
& \hspace{10mm} + W_k(x(s),t(s)-1,s) - V_k(x(s),t(s),s) + \sigma_k(s), \quad t(s) > 0 \\
\eal
\een
where
\ben \label{6.301}
\bal
V_k(x(s),t(s),s) = & \sum_{y(s) \in \m{B}(x(s))} u_k(x(s),y(s),t(s),s) \\
W_k(x(s),t(s),s) = & \sum_{y(s) \in \m{B}(x(s))} u_k(y(s),x(s),t(s),s) \\
\eal
\een
and $\sigma_k(s)$ is a reaction term that can be represented by
\be
\sigma_k(s) = \sum_{\substack{j=1 \\ j \neq k}}^m \sigma_{kj}(s)
\ee
where $\sigma_{kj}(s)=\sigma_{jk}(s)$ is a rate of reaction of the $k$th species with the $j$th species.

For the fully resolved system, $s=0$, the maximum amount of information contained at any node will be bound by the minimum amount of information needed to describe the individual particle of a given species.
Thus each state variable, $r_k$, at each node of the fully resolved system can only take on the values $[0~c_k]$, where $c_k$ is a positive integer value that represents the minimum amount of information needed to describe an individual particle of species $k$.
For $s>0$ each node can contain a mixture of particles of any species.

Following along the lines of the estimate given above for a single species we have a bound for the $k$th species on each scale, $s$, 
\ben
0 \leq r_k(x(s),t(s),s) \leq (2c_k)^{ps}
\een

\textbf{Scaling.} The aim is to obtain a discrete time evolution of the state variables $[r~u]$ on an isolated fixed scale for some $s>0$.
Typically the saturation value at each node, $r_{max}=2^{ps}$, is a very large integer.
One may prefer to work over $\mb{J}$ rather than $\mb{I}$ by introducing the scaled variables
\be
\bar{r}=r/r_0,~\bar{u}=u/r_0
\ee
for some suitable scaling parameter $r_0:\mb{I}_+$.
When we choose to  work over $\mb{J}$ instead of $\mb{I}$, by using the scaled state vector $[\bar{r}~\bar{u}]$, we are effectively redefining a single bit of information by the quantity $1/r_0$.
There is a point of caution here because $1/r_0$ cannot be less than the resolution $\epsilon$ of $\mb{J}$ so that underflows become an issue.  

\textbf{Notes.}

\begin{itemize}

\item The general filters (\ref{6.200})-(\ref{6.210}) can play an important role in the construction of closure models.
To get some idea of how this works one may consider the methodology to be partly related as a discrete analogy of the process employed in the construction of closure models for nonlinear continuum theories \cite{pan99}-\cite{pan09}.
This analogy is somewhat limited because there exist some properties of discrete systems that will require some special treatment.
This is a work in progress and will be documented elsewhere.

\item Of particular note is the observation that there exists an additional constraint imposed by (\ref{6.200})-(\ref{6.210}) that requires that the state variables $r(x,t)$ and $u(x,y,t)$ remain type $\mb{I}_+$.
This introduces an additional restriction on the choices of the constant coefficient arrays $a,b,c$ and $d$ that appear in (\ref{6.200})-(\ref{6.210}).  
The constraint $r(x,t),~u(x,y,t):\mb{I}_+$ is automatically satisfied when applying the cell sum.

\end{itemize}

\section{Closure.}
In light of the discussion presented so far we can regard (\ref{6.100})-(\ref{6.110}) as representing the laws governing the state vector $[r~u]$ on an isolated single scale slice $s > 0$.
With this understanding we set
\be
r_{max} = 2^{ps}
\ee
and remove the parameter $s$ by simply writing
\ben \label{6.400}
\bal
& r(x,t) = r(x,t-1) + W(x,t-1) - V(x,t) \\
\eal
\een
\ben \label{6.410}
r(x,t) \leq r_{max}
\een
\ben \label{6.411}
u(x,y,t) \leq r(x,t)
\een
\ben \label{6.420}
r(x,t) \geq 0,~u(x,y,t) \geq 0
\een
where
\ben \label{6.421}
\bal
V(x,t) = & \sum_{y \in \m{B}(x)} u(x,y,t) \\
W(x,t) = & \sum_{y \in \m{B}(x)} u(y,x,t) \\
\eal
\een
To this system we may include certain boundary conditions.
It is understood that on our lattice network the node to node interactions are local and are reflected in the statement
\ben \label{6.430}
u(x,y,t) = 0, \qquad y \notin \m{B}(x)
\een
The system (\ref{6.400})-(\ref{6.430}) can be regarded as the most fundamental structure at the core of finite dynamical systems on a lattice network.

Closure of the system (\ref{6.400})-(\ref{6.430}) is application specific and may require the introduction of additional laws along with additional state variables.
Closure can be said to be well posed if it automatically satisfies the constraints (\ref{6.410})-(\ref{6.420}).

\section{Discrete interval methods.}
In Section \ref{sln} we introduced the state vector $[r(x,t)~z(x,t)]$, (\ref{6.110b}), where $r(x,t)$ represents the amount of information at node, $x$, of the lattice and time, $t$, and $z(x,t)$ is the reduced list, (\ref{6.110a}), containing the nonzero elements of $u(x,y,t)$ that quantify the amount of information leaving node, $x$, to node, $y$, at time, $t$.
For instance on a $p$-dimensional Neumann lattice the length of the state vector $[r(x,t)~z(x,t)]$ is $2p+1$.
It may be convenient to work with arrays $\mb{A}(m)$ of rank $1$, where $m$ is not a dimension list but simply a scalar $m:\mb{I}_+$.
We introduce the system state vector\index{system state vector} $v^{(t)}:\mb{A}(m)$ that can be thought of as some ordered list of a concatenation of the $2p+1$ elements of $[r(x,t)~z(x,t)]$ at all lattice nodes $x \in \Omega(l)$ at time $t$.  

For each time, $t=1,\ldots,n$, the system state vector $v^{(t)}:\mb{A}(m)$ is evaluated by an assignment program $f~\la v^{(t-1)} \ra~\la v^{(t)} \ra$ that will include the conservation of information constraint (\ref{6.100}) at each node along with additional application specific constraints that have been introduced to obtain closure of the system.
Here we are using the angle brackets $\la ~ \ra$ that enclose the primary variables as shorthand notation for the I/O lists as outlined in the introduction of Chapter \ref{cfds}.

While the constraint (\ref{6.400}) is linear we can expect that typical closure models will contain additional constraints that are nonlinear, most likely in polynomial form.
The system can be defined as a finite dynamical system by constructing an associated iteration program $itf~\la v~n \ra~\la w \ra$, for some $n:\mb{I}_0$.
Following along similar lines to those outlined in Chapter \ref{cfds}\index{iteration program}, the iteration program $itf~\la v~n \ra~\la w \ra$ is defined as an atomic program that is constructed by some imperative language using an iteration loop as follows.
\ben \label{iterfn}
itf~\la v~n \ra~\la w \ra = \left \{
\begin{array}{l}
	n:\mb{I},~m:\mb{I}_+,~v:\mb{A}(m),~w:\mb{A}(m) \\
	\ldots \\
	t:\mb{I},~z:\mb{A}(m) \\
	le~[0~n]~[~] \\
	w:=v \\
	do~t=1,n \\
	~~~z:=w \\
	~~~f~\la z \ra~\la w \ra \\
	end~do \\
\end{array}
\right .
\een
Under the representation $itf~x~y = itf~\la v~n \ra~\la w \ra$, the second line indicated by the dots $\ldots$ is meant to represent the type checks of the parameters of $x \setminus [v~n]$ and $y \setminus w$.
These are the same parameter names not shown in the I/O lists of $f~\la z \ra~\la w \ra$.   

Since (\ref{iterfn}) represents a program constructed from an imperative language, each preceding solution of the iteration is discarded through the reassignment $z:=w$.
Here the variables $t:\mb{I}$ and $z:\mb{A}(m)$ are defined internally and are released from memory storage once the program has been executed.
The $do$-loop is not activated when $n=0$, in which case the value assignment $w:=v$ is returned as output.
If the user prescribes $n$ as a negative integer the atomic program $le~[0~n]~[~]$, and hence $itf~\la v~n \ra~\la w \ra$, will halt with an execution error.
Here $v=v^{(0)}$ is the initial system state vector and $w=v^{(n)}$ is the solution of $f~\la v^{(t-1)} \ra~\la v^{(t)} \ra$ after $n$ iterations.

To establish computability we may use the methods similar to those outlined in Chapter \ref{cfds}.
Since we are working with state variables of type $\mb{A}(m)$, $m:\mb{I}_+$, the theory based on discrete intervals needs to be modified by replacing the intervals with boxes, where each box is an $m$-dimensional interval defined as a cross product of $m$ discrete intervals associated with each dimension.

We can apply axioms axdi 1-16 and axoc 1-5 by using the following modifications.
Firstly, we make use of the non-atomic program for array inequality (see Section \ref{sai}) 
\ben
lea~[a~b]~[~], \quad a,b:\mb{A}(m)
\een
for some dimension list $m=[m_1 \ldots m_q]$ that checks for the array inequality
\ben
a \leq b
\een
with the component representation
\ben\label{lea}
a(x) \leq b(x), \quad x \in \Omega(m)
\een
In our application we are dealing with arrays of rank 1, namely objects of type $\mb{A}(m)$, so the dimension list $m=[m_1 \ldots m_q]$ reduces to the scalar $m:\mb{I}_+$, i.e. $q=1$.

A discrete interval\index{discrete interval} $p$ is assigned the type $\mb{B}$ and defined by the two element list, $p=[a~b]$, where $a,b:\mb{I}$, $a \leq b$, are the interval bounds.
An $m$-dimensional box\index{discrete box}, $p$, is assigned the type $\mb{B}(m)$ and defined by the two element list
\ben
p=[a~b], \quad a,b:\mb{A}(m),~a \leq b,~m:\mb{I}_+
\een
Here $a,b:\mb{A}(m)$ represent the bounds of the $m$-dimensional box $p:\mb{B}(m)$ that represents the cross product of $m$ intervals
\ben
p=[a(1)~b(1)] \times \ldots \times [a(m)~b(m)]
\een
where each $[a(i)~b(i)]:\mb{B}(1)$, $i=1,\ldots,m$, is a discrete interval or 1-dimensional box.
In this way all of the atomic programs defined in Section \ref{sintap} will now accept as input type $\mb{B}(m)$ objects instead of $\mb{B}(1)$ objects and the interval construction program, $int~[a~b]~[p]$, accepts as input the box bounds $a,b:\mb{A}(m)$ such that $a \leq b$, as defined component wise by (\ref{lea}).   
Axioms axdi 1-16 and axoc 1-5 can now be applied by replacing all occurrences of the program names $eqi$ with $eqa$ and $le$ with $lea$, with the exception of the program $le~[0~n]~[~]$ in axoc 5.
In a similar way, the axioms labeled by axdia associated with interval arithmetic can be modified for $m$-dimensional boxes.

Alternatively, explicit closure based on equality constraints may be abandoned by posing the problem as an iteration of a linear programming problem.
In this case the dynamical system $itf~\la v~n \ra~\la w \ra$ will be associated with the assignment program $f~\la v^{(t-1)} \ra~\la v^{(t)} \ra$ that solves, at each time step, the linear programming problem based on (\ref{6.400})-(\ref{6.420}). 
The program $f~\la v^{(t-1)} \ra~\la v^{(t)} \ra$ will include some linear objective function that will be subject to optimization.
To establish computability of the dynamical system $itf~\la v~n \ra~\la w \ra$ based on an extension of the methods of discrete intervals described above may not be the best approach.
In this case the bounds on the state variables are already set by the linear programming problem via (\ref{6.410})-(\ref{6.420}).
In other words the tightest enclosure $R(f,p)$ over any $m$-dimensional box, $p:\mb{B}(m)$, is already bounded within the program $f~\la v \ra~\la w \ra$ and hence the construction of a box enclosure $B(f,p)$ of $R(f,p)$ may not be required.
The task reduces to identifying, at each time, $t$, the required properties of a system state vector $v^{(t-1)}$ that would guarantee the existence of the system state vector $v^{(t)}$ evaluated from the linear programming algorithm of $f~\la v^{(t-1)} \ra~\la v^{(t)} \ra$.

In any event, it is evident that the methods required to establish computability of computer models based on dynamical systems on a working platform $\mathfrak{M}(\mathcal{K},\mathcal{L},\mathcal{M})$, whether they be based on boxed regions or otherwise, are largely underdeveloped.
This highlights one of the objectives pursued throughout this book, i.e. to initiate the construction of tools of analysis based on a language of programs and finite state arithmetic that directly address the issues encountered when working in an environment $\mathfrak{M}(\mathcal{K},\mathcal{L},\mathcal{M})$.

\chapter{Formal Systems in Science.}\label{cfsis}

\section{Introduction.}

In the early part of the 20th century mathematicians set about to finally settle the issues surrounding the foundations of mathematics.
In this way they aimed to remove the discourse out of the hands of the philosophers and by a process of self referencing carry out a formal study of the foundations of mathematics using the tools of mathematics itself.
Although early efforts can be traced back to the work of Frege and others, the program was largely initiated by a series of lectures given by Hilbert, culminating in the work of Godel's incompleteness theorem.

Scientist, on the other hand, continue to carry out research into their special subject area with an acceptance that there is a well defined scientific method that they have an intuitive grasp of.
It is uncommon to find a formal course on the scientific method offered to undergraduates in any branch of the sciences.
Students are expected to acquire the rules of conduct when carrying out scientific research through general guidelines offered in the coursework of the various science disciplines that they have elected as part of their major.
It is ironic, then, that scientist have left the in-depth investigations and interpretations of the scientific method to be carried out exclusively by philosophers.

In recent decades controversies over what actually constitutes scientific research have arisen with the ever increasing activity in peripheral areas such as the social sciences and related human and life sciences.  
It therefore seems timely that scientist make a similar effort to that made by mathematicians and examine the scientific method in some formal sense.
To understand that such a project is possible one needs to recognize that the scientific method is a recursive self-correcting process that is essentially a dynamical system and hence can be posed as a problem in computation subject to constraints imposed by empirical data.
    
Here an attempt will be made to initiate this project by introducing some preliminary ideas based upon the tools that have been developed in the previous chapters of this book.
In order that this make sense one must be receptive to the idea that scientific theories of the future will be expressed in a language of algorithms and programs.
Consequently the status of a fully discrete computer model that can be derived from a theory is raised from one that is not just a useful research tool but in itself is the language that is used to define the laws of the application.

It should be noted that in the current paradigm the language of science is developed in a separate discipline, namely mathematics.
By adopting a language based on algorithms and programs the validation of a scientific theory automatically includes the validation of the language and formal system on which it is based.     

\textbf{Complexity.}
The scientific method relies heavily on data obtained from real world observations against which simulation results of specific models are tested.
We will focus mainly on validating theories by way of empirical checks associated with the computability map (see Section \ref{sc}).
We do this because we are primarily interested here in employing some empirically based notion of soundness of applications in our formal system.    

In the sense of Chaitin-Kolmogorov complexity a major objective is to construct the shortest code that represents a computer model.
This needs to be assessed with respect to the scope of applicability of the computer model.
Roughly speaking, the scope of applicability can be defined as the model's ability to generate solutions that simulate real world observations to within the experimental errors and confidence intervals of the widest range of measured data.
If we are to regard the computer model as the primary descriptor of a theory then we are moving towards some quantifiable way of assessing the elegance of a theory.
The objective, then, is to construct theories with minimal program complexity while possessing a maximal scope of applicability.  

Real world measurements targeting specific models can be incorporated into the methods to be outlined.
In order to remain focused on the main thrust of the discussion that follows we omit details on how this could be done.
Because of this omission we are exploring a more general universe of valid computer models where the specific interpretation of a model and its scope of application is left unspecified.

\textbf{Empirical Computations.}
From a strict formalist point of view the semantics of statements in a formal system are less of a concern than that of consistency.
In a computer environment a formal statement is expressed as a program whose functionality is largely well defined.
In this context the semantics of formal statements is unambiguous.

We have constructed our formal system, PECR, to be compatible with the constraints imposed by a machine environment $\mathfrak{M}(\mathcal{K},\mathcal{L},\mathcal{M})$.
In PECR the well formed formulas of classical logic are replaced by functional programs and the classical notion of attaching a truth value to a formal statement is now replaced by the notion of computability.

Although our primary objective is to establish the computability of programs by way of inference based upon a collection of construction rules, we can also check the computability of a program by empirical means through the computability map.
This simply involves executing a program for a given value assigned input and observing whether it halts prematurely with an execution error or returns an output.
We shall call this process \emph{empirical computation}.

Of course, empirically checking for computability can only be useful if the program can be observed to either return an output or an error message in a reasonable time period.
Here we can be guided by a preliminary analysis of the algorithm of a program to establish whether it can be executed in polynomial time. 
Otherwise, what may be regarded as a reasonable time cannot be strictly defined and will be an arbitrary constraint imposed by a user.
  
For this reason establishing computability by inference is preferred because of its generality but there are situations where empirical computation will have an important role to play.

\textbf{Consistency and soundness.}
In conventional theories of logic, consistency is defined in terms of formal statements and their negations.
Applications of the formal system PECR in its most primitive form do not make much use of negations so consistency in the conventional sense is not appropriate.
We can, however, approach the conventional notion of soundness as follows.

Suppose $[p~c],~c:\mb{P}_{ext}(p)$, has either been supplied as an axiom or obtained by inference.
Suppose further that by empirical computation the program $p$ is found to be computable for a given assigned input.
Soundness will be violated if by empirical computation it is found that $[p~c]$ is not computable for the same value assigned input.

Similarly, suppose that the statement $p:\mb{P}_{false}$ has been inferred or simply supplied as an axiom of falsity.
Another form of a violation of soundness may occur if by empirical computation it is found that there exists an assigned valued input such that the program $p$ is computable.

We can rewrite the two conditions for violation of soundness of a theory, $S$, as follows.
\begin{itemize}

\item $s=[p~c],~c:\mb{P}_{ext}(p)$, is either an axiom or theorem of $S$ and we have by empirical computation $p:\mb{P}_{comp}(x_p)$, for some value assigned input list $x_p$ of $p$, and $s=[p~c]$ is not of type $\mb{P}_{comp}(x_s)$, where $x_s$ is the value assigned input list of $s$ that acquires its value through the identities $x_s = \bar{x}_s \setminus (x_c \cap y_p)$ and $\bar{x}_s \simeq [x_p~x_c]$.

\item $p:\mb{P}_{false}$ is an axiom or theorem of falsity of $S$ and we have by empirical computation $p:\mb{P}_{comp}(x_p)$ for some value assigned input $x_p$ of $p$.

\end{itemize}

By analogy with classical mathematical logic we are employing empirical computation to search for counter examples of a proposition asserting soundness.   
Hence the above conditions are weak in the sense that they only address violations of soundness in an empirical sense and do not provide a formal procedure from which we can establish that a formal system is sound.
On the other hand, computability is defined in an unambiguous way by empirical computation and can establish the computability of a program with respect to a given assigned input list with absolute certainty and hence requires no interpretation.
This reliance on empirical observations suggests a process closer to the scientific method rather than the higher goals demanded by conventional mathematics.
We are led to seriously consider the following.

\textbf{Iterated axiomatic method.}
The controversy surrounding the foundations of mathematics and formal systems in general are well known and remain a topic of serious debate.
Rather than attack this problem head on we may seek a path around it.
One approach is to accept a less ambitious form of inquiry that is closer to that found through the self correcting recursive process of the scientific method.
Consequently the axiomatic method is weakened to incorporate some procedures that may be empirical.

First of all one concedes to the notion that, like postulates in science, laying down a collection of axioms to define a specific theory may be a tentative process that is subject to modification.
It is through such a concession that an iterative mechanism is required for continual reevaluation and self correction.
One initiates an action of theorem mining by first laying down a collection of axioms for a theory, $S$.
By applying these axioms in conjunction with the construction rules, proofs are derived from which theorems are extracted as irreducible extended programs.

We concede that there may be irreducible extended programs of the theory $S$ that may be missed by this process, i.e. irreducible extended programs that cannot be derived under the current collection of axioms.
Irreducible extended programs that cannot be derived are potential candidates for new axioms of a theory.
If by some means outside of the action of theorem mining a new irreducible extended program is found that cannot be derived from the existing axioms then it can be appended to the collection of axioms.
In this way the theory under investigation is built up with increasing scope of its theorem mining capabilities.

There is a point of caution here in that simply appending a new irreducible extended program to the current list of axioms of a theory can radically change the whole dynamics of the system.
This is because such a procedure does not guarantee that the current list of axioms will remain non-derivable.  
Taking this into account, in our recursive self improving procedure we include the following two actions that run concurrent to the action of theorem mining.

\begin{itemize}

\item Axioms are assumed to be irreducible extended programs until such time that they are found to violate soundness.
Violations of soundness can be detected through empirical computation.
When this occurs the offending programs that are stored as axioms are removed from storage along with all theorems whose derivations depend directly or indirectly upon them.

\item If a derivation or proof is found for an axiom then it is accessed in the file \emph{axiom.dat} and relabeled as a theorem.
This situation may occur when a program was incorrectly identified as an axiom from the start or a new axiom is introduced into the current collection of axioms.

\end{itemize}

\textbf{Identifying new axioms.}
The actual task of identifying new axioms lies outside of the formal system in which they are employed.
At this stage such a task is largely a human enterprise but it is worthwhile to speculate that automation may be possible.

It is difficult to envisage a procedure of identifying axioms that can avoid some kind of empirical process.
This may involve a mechanism employing some kind of targeted pattern recognition on permutations of lists of atomic programs.
Immediate elimination of possible candidates can rely on the structural Condition 1 of Definition \ref{ce}.  
Each of the remaining candidates of programs lists will be subject to extensive testing with respect to a large range of prescribed value assigned inputs through empirical computation in combination with confidence valuation through statistical analysis.
This empirically based procedure will largely test for violations of Condition 2 of Definition \ref{ce}.
  
Identifying new axioms in this way is another action that could be conducted concurrent to the main action of generating proofs and theorems.
Since our formal system is constrained by the machine environment $\mathfrak{M}(\mathcal{K},\mathcal{L},\mathcal{M})$, we can expect that the empirical procedure just described may identify new axioms that are machine specific.
In a larger realm of investigation the machine specific parameters become variables that enter the self correcting recursive process.

\textbf{Premature derivation halting.}
In any theorem mining activity there will always be a lack of certainty that all programs that are of type $\mb{P}_{false}$ have been detected.
As a result we might extract theorems from derivations that have been halted prematurely with conclusions that do not state the falsity of their premises.
However, such theorems that have been stored in \emph{axiom.dat} are benign in the sense that any proof construction starting from a premise program that is computable will never access such theorems.
For reasons outlined in Section \ref{ce} we do not regard the storage of these benign programs to be in violation of the formal definition of a program extension. 

Once a proof of a new theorem of falsity has been obtained it is stored in the file \emph{axiom.dat}.
A search can then be conducted of all axioms and theorems currently stored in \emph{axiom.dat} whose premise programs contain, as a sublist, the new false program associated with the new theorem of falsity.
When these are identified they are simply removed from storage along with all theorems whose derivations directly or indirectly depend  on those programs that were stored as axioms/theorems.

\section{Theorem connection lists.}\label{thmconn}

In any proof, the connection list of each statement gives knowledge of the axiom or theorem used to infer that statement along with its dependence on the preceding statements of the proof.
Provided that there are no redundant premise statements, it is a straight forward matter to establish that each derived statement in a proof can be traced back to the premises of the proof program.

In a similar way we can define connection lists for theorems that can be employed to trace back dependencies to the axioms of the theory.
When a new theorem is extracted from a proof as an irreducible extended program it is stored in the file \emph{axiom.dat} along with a connection list that includes the labels of the axioms and theorems that were used in its proof.

Consider a proof program $[p~q]$, where $p=[p_i]_{i=1}^n$ is the list of premises of the proof and $q=[q_i]_{i=1}^m$ are the derived statements.
Suppose that a theorem $[p~q_m]$ has been extracted from the proof program $[p~q]$ and stored in the file \emph{axiom.dat} as a new theorem.
Suppose further that there are $\lambda-1$ axioms/theorems already stored in \emph{axiom.dat}.
Let $c_i$, $i=1,\ldots,\lambda-1$, be the labels of these axioms/theorems.
We label the new theorem $[p~q_m]$ as $c_\lambda$.

The connection list of each statement $q_i$ of the proof $[p~q]$ associated with the theorem $c_\lambda$ has the form
\be
 a_\lambda(i)~[l_\lambda(i,1)~\ldots~l_\lambda(i,k_\lambda)]
\ee
where $k_\lambda$ is the length of the axiom/theorem $a_\lambda(i)$ and $1 \leq l_\lambda(i,1),\ldots,l_\lambda(i,k_\lambda) \leq n+i-1$ are the labels of the programs of the sublist of $[p,[q_j]_{j=1}^{i-1}]$ that is program and I/O equivalent to the premise program of the axiom/theorem $a_\lambda(i)$ that is stored in \emph{axiom.dat}.
The theorem connection list for the theorem $c_\lambda$ can be defined by
\be
 c_\lambda~[a_\lambda(1)~\ldots~a_\lambda(m)]
\ee
Here $[a_\lambda(1)~\ldots~a_\lambda(m)]$ is the complete list of axioms/theorems that were used in the proof of theorem $c_\lambda$.
Each label $a_\lambda(i)$, $i=1,\ldots,m$, coincides with some axiom/theorem label $c_i$, $i=1,\ldots,\lambda-1$, that is currently stored in \emph{axiom.dat}.
Following an algorithm similar to that outlined in Section \ref{clist} we can trace each theorem dependency of $c_\lambda$ to the axioms of the theory.
The procedure starts with the list $[a_\lambda(1)~\ldots~a_\lambda(m)]$ and then tracing back all dependencies to arrive at a final list whose elements are axiom labels only.
We shall refer to this procedure as a \emph{theorem connection list reduction}.

\section{Iteration.}

The self correcting procedure of the iterated axiomatic method discussed earlier has some similarity with belief revision theory.
Belief revision theory began with the seminal paper \cite{agm} and remains the dominant paradigm of the subject to this present day.
The theory is based on the so called AGM postulates that reflect the minimal change of a rational agent's belief state through the acquisition of new information.
The three main actions of a change in a belief state are contraction, expansion and revision.

The AGM paradigm draws heavily on conventional theories of logic and set theory and is not readily adapted to our formal system.
Although the objectives of the AGM paradigm of belief revision appear to be related to the iterated axiomatic method there are properties of our formal system that require some significant departures.
Keeping with our motivation for practical implementation we will take a more constructive approach to the self correcting process for the iterated axiomatic method.

We need to distinguish between the construction rules, that can be regarded as the general inference rules, and the axioms associated with an application of a specific theory, $S$.
For each launching of an action of theorem mining the construction rules are fixed, with the starting hypothesis that the formal system based upon the construction rules is sound. 
The application specific axioms are supplied by the user and serve as input to the proof assistance software (in our case VPC).
In the iterated axiomatic scheme the application specific axioms, that define the theory under investigation, can be modified under the self correcting process through the procedures that will now be outlined.

Let $S$ be a theory accompanied by a finite collection of well defined atomic programs.
We denote by $\mb{P}_S$ the type program for theory $S$ that includes the atomic programs and all programs inductively constructed from them as lists and/or disjunctions.
Along with a collection of atomic programs, a theory $S$ is defined by a list of axioms
\ben\label{at01}
s_{axm} = [ s_{axm}(i) ]_{i=1}^{n_{axm}}
\een
for some $n_{axm}:\mb{I}_0$.
Here each $s_{axm}(i):\mb{P}_S$, $i=1,\ldots,n_{axm}$, is an axiom.
It is possible that a theory $S$ is defined by no axioms, in which case we set $n_{axm}=0$ and $s_{axm}$ is an empty list.
It is important to note that $s_{axm}$ is a list of axioms but is not in itself meant to represent a program, i.e. while each element, $s_{axm}(i)$, of the list $s_{axm}$ is of type $\mb{P}$, the list $s_{axm}$ is not itself of type $\mb{P}$.    

When we launch a theorem mining action for a theory, $S$, we explore a space of derivable irreducible extended programs along with their proofs.
We thus generate a time dependent list
\ben\label{at02}
\bal
s_{thm}^{(t)} = & [ s_{thm}(i) ]_{i=1}^t \\
\eal
\een
where $s_{thm}^{(t)}$ is the list of theorems, $s_{thm}(i):\mb{P}_S$, $i=1,\ldots,t$, that have been generated at time $t$.
We have defined the time $t:\mb{I}_0$ to be initially set at $t=0$ and advanced by one unit when a new theorem is added to the theorems list $s_{thm}^{(t)}$.
As with the list $s_{axm}$, $s_{thm}^{(t)}$ is a list of programs but is not itself meant to represent a program.

We can think of theorem mining for a theory, $S$, as a process of generating a sequence of sub theories $S(t)$, for $t=0,1,\ldots,\tau$, such that each $S(t)$ corresponds to the time dependent state of the theory $S$ with a list $s_{thm}^{(t)}$ of theorems that are obtained under the fixed list of axioms $s_{axm}$.
Here $\tau$ is the finite time such that the length of the theorems list of $S(\tau)$ contains all possible theorems that can be derived from the axioms contained in the list $s_{axm}$.
(Given that we are working on $\mathfrak{M}(\mathcal{K},\mathcal{L},\mathcal{M})$ there must exist a finite time $\tau$ for any given theory $S$).
We can regard the theory, $S$, with its fixed list of axioms $s_{axm}$ and the associated sequence of sub theories $S(t)$, $t=1,\ldots,\tau$, as a paradigm.

The irreducible extended programs of $S(\tau)$ can be partitioned into the three distinct subtypes
\ben\label{at03}
\mb{P}_S^{axm},~\mb{P}_S^{thm},~\mb{P}_S^{nd}
\een
The objects of type $\mb{P}_S^{axm}$ are the elements of the axioms list $s_{axm}$ and objects of type $\mb{P}_S^{thm}$ are irreducible extended programs that are derivable from the axioms $s_{axm}$.
Objects of type $\mb{P}_S^{nd}$ are irreducible extended programs that cannot be derived from the axioms $s_{axm}$ and are a potential source of new axioms for a modified theory, $S^\prime$, of $S$.
If for a given theory, $S$, there are no objects of $S(\tau)$ that are type $\mb{P}_S^{nd}$ then we say that the paradigm $S$ is complete with respect to the list of axioms $s_{axm}$. 

Starting with a prescribed list of axioms, $s_{axm}$, for an initial paradigm, $S$, an action of theorem mining is launched from which theorems are extracted from proofs as irreducible extended programs.
The following concurrent actions are as follows.

\begin{itemize}

\item \textbf{TM}. Theorem mining. The search for theorems extracted from proofs as irreducible extended programs. 

\item \textbf{AXT}. Test current list of axioms for violations of soundness.

\item \textbf{AXS}. Search for new axioms.

\end{itemize}
The iteration process starts with the action \textbf{TM} of the current theory, $S$, that runs independently until such time that a violation of soundness or a new axiom is found by the concurrent actions of \textbf{AXT} and \textbf{AXS}.
When this occurs, \textbf{TM} is halted and the current list of axioms and theorems of the theory $S$ are modified.
The action of theorem mining, \textbf{TM}, is restarted generating theorems of the revised theory $S^\prime$.

An ideal situation is one in which the actions of \textbf{TM}, \textbf{AXT} and \textbf{AXS} are fully automated so that we are confined entirely within a machine environment $\mathfrak{M}(\mathcal{K},\mathcal{L},\mathcal{M})$.
A description of what these fully automated procedures might look like is beyond the scope of this book and may be better investigated within the wider discipline of artificial intelligence research.
Therefore, since we cannot claim that these automated procedures have at present been fully developed we will require, within each action, the intervention of an external agent.
By the use of the expression external agent we will always mean a human.
In any event, whether they are fully automated or largely managed by the intervention of an external agent, we will work with the starting assumption that the actions of \textbf{TM}, \textbf{AXT} and \textbf{AXS}, as they are defined in the above items list, will be of a sufficiently high level to perform their designated tasks. 

The action \textbf{TM} involves the search for theorems, expressed in terms of the atomic programs of the theory.
The process may involve some type of goal oriented conjecturing followed by a proof construction.
The proof construction and theorem extraction largely involves the application of VPC as it has already been demonstrated in the previous chapters.
Once a proof is completed the theorem is extracted from the proof as an irreducible extended program by employing the algorithm outlined in Section \ref{connection}.
We can regard this task to be largely managed by an external agent, although some internal automated reasoning could also be included to assist in some aspects of the external agent's strategic decision making, i.e. targeted conjecturing. 

The action of \textbf{AXT} involves testing the soundness of the current list of axioms.
This task would involve a process of confidence building based on some kind of statistical analysis of data supplied by empirical computations.
These procedures could also include some internal automated reasoning along with the interaction of an external agent to find shortcuts in the raw testing process.
This may include targeting specific assigned value input lists.
If the current complete list of axioms is not too large it would appear reasonable that the action of \textbf{AXT} would run the tests on all axioms in parallel.

At this stage the action \textbf{AXS} is much less developed than the previous two actions.
Constructing a fully automated procedure for \textbf{AXS} is still a long way off and requires considerably more work.
As a consequence, we may regard, based upon current level of development, that the action \textbf{AXS} is largely managed by the intervention of an external agent.

The recursive application of the actions \textbf{TM}, \textbf{AXT} and \textbf{AXS} can be run indefinitely until such time that the external agent managing the actions \textbf{AXT} and \textbf{AXS} intervenes and halts the whole process.
The overall self correcting procedure means that, for any paradigm $S$, we must regard an assignment of an object as type $\mb{P}_S^{axm}$ or type $\mb{P}_S^{thm}$ as tentative.
By this we mean that an object will be assigned a type $\mb{P}_S^{axm}$ or type $\mb{P}_S^{thm}$ until such time that an axiom is found to violate soundness.

In practice the actions of \textbf{AXT} and \textbf{AXS} will often halt the current theorem generation of \textbf{TM} in some time well before $t=\tau$ is reached.
When the action $\textbf{TM}$ is halted in this way the current lists of axioms $s_{axm}$ and theorems $s_{thm}^{(t)}$ are modified, the clock speed is reset to $t=0$ and the action \textbf{TM} is restarted, generating a revised sequence of sub theories $S^\prime(t)$, $t=1,\dots,\tau^\prime$, for a new paradigm $S^\prime$. 
   
We start the entire process by prescribing a list of axioms, $s_{axm}$, for an initial paradigm, $S$.
We launch the action \textbf{TM} that generates the sequence of sub theories $S(t)$, for $t=0,1,\ldots$, until such time that $t=\tau$ is reached or the action of \textbf{TM} is interrupted by the concurrent action of \textbf{AXT} or \textbf{AXS}.
There are three main procedures that result in the halting and restarting of \textbf{TM}. 

\textbf{Contraction.} 

\begin{itemize}

\item By action \textbf{AXT} halt \textbf{TM} at time $t$: if for some $r \in s_{axm}$, where $r=[p~c],~c:\mb{P}_{iext}(p)$, it is found that $p:\mb{P}_{S,comp}(x_p)$, for some value assigned input $x_p$, and $r$ is not of type $\mb{P}_{S,comp}(x_r)$, where $x_r$ is the value assigned input list of $r$ that acquires its value through the identities $x_r = \bar{x}_r \setminus (x_c \cap y_p)$ and $\bar{x}_r \simeq [x_p~x_c]$. \\
Extract the list $u \subseteqq s_{thm}^{(t)}$ such that the derivation of the proof of each element of $u$ depends on the axiom $r$ as identified through its theorem connection list reduction. \\
Construct $s_{axm}^\prime = s_{axm} \setminus [r]$. \\
Construct $s_{thm}^{\prime~(0)}=s_{thm}^{(t)} \setminus u$.

\item By action \textbf{AXT} halt \textbf{TM} at time $t$: if for some axiom of falsity $r=false~[p]~[~]$, for some $p:\mb{P}_{S}$, it is found that $p:\mb{P}_{S,comp}(x_p)$, for some value assigned input $x_p$. \\
Extract the list $u \subseteqq s_{thm}^{(t)}$ such that the derivation of the proof of each element of $u$ depends on the axiom $r$ as identified through its theorem connection list reduction ($u$ is a list of theorems of falsity). \\
Construct $s_{axm}^\prime = s_{axm} \setminus [r]$. \\
Construct $s_{thm}^{\prime~(0)}=s_{thm}^{(t)} \setminus u$.

\item Reset the clock at $t=0$.
Restart \textbf{TM} and generate a revised sequence of sub theories $S^\prime(t)$, $t=1,\ldots,\tau^\prime$, for a new paradigm, $S^\prime$, starting with the contracted lists of axioms $s_{axm}^\prime$ and theorems $s_{thm}^{\prime~(0)}$.

\end{itemize}

\textbf{Expansion.} 

\begin{itemize}

\item By action \textbf{AXS} halt \textbf{TM} at time $t$: if a new program $s$ is identified as an axiom, i.e. $s:\mb{P}_S^{nd}$. \\
Construct $s_{axm}^\prime = [s_{axm}~s]$. \\
Set $s_{thm}^{\prime~(0)}=s_{thm}^{(t)}$.

\item Reset the clock at $t=0$.
Restart \textbf{TM} and generate a revised sequence of sub theories $S^\prime(t)$, $t=1,\ldots,\tau^\prime$, for a new paradigm, $S^\prime$, starting with the lists of axioms $s_{axm}^\prime$ and theorems $s_{thm}^{\prime~(0)}$.

\end{itemize}

We include an additional action \textbf{Modify}.    

\textbf{Modify.} 

\begin{itemize}

\item By action \textbf{TM} halt \textbf{TM} at time $t$: if a new theorem, $r:\mb{P}_S^{thm}$, is found and there exists $q \in s_{axm}$ such that $r$ coincides with $q$ and the proof of $r$ is independent of $q$ as identified through its theorem connection list reduction. \\
Construct $s_{axm}^\prime = s_{axm} \setminus [q]$. \\
Construct $s_{thm}^{\prime~(0)}=[s_{thm}^{(t)}~q]$.

\item By action \textbf{TM} halt \textbf{TM} at time $t$: if for some $p:\mb{P}_S$ a new theorem of falsity $r=false~[p]~[~]$ is identified. \\
Extract the list $a \subseteqq s_{axm}$ such that a sublist of the premise of each element of $a$ is program and I/O equivalent to $p$. \\
Extract the list $b \subseteqq s_{thm}^{(t)}$, such that a sublist of the premise of each element of $b$ is program and I/O equivalent to $p$. \\
Extract the list $c \subseteqq s_{thm}^{(t)}$, such that the derivation of the proof of each element of $c$ depends on any element of $a$ and/or $b$. \\
If $a$, $b$ and $c$ are empty lists then set $s_{thm}^{(t+1)}=[s_{thm}^{(t)}~r]$, exit \textbf{Modify} and continue the action \textbf{TM} for the current paradigm $S$.
Otherwise, \\
Construct $s_{axm}^\prime = s_{axm} \setminus a$. \\
Construct $s_{thm}^{\prime~(0)}=[s_{thm}^{(t)}~r] \setminus [b~c]$.

\item Reset the clock at $t=0$.
Restart \textbf{TM} and generate a revised sequence of sub theories $S^\prime(t)$, $t=1,\ldots,\tau^\prime$, for a new paradigm, $S^\prime$, starting with the lists of axioms $s_{axm}^\prime$ and theorems $s_{thm}^{\prime~(0)}$.

\end{itemize}

\section{The scientific method as a dynamical system.}

We have partitioned the irreducible extended programs of a theory, $S$, into the three distinct subtypes of (\ref{at03}), where objects of type $\mb{P}_S^{axm}$ are elements of the axioms list $s_{axm}$ and objects of type $\mb{P}_S^{thm}$ are irreducible extended programs that are derivable from the axioms.
Objects of type $\mb{P}_S^{nd}$ are irreducible extended programs that cannot be derived from the axioms of the list $s_{axm}$.
We can expect that it is not uncommon that the total number of objects of type $\mb{P}_S^{nd}$ of a theory, $S$, will be very much larger than the combined number of objects of type $\mb{P}_S^{axm}$ and $\mb{P}_S^{thm}$.
Objects of type $\mb{P}_S^{nd}$ are a source of new axioms for a modified theory $S^\prime$ (subject to the action \textbf{Modify} as outlined in the previous section). 

This stresses that in any scientific study the acquisition of knowledge for a given theory is not necessarily dominated by the formal deductive methods associated with the action \textbf{TM} but instead by the search for non-derivable objects of type $\mb{P}_S^{nd}$.
This would essentially involve the same procedures associated with the search for new axioms.
It can then be argued that the action of \textbf{AXS} should be regarded as the most important component of any scientific research.
This is very much reflected by a recent movement of mathematicians who are proponents of the idea that mathematics should place less emphasis on the axiomatic method and concentrate more on experimental mathematics largely by making use of various available tools in the form of specialized computer software (see for example \cite{zeil02}).

As has already been discussed, the process of identifying new axioms by way of the action of \textbf{AXS} is still far from being understood and is often attributed to some vague notion of a creative process possessed by the human mind.
If we are to elevate the importance of the action of \textbf{AXS} in the sciences much more effort needs to be directed into understanding this process.
The hope is that the process can be ultimately described by algorithms.      
This is very much aligned with research efforts in artificial intelligence and the related areas such as machine learning and Bayesian inference methods but it is possible that some of the formal deductive methods associated with the action \textbf{TM} may also be involved.

This book has focused on the mechanical aspects of deduction that are largely associated with the action \textbf{TM}.
In light of the above comments we could argue that this is perhaps the easiest part.
The development of methods that effectively move towards a goal of fully automating the action of \textbf{AXS} and its allied action \textbf{AXT} is an effort that would be an essential part of scientific research in the future.  

Ideally, the user has prescribed an initial collection of axioms for a given theory that is as concise as possible.
While soundness is the major objective, the hope that such a collection of axioms is exhaustive must often be abandoned.
As such there is the concern that an iteration of the self correcting recursive process of the iterated axiomatic method would result in a system that will generate a collection of axioms that is too large to manage.
Other phenomena such as cycling may also be encountered.
Thus the selection of the initial list of axioms will be crucial.

From these observations it becomes evident that we are dealing with a dynamical system.
At this stage it is uncertain that our constructive procedure of the recursive process of generating a sequence of revised paradigms will approach a limiting paradigm $\bar{S}$, where $\bar{S}$ is sound and complete.
The best that one can hope to achieve is that, given a good choice of the initial list of axioms and a sufficiently large time for the extensive checks of soundness violations through the action \textbf{AXT} and the search for new axioms through the action \textbf{AXS}, a sequence of paradigms will evolve with increasing scope and reliability.
To be more precise we wish to generate a sequence of paradigms such that the lengths of the lists of axioms and non-derivable irreducible extended programs are minimized.

All of the above suggest that a new subject area will emerge that is aimed at investigating the properties associated with the iterated axiomatic method as a dynamical system in itself.
Such an area will be useful in uncovering such behavior as stability, sensitivity to initial conditions and other phenomena associated with dynamical systems.
Any knowledge obtained from this study will yield vital feedback on what limitations a user might expect from the initial data that is supplied to define a specific theory and possible procedures that could be applied in their selection that will ensure the best results.

We may formulate the dynamical system as an incorporation of the combined actions of \textbf{TM}, \textbf{AXT} and \textbf{AXS}.
Alternatively, we may regard the action of theorem mining, \textbf{TM}, as the sole defining process of the dynamical system and the actions of \textbf{Contraction}, \textbf{Expansion} and \textbf{Modify} as external sources of perturbations to \textbf{TM}.
We can think of the dynamical system based on the action of \textbf{TM} as a map that generates, for each paradigm, $S$, a sequence of sub theories $S(t)$, in the discrete time $t=0,\ldots,\tau$.
The global target space of this map is a discrete space encompassing all possible paradigms where each point of this space is a sub-theory of a distinct paradigm.  
A local fixed point of each paradigm, $S$, can be defined by $S(\tau)$, for some usually large but finite time $t=\tau$, that contains all of the theorems that can be derived from the fixed list of axioms, $s_{axm}$, associated with $S$.

One may be interested in observing the dynamic behavior in a region containing the fixed point, $S(\tau)$, that takes on the properties of a local basin of attraction.
The actions of \textbf{Contraction}, \textbf{Expansion} and \textbf{Modify} will provide a potential source of perturbations to our dynamical system that could knock a trajectory out of its current basin of attraction to a new paradigm.
Thus the traditional notions of attractors and other well known phenomena of dynamical systems will acquire a useful interpretation in this application area.
We can anticipate that such a field of study will provide very useful insights that will eventually lay to rest many philosophical debates that currently surround the scientific method itself.

\backmatter

\newpage

\printindex

\end{document}